% physrep.tex
%
% this is the master_file for the PhysRep:
% "W- symmetry in conformal field theory," CERN-TH.6583/92
%
% Needs input: pkjmac, harvmac (version 9/91), chfront, chtoc,
% ch12345, ch67, ch89, chrefs.
%
% DO NOT CHANGE PAGE-LAYOUT, OTHERWISE TABLE OF CONTENTS WILL BE
% MESSED UP !!!!!
%
%          last updated 7/31/92
%
\message{Please, respond `b' to the next query!}
\input harvmac
%%%%%%%%%%%%%%%%%%%%%%%%%%%%%%%%%%%%%%%%%%%%%%%%%%%%%%%%%%%%%%%%%%%%%%
%%
%%               pkjmac.tex         last updated 06/16/91
%%
%%%%%%%%%%%%%%%%%%%%%%%%%%%%%%%%%%%%%%%%%%%%%%%%%%%%%%%%%%%%%%%%%%%%%%%
%
% \input harvmac.tex    % version 9/91 required
%
%%
%%  Some additions/modifications to harvmac
%%
\global\newcount\mthno \global\mthno=1
\global\newcount\mexno \global\mexno=1
\global\newcount\mquno \global\mquno=1
\def\newsec#1{\global\advance\secno by1\message{(\the\secno. #1)}
\global\subsecno=0\xdef\secsym{\the\secno.}\global\meqno=1\global\mthno=1
\global\mexno=1\global\mquno=1
%\ifx\answ\bigans \vfill\eject \else \bigbreak\bigskip \fi  %if desired
\bigbreak\medskip\noindent{\bf\the\secno. #1}\writetoca{{\secsym} {#1}}
\par\nobreak\medskip\nobreak}
\xdef\secsym{}
\global\newcount\subsecno \global\subsecno=0
\def\subsec#1{\global\advance\subsecno by1\message{(\secsym\the
\subsecno. #1)}
\bigbreak\noindent{\it\secsym\the\subsecno. #1}\writetoca{\string\quad
{\secsym\the\subsecno.} {#1}}\par\nobreak\medskip\nobreak}
\def\appendix#1#2{\global\meqno=1\global\mthno=1\global\mexno=1%
\global\mquno=1
\global\subsecno=0
\xdef\secsym{\hbox{#1.}}
\bigbreak\bigskip\noindent{\bf Appendix #1. #2}\message{(#1. #2)}
\writetoca{Appendix {#1.} {#2}}\par\nobreak\medskip\nobreak}
%
%          theorems and examples and questions
%
\def\thm#1{\xdef #1{\secsym\the\mthno}\writedef{#1\leftbracket#1}%
\global\advance\mthno by1\wrlabeL#1}
\def\que#1{\xdef #1{\secsym\the\mquno}\writedef{#1\leftbracket#1}%
\global\advance\mquno by1\wrlabeL#1}
\def\exm#1{\xdef #1{\secsym\the\mexno}\writedef{#1\leftbracket#1}%
\global\advance\mexno by1\wrlabeL#1}
%
%     \ref\label{text}
% generates a number, assigns it to \label, generates an entry.
% To list the refs on a separate page,  \listrefs
%            on the same page,  \listrefsnoskip
%
\def\ref{\the\refno\nref}
\def\nref#1{\xdef#1{\the\refno}\writedef{#1\leftbracket#1}%
\ifnum\refno=1\immediate\openout\rfile=\jobname.refs\fi
\global\advance\refno by1\chardef\wfile=\rfile\immediate
\write\rfile{\noexpand\item{[#1]\ }\reflabeL{#1\hskip.31in}\pctsign}
\findarg}
\def\bref{\nref}
%

%
%%%%%%%%%%%%%%%%%%%%%%%%%%%%%%%%%%%%%%%%%%%%%%%%%%%%%%%%%%%%%%%%%%%%%%%%
%
%    Here our macros start
%
%%%%%%%%%%%%%%%%%%%%%%%%%%%%%%%%%%%%%%%%%%%%%%%%%%%%%%%%%%%%%%%%%%%%%%%%

\def\p{\partial}
%
%  Greek abbreviations (to be used within $....$)
%
\def\al{\alpha} \def\be{\beta} \def\ga{\gamma}  \def\Ga{\Gamma}
\def\de{\delta}  \def\De{\Delta} \def\ep{\epsilon}
  \def\et{\eta} \def\th{\theta}
\def\Th{\Theta}    \def\ka{\kappa}
\def\la{\lambda} \def\La{\Lambda} \def\rh{\rho} 
 \def\ta{\tau}   
\def\ph{\phi}  \def\Ph{\Phi}  \def\vph{\varphi} \def\ch{\chi}
\def\ps{\psi}     
%
%  boldface abbreviations  (to be used within $..$)
%

%
\def\bGa{{\bf\Ga}} 
%
% Calligraphic abbreviations (to be used within $..$)
%
  \def\cF{{\cal F}} 
 \def\cM{{\cal M}}  
  \def\cW{{\cal W}} 
\def\cC{{\cal C}} \def\cH{{\cal H}}
%
%
%  definition of \underhook{...}
%
\def\lefthook{{\vrule height5pt width0.4pt depth0pt}}
\def\righthook{{\vrule height5pt width0.4pt depth0pt}}
\def\leftrighthookfill{$\mathsurround=0pt \mathord\lefthook
     \hrulefill\mathord\righthook$}
\def\underhook#1{\vtop{\ialign{##\crcr$\hfil\displaystyle{#1}\hfil$\crcr
      \noalign{\kern-1pt\nointerlineskip\vskip2pt}
      \leftrighthookfill\crcr}}}
%
%  other abbreviations
%
\def\dCo{{\rm h}^{\vee}}

\def\hb{\hfill\break}
\def\ie{{\it i.e.\ }}
\def\eg{{\it e.g.\ }}

\def\ZZ{Z\!\!\!Z}           %%%%%%%%%%%%%
\def\RR{I\!\!R}             % produces reasonably looking math. Z,N,etc.
\def\CC{I\!\!\!\!C}         %
\def\QQ{I\!\!\!\!Q}         %%%%%%%%%%%%%

%%%%%%%%%%%%%%%%%%%%%%%%%%%%%%%%%%%%%%%%%%%%%%%%%
%% some Lie algebra macros
%%%%%%%%%%%%%%%%%%%%%%%%%%%%%%%%%%%%%%%%%%%%%%%%%%

\def\bfg{{\bf g}}
\def\bfh{{\bf h}}

\def\bfnp{{\bf n}_+}

\def\bfk{{\bf k}}
\def\whg{\widehat{\bf g}}
\def\whk{{\widehat{\bf k}}}
\def\rhv{{\rh^\vee}}
\def\Lap{\La^{(+)}}
\def\Lam{\La^{(-)}}

%%%%%%%%%%%%%%%%%%%%%%

\def\bs{\bigskip}

    % Use within $..$

%%%
%%%
%%% some macros for making commutative diagrams
%%%

\def\comdiag{
\def\normalbaselines{\baselineskip15pt
\lineskip3pt \lineskiplimit3pt }}

\def\mapright#1{\smash{\mathop{\longrightarrow}\limits^{#1}}}

\def\mapdown#1{\Big\downarrow\rlap{$\vcenter{
                \hbox{$\scriptstyle#1$}}$}}

%%%%%%%%%%%%%%%%%%%%%%%%%%%%%%%%%%%%%%%%%%%
%%    some additional macros for ch67
%%%%%%%%%%%%%%%%%%%%%%%%%%%%%%%%%%%%%%%%%%%%%%

\def\fhg{{{\hbox{{\hbox{$\scriptstyle\bf g$}}\kern-.8em\lower.5ex%
\hbox{$\scriptstyle\widehat{\phantom{\bf g}}$}}}}}

\def\cite#1{[#1]}
\def\onethird{\textstyle{{1\over3}}}
\def\zb{\bar{z}}
\def\del{\partial}
\def\delb{\bar\partial}

\def\wLa{{\widehat{\La}}}

\def\third{{\textstyle{1\over3}}}

%%%%%%%%%%%some additional used in ch1234589

\def\sW{{\cal SW}}
\def\frac#1#2{{#1\over #2}}  % LaTeX remnant

\def\zw{(z-w)}
\def\Th{\widehat{T}}
\def\Wh{\widehat{W}}
\def\del{\partial}
\def\hbar{\bar{h}}
\def\htil{\dCo}
\def\zb{\bar{z}}

\def\kb{\bar{k}}
\def\Lb{\overline{L}}

\def\whso{\widehat{so}}
\def\whsu{\widehat{su}}
\def\whW{\widehat{\cW}}
%%%%%%%%%%%%%%%%%%%%%%%%%%%%%%%%%%%%%%%%%%%%%%%%%%%%%%%%%%%%%%%%%%%%%%%%%%
%%%%
%%%%   END MACRO FILE
%%%%
%%%%%%%%%%%%%%%%%%%%%%%%%%%%%%%%%%%%%%%%%%%%%%%%%%%%%%%%%%%%%%%%%%%%%%%%%%

% chrefs.tex, last updated 12/22/92 (3 items added)
%\pageno=123
%\writelabels
%
% This file contains a list of references in alphabetical order.
%
% The principal rule for label-assignment is that the label consists of
% the combination of (upper-case) first letters of each authors' surname,
% to which (lower-case) letters (a,b,c,...) are appended if there are
% several papers with the same author(s). In the case of one author or
% other confusing cases an additional (lower-case) letter is added.
%
%
\def\NP{Nucl. Phys.\ }
\def\PL{Phys. Lett.\ }
\def\CMP{Comm. Math. Phys.\ }
\def\PR{Phys. Rev.\ }
\def\IJMP{Int. Journ. Mod. Phys.\ }
\def\MPL{Mod. Phys. Lett.\ }
\def\AP{Ann. Phys.\ }
\def\PRL{Phys. Rev. Lett.\ }
\def\IM{Inv. Math.\ }

\def\sW{{\cal SW}}

\bref\Ad{
  M. Ademollo {\it et al.}, {\it Supersymmetric strings and
  colour confinement}, \PL {\bf 62B} (1976) 105.}
\bref\Adl{
  M. Adler, {\it On a trace functional for formal pseudo-differential
  operators and the symplectic structure of the Korteweg-de Vries
  type equations}, Inv. Math. {\bf 50} (1979) 219.}
\bref\Aa{
  C. Ahn, {\it $c=5/2$ Free fermion model of $\cW B_2$ algebra},
  preprint ITP-SB-91-61.}  % hepth@xxx/9111061
\bref\ASS{
  C. Ahn, K. Schoutens and A. Sevrin, {\it The full structure
  of the super-$\cW_3$ algebra}, \IJMP {\bf A6} (1991) 3467.}
\bref\Al{
  D. Altschuler, {\it Quantum equivalence of coset space models},
  \NP {\bf B313} (1989) 293.}
\bref\AM{
  G. Anderson and G. Moore, {\it Rationality in conformal field
  theory}, \CMP {\bf 117} (1988) 441.}
% \bref\Ap{
%   S. A. Apikyan, {\it Infinite additional symmetries in the two
%   dimensional conformal quantum field theory}, \MPL
%   {\bf A2} (1987) 317.}
\bref\AQ{
  M. Awada and Z. Qiu, {\it 2D Higher-spin gravity and the
  multimatrix models}, \PL {\bf 245B} (1990) 85.}
%\bref\BCR{
%  M. Baake, P. Christe and V. Rittenberg, {\it Higher spin conserved
% currents in $c=1$ conformally invariant systems}, \NP {\bf B300} [FS22]
% (1988) 637.}
\bref\BNY{
  J. Bagger, D. Nemeschansky and S. Yankielowicz, {\it Virasoro
  algebras with central charge $c>1$}, \PRL {\bf 60}
  (1988) 389.}
\bref\BN{
  J. Bagger and D. Nemeschansky,
  {\it Coset construction of chiral algebras},
  in `College Park Workshop 1988', Proc. of the Maryland Superstring
  Workshop, May 24-28, 1988.}
\bref\BBa{
  F.A. Bais and P. Bouwknegt, {\it A classification of
  subgroup truncations of the bosonic string},
  \NP {\bf B279} (1987) 561.}
\bref\BBSSa{
  F.A. Bais, P. Bouwknegt, K. Schoutens and M. Surridge,
  {\it Extensions of the Virasoro algebra constructed from Kac-Moody
  algebras using higher order Casimir invariants}, \NP {\bf
  B304} (1988) 348.}
\bref\BBSSb{
  F.A. Bais, P. Bouwknegt, K. Schoutens and M. Surridge,
  {\it Coset construction for extended Virasoro algebras},
  \NP {\bf B304} (1988) 371.}
\bref\BBSSc{
  F.A. Bais, P. Bouwknegt, K. Schoutens and M. Surridge,
  {\it Extended Virasoro algebras}, in Proc. of `Perspectives in
  String Theory', Niels Bohr Institute/Nordita, Copenhagen, Oct.'87,
  ed.\ P. DiVecchia and J.L. Petersen (World Scientific, Singapore, 1988).}
\bref\BTV{
  F.A. Bais, T. Tjin and P. van Driel, {\it Coupled chiral algebras
  obtained by reduction of WZNW theories}, \NP {\bf B357} (1991) 632.}
\bref\Baa{
  I. Bakas, {\it Conformal invariance, the KdV equation and coadjoint
  orbits of the Virasoro algebra}, \NP {\bf B302} (1988) 189.}
\bref\Bab{
  I. Bakas, {\it The hamiltonian structure of the spin-4
  operator algebra}, \PL {\bf 213B} (1988) 313.}
\bref\BabII{
  I. Bakas, {\it The large-$N$ limit of extended conformal systems},
  \PL {\bf 228B} (1989) 57.}
\bref\Bac{
  I. Bakas, {\it Higher spin fields and the Gel'fand-Dickey algebra},
  \CMP {\bf 123} (1989) 627.}
\bref\Bad{
  I. Bakas, {\it The structure of the $\cW_\infty$ algebra},
  \CMP {\bf 134} (1990) 487.}
\bref\BKa{
  I. Bakas and E. Kiritsis, {\it Bosonic realization of a universal
  $\cW$-algebra and $Z_\infty$ para\-fermions},
  \NP {\bf B343} (1990) 185; erratum \NP {\bf B350} (1991) 512.}
\bref\BKb{
  I. Bakas and E. Kiritsis,
  {\it Grassmannian coset models and unitary representations of
  $\cW_\infty$}, \MPL {\bf 5} (1990) 2039.}
\bref\BKe{
  I. Bakas and E. Kiritsis, {\it Beyond the large-$N$ limit: non-linear
  $\cW_\infty$ as symmetry of the $SL(2,R)/U(1)$ coset model},
  preprint UCB-PTH-91/44.}    % hepth@xxx/9109029
\bref\BFFOWa{
  J. Balog, L. Feh\'er, P. Forg\'acs, L. O'Raifeartaigh and
  A. Wipf, {\it Liouville and Toda theories as conformally reduced WZNW
  theories}, \PL {\bf 227B} (1989) 214.}
\bref\BFFOWb{
  J. Balog, L. Feh\'er, L. O'Raifeartaigh, P. Forg\'acs and A. Wipf,
  {\it Toda theory and $\cW$-algebra from a gauged WZNW point of view},
  \AP {\bf 203} (1990) 76.}
\bref\BFFOWc{
  J. Balog, L. Feh\'er, L. O'Raifeartaigh, P. Forg\'acs and A. Wipf,
  {\it Kac-Moody realization of $\cW$-algebras},
  \PL {\bf 244B} (1990) 435.}
\bref\BDFM{
  T. Banks, L.J. Dixon, D. Friedan and E. Martinec, {\it Phenomenology
  and conformal field theory, or Can string theory predict the weak
  mixing angle}, \NP {\bf B299} (1988) 613.}
\bref\BHN{
  T. Banks, D. Horn and H. Neuberger, {\it Bosonization of the
  $SU(N)$ Thirring models}, \NP {\bf B108} (1976) 119.}
\bref\BH{
  K. Bardakci and M.B. Halpern, {\it New dual quark models},
  \PR {\bf D3} (1971) 2493.}
%\bref\BRSa{
%  C. Becchi, A. Rouet and R. Stora, {\it The abelian Higgs-Kibble
%  model, unitarity and the $S$-operator}, \PL {\bf 52B} (1974) 344.}
%\bref\BRSb{
%  C. Becchi, A. Rouet and R. Stora,
%  {\it Renormalization of gauge theories}, \AP {\bf 98} (1976) 287.}
\bref\BPZ{
  A.A. Belavin, A.M. Polyakov and A.B. Zamolodchikov,
  {\it Infinite conformal symmetry in two dimensional quantum field
  theory}, \NP {\bf B241} (1984) 333.}
\bref\Be{
  A. Belavin, {\it KdV-type equations and $\cW$-algebras},
  Adv. Stud. in Pure Math. {\bf 19} (1989) 117.}
\bref\Bera{
  E. Bergshoeff, {\it Non-linear super-$\cW$ algebras at fixed
  central charge}, preprint UG-50/91.}
\bref\Berb{
  E. Bergshoeff, {\it Super-$\cW$ algebras}, preprint UG-63/91,
  to be published in Proc. of the 1991 Trieste Workshop on Superstrings
  and Related Topics.}
\bref\BBS{
  E. Bergshoeff, A. Bilal and K.S. Stelle, {\it $\cW$ symmetries:
  gauging and geometry}, \IJMP {\bf A6} (1991) 4959.}
\bref\BdWVa{
  E. Bergshoeff, B. de Wit and M. Vasiliev, {\it The
  super-$\cW_\infty$ algebra}, \PL {\bf 256B} (1991) 199.}
\bref\BdWVb{
  E. Bergshoeff, B. de Wit and M. Vasiliev, {\it The structure of the
  super-$\cW_\infty$ algebra}, preprint CERN-TH.6021/91.}
\bref\BHPSSS{
  E. Bergshoeff, P.S. Howe, C.N. Pope, E. Sezgin, X. Shen and K.S. Stelle,
  {\it Quantisation deforms $w_\infty$ to $\cW_\infty$ gravity},
  \NP {\bf B363} (1991) 163.}
\bref\BPRSS{
  E. Bergshoeff, C.N. Pope, L.J. Romans, E. Sezgin and X. Shen,
  {\it The super-$\cW_\infty$ algebra}, \PL {\bf 245B} (1990) 447.}
\bref\BPRSSS{
  E. Bergshoeff, C.N. Pope, L.J. Romans, E. Sezgin, X. Shen and
  K.S. Stelle, {\it $w_\infty$ gravity}, \PL {\bf 243B} (1990) 350.}
\bref\BTh{
  D. Bernard and J. Thierry-Mieg, {\it Level one representations
  of the simple affine Kac-Moody algebras in their homogeneous
  gradations}, \CMP {\bf 111} (1987) 181.}
\bref\Bea{
  M. Bershadsky, {\it Superconformal algebras in two dimensions
  with arbitrary $N$}, \PL {\bf 174B} (1986) 285.}
\bref\Beb{
  M. Bershadsky, {\it Conformal field theories via hamiltonian
  reduction}, \CMP {\bf 139} (1991) 71.}
\bref\BOa{
  M. Bershadsky and H. Ooguri, {\it Hidden $SL(n)$ symmetry in
  conformal field theories}, \CMP {\bf 126} (1989) 49.}
\bref\BOb{
  M. Bershadsky and H. Ooguri, {\it Hidden $Osp(N,2)$ symmetries
  in superconformal field theories}, \PL {\bf 229B} (1989) 374.}
\bref\Bia{
  A. Bilal,
  {\it A note on super $\cW$-algebras}, \PL {\bf 238B} (1990) 239.}
\bref\Bib{
  A. Bilal,
  {\it A remark on the $N\to \infty$ limit of $\cW_N$-algebras},
  \PL {\bf 227B} (1989) 406.}
\bref\Biba{
  A. Bilal, {\it What is $\cW$-geometry?}, \PL {\bf 249B} (1990) 56.}
\bref\Bic{
  A. Bilal, {\it Introduction to $\cW$-algebras}, presented at the
  Spring School on String Theory and Quantum Gravity, Trieste, Italy,
  April 1991, preprint CERN-TH.6083/91.}
%\bref\Bid{
%  A. Bilal, {\it Fusion and braiding in $\cW$-algebra extended conformal
%  theories}, \NP {\bf B330} (1990) 399.}
%\bref\Bie{
%  A. Bilal, {\it Fusion and braiding in $\cW$-algebra extended conformal
%  theories, II:
%  Generalization to chiral screened vertex operators labelled
%  by arbitrary Young tableaux}, \IJMP {\bf A5} (1990) 1881.}
%\bref\Bif{
%  A. Bilal,
%  {\it Braiding and fusion properties of $\cW$-algebras and their
%  relation with integrable lattice models}, in Proc. of `Strings '90',
%  Texas A \& M, 1990, ed.\ R. Arnowitt {\it et al.} (World Scientific,
%  Singapore, 1991).}
\bref\BFK{
  A. Bilal, V.V. Fock and I.I. Kogan, {\it On the origin of
  $\cW$-algebras}, \NP {\bf B359} (1991) 635.}
\bref\BGa{
  A. Bilal and J.-L. Gervais, {\it Systematic approach to conformal
  systems with extended Virasoro symmetries}, \PL {\bf 206B} (1988) 412.}
\bref\BGb{
  A. Bilal and J.-L. Gervais,
  {\it Extended $c=\infty$ conformal systems from classical Toda
  field theories}, \NP {\bf B314} (1989) 646.}
\bref\BGc{
  A. Bilal and J.-L. Gervais,
  {\it Systematic construction of conformal theories with higher-spin
  Virasoro symmetries}, \NP {\bf B318} (1989) 579.}
\bref\BGd{
  A. Bilal and J.-L. Gervais,
 {\it Conformal theories with non-linearly extended Virasoro symmetries
  and Lie algebra classification},
  in `Infinite dimensional Lie algebras and Lie groups', ed. V. Kac,
  Proc. CIRM-Luminy conf., 1988 (World Scientific, Singapore, 1989),
  Adv. Ser. Math. Phys. {\bf 7} (1988) 483.}
\bref\BGe{
  A. Bilal and J.-L. Gervais, {\it Nonlinearly extended Virasoro algebras:
  new prospects for building string theories}, \NP {\bf B326} (1989) 222.}
\bref\BY{
  B. Blok and S. Yankielowicz, {\it Extended current algebras and the coset
  construction of conformal field theories}, \NP {\bf B315} (1989) 25.}
\bref\Bl{
  R. Blumenhagen, {\it Covariant construction of $N=1$ super $\cW$-algebras},
  preprint BONN-HE-91-20.}
\bref\Blb{
  R. Blumenhagen, {\it $\cW$-algebras in conformal quantum field theory},
  Diplomarbeit (Master's thesis), preprint BONN-IR-91-06 (in German).}
\bref\BEHH{
  R. Blumenhagen, W. Eholzer, A. Honecker and R. H\"ubel,
  {\it New $N=1$ extended superconformal algebras with two and
  three generators}, preprint BONN-HE-92-02.}
\bref\BFKNRV{
  R. Blumenhagen, M. Flohr, A. Kliem, W. Nahm, A. Recknagel and
  R. Varnhagen, {\it $\cW$-algebras with two and three generators},
  \NP {\bf B361} (1991) 255.}
\bref\BT{
  R. Bott and L.W. Tu, {\it Differential forms in algebraic topology},
  Springer Verlag, New-York, 1982.}
\bref\Boa{
  P. Bouwknegt, {\it On the construction of modular invariant partition
  functions}, \NP {\bf B290} [FS20] (1987) 507.}
\bref\Bob{
  P. Bouwknegt,
  {\it Extended conformal algebras}, \PL {\bf 207B} (1988) 295.}
\bref\Boc{
  P. Bouwknegt,
  {\it Extended conformal algebras from Kac-Moody algebras},
  in `Infinite-dimensional Lie algebras and Lie groups',
  ed. V. Kac, Proc.
  CIRM-Luminy conf., 1988 (World Scientific, Singapore, 1989).}
\bref\Bod{
  P. Bouwknegt, {\it An algebraic approach towards the classification
  of two dimensional conformal field theories}, Univ.\ of Amsterdam
  PhD thesis, 1988.}
%\bref\BCMNa{
%  P. Bouwknegt, A. Ceresole, J. McCarthy and P. van Nieuwenhuizen,
%  {\it Extended Sugawara construction for the superalgebras
%  $SU(M+1|N+1)$,
%  I : free field representation and bosonization of super
%  Kac-Moody currents}, \PR {\bf D39} (1989) 2971.}
%\bref\BCMNb{
%  P. Bouwknegt, A. Ceresole, J. McCarthy and P. van Nieuwenhuizen,
%  {\it Extended Sugawara construction for the superalgebras
%  $SU(M+1|N+1)$, II : the third order Casimir algebra},
%  \PR {\bf D40} (1989) 415.}
\bref\BMPa{
  P. Bouwknegt, J. McCarthy and K. Pilch, {\it Quantum group structure
  in the Fock space resolutions of $\widehat{sl}(n)$ representations},
  \CMP {\bf 131} (1990) 125.}
\bref\BMPb{
  P. Bouwknegt, J. McCarthy and K. Pilch, {\it Free field approach
  to 2-dimensional conformal field theories}, Prog. Theor. Phys. Suppl.
  {\bf 102} (1990) 67.}
\bref\BMPc{
  P. Bouwknegt, J. McCarthy and K. Pilch, {\it On the free field
  resolutions for coset conformal field theories}, \NP {\bf B352}
  (1991) 139. }
\bref\BMPd{
  P. Bouwknegt, J. McCarthy and K. Pilch, {\it Some aspects of free
  field resolutions in 2D CFT with application to the quantum
  Drinfeld-Sokolov reduction}, preprint CERN-TH.6267/91.}
\bref\Bow{
  P. Bowcock, {\it Quasi-primary fields and associativity of
  chiral algebras}, \NP {\bf B356} (1991) 367.}
\bref\Bowb{
  P. Bowcock, {\it Exceptional superconformal algebras},
  preprint EFI 92-09.}
\bref\BoGa{
  P. Bowcock and P. Goddard, {\it Virasoro algebras with central
  charge $c<1$}, \NP {\bf B285} [FS19] (1987) 651.}
\bref\BoGb{
  P. Bowcock and P. Goddard, {\it Coset constructions and extended
  conformal algebras}, \NP {\bf B305} [FS23] (1988) 685.}
\bref\BW{
  P. Bowcock and G.M.T. Watts, {\it On the classification of
  quantum $\cW$-algebras}, preprint EFI 91-63, DTP-91-63.}
%\bref\BYZ{
%  R. Brustein, S. Yankielowicz and J.-B. Zuber, {\it Factorization
%  and selection rules of operator algebras in conformal field
%  theories}, \NP {\bf B313} (1989) 321.}
\bref\Bu{
  N.J. Burroughs, {\it Coadjoint orbits of the generalised $Sl(2)$,
  $Sl(3)$ KdV hierarchies}, preprint IASSNS-HEP-91/67.}
\bref\BDHM{
  N.J. Burroughs, M.F. De Groot, T.J. Hollowood and J.L. Miramontes, {\it
  Generalized Drinfel'd-Sokolov hierarchies II: The Hamiltonian
  structures}, preprint IASSNS-HEP-91/42.}
\bref\Ca{
  A. Cappelli, {\it Modular invariant partition functions of
  superconformal theories}, \PL {\bf 185B} (1987) 82.}
\bref\CIZa{
  A. Cappelli, C. Itzykson and J.-B. Zuber, {\it Modular invariant
  partition functions in two dimensions}, \NP {\bf B280} [FS18] 445.}
\bref\CIZb{
  A. Cappelli, C. Itzykson and J.-B. Zuber,
  {\it The ADE classification of minimal and $A_1^{(1)}$ conformal
  invariant theories}, \CMP {\bf 113} (1987) 1.}
\bref\Caa{
  J.L. Cardy, {\it Conformal invariance}, in `Phase Transitions and
  Critical Phenomena', vol.\ 11,
  ed.\ C. Domb and J. Lebowitz (Academic Press, 1986).}
\bref\Cab{
  J.L. Cardy, {\it Operator content of two dimensional conformally
  invariant theories}, \NP {\bf B270} [FS16] (1986) 186.}
\bref\Cac{
  J.L. Cardy, {\it Conformal invariance and statistical mechanics},
  in Proc. of `Fields, Strings and Critical Phenomena,' Les Houches
  1988, ed. E.~Br\'ezin and J.~Zinn-Justin
  (North Holland, 1990).}
\bref\Cas{
  M. Caselle, {\it Character identities implied by super-$\cW_3$
  invariance and generalized Coulomb gas descriptions}, preprint
  DFTT 2/91.}
%\bref\CK{
%  D. Chang and A. Kumar, {\it Becchi-Rouet-Stora-Tyutin
%  quantization of superconformal theories}, \PR
%  {\bf D35} (1987) 1388.}
%\bref\CKW{
%  D. Chang, A. Kumar and J. Wang, {\it Search for extended
%  conformal algebras}, \MPL {\bf 5} (1990) 1511.}
%\bref\CT{
%  A. Chodos and C.B. Thorn, {\it Making the massless string
%  massive}, \NP {\bf B72} (1974) 509.}
\bref\CRa{
  P. Christe and F. Ravanini, {\it $G_N\oplus G_L/G_{N+L}$ conformal
  field theories and their modular invariant partition functions},
  \IJMP {\bf A4} (1989) 897.}
%\bref\CRb{
%  P. Christe and F. Ravanini, {\it A new tool in the classification
%  of rational conformal field theories}, \PL {\bf 217B} (1989) 252.}
%\bref\CG{
%  E. Cremmer and J.-L. Gervais, {\it The quantum group structure
%  associated with nonlinearly extended Virasoro algebras},
%  \CMP {\bf 134} (1990) 619.}
%\bref\CPS{
%  C. Crnkovic, R. Paunov and M. Stanishkov, {\it Fusions of conformal
%  models}, \NP {\bf B336} (1990) 637.}
\bref\DHR{
  A. Das, W.-J. Huang and S. Roy, {\it Zero curvature condition and 2D
  gravity theories}, preprint UR-1197 (1991).}
\bref\DDR{
  S.R. Das, A. Dhar and K.S. Rama, {\it Physical states and scaling
  properties of $\cW$-gravities and $\cW$-strings}, preprint
  TIFR/TH/91-20.}
\bref\Db{
  B.V. de Bakker, {\it $\cW$-algebras in the coset model $SU(2)/U(1)$},
  Univ.\ of Amsterdam Master's thesis, 1991.}
\bref\dBGa{
  J. de Boer and J. Goeree, {\it The covariant $\cW_3$ action},
  \PL {\bf 274B} (1992) 289.}
\bref\dBGb{
  J. de Boer and J. Goeree, {\it $\cW$-gravity form Chern-Simons theory},
  preprint THU-91/24.}
\bref\DSc{
  A. Deckmyn and S. Schrans, {\it $\cW_3$ constructions on affine Lie
  algebras}, \PL {\bf 274B} (1992) 186.}
%\bref\Deg{
%  P. Degiovanni, {\it $\ZZ/N\ZZ$ Conformal field theories},
%  \CMP {\bf 127} (1990) 71.}
\bref\DHM{
  M.F. De Groot, T.J. Hollowood and J.L. Miramontes, {\it Generalized
  Drinfel'd-Sokolov hierarchies}, \CMP {\bf 145} (1992) 57.}
\bref\DRSa{
  F. Delduc, E. Ragoucy and P. Sorba, {\it Super-Toda theories
  and $\cW$-algebras from superspace Wess-Zumino-Witten models},
  preprint ENSLAPP-AL-352/91.}
%\bref\DRSb{
%  F. Delduc, E. Ragoucy and P. Sorba, {\it Towards a classification
%  of $\cW$-algebras arising from non-abelian Toda theories},
%  preprint ENSLAPP-AL-362/92.}
\bref\DM{
  D.A. Depireux and P. Mathieu, {\it On the classical $\cW_N^{(l)}$
  algebras}, preprint LAVAL PHY-27/91.}
\bref\DF{
  A.H. D\'iaz and J.M. Figueroa-O'Farrill,
  {\it A new explicit construction of $\cW_3$ from the affine algebra
  $A_2^{(1)}$}, \NP {\bf B349} (1991) 237.}
\bref\DIZ{
  P. Di Francesco, C. Itzykson and J.-B. Zuber, {\it Classical
  $\cW$-algebras}, \CMP {\bf 140} (1991) 543.}
\bref\DV{
  R. Dijkgraaf and E. Verlinde, {\it Modular invariance and the
  fusion algebra}, \NP [Proc.Suppl.]
  {\bf 5B} (1988) 87.}
\bref\DVVV{
  R. Dijkgraaf, C. Vafa, E. Verlinde and H. Verlinde, {\it The operator
  algebra of orbifold models}, \CMP {\bf 123} (1989) 485.}
%\bref\DVV{
%  R. Dijkgraaf, E. Verlinde and H. Verlinde, {\it $c=1$ Conformal
%  field theories on Riemann surfaces}, \CMP {\bf 115} (1988) 649.}
\bref\DVVb{
  R. Dijkgraaf, E. Verlinde and H. Verlinde, {\it Loop equations
  and Virasoro constraints in non-perturbative 2-D quantum
  gravity}, \NP {\bf B348} (1991) 435.}
%\bref\DGH{
%  L. Dixon, P. Ginsparg and J. Harvey, {\it $\hat{c}=1$ Conformal
%  field theory}, \NP {\bf B306} (1988) 470.}
%\bref\DFa{
%  V.S. Dotsenko and V.A. Fateev, {\it Conformal algebra and multipoint
%  correlation functions in 2D statistical models}, \NP {\bf B240} [FS12]
%  (1984) 312.}
%\bref\DFb{
%  V.S. Dotsenko  and V.A. Fateev,
%  {\it Four-point correlation functions and operator algebra
%  in 2D conformal invariant theories with central charge $c \leq 1$},
%  \NP {\bf B251} [FS13] (1985) 691.}
\bref\Doa{
  M.R. Douglas, {\it $G/H$ conformal field theory},
  Caltech preprint 68-1453, unpublished.}
\bref\Dob{
  M.R. Douglas,
  {\it $G/H$ conformal field theory}, Caltech PhD thesis, 1988.}
\bref\Doc{
  M. Douglas, {\it Strings in less than one dimension and the
  generalized KdV hierarchies}, \PL {\bf 238B} (1990) 176.}
\bref\DS{
  V. Drinfeld and V. Sokolov, {\it Lie algebras and equations of
  Korteweg-de Vries type}, J. Sov. Math. {\bf 30} (1984) 1975.}
\bref\DHS{
  G. Dunne, I. Halliday and P. Suranyi, {\it Bosonization
  of parafermionic conformal field theories}, \NP {\bf B325}
  (1989) 526.}
\bref\Dy{
  E. Dynkin, {\it Semi-simple subalgebras of semi-simple Lie algebras},
  Amer. Math. Soc. Transl. Ser. 2, {\bf 6} (1957) 111.}
%\bref\ET{
%  T. Eguchi and A. Taormina, {\it Extended superconformal algebras
%  and string compactifications}, preprint CERN-TH.5123/88.}
\bref\EY{
  T. Eguchi and S.-K. Yang, {\it $N=2$ superconformal models
  as topological field theories}, \MPL {\bf A5} (1990) 1693.}
\bref\EFHHNV{
  W. Eholzer, M. Flohr, A. Honecker, R. H\"ubel,
  W. Nahm and R. Varnhagen, {\it Representations
  of $\cW$-algebras with two generators and new rational
  models}, preprint BONN-HE-91-22.}
\bref\EMNa{
  J. Ellis, N. Mavromatos and D.V. Nanopoulos, {\it Quantum coherence
  and two-dimensional black holes}, \PL {\bf 267B} (1991) 465.}
\bref\EMNb{
  J. Ellis, N. Mavromatos and D.V. Nanopoulos, {\it On the connection
  between quantum mechanics and the geometry of two-dimensional strings},
  \PL {\bf 272B} (1991) 261.}
\bref\EH{
  M. Evans and T. Hollowood, {\it Supersymmetric Toda field theories},
  \NP {\bf B352} (1991) 723.}
\bref\Fa{
  V.A. Fateev, unpublished.}
\bref\FLa{
  V.A. Fateev and S.L. Lykyanov, {\it The models of two dimensional
  conformal quantum field theory with $\ZZ_n$ symmetry}, \IJMP
  {\bf A3} (1988) 507.}
\bref\FLb{
  V.A. Fateev and S.L. Lykyanov,
  {\it Conformally invariant models of two-dimensional
  quantum field theory with $\ZZ_n$-symmetry}, Sov. Phys. JETP {\bf 67}
  (1988) 447.}
\bref\FLc{
  V.A. Fateev and S.L. Lykyanov,
  {\it Additional symmetries and exactly-soluble models
       in two-dimensional conformal field theory},
  parts I, II and III, Sov. Sci. Rev. A. Phys. {\bf 15} (1990) 1.}
\bref\FLd{
  V.A. Fateev and S.L. Lukyanov, {\it Exactly soluble models of
  conformal quantum field theory associated with the simple Lie algebra
  $D_n$}, Sov. J. Nucl. Phys. {\bf 49} (1989) 925.}
\bref\FZa{
  V.A. Fateev and A.B. Zamolodchikov,
  {\it Nonlocal (parafermion) currents in two-dimensional conformal
  quantum field theory and self-dual critical points in $Z_N$-symmetric
  statistical systems}, Sov. Phys. JETP {\bf 62} (1985) 215.}
\bref\FZb{
  V.A. Fateev and A.B. Zamolodchikov, {\it Conformal quantum field
  theory models in two dimensions having $\ZZ_3$ symmetry}, \NP
  {\bf B280} [FS18] (1987) 644.}
\bref\FZc{
  V.A. Fateev and A.B. Zamolodchikov, {\it Spin $4/3$ ``parafermionic''
  currents in two-dimensional conformal field theory. Minimal models
  and tricritical $Z_3$ Potts model},
  Theor. Math. Phys. {\bf 71} (1988) 451.}
\bref\FORTW{
  L. Feh\'er, L. O'Raifeartaigh, P. Ruelle, I. Tsutsui and A. Wipf,
  {\it On the general structure of Hamiltonian reductions
  of the WZNW theory}, preprint DIAS-STP-91-29.}
\bref\FFrc{
  B.L. Feigin and E. Frenkel, {\it The family of representations of
  affine Lie algebras}, Russ. Math. Surv. {\bf 43} (1989) 221.}
\bref\FFra{
  B.L. Feigin and E. Frenkel, {\it Quantization of the Drinfeld-Sokolov
  reduction}, \PL {\bf 246B} (1990) 75.}
\bref\FFrb{
  B.L. Feigin and E. Frenkel, {\it Affine Kac-Moody algebras at the
  critical level and Gelfand-Dikii algebras}, preprint MSRI 04029-91.}
\bref\FFa{
  B.L. Feigin and D.B. Fuchs, {\it Invariant skew-symmetric differential
  operators on the line and Verma modules over the Virasoro algebra},
  Funct. Anal. Appl. {\bf 16} (1982) 114.}
\bref\FFb{
  B.L. Feigin and D.B. Fuchs,
  {\it Verma modules over the Virasoro algebra},
  Funct. Anal. Appl. {\bf 17} (1983) 241.}
\bref\FFc{
  B.L. Feigin and D.B. Fuchs,
  {\it Verma modules over the Virasoro algebra},
  Lect. Notes Math. {\bf 1060} (1984) 230.}
\bref\FFd{
  B.L. Feigin and D.B. Fuchs,
  {\it Representations of the Virasoro algebra},
  in `Representations of Lie Groups and Related Topics',
  ed. A.M.~Vershik and D.P.~Zhelobenko
  (Gordon and Breach, New York, 1990).}
\bref\Fe{
  G. Felder, {\it BRST approach to minimal models}, \NP {\bf B317} (1989)
  215; erratum \NP {\bf B324} (1989) 548.}
%\bref\FGK{
%  G. Felder, K. Gawedzki and A. Kupiainen, {\it The spectrum of
%  Wess-Zumino-Witten models}, \NP {\bf B299} (1988) 355.}
\bref\FGRZ{
  L.A. Ferreira, J.F. Gomes, R.M. Ricotta and A.H. Zimerman,
  {\it Supersymmetric construction of $\cW$-algebras from super Toda
  and WZNW theories}, preprint IFT-P.45/91.}
\bref\Fi{
  J.M. Figueroa-O'Farrill,
  {\it On the homological construction of Casimir algebras},
  \NP {\bf B343} (1990) 450.}
\bref\FMR{
  J.M. Figueroa-O'Farrill, J. Mas and E. Ramos, {\it Bihamiltonian
  structure of the KP hierarchy and the $\cW_{KP}$ algebra},
  \PL {\bf 266B} (1991) 298.}
\bref\FRa{
  J.M. Figueroa-O'Farrill and E. Ramos, {\it $\cW$-Superalgebras from
  supersymmetric Lax operators}, \PL {\bf 262B} (1991) 265.}
\bref\FRb{
  J.M. Figueroa-O'Farrill and E. Ramos, {\it Classical $N=2$
  $\cW$-superalgebras and supersymmetric Gel'fand-Dickey brackets},
  \NP {\bf B368} (1992) 361.}
\bref\FRc{
  J.M. Figueroa-O'Farrill and E. Ramos, {\it Classical $N=1$
  $\cW$-superalgebras from hamiltonian reduction},
  \CMP {\bf 145} (1992) 43.}
%\bref\FRd{
%  J.M. Figueroa-O'Farrill and E. Ramos, {\it Existence and uniqueness
%  of the universal $w$-algebra}, preprint KUL-TF-91/27.}
%\bref\FRe{
%  J.M. Figueroa-O'Farrill and E. Ramos, {\it The classical limit
%  of $\cW$-algebras}, \PL {\bf 282B} (1992) 357.}
\bref\FSa{
  J.M. Figueroa-O'Farrill and S. Schrans, {\it The spin-6 extended
  conformal algebra}, \PL {\bf 245B} (1990) 471.}
\bref\FSb{
  J.M. Figueroa-O'Farrill and S. Schrans, {\it The conformal bootstrap
  and super-$\cW$ algebras}, \IJMP {\bf A7} (1992) 591.}
\bref\FSc{
  J.M. Figueroa-O'Farrill and S. Schrans, {\it Extended superconformal
  algebras}, \PL {\bf 257B} (1991) 69.}
\bref\FST{
  J.M. Figueroa-O'Farrill, S. Schrans and K. Thielemans, {\it On the
  Casimir algebra of $B_2$}, \PL {\bf 263B} (1991) 378.}
\bref\Flo{
  M. Flohr, {\it $\cW$-algebra quasiprimary fields and non-minimal
  models}, Diplomarbeit (Master's thesis), preprint BONN-IR-91-30
  (in German).}
\bref\FrLa{
  E.S. Fradkin en V.Ya. Linetsky, {\it Result of the classification
  of superconformal algebras in two dimensions}, preprint ETH-TH/92-6.}
\bref\Frea{
  E. Frenkel, {\it Affine Kac-Moody algebras at the critical level and
  quantum Drinfeld-Sokolov reduction}, Harvard University PhD thesis, 1991.}
\bref\Freb{
  E. Frenkel, {\it $\cW$-algebras and Langlands-Drinfeld correspondence},
  Harvard preprint (Nov. '91).}
\bref\FKW{
  E. Frenkel, V.G. Kac and M. Wakimoto, {\it Characters and fusion
  rules for $\cW$-algebras via quantized Drinfeld-Sokolov reductions},
  preprint (1991).}
\bref\Fr{
  I.B. Frenkel, {\it Representations of Kac-Moody algebras and dual
  resonance models}, in ``Applications of group theory in physics and
  mathematical physics'', Lect. in Appl. Math. {\bf 21} (1985) 325.}
\bref\FK{
  I.B. Frenkel and V.G. Kac, {\it Basic representations of affine Lie
  algebras and dual resonance models}, Inv. Math. {\bf 62} (1980) 23.}
\bref\Fra{
  D. Friedan, {\it Notes on string theory and two-dimensional conformal
  field theory}, in Proc.\ Santa Barbara Workshop on Unified String
  Theories, ed.\ M.B. Green and D. Gross (World Scientific, Singapore,
  1986).}
\bref\Frb{
  D. Friedan, {\it The space of conformal field theories and the space
  of classical string ground states}, in `Physics and Mathematics
  of Strings', V. Knizhnik Memorial Volume, ed.\ L. Brink {\it et al}
  (World Scientific, Singapore, 1989).}
%\bref\Frc{
%  D. Friedan,
%  {\it A new formulation of string theory},
%  Phys. Scripta {\bf T15} (1987) 78.}
\bref\FKSW{
  D. Friedan, A. Kent, S. Shenker and E. Witten, unpublished (1987).}
\bref\FMS{
  D. Friedan, E. Martinec and S. Shenker, {\it Conformal invariance,
  supersymmetry and string theory}, \NP {\bf B271} (1986) 93.}
\bref\FQSa{
  D. Friedan, Z. Qiu and S. Shenker, {\it Conformal invariance,
  unitarity and critical exponents in two dimensions}, Phys. Rev. Lett.
  {\bf 52} (1984) 1575.}
\bref\FQSb{
  D. Friedan, Z. Qiu and S. Shenker, {\it Conformal invariance, unitarity
  and two-dimensional critical exponents},
  in `Vertex Operators in Mathematics and Physics', ed.\
  J. Lepowski {\it et al.}, (Springer Verlag, 1984) 419.}
\bref\FQSc{
  D. Friedan, Z. Qiu and S. Shenker,
  {\it Details of the non-unitarity proof for highest weight
  representations of the Virasoro algebra}, \CMP {\bf 107}
  (1986) 535.}
%\bref\FS{
%  D. Friedan and S. Shenker, unpublished (1987).}
\bref\Fu{
  J. Fuchs, {\it Bosonic superconformal algebras from hamiltonian
  reduction}, preprint CERN-TH.6040/91.}
\bref\FKNa{
  M. Fukuma, H. Kawai, and R. Nakayama {\it Continuum Schwinger-Dyson
  equations and universal structures in two-dimensional quantum
  gravity}, Int. Jour. Mod. Phys. {\bf A6} (1991) 1385.}
\bref\FKNb{
  M. Fukuma, H. Kawai, and R. Nakayama, {\it Infinite dimensional
  grassmannian structure of two-dimensional quantum gravity},
  preprint UT-572, KEK-TH-272.}
\bref\GaSa{
  B. Gato-Rivera and A.N. Schellekens, {\it Complete classification
  of simple current automorphisms}, \NP {\bf B353}
  (1991) 519.}
\bref\GaSb{
  B. Gato-Rivera and A.N. Schellekens, {\it Complete classification
  of simple current modular invariants for $(\ZZ_p)^k$},
  \CMP {\bf 145} (1992) 85.}
\bref\GaSc{
  B. Gato-Rivera and A.N. Schellekens, {\it Classification of simple
  current invariants}, preprint CERN-TH.6246/91.}
\bref\GD{
  I.M. Gel'fand and L.A. Dickey, {\it A family of Hamiltonian structures
  connected with integrable non-linear differential equations}, preprint
  IPM AN SSSR, Moscow (1978).}
\bref\GDb{
  I.M. Gel'fand and L.A. Dickey, Annals of the New York
  Academy of Science {\bf 491} (1987) 131.}
\bref\Gea{
  D. Gepner, {\it On the spectrum of 2D conformal field theories},
  \NP {\bf B287} (1987) 111.}
\bref\Geb{
  D. Gepner, {\it New conformal field theories associated with Lie
  algebras and their partition functions}, \NP {\bf B290} (1987) 10.}
\bref\Gec{
  D. Gepner,
  {\it Space-time supersymmetry in compactified string theory and
  superconformal models}, \NP {\bf B296} (1988) 757.}
\bref\GQ{
  D. Gepner and Z. Qiu, {\it Modular invariant partition functions
  for parafermionic field theories}, \NP {\bf B285} [FS19] (1987) 423.}
\bref\GW{
  D. Gepner and E. Witten, {\it String theory on group manifolds},
  \NP {\bf B278} (1986) 493.}
\bref\GLM{
  A. Gerasimov, A. Levin and A. Marshakov, {\it On $\cW$-gravity
  in two dimensions}, \NP {\bf B360} (1991) 537.}
%\bref\GMM{
%  A. Gerasimov, A. Marshakov and A. Morozov, {\it Hamiltonian reduction
%  of Wess-Zumino-Witten theory from the point of view of bosonization},
%  \PL {\bf 236B} (1990) 269.}
\bref\Ger{
  J.-L. Gervais, {\it Infinite family of polynomial functions of
  the Virasoro generators with vanishing Poisson brackets},
  \PL {\bf 160B} (1985) 277.}
\bref\GM{
  J.-L. Gervais and Y. Matsuo, {\it Classical $A_n$ $\cW$-geometry},
  preprint LPTENS-91/35, NBI-HE-91-50.}
\bref\GN{
  J.-L. Gervais and A. Neveu, {\it Dual string spectrum in Polyakov's
  quantization. (II). Mode separation}, \NP {\bf B209} (1982) 125.}
%\bref\Gia{
%  P. Ginsparg, {\it Curiosities at $c=1$}, \NP {\bf B295} [FS21] (1988)
%  153.}
\bref\Gib{
  P. Ginsparg, {\it Applied conformal field theory},
  in Proc. of `Fields, Strings and Critical Phenomena,' Les Houches
  1988, ed. E.~Br\'ezin and J.~Zinn-Justin
  (North Holland, 1990).}
\bref\Go{
  P. Goddard, {\it Meromorphic conformal field theory}, in `Infinite
  dimensional Lie algebras and Lie groups', ed. V. Kac, Proc. CIRM-Luminy
  conf., 1988 (World Scientific, Singapore, 1989).}
\bref\GKOa{
  P. Goddard, A. Kent and D. Olive, {\it Virasoro algebras
  and coset space models}, \PL {\bf 152B} (1985) 88.}
\bref\GKOb{
  P. Goddard, A. Kent and D. Olive,
  {\it Unitary representations of the Virasoro and super-Virasoro
  algebras},  \CMP {\bf 103} (1986) 105.}
\bref\GNOS{
  P. Goddard, W. Nahm, D. Olive and A. Schwimmer, {\it Vertex operators
  for non-simply-laced algebras}, \CMP {\bf 107} (1986) 179.}
\bref\GO{
  P. Goddard and D.I. Olive, {\it Kac-Moody and Virasoro algebras
  in relation to quantum physics}, \IJMP {\bf A1}
  (1986) 303.}
\bref\GSa{
  P. Goddard and A. Schwimmer, {\it Unitary constructions of extended
  conformal algebras}, \PL {\bf 206B} (1988) 62.}
\bref\GSb{
  P. Goddard and A. Schwimmer, {\it Factoring out free fermions and
  superconformal algebras}, \PL {\bf 214B} (1988) 209.}
\bref\Goe{
  J. Goeree, {\it $\cW$-Constraints in 2-D quantum gravity},
  \NP {\bf B358} (1991) 737.}
\bref\GoW{
  R. Goodman and N.R. Wallach, {\it Higher order Sugawara operators
  for affine Lie algebras}, Trans. Amer. Math. Soc. {\bf 315} (1989) 1.}
\bref\GSW{
  M.B. Green, J.H. Schwarz and E. Witten, {\it Superstring theory},
  vol.\ 1 \& 2 (Cambridge University Press, 1987).}
\bref\GrN{
  P. Griffin and D. Nemeschansky, {\it Bosonization of $\ZZ_N$
  parafermions}, \NP {\bf B323} (1989) 545.}
\bref\GvN{
  M. Grisaru and P. van Nieuwenhuizen, {\it Loop calculations in
  two-dimensional nonlocal field theories}, preprint CERN-TH.63898/92.}
\bref\Halc{
  M.B. Halpern, {\it The two faces of a dual pion-quark model},
  Phys. Rev. {\bf D4} (1971) 2398.}
\bref\Halb{
  M.B. Halpern, {\it Quantum `solitons' which are $SU(N)$ fermions},
  Phys. Rev. {\bf D12} (1975) 1684.}
\bref\Hal{
  M.B. Halpern, {\it Recent developments in the Virasoro master
  equation}, in Proc. of `Strings and Symmetries 1991', Stony Brook,
  May 1991, ed.\ N. Berkovits {\it et al.} (World Scientific,
  Singapore, 1992).}
\bref\HT{
  K. Hamada and M. Takao, {\it Spin-4 current algebra}, \PL {\bf 209B}
  (1988) 247; erratum \PL {\bf 213B} (1988) 564.}
%\bref\Ha{
%  Q. Han-Park, {\it Extended conformal symmetries in real heavens},
%  preprint UMDEPP 90-037.}
\bref\Hay{
  T. Hayashi, {\it Sugawara operators and the Kac-Kazhdan conjecture},
  Inv. Math. {\bf 94} (1988) 13.}
%\bref\HY{
%  Z. Hlousek and K. Yamagishi, {\it An approach to BRST formulation
%  of Kac-Moody algebra}, \PL {\bf 173B} (1986) 65.}
\bref\HZa{
  Q. Ho-Kim and H.B. Zheng, {\it Twisted conformal field theories
  with $Z_3$ invariance}, \PL {\bf 212B} (1988) 71.}
\bref\HZb{
  Q. Ho-Kim and H.B. Zheng,
  {\it Twisted structures of extended Virasoro algebras},
  \PL {\bf 223B} (1989) 57.}
\bref\HZc{
  Q. Ho-Kim and H.B. Zheng,
  {\it Twisted characters and partition functions in extended
  Virasoro algebras}, \MPL {\bf A5} (1990) 1181.}
\bref\Hoa{
  K. Hornfeck, {\it Supersymmetrizing the $\cW_4$ algebra},
  \PL {\bf 252B} (1990) 357.}
\bref\Hob{
  K. Hornfeck, {\it On the central charge of $\cW$ algebras},
  \PL {\bf 255B} (1991) 337.}
\bref\Hoc{
  K. Hornfeck, {\it Realizations for the $\sW (7/2)$-algebra
  and the minimal supersymmetric extension of $\cW A_3$},
  Kings College preprint (July 1991).}
\bref\Hod{
  K. Hornfleck, {\it The minimal supersymmetric extensions of $\cW A_{n-1}$},
  \PL {\bf 275B} (1992) 355.}
\bref\HR{
  K. Hornfeck and E. Ragoucy, {\it A coset construction for the
  super-$\cW_3$ algebra}, \NP {\bf B340} (1990) 235.}
%\bref\HZ{
%  C.S. Huang and D.H. Zhang, {\it Spin 4 and $7/2$ extended conformal
%  algebra}, IC/90/211.}
%\bref\HN{
%  K. Huitu and D. Nemeschansky, {\it Supersymmetric Gelfand-Dickey
%  algebra}, \MPL {\bf A6} (1991) 3179.}
\bref\Hua{
  C.M. Hull, {\it Gauging the Zamolodchikov $\cW$-algebra},
  \PL {\bf 240B} (1990) 110.}
\bref\Hub{
  C.M. Hull, {\it Higher-spin extended conformal algebras and
  $\cW$-gravities}, \NP {\bf B353} (1991) 707.}
\bref\Huba{
  C.M. Hull, {\it The geometry of $\cW$-gravity}, \PL {\bf 269B}
  (1991) 257.}
\bref\Hubb{
  C.M. Hull, {\it $\cW$-gravity anomalies 1: induced quantum
  $\cW$-gravity}, \NP {\bf B367} (1991) 731.}
\bref\Huc{
  C.M. Hull, {\it Classical and quantum $\cW$-gravity},
  in Proc.\  of `Strings and Symmetries 1991',
  Stony Brook, May 1991, ed. N. Berkovits {\it et al.}
  (World Scientific, Singapore, 1992).}
\bref\Ina{
  T. Inami, {\it Super-$\cW$ algebras and generalized super-KdV
  equations}, in Proc.\ of `Strings '90', Texas A \& M,
  ed. R. Arnowitt {\it et al.}\ (World Scientific, Singapore, 1991).}
\bref\II{
  T. Inami and K.-I. Izawa, {\it Super Toda theory from WZNW models},
  \PL {\bf 255B} (1991) 521.}
\bref\IKa{
  T. Inami and H. Kanno, {\it $N=2$ Super KdV and super sine-Gordon
   equations based on Lie superalgebra $A(1,1)^{(1)}$},
   \NP {\bf B359} (1991) 201.}
\bref\IKb{
  T. Inami and H. Kanno, {\it $N=2$ Super $\cW$ algebras and generalized
  $N=2$ super KdV hierarchies based on Lie superalgebras},
  preprint YITP/K-928.}
%\bref\IKc{
%  T. Inami and H. Kanno, {\it Generalized $N=2$ super KdV hierarchies:
%  Lie superalgebraic methods and scalar super Lax formalism},
%  preprint YITP/K-929.}
\bref\IMYa{
  T. Inami, Y. Matsuo and I. Yamanaka, {\it Extended conformal algebras
  with $N=1$ supersymmetry}, \PL {\bf 215B} (1988) 701.}
\bref\IMYb{
  T. Inami, Y. Matsuo and I. Yamanaka, {\it Extended Conformal Algebra
  with $N=2$ Supersymmetry}, \IJMP {\bf A5} (1990) 4441.}
\bref\In{
  K. Intriligator, {\it Bonus symmetry in conformal field theory},
  \NP {\bf B332} (1990) 541.}
\bref\Ita{
  K. Ito, {\it Quantum hamiltonian reduction and $\cW B$ algebra},
  preprint YITP/K-885 (Sept. '90).}
\bref\Itb{
  K. Ito, {\it Quantum Hamiltonian reduction and $N=2$ coset models},
  \PL {\bf 259B} (1991) 73.}
\bref\Itc{
  K. Ito, {\it $N=2$ superconformal $CP_n$ model},
  \NP {\bf B370} (1992) 123.}
\bref\IM{
  K. Ito and J.O. Madsen, {\it Hamiltonian reduction and classical
  extended superconformal algebras}, preprint NBI-HE-92-12.}
\bref\ISZ{
  C. Itzykson, H. Saleur and J.-B. Zuber eds., {\it Conformal invariance
  and applications to statistical mechanics}, (World Scientific,
  Singapore, 1988).}
\bref\JMa{
  M. Jimbo and T. Miwa, Adv. Stud. Pure Math. {\bf 4} (1984) 97.}
\bref\JMb{
  M. Jimbo and T. Miwa, Adv. Stud. Pure Math. {\bf 6} (1985) 17.}
%\bref\JMOa{
%  M. Jimbo, T. Miwa and M. Okado, {\it Local state probabilities
%  of solvable lattice models, a $A_{n-1}^{(1)}$ family}, \NP {\bf B300}
%  [FS22] (1988) 74.}
%\bref\JMOb{
%  M. Jimbo, T. Miwa and M. Okado, {\it Solvable lattice models related
%  to the vector representation of classical simple Lie algebras},
%  \CMP {\bf 116} (1988) 507.}
\bref\Kaa{
  V.G. Kac, {\it Simply graded Lie algebras of finite growth},
  Funct. Anal. Appl. {\bf 1} (1967) 328.}
%\bref\Kab{
%  V.G. Kac, {\it Infinite-dimensional Lie algebras and Dedekind's
%  $\eta$ function}, Funct. Anal. Appl. {\bf 8} (1974) 68.}
\bref\Kac{
  V.G. Kac, {\it Contravariant form for infinite-dimensional Lie
  algebras and superalgebras}, Lect. Notes in Phys. {\bf 94}
  (1979) 441.}
\bref\Kae{
  V.G. Kac, {\it Lie superalgebras}, Adv. Math. {\bf 26} (1977) 8.}
%
% \Kac has the Kac-determinant
%
\bref\Kad{
  V.G. Kac, {\it Infinite dimensional Lie algebras} (Cambridge
  University Press, 1985).}
\bref\KK{
  V.G. Kac and D.A. Kazhdan, {\it Structure of representations
  with highest weight of infinite dimensional Lie algebras},
  Adv. Math. {\bf 34} (1979) 97.}
\bref\KP{
  V.G. Kac and D.H. Peterson, {\it Infinite dimensional Lie algebras,
  theta functions and modular forms}, Adv. Math {\bf 53} (1984) 125.}
\bref\KaWa{
  V.G. Kac and M. Wakimoto, {\it Modular invariant representations
  of infinite dimensional Lie algebras and superalgebras},
  Proc. Natl. Acad. Sci. USA {\bf 85} (1988) 4956.}
\bref\KaWc{
  V.G. Kac and M. Wakimoto, {\it Modular and conformal invariance
  constraints in representation theory of affine Lie algebras},
  Lect. Notes in Phys. {\bf 261}, 345.}
\bref\KaWb{
  V.G. Kac and M. Wakimoto, {\it Branching functions for winding
  subalgebras and tensor products}, MIT preprint (1990).}
\bref\KTN{
  M. Kaku, P.K. Townsend and P. van Nieuwenhuizen, {\it Properties
  of conformal supergravity}, Phys. Rev. {\bf D17} (1987) 3179.}
\bref\Ka{
  D. Kastor, {\it Modular invariance in superconformal models},
  \NP {\bf B280} [FS18] (1987) 304.}
\bref\KMQ{
  D. Kastor, E. Martinec and Z. Qiu, {\it Current algebra
  and conformal discrete series}, \PL {\bf 200B} (1988) 434.}
\bref\Kat{
  A. Kato, {\it Classification of modular invariant partition
  functions in 2 dimensions}, \MPL {\bf A2} (1987) 585.}
%\bref\KO{
%  M. Kato and K. Ogawa, {\it Covariant quantization of string
%  based on BRS invariance}, \NP {\bf B212} (1983) 443.}
\bref\Kau{
  H.G. Kausch, {\it Extended conformal algebras generated by a
  multiplet of primary fields}, \PL {\bf 259B} (1991) 448.}
\bref\KWa{
  H.G. Kausch and G.M.T. Watts, {\it A study of $\cW$-algebras
  using Jacobi identities}, \NP {\bf B354} (1991) 740.}
\bref\KWb{
  H.G. Kausch and G.M.T. Watts, {\it Quantum Toda theory and the Casimir
  algebra of $B_2$ and $C_2$}, preprint DAMTP 91-24, DTP 91-35.}
\bref\KWc{
  H.G. Kausch and G.M.T. Watts, {\it Duality in quantum Toda theory
  and $\cW$-algebras}, preprint Imperial/ TH/ 91-92/17.}
\bref\KN{
  T. Kawai and T. Nakatsu, {\it Comments on generalized quantum
  Hamiltonian reductions}, \MPL {\bf A6} (1991) 3557; erratum
  \MPL {\bf A7} (1992) 267.}
%\bref\KS{
%  Y. Kazama and H. Suzuki, {\it Bosonic construction of conformal field
%  theories with extended supersymmetry}, preprint UT-Komaba 88-13.}
%\bref\Ke{
%  A. Kent {\it Normal ordered Lie algebras}, preprint
%  IASSNS-HEP-88/04.}
\bref\Ki{
  E.B. Kiritsis, {\it Structure of $N=2$ superconformally invariant
  unitary `minimal' theories: Operator algebra and correlation
  functions}, \PR {\bf D36} (1987) 3048.}
\bref\KPa{
  I.R. Klebanov and A.M. Polyakov, {\it Interaction of
  discrete states in two-dimensional string theory},
  \MPL {\bf A6} (1991) 3273.}
\bref\Kli{
  A. Kliem, {\it Construction of $\cW$-algebras}, Diplomarbeit (Master's
  thesis), preprint BONN-IR-91-46 (in German).}
\bref\Kn{
  V.G. Knizhnik, {\it Superconformal algebras in two dimensions},
  Theor. Math. Phys. {\bf 66} (1986) 68.}
\bref\KPZ{
  V.G. Knizhnik, A.M. Polyakov and A.B. Zamolodchikov,
  {\it Fractal structure of 2d-quantum gravity},
  \MPL {\bf A3} (1988) 819.}
\bref\KZ{
  V.G. Knizhnik and A.B. Zamolodchikov, {\it Current algebra and
  Wess-Zumino model in two dimensions}, \NP {\bf B247} (1984) 83.}
\bref\KMN{
  S. Komata, K. Mohri and H. Nohara, {\it Classical and quantum
  extended superconformal algebra}, \NP {\bf B359} (1991) 168.}
%\bref\Ko{
%  I.K. Kostov, {\it Free field representation of the $A_n$ coset
%  models on the torus}, \NP {\bf B300} (1988) 559.}
\bref\Ku{
  B.A. Kupershmidt, {\it Super Korteweg-de Vries equations
  associated to super extensions of the Virasoro algebra},
  \PL {\bf 109A} (1985) 417.}
\bref\LPRSW{
  H. Lu, C.N. Pope, L.J. Romans, X. Shen and X.-J. Wang,
  {\it Polyakov construction of the $N=2$ super-$\cW_3$ algebra},
  \PL {\bf 264B} (1991) 91.}
\bref\LPScW{
  H. Lu, C.N. Pope, S. Schrans and X.W. Wang, {\it On sibling and
  exceptional $\cW$-strings}, preprint CTP TAMU-10/92.}
\bref\LPSX{
  H. Lu, C.N. Pope, S. Schrans and K.W. Xu, {\it The complete spectrum
  of the $\cW_N$ string}, preprint CTP TAMU-5/92.}
\bref\LPSW{
  H. Lu, C.N. Pope, X. Shen and X.-J. Wang, {\it The complete
  structure of $\cW_N$ form $\cW_{\infty}$ at $c=-2$},
  \PL {\bf 267B} (1991) 356.}
\bref\Lu{
  S.L. Lukyanov, {\it Quantization of the Gelfan'd Dikii brackets},
  Funct. Anal. Appl. {\bf 22} (1990) 1.}
\bref\MSW{
  A.J. Macfarlane, A. Sudbery and P.H. Weisz, {\it On Gell-Mann's $\la$
  matrices, $d$- and $f$-tensors, octets,
  and parametrizations of $SU(3)$}, \CMP {\bf 11} (1968) 77.}
\bref\Mag{
  F. Magri, {\it A simple model of the integrable hamiltonian
  equation}, Journ. Math. Phys. {\bf 19} (1978) 1156.}
\bref\Ma{
  D. Malikov, {\it Special vectors in Verma modules over affine
  algebras}, Funct. Anal. Appl. {\bf 23} (1989) 76.}
\bref\Mana{
  P. Mansfield, {\it Light-cone quantization of the Liouville and
  Toda field theories}, \NP {\bf B222} (1983) 419.}
\bref\Manb{
  P. Mansfield, {\it Conformally extended Toda theories},
  \PL {\bf 242B} (1990) 387.}
\bref\MaSp{
  P. Mansfield and B. Spence, {\it Toda theories, the geometry
  of $\cW$-algebras and minimal models}, \NP {\bf B362} (1991) 294.}
%\bref\MM{
%  A. Marshakov and A. Morozov, {\it A note on $\cW_3$ algebra},
%  \NP {\bf B339} (1990) 79.}
\bref\Maa{
  P. Mathieu, {\it Extended classical conformal algebras and the
  second hamiltonian structure of Lax equations}, \PL {\bf 208B}
  (1988) 101.}
\bref\Mab{
  P. Mathieu, {\it Representation of the $so(N)$ and $u(N)$
  superconformal algebras via Miura transformations}, \PL
  {\bf 218B} (1989) 185.}
\bref\MO{
  P. Mathieu and W. Oevel, {\it The $\cW_3^{(2)}$ conformal
  algebra and the Boussinesq hierarchy}, \MPL {\bf A6} (1991) 2397.}
%\bref\MMS{
%  S. Mathur, S. Mukhi and A. Sen, {\it Differential equations
%  for correlators and characters in arbitrary rational conformal
%  field theories}, \NP {\bf B312} (1989) 15.}
\bref\Mat{
  Y. Matsuo, {\it Remarks on fractal $\cW$-gravity}, \PL {\bf 227B}
  (1989) 209.}
\bref\Mik{
  A. Mikovic, {\it Hamiltonian construction of $\cW$-gravity
  actions}, preprint QMW/PH/91/13.}
\bref\Mia{
  S. Mizoguchi, {\it Determinant formula and unitarity for the
  $\cW_3$ algebras}, \PL {\bf 222B} (1989) 226.}
\bref\Mib{
  S. Mizoguchi, {\it Non-unitarity theorem for the $A$-type $\cW_{(n)}$
  algebras}, \PL {\bf 231B} (1989) 112.}
\bref\Mic{
  S. Mizoguchi,
  {\it The structure of representation for the $\cW_3$ minimal model},
  \IJMP {\bf A6} (1991) 133.}
\bref\Moo{
  R.V. Moody, {\it Lie algebras associated with generalized Cartan
  matrices}, Bull. Am. Mat. Soc. {\bf 73} (1967) 217.}
\bref\MSa{
  G. Moore and N. Seiberg, {\it Polynomial equations for
  rational conformal field theories}, \PL {\bf 212B} (1988) 451.}
\bref\MSb{
  G. Moore and N. Seiberg,
  {\it Naturality in conformal field theory}, \NP {\bf B313}
  (1989) 16.}
\bref\MSc{
  G. Moore and N. Seiberg,
  {\it Classical and quantum conformal field theory},
  \CMP {\bf 123} (1989) 177.}
%\bref\MSd{
%  G. Moore and N. Seiberg, {\it Taming the conformal zoo},
%  \PL {\bf 220B} (1989) 422.}
\bref\Mo{
  A. Morozov, {\it On the concept of universal $\cW$-algebra},
  \NP {\bf B357} (1991) 619.}
\bref\MP{
  S. Mukhi and S. Panda, {\it Fractional level current algebras
  and the classification of characters}, \NP {\bf B338} (1990) 263.}
\bref\Na{
  W. Nahm, {\it Conformal quantum field theories in two dimensions},
  World Scientific, to be published.}
\bref\Naa{
  F.J. Narganes Quijano, {\it On the parafermionic $\cW_N$ algebra},
  \IJMP {\bf A6} (1991) 2611.}
\bref\Nab{
  F.J. Narganes Quijano, {\it Bosonization of parafermions
  and related conformal models: $\cW_N$-algebras},
  \AP {\bf 206} (1991) 494.}
\bref\NY{
  D. Nemeschansky and S. Yankielowicz, {\it $N=2$ $\cW$-algebra,
  Kazama-Suzuki models and Drinfeld-Sokolov reduction}, preprint
  USC-91/005}
\bref\NS{
  A. Neveu and J.H. Schwarz, {\it Factorizable dual model of pions},
  \NP {\bf B31} (1971) 86.}
%\bref\Nia{
%  M. Niedermaier, {\it Quantum integrability of perturbed Virasoro
%  degenerate conformal models}, \PL {\bf 243B} (1990) 73.}
%\bref\Nib{
%  M. Niedermaier, {\it Infinite abelian subalgebra of $\cW(sl(n))$},
%  \NP {\bf B364} (1991) 165.}
\bref\Nic{
  M. Niedermaier, {\it Irrational free field resolutions for
  $\cW(sl(n))$ and extended Sugawara construction}, preprint DESY 90-111.}
\bref\Nid{
  M. Niedermaier, {\it $\cW(sl(n))$: Existence, Cartan basis and infinite
  abelian subalgebras}, preprint DESY 91-148.}
\bref\Noh{
  H. Nohara, {\it Extended superconformal algebra as symmetry
  of super Toda field theory}, \AP {\bf 214} (1992) 1.}
\bref\NM{
  H. Nohara and K. Mohri, {\it Extended Superconformal Algebras
  from Super Toda Field Theory}, \NP {\bf B349} (1991) 253.}
\bref\Od{
  S. Odake, {\it Superconformal algebras and their extensions}, Soryushiron
  Kenkyu (Kyoto) {\bf 78} (1989) 201 (in Japanese).}
\bref\Odb{
  S. Odake, {\it Unitary representations of $\cW$-infinity
  algebras}, preprint YITP/K-955.}
\bref\OS{
  S. Odake and T. Sano, {\it $\cW_{1+\infty}$ and super-$\cW_{1+\infty}$
  algebras with $SU(N)$ symmetry}, \PL {\bf 258B} (1991) 369.}
\bref\OSSvN{
  H. Ooguri, K. Schoutens, A. Sevrin and P. van Nieuwenhuizen,
  {\it The induced action of $\cW_3$ gravity},
  \CMP {\bf 145} (1992) 515.}
\bref\ORTW{
  L. O'Raifeartaigh, P. Ruelle, I. Tsutsui and A. Wipf,
  {\it $\cW$-algebras for generalized Toda theories},
  \CMP {\bf 143} (1992) 333.}
\bref\Paa{
  V. Pasquier, {\it Operator content of the ADE lattice models},
  J. Phys. {\bf A20} (1987) 5707.}
\bref\Pab{
  V. Pasquier, {\it Continuum limit of lattice models built on
  quantum groups}, \NP {\bf B295} [FS21] (1988) 491.}
%\bref\Pac{
%  V. Pasquier, {\it Etiology of IRF models}, \CMP {\bf 118}
% (1988) 355.}
%\bref\PSY{
%  S.K. Paul, R. Sasaki and H. Yoshii, {\it Supersymmetric generalization
%  of extended conformal algebras}, preprint RRK 88-22.}
%\bref\PT{
%  R. Paunov and I. Todorov, {\it Local quantum field theories
%  involving the $U(1)$ current algebra on the circle}, preprint
%  SISSA 95 EP (July 1988).}
\bref\Poa{
  A.M. Polyakov, {\it Quantum gravity in two dimensions},
  \MPL {\bf A2} (1987) 893.}
\bref\Pob{
  A.M. Polyakov, {\it Two-dimensional quantum gravity. Superconductivity
  at high $T_c$},
  in Proc. of `Fields, Strings and Critical Phenomena,' Les Houches
  1988, ed. E.~Br\'ezin and J.~Zinn-Justin
  (North Holland, 1990).}
\bref\Poc{
  A.M. Polyakov, {\it Gauge transformations and diffeomorphisms},
  \IJMP {\bf A5} (1990) 833.}
\bref\Pop{
  C.N. Pope, {\it Lectures on $\cW$ algebras and $\cW$ gravity},
  in Proc. of the 1991 Trieste Summer School in High-Energy
  Physics (World Scientific, Singapore, 1992).}
\bref\PRSa{
  C.N. Pope, L.J. Romans and X. Shen, {\it The complete structure of
  $\cW_\infty$}, \PL {\bf 236B} (1990) 173.}
\bref\PRSb{
  C.N. Pope, L.J. Romans and X. Shen, {\it $\cW_\infty$ and the Racah-Wigner
  algebra}, \NP {\bf B339} (1990) 191.}
\bref\PRSc{
  C.N. Pope, L.J. Romans and X. Shen, {\it A new higher-spin algebra
  and the lone-star product}, \PL {\bf 242B} (1990) 401.}
\bref\PRSd{
  C.N. Pope, L.J. Romans and X. Shen, {\it A brief history
  of $\cW_\infty$}, in Proc. of `Strings '90', Texas A \& M,
  ed. R. Arnowitt {\it et al.}\ (World Scientific, Singapore, 1991).}
%\bref\PRSe{
%  C.N. Pope, L.J. Romans and X. Shen, {\it Conditions for anomaly-free
%  $\cW$ and super-$\cW$ algebras}, \PL {\bf 254B} (1991) 401.}
\bref\PRSta{
  C.N. Pope, L.J. Romans and K.S. Stelle, {\it Anomaly-free $\cW_3$
  gravity and critical $\cW_3$ strings}, \PL {\bf 268B} (1991) 167.}
\bref\PRStb{
  C.N. Pope, L.J. Romans and K.S. Stelle, {\it On $\cW_3$ strings},
  \PL {\bf 269B} (1991) 287.}
%\bref\Rad{
%  O. Radul, {\it Central expansion of Lie algebra of differential
%  operators on a circle and $\cW$-algebras},
%  JETP Lett. {\bf 50} (1989) 371.}
\bref\Ram{
  P. Ramond, {\it Dual theory for free fermions},
  \PR {\bf D3} (1971) 2415.}
%\bref\RS{
%  P. Ramond and J.H. Schwarz, {\it Classification of dual model
%  gauge algebras}, \PL {\bf 64B} (1976) 75.}
\bref\Raa{
  F. Ravanini, {\it An infinite class of new conformal field
  theories with extended algebras}, \MPL {\bf A3} (1988) 397.}
\bref\Rac{
  F. Ravanini, {\it Modular invaraince in $S_3$ symmetric conformal field
  theories}, Mod. Phys. Lett. {\bf A3} (1988) 271.}
\bref\Rab{
  F. Ravanini, {\it Informal introduction to extended algebras
  and conformal field theories with $c\geq 1$}, lectures at the
  `Program on String and Conformal Field Theory', Copenhagen
  1989, preprint NORDITA 89/21 P.}
\bref\RC{
  A. Rocha-Caridi, {\it Vacuum vector representations of the
  Virasoro algebra},
  in `Vertex Operators in Mathematics and Physics', ed.\
  J. Lepowski {\it et al.}, (Springer Verlag, 1984) 451.}
\bref\Roa{
  L.J. Romans, {\it Realizations of classical and quantum $\cW_3$
  symmetry}, \NP {\bf B352} (1991) 829.}
\bref\Rob{
  L.J. Romans, {\it Quasi-superconformal algebras in two dimensions
  and hamiltonian reduction}, preprint USC-90/ HEP-30.}
\bref\Roc{
  L.J. Romans, {\it The $N=2$ super-$\cW_3$ algebra}, \NP {\bf B369}
  (1992) 403.}
\bref\SSY{
  Y. Saitoh, N. Sakai and N. Yugami, {\it Bosonic construction of
  $\cW_3$ algebra}, Prog. Theor. Phys. {\bf 83} (1990) 621.}
%\bref\Sa{
%  R. Sasaki, {\it Notes on extended conformal algebras},
%  preprint RRK 88-21.}
\bref\SW{
  A.N. Schellekens and N.P. Warner, {\it Conformal subalgebras
  of Kac-Moody algebras}, \PR {\bf D34} (1986) 3092.}
\bref\SYa{
  A.N. Schellekens and S. Yankielowicz, {\it Extended chiral
  algebras and modular invariant partition functions},
  \NP {\bf B327} (1989) 673.}
\bref\SYb{
  A.N. Schellekens and S. Yankielowicz,
  {\it Modular invariants from simple currents. An explicit proof},
  \PL {\bf 227B} (1989) 387.}
\bref\Sca{
  K. Schoutens, {\it $O(N)$-extended superconformal field theory
  in superspace}, \NP {\bf B295} [FS21] (1988) 634.}
%\bref\Scb{
%  K. Schoutens, {\it Representation theory for a class of $so(N)$
%  extended superconformal operator algebras},
%  \NP {\bf B314} (1989) 519.}
\bref\Scc{
  K. Schoutens, {\it Extensions of conformal symmetry in two dimensional
  quantum field theory}, Univ. of Utrecht PhD thesis, 1989.}
\bref\SS{
  K. Schoutens and A. Sevrin, {\it Minimal super $\cW_N$ algebras
  in coset conformal field theories}, \PL {\bf 258B} (1991) 134.}
\bref\SSvNa{
  K. Schoutens, A. Sevrin and P. van Nieuwenhuizen, {\it Quantum BRST
  charge for quadratically nonlinear Lie algebras},
  \CMP {\bf 124} (1989) 87.}
\bref\SSvNb{
  K. Schoutens, A. Sevrin and P. van Nieuwenhuizen,
  {\it A new gauge theory for $\cW$-type algebras},
  \PL {\bf 243B} (1990) 245.}
\bref\SSvNc{
  K. Schoutens, A. Sevrin and P. van Nieuwenhuizen,
  {\it Covariant formulation of classical $\cW$-gravity},
  \NP {\bf B349} (1991) 791.}
\bref\SSvNca{
  K. Schoutens, A. Sevrin and P. van Nieuwenhuizen, {\it Covariant
  $w_\infty$ gravity and its reduction to $\cW_N$ gravity},
  \PL {\bf 251B} (1990) 355.}
\bref\SSvNd{
  K. Schoutens, A. Sevrin and P. van Nieuwenhuizen, {\it Quantum
  $\cW_3$ gravity in the chiral gauge}, \NP {\bf B364} (1991) 584.}
\bref\SSvNe{
  K. Schoutens, A. Sevrin and P. van Nieuwenhuizen, {\it On the effective
  action of chiral $\cW_3$ gravity}, \NP {\bf B371} (1992) 315.}
\bref\SSvNf{
  K. Schoutens, A. Sevrin and P. van Nieuwenhuizen, {\it Induced gauge
  theories and $\cW$-gravity}, in Proc. of `Strings and Symmetries 1991',
  Stony Brook, May 1991, ed.\ N. Berkovits {\it et al.}\
  (World Scientific, Singapore, 1992).}
\bref\Sc{
  S. Schrans, {\it Extensions of conformal invariance in two-dimensional
  quantum field theory}, Univ. of Leuven PhD thesis, 1991.}
%\bref\ScS{
%  A. Schwimmer and N. Seiberg, {\it Comments on the $N=2,3,4$
% superconformal algebras in two dimensions}, \PL {\bf 184B} (1987) 191.}
\bref\Seg{
  G. Segal, {\it Unitary representations of some infinite dimensional
  groups}, Comm. Math. Phys. {\bf 80} (1981) 301.}
%\bref\Se{
%  A. Sevrin, {\it Superconformal algebras and supersymmetric
%  nonlinear $\sigma$-models}, Univ. of Leuven PhD thesis, 1988.}
\bref\STPS{
  A. Sevrin, W. Troost and A. Van Proeyen, {\it Superconformal
  algebras in two dimensions with $N=4$}, \PL {\bf 208B} (1988) 447.}
\bref\Sh{
  X. Shen, {\it $\cW$-infinity and string theory},
  preprint CERN-TH.6404/92.}
\bref\SY{
  J. Soda and H. Yoshii, {\it Kac formulas for the extended Virasoro
  algebras}, preprint RRK-88-5.}
\bref\SoS{
  G.M. Sotkov and M. Stanishkov, {\it Affine Geometry and
  $\cW_n$-gravities}, \NP {\bf B356} (1991) 439.}
\bref\SSZ{
  G.M. Sotkov, M. Stanishkov and C. Zhu, {\it Extrinsic Geometry
  of Strings and $\cW$-gravities}, \NP {\bf B356} (1991) 245.}
\bref\Su{
  H. Sugawara, {\it A field theory of currents}, \PR {\bf 170}
  (1968) 1659.}
\bref\Thi{
  K. Thielemans, {\it A Mathematica package for computing operator
  product expansions}, \IJMP {\bf C2} (1991) 787.}
\bref\Tha{
  J. Thierry-Mieg, {\it BRS-analysis of Zamolodchikov's spin 2
  and 3 current algebra}, \PL {\bf 197B} (1987) 368.}
\bref\Thb{
  J. Thierry-Mieg, {\it Generalizations of the Sugawara Construction},
  in `Nonperturbative
  Quantum Field Theory', ed. G. 't Hooft {\it et al.},
  Proc. Cargese School 1987 (Plenum Press, New York, 1988), 567.}
\bref\Tho{
  C. Thorn, {\it Computing the Kac determinant using dual model
  techniques and more about the no-ghost theorem}, \NP {\bf B248}
  (1984) 551.}
\bref\TV{
  T. Tjin and P. Van Driel, {\it Coupled WZNW-Toda models and covariant
  KdV hierarchies}, preprint ITFA-91-04.}
%\bref\Ty{
%  I.V. Tyutin, unpublished (1975).}
%\bref\Va{
%  C. Vafa, {\it Toward classification of conformal theories},
%  \PL {\bf 206B} (1988) 421.}
%\bref\Var{
%  V.S. Varadarajan, {\it Lie groups, Lie algebras and their
%  representations} (Prentice-Hall inc., 1974).}
\bref\Varn{
  R. Varnhagen, {\it Characters and representations
  of new fermionic $\cW$-algebras}, BONN-HE-91-16.}
\bref\Vea{
  E. Verlinde, {\it Fusion rules and modular transformations in
  2D conformal field theory}, \NP {\bf B300} [FS22] (1988) 360.}
%\bref\Veb{
%  E. Verlinde, {\it Conformal field theory and its application to string
%  theory}, Univ. of Utrecht PhD thesis, 1988.}
\bref\Wak{
  M. Wakimoto, {\it Fock representations of the affine Lie algebra
  $A_1^{(1)}$}, \CMP {\bf 104} (1986) 605.}
\bref\Waa{
  G.M.T. Watts, {\it Determinant formulae for extended algebras in two
  dimensional conformal field theory}, \NP {\bf B326} (1989) 648;
  erratum \NP {\bf B336} (1990) 720.}
\bref\Wab{
  G.M.T. Watts, {\it $\cW B$ algebra representation theory},
  \NP  {\bf B339} (1990) 177.}
\bref\Wac{
  G.M.T. Watts, {\it $\cW$-algebras and coset
  models}, \PL {\bf 245B} (1990) 65.}
\bref\Wad{
  G.M.T. Watts, {\it Extended algebras in conformal field theory},
  Trinity college PhD thesis, 1990.}
\bref\Wae{
  G.M.T. Watts, {\it $\cW B_n$ symmetry, Hamiltonian reduction and
  $B(0,n)$ Toda theory}, \NP {\bf B361} (1991) 311.}
\bref\WZ{
  J. Wess and B. Zumino, {\it Consequences of anomalous
  Ward identities}, \PL {\bf 37B} (1971) 95.}
\bref\Wia{
  E. Witten, {\it Nonabelian bosonization}, \CMP {\bf 92} (1984) 455.}
\bref\Wiaa{
  E. Witten, {\it On the structure of the topological phase
  of two-dimensional gravity}, \NP {\bf B340} (1990) 281.}
\bref\Wiab{
  E. Witten, {\it Topological sigma models},
  \CMP {\bf 118} (1988) 411.}
%\bref\Wib{
%  E. Witten, {\it Quantum field theory and the Jones polyniomial},
%  \CMP {\bf 121} (1989) 351.}
\bref\Wic{
  E. Witten, {\it On string theory and black holes},
  Phys. Rev. {\bf D44} (1991) 314.}
\bref\Wid{
  E. Witten, {\it Ground ring of two-dimensional string
  theory}, \NP {\bf B373} (1992) 187.}
\bref\WYa{
  Y.-S. Wu and F. Yu, {\it Hamiltonian structure, (anti-)self-adjoint
  flows in KP hierarchy and the $\cW_{1+\infty}$
  and $\cW_\infty$ algebras},
  \PL {\bf 263B} (1991) 220.}
\bref\WYb{
  Y.-S. Wu and F. Yu, {\it Nonlinearly deformed
  $\widehat{W}_\infty$ algebra and second hamiltonian structure
  of KP hierarchy}, \NP {\bf B373} (1992) 713.}
\bref\Ya{
  K. Yamagishi, {\it The KP hierarchy and extended Virasoro
  algebras}, \PL {\bf 205B} (1988) 466.}
\bref\Yab{
  K. Yamagishi, {\it $\widehat{W}_\infty$ algebra is anomaly free
  at $c=-2$}, \PL {\bf 266B} (1991) 370.}
\bref\Yac{
  K. Yamagishi, {\it A hamiltonian structure of KP hierarchy,
  $\cW_{1+\infty}$ algebra, and self-dual gravity},
  Phys. Lett. {\bf 259B} (1991) 436.}
\bref\Za{
  A.B. Zamolodchikov, {\it Infinite additional symmetries in two
  dimensional conformal quantum field theory}, Theor. Math. Phys.
  {\bf 65} (1985) 1205.}
\bref\Zaaa{
  A.B. Zamolodchikov, {\it Integrals of motion in scaling 3-state
  Potts model field theory}, \IJMP {\bf A3} (1988) 743.}
\bref\Zab{
  A.B. Zamolodchikov, {\it Integrable field theory from CFT},
  Advanced Studies in Pure Mathematics {\bf 19} (1989) 1.}
\bref\Zh{
  D.H. Zhang, {\it Spin-4 extended conformal algebra}, \PL {\bf 232B}
  (1989) 323.}

\magnification=1200
\hfuzz=7pt
%---------------------------------------------------------------
\pageno=0 \nopagenumbers
% chfront.tex
%---------------frontpage-----------------------------------------
\baselineskip=.8\baselineskip
\line{\hfill CERN-TH.6583/92}
\line{\hfill ITP-SB-92-23}
\vskip1cm

\centerline{\bf $\cW$-SYMMETRY IN CONFORMAL FIELD THEORY}
\vskip1.5cm

\centerline{{Peter Bouwknegt}\footnote{$^*$}{email:
bouwkneg@cernvm.cern.ch}}\smallskip

\centerline{\sl CERN - Theory Division}
\centerline{\sl CH-1211 Gen\`eve 23}
\centerline{\sl SWITZERLAND}
\vskip1cm

\centerline{{Kareljan Schoutens}\footnote{$^{**}$}{email:
schouten@max.physics.sunysb.edu}}\smallskip

\centerline{\sl Institute for Theoretical Physics}
\centerline{\sl State University of New York}
\centerline{\sl Stony Brook, NY 11794-3840}
\centerline{\sl USA}
\vskip2cm

\centerline{\bf Abstract}\smallskip

We review various aspects of $\cW$-algebra symmetry in two-dimensional
conformal field theory and string theory.
We pay particular attention to the construction of $\cW$-algebras
through the quantum Drinfeld-Sokolov reduction and through the coset
construction.
\vskip1cm

\centerline{To be published in: {\sl Physics Reports}}

\vfill
\line{CERN-TH.6583/92\hfil}
\line{ITP-SB-92-23\hfil}
\line{July 1992\hfil}
\baselineskip=1.25\baselineskip
\eject

\footline={\hss\tenrm\folio\hss}
% table of contents for PhysRep,.....chtoc.tex
% last updated 12/22/92

\def\regel{\leaders\hbox to 1em{\hss.\hss}\hfill}
\def\tab{\hskip 5mm}
\def\tabtab{\hskip 12mm}

\centerline{\bf $\cW$-Symmetry in Conformal Field Theory}

\vskip 3mm

\centerline{P. Bouwknegt and K. Schoutens}

\vskip 3mm

\tenpoint

\line{1. Introduction \regel 3}

\line{\tab 1.1. Extensions of conformal symmetry \regel 3}

\line{\tab 1.2. Studying extended symmetries \regel 6}

\line{\tab 1.3. Outline of the paper \regel 10}

\line{2. Preliminaries \regel 11}

\line{\tab 2.1. Conformal invariance: basic notions \regel 11}

\line{\tab 2.2. OPE's, normal ordered
products and associativity \regel 17}

\line{\tab 2.3. Auxiliary field theories \regel 20}

\line{\tabtab 2.3.1. Free fields \regel 20}

\line{\tabtab 2.3.2. Affine Kac-Moody algebras and WZW models \regel 23}

\line{3. $\cW$-algebras and Casimir algebras \regel 28}

\line{\tab 3.1 $\cW$-algebras: definitions and
               the example of $\cW_3$ \regel 28}

\line{\tab 3.2 Casimir algebras \regel 33}

\line{\tab 3.3 $\cW$-superalgebras;
the example of super-$\cW_3$ \regel 35}

\line{4. $\cW$-algebras and RCFT \regel 38}

\line{\tab 4.1 The chiral algebra in RCFT's \regel 38}

\line{\tab 4.2 Examples: $\cW_3$ and super-$\cW_3$ minimal
models \regel 40}

\line{5. Classification through direct construction \regel 44}

\line{\tab 5.1. The method \regel 44}

\line{\tab 5.2. Overview of results \regel 45}

\line{\tabtab 5.2.1. Generic, linear algebras \regel 46}

\line{\tabtab 5.2.2. Generic, non-linear algebras \regel 48}

\line{\tabtab 5.2.3. Exotic algebras \regel 51}

\line{\tab 5.3. Relating various $\cW$-algebras \regel 52}

\line{\tabtab 5.3.1. Factoring out spin-${1 \over 2}$ fermions \regel 52}

\line{\tabtab 5.3.2. Twisted and projected $\cW$-algebras \regel 53}

\line{\tabtab 5.3.3. Relations with parafermion algebras \regel 54}

\line{\tabtab 5.3.4. $\whW_\infty$ as a universal structure \regel 55}

\line{6. Quantum Drinfeld-Sokolov reduction \regel 57}

\line{\tab 6.1. Introduction \regel 57}

\line{\tab 6.2. Lagrange approach:
                constrained WZW and Toda field theories \regel 58}

\line{\tab 6.3. Algebraic approach to DS reduction \regel 60}

\line{\tabtab 6.3.1. $\cW$-algebras from DS reduction \regel 60}

\line{\tabtab 6.3.2. Character technique \regel 68}

\line{\tabtab 6.3.3. Examples \regel 71}

\line{\tab 6.4. Representation theory \regel 79}

\line{\tabtab 6.4.1. The Kac determinant \regel 79}

\line{\tabtab 6.4.2. Completely degenerate representations
            and minimal models \regel 81}

\line{\tabtab 6.4.3. Character formulae \regel 83}

\line{7. Coset constructions \regel 85}

\line{\tab 7.1. Introduction \regel 85}

\line{\tab 7.2. Casimir algebras \regel 87}

\line{\tabtab 7.2.1. Generalities \regel 87}

\line{\tabtab 7.2.2. Character technique \regel 89}

\line{\tabtab 7.2.3. 3rd Order Casimir example \regel 91}

\line{\tab 7.3. $G \times G / G$ coset conformal field theories \regel 95}

\line{\tabtab 7.3.1. Deforming the singlet algebra \regel 95}

\line{\tabtab 7.3.2. Concrete example \regel 97}

\line{\tabtab 7.3.3. Representation theory \regel 99}

\line{\tabtab 7.3.4. The Kac determinant \regel 100}

\line{\tabtab 7.3.5. The limiting $\cW$-algebra of
                     diagonal coset models  \regel 101}

\line{\tab 7.4. Other cosets \regel 103}

\line{\tabtab 7.4.1. $G\times G'/G'$ coset conformal field theories
\regel 103}

\line{\tabtab 7.4.2. Dual coset pairs \regel 104}

\line{8. Further developments \regel 107}

\line{\tab 8.1. $\cW$-gravity \regel 107}

\line{\tab 8.2. $\cW$-symmetry in string theory \regel 111}

\line{9. Appendices \regel 116}

\line{\tab A. Lie algebra conventions \regel 116}

\line{\tab B. $\cW$-algebra nomenclature \regel 118}

\line{References \regel 120}

\vfil\eject

% ch12345.tex, last updated 12/22/92
\secno=0
%-----------------------------------------------------------------------

\newsec{Introduction}

\subsec{Extensions of conformal symmetry}

Conformal invariance in two dimensions is a spectacularly powerful
symmetry. Two-dimensional quantum field theories that possess conformal
symmetry, which are called conformal field theories, can be solved exactly
by exploiting the conformal symmetry. This fact, and the circumstance
that conformal field theories have found remarkable applications in
string theory (see \cite{\GSW}) and in the study of critical phenomena in
statistical mechanics (see \cite{\ISZ} for a collection of reprints),
has resulted in a large-scale study of conformal field theories in
recent years.

{}From the mathematical point of view,
the main reason why conformal symmetry
is so powerful, is the fact that the corresponding symmetry algebra,
which is the product of two copies of the Virasoro algebra, is
infinite-dimensional. In a quantum field theory the conformal symmetry
gives rise to Ward identities that interrelate various correlation
functions. In certain special theories (so-called minimal models)
these relations take
the form of differential equations whose solutions provide an explicit
solution of the theory \cite{\BPZ}.

In string theory, conformal symmetry arises as a remnant of the
reparametrization invariance of the string world-sheet, which guarantees
that the physics of a string theory does not depend on the choice of
coordinates on the world-sheet. Due to this, the conformal symmetry
in a string theory is `gauged', which implies that the physical states
satisfy a number of constraints, the so-called Virasoro constraints,
which can be compared to the Gauss-law in quantum electro-dynamics.
In the modern formalism these constraints are implemented in a BRST
quantization procedure. Consistency of the quantization requires
that the conformal invariance is not affected at the quantum level,
which leads to important conditions on the space-time backgrounds
in which a quantum string can propagate. Through this mechanism, string
theory makes contact with general relativity, thus raising the hope
that quantum strings may teach us about a consistent theory of quantum
gravity.

The applications of conformal field theory to statistical mechanics
have led to exact results for critical exponents and finite-size
corrections for statistical systems (two-dimensional classical lattice
models or one-dimensional quantum chains) at a second order phase
transition point. The classical example of the Ising model (corresponding
to a conformal field theory of central charge $c=1/2$) has been
generalized in many directions, which has led to a wealth of systems for
which the critical behavior is known exactly.

Since the early days of the massive attention for conformal field theory,
several fields have been explored which are closely related to conformal
field theory, and in which conformal field theory `technology' is heavily
used. For example, the study of so-called perturbed
conformal field theories
has given rise to surprising new results for certain massive integrable
quantum field theories \cite{\Zab}. The study of two-dimensional gravity
(pure or coupled to matter fields) relies heavily on conformal field
theory techniques and the same is true for two-dimensional topological
quantum field theories. These applications provide additional motivation
for a detailed study of the structure of conformal field theories.

\vskip 6mm

In a systematic study of $D=2$ conformal quantum field theory extensions
of the conformal symmetry play an important role. The algebraic structures
that emerge in the study of bosonic extended symmetry are higher-spin
extensions of the Virasoro algebra, which are commonly called $\cW$-algebras.
These algebras, and the associated $\cW$-symmetry in conformal field theory,
are the main topic of this review.

There are two main reasons for studying extended symmetries in conformal
field theory. The first is that certain applications of conformal field
theory (in string theory or statistical mechanics) require some extra
symmetry in addition to conformal invariance. The second reason is that
extended symmetries can be used to facilitate the analysis of a large
class of conformal field theories (called rational conformal field theories)
and, eventually, to classify certain types of conformal field theories.
In the remainder of this introductory section we shall briefly discuss
these two aspects.

We first take a look at the role played by extended symmetries
in applications of conformal field theory. An important example of
this are applications in string theory, where extensions of the worldsheet
conformal symmetry have been very important.
Probably the best known example
for this is the $N=1$ supersymmetric extension of conformal symmetry,
called superconformal symmetry, which promotes strings to superstrings,
thereby improving their properties. Compactified superstrings that are
supersymmetric in space-time require $N=2$ extended
superconformal invariance on the string world-sheet
\cite{\FKSW,\BDFM,\Gec}. Tentative extensions of
string-theory based on extra bosonic symmetry ($\cW$-symmetry) on the
worldsheet have been proposed and are called $\cW$-strings
\cite{\BGe,\DDR,\PRStb}.

Symmetries of lattice models in statistical mechanics are necessarily
finite-dimensional or discrete. However, in the conformal field theory
that describes their scaling limit at criticality such symmetries may
give rise to continuous extensions of the conformal symmetry in the field
theory. As an example we mention the $\ZZ_N$ symmetric lattice model,
$N=2,3,\ldots$, of \cite{\FZa}, whose scaling limit gives rise to a
conformal field theory of central charge $c_N=2(N-1)/(N+2)$.
It has been found that this conformal field theory is invariant under
the so-called $\cW_N$-algebra, which is an extension of the Virasoro
algebra with extra generators of spin $3,4,\ldots,N$. Another example
is the XXX spin-1/2 Heisenberg spin-chain which is invariant under
$A_1=SU(2)$, and for which the associated $c=1$ conformal field theory is
invariant under the semi-direct product of the Virasoro algebra with the
level-1 affine Kac-Moody algebra $A_1^{(1)}$.

In related fields which employ conformal field theory techniques,
extended symmetries are equally important. In perturbed conformal
field theories, the presence of $\cW$-symmetries in the original
conformal field theory may lead to additional integrals of motion in
the perturbed theory \cite{\Zaaa}. For the relation with topological
field theories, $N=2$ extended superconformal symmetry is essential
\cite{\Wiaa,\Wiab,\EY}.

Extended symmetries appear to be particularly important for the coupling
of conformal field theory `matter systems' to two-dimensional gravity. It
has been found \cite{\KPZ} that there is a threshold value $c=1$
for the central charge $c$ of the matter system, above which the coupling
of conformal matter to gravity runs into strong-coupling problems.
These problems can sometimes be cured by replacing gravity
by an appropriate
extension, namely $\cW$-gravity. Classical and
quantum $\cW$-gravity, in particular $\cW_3$ gravity, have
recently been studied by various groups (see Section 8.1).

We finally mention applications of extended conformal algebras that
contain an {\it infinite} number of independent higher spin generators,
such as $w_\infty$ \cite{\BabII} and $\cW_\infty$ \cite{\PRSa,\PRSb}.
Some of these algebras have a clear geometrical meaning, and they play a
role in field theories that are not strictly two-dimensional, such as
membrane theories and the theory of self-dual $D=4$ gravity. They have
also become an important tool in the study of string theory in
two-dimensional target spaces and the associated matrix models.
In this context it is expected that an infinite $\cW$-algebra will
be part of a universal symmetry structure underlying two-dimensional
string field theory.

\vskip 6mm

At this point we come to the second reason for studying extended
symmetries, which is the role they play in a systematic analysis of
so-called rational conformal field theories.
The description of a conformal field theory that is invariant under
an extension of the conformal algebra can be considerably improved
by exploiting the extra symmetry. For example, degeneracies in the
spectrum of conformal dimensions can be resolved by the quantum numbers
that correspond to the additional symmetry. Furthermore, in rational
conformal field theories the presence of extra symmetry makes it possible
to have a finite decomposition of the Hilbert space of physical states
in terms of irreducible representations of the extended algebra (see
Section 4.1 for a discussion). If one does not extend the conformal
algebra, such a finite decomposition can only be made for the minimal
models of central charge $c<1$.

Once the existence of a certain extension of the conformal
algebra has been established, one can try to identify conformal
field theories that realize that symmetry. Such an analysis
will typically involve a study of the representation theory of
the extended algebra and of the properties of their
characters under modular transformations. By exploiting the requirement
of modular invariance of the torus partition function (see Section 2.1),
one can completely determine the possible operator contents of
the conformal
field theory. If the extended algebra contains fermionic currents
(of half-odd-integer spin), the representations that enter the torus
partition function will be subject to Gliozzi-Scherk-Olive (GSO)
projections,
which are familiar from the context of the superstring \cite{\GSW}.

Some of the issues concerning the role of extended symmetries in
conformal field theory have been clarified by the work on the structure
theory of general rational conformal field theories, see {\it e.g.}\
\cite{\Vea,\MSa,\DV,\MSc}.
In the systematic analysis of rational conformal
field theories a central role is played by the so-called chiral algebra,
which is a bosonic extension of the Virasoro algebra.
There exists a precise
characterization \cite{\DV} of all possible rational conformal field
theories with a given chiral algebra.
With that result, the classification of all rational conformal
field theories has formally been reduced to the study of chiral
algebras and of the automorphisms of the associated fusion
rules. However, since the chiral algebras of rational conformal
field theories
are in most cases so-called exotic or non-deformable $\cW$-algebras,
which appear to escape a systematic classification, the practical
applicability of these results to a realistic classification program
seems to be limited.

\vskip 8mm

\subsec{Studying extended symmetries}

Some examples of extended conformal symmetries in string theory
and conformal field theory have been known for a long time. Examples are
the semidirect products of the Virasoro algebra with affine Kac-Moody Lie
algebras \cite{\Kaa,\Moo,\BH,\GO}, which have for example been used to
describe the propagation of strings on group manifolds \cite{\GW}.
Superconformal extensions go back to \cite{\NS,\Ram}; some $N$-extended
superconformal algebras have been known since 1976 \cite{\Ad}. These
algebras all have linear defining relations.

Shortly after the 1984 paper by Belavin, Polyakov and Zamolodchikov
\cite{\BPZ}, it was realized by A.B. Zamolodchikov \cite{\Za} that the
extended symmetries in conformal field theory in general do not give rise
to (super)algebras with {\it linear} defining relations. Algebras of a
more general type are perfectly viable in this context.
A typical feature of
the more general algebras is that operator products (or, equivalently,
(anti)commutators of Laurent modes) are expressed as {\it multilinear}
expressions in the generating currents. This happens in all higher-spin
bosonic extensions of the Virasoro algebra, the $\cW$-algebras, but for
example also in certain extended superconformal algebras
\cite{\Kn,\Bea,\Bowb}. Since non-linear extensions
of the Virasoro algebra had not been studied in
a systematic way in the mathematics literature,
the challenge for physicists has been to understand these algebras and to
employ them for their study of conformal field theory and string theory.

The fact that $\cW$-algebras in general have non-linear
defining relations
puts them outside of the direct scope of Lie algebra theory. However, in
recent years the structure of these algebras has been clarified to
a large extent.
One of the most effective tools in their study is the technique
of so-called
Drinfeld-Sokolov reduction, which relates $\cW$-algebras to Lie
algebras. This reduction also explains the fact that $\cW$-algebras are
related to certain hierarchies of differential equations. In fact, it was
in this context that structures related to $\cW$-algebras made their
first appearance in the literature \cite{\GD}.
The simplest example of this relation is the connection of the
Virasoro algebra with the second hamiltonian
structure of the KdV hierarchy \cite{\Mag,\GN,\Ger,\Ku,\Baa}.
This then generalizes to a connection of the so-called $\cW_N$
algebras to the second hamiltonian structure of the generalized
KdV hierarchies \cite{\Adl,\GD,\Ya,\Maa,\Bab,\Bac}, and
of a certain non-linear infinite $\cW$-algebra, called $\cW_{KP}$, to the
second hamiltonian structure of the KP hierarchy \cite{\GDb,\WYb,\FMR}
(see Section 5.3.4).

\vskip 6mm

In order to organize our discussion in this report, we would like to
distinguish the following three approaches, which have been employed in
the study of extended symmetries in conformal field theory, and
of the corresponding $\cW$-algebras. Although they are often pursued in
parallel, they are of a rather different nature.

\item{1.} The first approach is to try to write down an extended
algebra by proposing a number of extra generators (characterized by
their spins which are usually chosen to be integer or halfinteger)
and closing the algebra. (The form of this algebra is to a large extent
fixed by the Ward identities arising from the conformal symmetry,
which dictate a specific form for the operator product of two primary
currents.) The difficult step is to guarantee that the proposed algebra
will actually be associative (see Section 2.2). Once a consistent algebra
has been found, one can try to find representations and, eventually,
conformal field theory models realizing the symmetry.

\item{2.} Extended conformal algebras can be obtained in a more systematic
way by employing the Drinfeld-Sokolov reduction procedure on the degrees
of freedom of a theory whose structure is based on a Lie algebra (or Lie
superalgebra). This approach is closely related to the study of
$\cW$-symmetries in Toda conformal field theories. When performed at
the classical level, Drinfeld-Sokolov reduction leads to a so-called
classical $\cW$-algebra, which would be expressed in terms of Poisson
brackets rather than commutator brackets (the two are different for
non-linear algebras!). It is also possible to do the Drinfeld-Sokolov
reduction at the quantum level, where it directly leads to a quantum
$\cW$-algebra.

\item{3.} A third approach is to take a known model of conformal field
theory and to see if there are extended symmetries in that model.
This means that one tries to actually construct additional currents
beyond the stress-energy tensor (which corresponds to the Virasoro
algebra) from the fields in the model. In the case of free fields, such
constructions often turn out to be related to the Lie algebra reductions
cited under 2. The most far-reaching construction of this type is the
so-called coset construction, which starts from the degrees of freedom of
a Wess-Zumino-Witten conformal field theory. The extended symmetries
that exist in coset conformal field theories can be studied in a
systematic
fashion. Of course, the resulting extended algebras are associative by
construction; the problem is to identify a complete set of independent
generating currents.

\vskip 6mm

Historically, the approach 1 was first employed by A.B. Zamolodchikov
in the pioneering paper \cite{\Za}. In this paper the $\cW_3$ algebra was
presented, and the phenomenon of `exotic' $\cW$-algebras (which are only
consistent for some isolated values of the central charge) was first
observed. In later papers, the techniques needed for direct
constructions of
extended conformal algebras have been refined and, with some help
of computer power, many more examples have been generated.

The approach 2, which goes back to the work of Gel'fand and Dickey
\cite{\GD} and of Drinfeld and Sokolov \cite{\DS}, was first
worked out in detail by Fateev and Lykyanov \cite{\FLa,\FLb,\FLc},
who in the course of their
work proposed three series of $\cW$-algebras based on the classical Lie
algebras $A_\ell$, $B_\ell$ and $D_\ell$. In this work, the construction
of these quantum $\cW$-algebras was based on the quantization of the
so-called Miura transformation. Later it was realized
\cite{\BOa,\DF,\Fi,\FFra,\FFrb,\Frea,\Freb}
that a more direct construction, called the quantum Drinfeld-Sokolov
reduction, of quantum $\cW$-algebras is possible.

On the level of Lagrange field theory, the theories associated with
Drinfeld-Sokolov reduction are constrained Wess-Zumino-Witten theories.
After implementing the constraints, these reduce to
(classical or quantum) Toda field theories, where the scalar
Toda fields are related to the Cartan
subalgebra of the original Lie algebra. $\cW$-symmetries in Toda field
theories were first studied in \cite{\BGa,\BGb,\BGc,\BGd}. The Lagrange
formulation of the Drinfeld-Sokolov reduction scheme has been
further worked out in \cite{\BFFOWa, \BFFOWb, \BFFOWc}.

The approach 3 to the study of $\cW$-symmetries was first employed
by Bais {\it et al.}\ in \cite{\BBSSa,\BBSSb}.
In these papers, the quantum
$\cW$-algebras based on the Lie algebras $A_\ell$, $D_\ell$
and $E_\ell$ were
independently proposed on the basis of the so-called Casimir construction
for level-1 Wess-Zumino-Witten models for simply laced Lie algebras. The
extension in \cite{\BBSSb} to a coset construction made it
possible to obtain detailed information about the representation
theory and the construction of modular invariant partition functions
for these $\cW$-algebras. In later work a so-called character
technique was developed \cite{\Boc}, which has made it
possible to prove the existence of $\cW$-symmetry in coset
conformal field theories \cite{\Wac} and to determine the spins
of the generating currents of the extended algebra.

\vskip 6mm

In this report we give a comprehensive review of results obtained
in the area of $\cW$-symmetry in conformal field theory. We explain
the basic approaches mentioned above, and discuss how these have
been worked out further in recent years. Where possible we emphasize
the relations that exist between different approaches.

We should stress that our selection of the material presented largely
reflects our personal preferences and own contributions to the field.
We have tried to avoid too much overlap with existing reviews.
Our discussion in Chapters 6 and 7 is somewhat more general than the
existing literature and contains original results.
The style of our presentation is descriptive,
reflecting a compromise between mathematical rigor and readability.
Many of
the fine details are omitted; for those we refer to the original
literature. We provide an extensive list of references, we hope
without too
many omissions. We apologize for leaving out references that would have
deserved to be mentioned but that for some reason escaped our attention.

\vskip 8mm

\subsec{Outline of the paper}

We give a brief outline on how the material in this review is organized.

We start with some preliminaries in Chapter 2. They
include a brief discussion of conformal symmetry and a discussion of some
technicalities concerning operator product expansions and normal
ordered products. We also introduce affine Kac-Moody algebras and
recall some basics of free field and Wess-Zumino-Witten conformal field
theories.

In Chapter 3 we will give a definition of (quantum) $\cW$-algebras.
We will show the example of the $\cW_3$ algebra and then present a
specific
class of $\cW$-algebras, which are the so-called Casimir algebras
for the simply laced algebras $A_\ell$, $D_\ell$ and $E_\ell$.
We will also briefly introduce the notion of $\cW$-superalgebras and
discuss the example of the super-$\cW_3$ algebra.

In Chapter 4 we will show how $\cW$-algebras arise in a natural way
in rational conformal field theories. We will illustrate this
by giving some examples, and by briefly reviewing the structure theory
of rational conformal field theories.

The main body of this paper is presented in the Chapters 5, 6 and
7, and is ordered according to the different approaches to
$\cW$-algebras that we listed in Section 1.2. In Chapter 5 we
discuss $\cW$-algebras that have been obtained through what can be
called `direct construction'. We will discuss the method and
give an extensive list of examples. In Chapter 6 we discuss the
constructions based on Drinfeld-Sokolov reduction from Lie algebra
valued theories. In Chapter 7 we then discuss the coset construction
and show how it leads to detailed information about representation
theory and modular invariants for $\cW$-algebras.

In Chapter 8 we briefly discuss $\cW$-gravity and we come back to
applications of $\cW$-symmetry in string theory.
Appendix A in Chapter 9 contains
our Lie algebra conventions and in Appendix B we
propose a nomenclature for $\cW$-algebras, trying to set a standard
for later works.

\vfill\eject

%----------------------------------------------------------------------

\newsec{Preliminaries}

\subsec{Conformal invariance: basic notions}

In this section we discuss, following \cite{\BPZ}, some of the very
basics of $D=2$ conformal field theory (CFT).
We will not attempt to give a self-contained account, because this alone
could easily fill an issue of this journal. Some additional background
can be found in Section 4.1. We refer to the literature, where
excellent introductions to CFT are available \cite{\Caa,\Cac,\Gib}, see
also the collection of reprints \cite{\ISZ} and references therein.
Some introductory papers with particular attention for $\cW$-algebras
are \cite{\Boc,\Scc,\Go,\BGd,\Rab,\Bic,\Pop,\Sc}.

The main object in a $D=2$ CFT is the stress-energy tensor
$T_{\mu\nu}(\vec{x})$, satisfying the conservation law
\eqn\chBa{
\nabla^{\mu}\ T_{\mu\nu}(\vec{x}) = 0 \ .
}
Because of local scale invariance it also satisfies the trace condition
\eqn\chBb{
T^{\mu}_{\mu} (\vec{x}) =0   \ .
}
It is convenient to choose a conformal gauge
$g_{\mu\nu}(\vec{x}) = \rho (\vec{x})\de_{\mu\nu}$.
(We work in a 2-dimensional Euclidean spacetime; most issues, however,
easily carry over to the Minkowskian domain.)

We introduce complex coordinates
$z=x_1+ix_2, \bar{z}=x_1-ix_2$ ({\it i.e.}\ light-cone coordinates in a
Minkowski space-time and therefore also often referred to as left and
right moving coordinates, respectively). In terms of these
coordinates conformal transformations are just analytic transformations
$z\rightarrow w(z)$, $\bar{z}\rightarrow \bar{w}(\bar{z})$
of the coordinates $z$ and $\bar{z}$.
The stress energy tensor splits into two components
$T\equiv T_{zz}=T_{11}-T_{22}+2iT_{12}$, $\bar{T}\equiv
T_{\bar{z}\bar{z}}= T_{11}-T_{22}-2iT_{12}$ which,  due to the
conservation law \chBa , only depend on $z$ and $\bar{z}$,
respectively. Because both components can be treated on equal
footing we will often restrict the discussion to the left moving
components only.

The short-distance operator product expansion (OPE)
for $T(z)$ can be argued to be
\eqn\chBc{
T(z)T(w) = {c/2 \over (z-w)^4} + {2T(w) \over (z-w)^2}+
{\partial T(w) \over z-w} + \ldots \ .
}
This relation should be understood as an identity which holds
within arbitrary correlation functions. As such it is independent
of the quantization scheme adopted. When using the operator
formalism, where the field $T(z)$ is represented by a field-operator
$T^{op}(z)$, one should keep in mind that the field product $T(z)T(w)$
in \chBc\ is represented by a radially ordered operator
product, which is $T^{op}(z)T^{op}(w)$ if $|z|>|w|$ and
$T^{op}(w)T^{op}(z)$ if $|z|<|w|$. Of course, this radial ordering
is nothing else than the time ordering, which is familiar in
the operator formalism for quantum field theories.
In \chBc\ the c-number $c$ is called the central charge and the dots
stand for the terms regular in the limit $z\rightarrow w$.

Among the fields in the theory there exists a preferred set which
transform as tensors of weight $(h,\bar{h})$ under conformal
transformations $z\rightarrow w(z)$, $\bar{z}\rightarrow \bar{w}(\bar{z})$
\eqn\chBd{
\phi^\prime_{h,\bar{h}}(z,\bar{z}) =\phi_{h,\bar{h}}
(w(z),\bar{w}(\bar{z})) \left( {dw \over dz}\right) ^h \left(
{d\bar{w} \over d\bar{z}}\right) ^{\bar{h}} \ .
}
These are called primary fields of conformal dimension
$(h,\bar{h})$. The property that the stress energy tensor $T(z)$ is
the generator of local scale transformations yields the following
OPE
\eqn\chBe{
T(z)\phi_h(w) = {h \, \phi_h(w) \over (z-w)^2} +
{\partial\phi_h(w) \over z-w}+ \ldots \ .
}
(The dependence on the right-moving coordinate is suppressed here.
We will often do this in the sequel.) The fields in the theory which are
not primary are called secondary or descendant fields. They can be obtained
by taking successive operator products with $T(z)$.

In the operator formalism, it is often convenient to work with
the Laurent modes $L_m$ of the stress-energy tensor $T(z)$, which
are defined by
\eqn\chBg{
T(z)= \sum_{m\in\ZZ} L_m z^{-m-2}, \quad\quad
L_m=\oint_{\cC_0} \frac{dz}{2\pi i}\ z^{m+1} T(z)\ ,
}
where the contour $\cC_0$ surrounds the origin counterclockwise.
The OPE relation \chBc\ then translates into a commutation relation
for the modes $L_m$
\eqn\chBh{
\left[ L_m,L_n\right] = (m-n)L_{m+n} + {c \over 12}
m(m^2-1)\de_{m+n,0}\ .
}
This is the Virasoro algebra. It is the algebra of analytic
transformations of $z$ (generated by $l_m = -z^{m+1} {d \over dz}$),
{\it i.e.}\ the 2-dimensional conformal group, together with a central
extension. The set $\{ L_{-1},L_0,L_1\}$ generates the $sl(2,\RR )$
subalgebra of translations, global scale transformations and special
conformal transformations. The vacuum $|0\rangle$ is a singlet
under this subalgebra.

The OPE \chBe\ for a primary field $\phi_h(z)$ translates into
\eqn\chBi{
[L_m,\phi_h(z)] = (m+1)z^{m}h\, \phi_h(z)+z^{m+1}\partial\phi_h(z)\ .
}
for all integers $m$. At this point it is useful to introduce the notion
of a quasi-primary field, which is defined by the relation \chBi\ for
$m=-1,0,1$ only. This notion is thus weaker than that of a primary field.
Examples of fields that are quasi-primary but not primary are
\eqn\chBiII{
T(z)\ , \qquad \La(z) = (TT)(z) - {3 \over 10} \del^2 T(z)\ , \quad \ldots\,
}
where the operator product $(TT)(z)$ is normal ordered (see Section 2.2).

\vskip 6mm

Let us now say a few words about the operator product algebras of
primary and quasi-primary fields. The operator product of two primary
fields can be decomposed as a linear combination of other primary fields
and their descendant fields
\eqn\chBeII{
\eqalign{
& \phi_n(z,\zb) \phi_m(0,0) = \cr
& \quad \sum_p \sum_{\{k\}} \sum_{\{\bar{k}\}}
C^{p;\{k\},\{\bar{k}\}}_{nm}\,
z^{h_p-h_n-h_m+\sum_i k_i}\,
\zb^{\hbar_p-\hbar_n-\hbar_m+\sum_i \kb_i}\,
\phi_p^{\{k\}\{\kb\}}(0,0) \ . \cr}
}
In this relation the index $p$ runs over the primary fields
that occur on the right hand side. The multi-indices $\{k\}$ and
$\{\kb\}$ label the descendants of the primary fields $\phi_p(z,\zb)$,
which are given by
\eqn\chBeIIa{
\phi_p^{\{k\}\{\kb\}}(z,\zb) =
L_{-k_1} \ldots L_{-k_N} \Lb_{-\kb_1} \ldots \Lb_{-\kb_M}
\phi_p(z,\zb).
}
Conformal invariance implies a factorization of the constants
$C^{p;\{k\},\{\kb\}}_{nm}$ according to
\eqn\chBeIII{
C^{p;\{k\},\{\kb\}}_{nm} = C_{nm}^p
\beta^{p,\{k\}}_{nm} \bar{\beta}^{p,\{\kb\}}_{nm}.
}
The coefficients $\beta$, $\bar{\beta}$ are trivial in the sense
that they can be expressed in terms of the conformal dimensions
$h$, $\hbar$, respectively. It can easily be seen that a three-point
function involving fields in the conformal families of $\phi_n$,
$\phi_m$ and $\phi_p$ can only be nonvanishing if $C_{nm}^p \neq 0$.
The coefficients $C_{nm}^p$ can thus be viewed as three-point vertices
describing the interactions in the theory. Schematically one writes
this OPE as
\eqn\chBf{
[\phi_n] \cdot [\phi_m] = \sum_p \,C_{nm}^p \,[\phi_p] \,,
}
where $[\phi_p]$ denotes the conformal family associated with the
primary field $\phi_p(z,\zb)$.

The descendant field structure in the operator product algebra
of quasi-primary fields is much simpler than it is for the primary
fields: the only descendant fields to be considered are (multiple)
derivatives of the quasi-primary fields. The OPE of two chiral
quasi-primary fields $\phi^i$ and $\phi^j$ of integer conformal
dimensions $h_i$ and $h_j$ takes the general form \cite{\Bow}
\eqn\chBfII{
\phi^i(z) \phi^j(0) = \sum_k C^{ij}{}_k \sum_{n=0}^\infty
{a^{(ijk)}_n \over n!} {\del^n \phi^k(0) \over z^{h_i+h_j-h_k-n}}
+ \gamma^{ij} {1 \over z^{h_i+h_j}}\ ,
}
where $k$ labels the quasi-primary fields occuring in the
right hand side,
$\gamma^{ij}$ plays the role of a metric on the space of quasi-primaries
and the coefficients $a^{(ijk)}_n$ are given by
\eqn\chBfIII{
a^{(ijk)}_n = { (h_i-h_j+h_k)_n \over (2 h_k)_n } \ ,
}
with the notation $(x)_n = \Ga(x+n) / \Ga(x)$.
In later chapters we will use the relation \chBfII\ as a fundamental
building block in the construction of $\cW$-algebras.

\vskip 6mm

Let us now focus on the Hilbert space of physical states of a CFT.
Conformal invariance implies that these states
assemble into representations of the
Virasoro algebra. The relevant representations are those for which
the (left) Hamiltonian $L_0$ is bounded from below. They are, by
convention, called highest weight modules (HWM's). The highest weight
vector $|h,c\rangle$, {\it i.e.}\ the state with the lowest $L_0$-eigenvalue,
is characterized by the properties
\eqn\chBj{
L_0 |h,c\rangle  =  h \, |h,c\rangle \ , \quad
L_n |h,c\rangle  =  0 \ , \quad n>0\ .
}
Before we further describe the structure of the HWM's, we mention
that there exists a 1--1 correspondence between states $|\phi \rangle$
in the Hilbert space and fields $\phi (z,\bar{z})$,
called vertex operators for the state $|\phi \rangle$. The correspondence
is given by
\eqn\chBjII{
|\phi \rangle = \lim_{z,\bar{z}\rightarrow 0}\  \phi (z,\bar{z}) |0\rangle\ .
}
For a primary field $\phi_{h,\bar{h}}(z,\bar{z})$ one easily shows,
using \chBi , that the associated state $|\phi \rangle$ defined by
\chBjII\ satisfies the conditions \chBj\ of a highest
weight vector $|h,c\rangle_{\rm left}\times |\bar{h},c\rangle_{\rm right}$.
So the primary fields in the theory are in 1-1 correspondence
with the highest weight vectors.

The module consisting of (finite) linear combinations of the states
\eqn\chBk{
L_{-k_1}L_{-k_2}\ldots L_{-k_m} |h,c\rangle\ ,\qquad k_i>0\ ,
}
is called the Verma module $M(h,c)$. The Verma module $M(h,c)$
admits an $L_0$-eigenspace decomposition
\eqn\chBl{
M(h,c)= \bigoplus_{N\geq 0}\ M(h,c)_{(N)}\ ,
}
where
\eqn\chBlII{
M(h,c)_{(N)} = \left\{ v\in M(h,c) | L_0v = (h+N)v \right\}\ .
}
A basis for the eigenspace $M(h,c)_{(N)}$ is given by the states
\eqn\chBm{
L_{-k_1}\ldots L_{-k_m} |h,c\rangle\ ,\qquad \sum_{i=1}^{m}k_i=N\ ,
\quad \quad k_1\geq k_2\geq\ldots k_m > 0\ .
}
The dimension of the eigenspace $M(h,c)_{(N)}$ is given by `Euler's
partition function' $p(N)$, {\it i.e.}\ the number of ways of partitioning
$N$ into a set of positive integers.

The hermiticity conditions
\eqn\chBn{
L_n^{\ \dagger} = L_{-n}\ , \quad \quad n\in\ZZ\ ,
}
which follow from the self-adjointness of $T(z)$, together with the
normalization
$\langle h,c|h,c\rangle=1$, uniquely define a symmetric bilinear form
$\vev{\ |\ }$ on the Verma module $M(h,c)$. It is easily seen that the
eigenspace decomposition \chBl\ is orthogonal with respect
to this bilinear form. Let $\cM (h,c)_{(N)}$ be the $p(N)\times p(N)$
matrix of inner products of a set of basis vectors of $M(h,c)_{(N)}$.
The determinant of this matrix, which is independent of the choice of
basis upto a multiplicative constant, is called the Kac determinant.
It is given by \cite{\Kac,\FFa}
\eqn\chBo{
{\rm det}\ \cM (h,c)_{(N)} = \prod_{k=1}^{N}\ \prod_{rs=k} \left(
h-h(r,s) \right) ^{p(N-k)} \ ,
}
where $r,s \in \ZZ_{>0}$ and
\eqn\chBp{
h(r,s) = {1 \over 48}\left( (13-c)(r^2+s^2)-24rs-2(1-c)+
\sqrt{(1-c)(25-c)}(r^2-s^2) \right)\ .
}

In general, the Verma module $M(h,c)$ is not irreducible, {\it i.e.}\
it contains invariant subspaces. It is easily seen that the radical of
$\vev{\ |\ }$, consisting of the so-called null-states $v\in M(h,c)$ which
are orthogonal to every state $w\in M(h,c)$, is such an invariant
subspace. One can prove that Rad$(\vev{\ |\ })$ is the unique maximal ideal
in $M(h,c)$, implying that the coset vectorspace
$L(h,c)\equiv M(h,c)/{\rm Rad}(\vev{\ |\ })$ is an irreducible HWM.
In physical terms: states which are orthogonal to every other state
decouple from all the correlation functions and can therefore be omitted
altogether. Physical spectra consist therefore of irreducible HWM's $L(h,c)$.

It is clear that states in Rad$(\vev{\ |\ })$
are in 1-1 correspondence with zeros of the Kac determinant. By
analyzing the Kac determinant one can therefore determine which
Verma modules are reducible (also called degenerate).
In particular it is easily seen that for central charges $c>1$ and $h>0$
the Verma module is irreducible.

Another important issue is unitarity.
A HWM $L(h,c)$ is called unitary if all the states have positive norm.
The question which irreducible HWM's $L(h,c)$ are unitary can also be
analyzed by means of the Kac determinant \chBo .
It has been shown  \cite{\FQSa,\FQSb,\FQSc,\GKOa,\GKOb}
that the requirement of unitarity restricts the possible $h$ and $c$
values of an irreducible HWM $L(h,c)$ to either
\eqn\chBq{
c\geq 1\ ,\qquad h\geq 0\ ,
}
or
\eqn\chBr{
c=c(m)= 1-\frac{6}{m(m+1)}\ ,\qquad m=2,3,4,\ldots\ ,
}
in which case there are only a finite number of allowed
$h$-values given by
\eqn\chBs{
h=h^{(m)}(r,s) = {( (m+1)r-ms)^2-1 \over 4m(m+1)}\ , \quad
1\leq r\leq m-1, \quad 1\leq s \leq m\ .
}
Information on the multiplicity of states in a HWM $V$ is contained in
the character $\ch_V$ of $V$. This is the holomorphic function on the
complex upper half plane $\{ \ta\in\CC\ ,{\rm Im}(\ta) >0\}$, defined by
\eqn\chBt{
\ch_V (\ta) = {\rm Tr}_V \left( q^{L_0-{c \over 24}} \right)\ ,
\qquad  q={\rm e}^{2\pi i\ta} \ .
}

{}From the discussion above, the character of the Verma module $M(h,c)$ is
found to be
\eqn\chBu{
\ch_{M(h,c)}(\ta ) = q^{h-\frac{c}{24}} \sum_{N\geq 0} p(N)q^N
= \frac{q^{h-\frac{c}{24}}}{\prod_{n\geq 1}(1-q^n)} \ .
}
Explicit expressions for the characters of the irreducible HWM's $L(h,c)$
can be found in \cite{\RC}.

Our last piece of introduction in this section concerns CFT's
defined on a two-dimensional torus, which can be characterized by its
modular parameter $\tau$. It has proved interesting to study the dependence
of various quantities in the CFT on $\tau$. One such quantity is the
partition function, which is formally defined as
\eqn\chBv{
Z(q) = {\rm Tr}
\left( q^{-\frac{c}{24}+L_0} \bar{q}^{-\frac{c}{24}+\bar{L}_0} \right),
}
where $q=e^{2\pi i\tau}$. Conformal invariance implies that the Hilbert
space of the CFT splits as a sum of representations (irreducible HWM's) of
the conformal algebra. Accordingly, the torus partition function has the
following form
\eqn\chBw{
Z(q) = \sum_{h,\hbar} \, \chi_h(q) {\cal N}_{h\hbar} \,
\chi_{\hbar}(\bar{q}),
}
where $\chi_h(q)$ is the character for the irreducible representation of
the Virasoro algebra with highest weight $h$ as in \chBt. The coeficients
${\cal N}_{h\hbar}$ are all integers and ${\cal N}_{00}=1$.
Modular invariance,
which expresses the fact that different values for $\tau$ can give rise
to the same torus, leads to the statement that the partition sum $Z$ is
invariant under the following transformations
\eqn\chBx{
T: \tau \rightarrow \tau+1 \ , \qquad
S: \tau \rightarrow -\frac{1}{\tau}  \quad \,.
}
The modular invariant torus partition functions for the minimal series
\chBr\ of central charges $c$ have been classified in
\cite{\Gea,\CIZa,\CIZb,\Kat}
(this is the so-called ADE classification). With this result, all unitary
models of CFT with $c<1$ are explicitly known.

At this point we stop our discussion of CFT basics. We have put some
emphasis on the algebraic aspects of CFT's and on the representation
theory of the Virasoro algebra, since these will be generalized by the
introduction of $\cW$-symmetry.

\vskip 8mm

\subsec{OPE's, normal ordered products and associativity}

As a preparation for our discussion of $\cW$-symmetry in later
chapters, we will now discuss various aspects of a special subset
of fields in a CFT, which are the chiral fields of integer conformal
dimension, which we will call `currents'.
(Whenever needed, a `graded' extension to a situation
including chiral fields of half-integer dimension can easily be made.)
For such fields the notions
conformal dimension and conformal spin are the same, and we will be using
both terms. It is useful to realize that the complete set of currents
can be split into quasi-primary currents and currents that are (multiple)
derivatives thereof.

The OPE of any two currents $A(z)$ and $B(w)$ can be written as
\eqn\chBaa{
A(z)B(w) = B(w)A(z) = \sum_{r=r_0}^\infty \{ AB \}_r(w) (z-w)^r \ ,
}
where $r$ runs over integer values. In general, $r_0$ is negative so that
the OPE contains a finite number of singular terms. The singular OPE or
{\it contraction}, which will be denoted by a hook, is given by
\eqn\chBab{
A\underhook{(z)B}(w) = \sum_{r<0} \{ AB \}_r(w) (z-w)^r \,.
}
The OPE $B(z)A(w)$ is determined by \chBaa\ through
a formal Taylor expansion.

Following \cite{\BBSSa}, we define the normal ordered field product,
to be denoted by $(AB)(z)$, of $A(z)$ and $B(z)$ to be the constant
term in the OPE \chBaa
\eqn\chBac{
(AB)(z) \equiv \{ AB \}_0(z)
= \frac{1}{2 \pi i} \oint_{\cC_z} \frac{dw}{w-z} A(w)B(z) \ ,
}
where the contour $\cC_z$ encloses the point $z$.
The normal ordered field commutator is defined as
$[A,B](z) \equiv (AB)(z)-(BA)(z)$. It can be expressed
in the fields occuring in the singular OPE according to
\eqn\chBad{
[A,B](z) = \sum_{r<0} (-1)^{r+1} {1 \over r!} \partial^r \{ AB \}_r(z) \ .
}

There exists a simple calculus, which allows one to compute with
contractions and normal ordered products of composite fields.
In particular, there is the Wick theorem for the contraction of $A(z)$
with the composite field $(BC)(w)$
\eqn\chBae{
A\underhook{(z)(B}C)(w) = \frac{1}{2 \pi i} \oint_{\cC_w} \frac{dx}{x-w}
\{ A\underhook{(z)B}(x)C(w) + B(x)A\underhook{(z)C}(w) \} \ .
}

We remark that the normal ordered product as defined in \chBac\ is
neither commutative (this we saw in \chBad) nor associative. Using
\eqn\chBaf{
(A(BC))(z) - (B(AC))(z)  = ([A,B]C)(z)
}
one can compute the associator $(A(BC))(z)-((AB)C)(z)$ and find it
to be nonvanishing in general. (Let us stress that this fact is not
at all in conflict with the property that the full operator product
algebra (OPA) is required to be associative.) More complicated
rearrangement lemmas, for example for the difference between
$((AB)(CD))(z)$ and $(A(B(CD)))(z)$,
have been given in appendix A of \cite{\BBSSa}. The rules for calculating
operator product expansions and rearranging normal ordered expressions
have been implemented in a Mathematica package \cite{\Thi}.

We define the modes $A_m$ of a current $A(z)$ of conformal
dimension $h_A$ by
\eqn\chBag{
A_m = \frac{1}{2 \pi i} \oint_{\cC_0} dz \, A(z) z^{m+h_A-1}, \quad
A(z) = \sum_{m \in \ZZ} A_m z^{-m-h_A}
}
The contraction $A(z)B(w)$ can then be translated into a commutation
relation $ [A_m,B_n] $ for the modes $A_m$ and $B_n$. In particular,
the OPE \chBe\ translates into
\eqn\chBagII{
[ L_m , A_n ] = ((h_A-1)m - n) A_{m+n} \ .
}

An important property is the {\it associativity}\ of the commutator
algebra of the modes of all currents in a CFT.
It is expressed by the Jacobi identity,
\eqn\chBah{
[ A_m, [ B_n, C_r]] + [ C_r, [ A_m, B_n]] +[ B_n, [ C_r, A_m]] = 0 \,.
}

At the level of currents, the property of associativity is most
easily expressed by considering four-point correlation functions.
If one denotes the currents by $A^I=A^i(z_i)$ and formally writes the
OPE's as $A^I A^J = \sum_K C^{IJ}{}_K A^K$, then the associativity
condition can be written as
\eqn\chBai{
\sum_M C^{IJ}{}_M C^{MK}{}_L = \sum_M C^{IM}{}_L C^{JK}{}_M
}
As explained in App B. of \cite{\BPZ}, this property precisely
corresponds to the so-called crossing symmetry of the 4-point
functions involving the currents $A^I$, $A^J$, $A^K$ and $A^L$.
Crossing symmetry simply means that the four-point function
$ \langle A^i(z_i)A^j(z_j)A^k(z_k)A^l(z_l) \rangle $ is
invariant under all permutations of the labels $i,j,k$ and $l$.
We can easily work out this condition in the case where all
four currents are quasi-primary: using the $sl(2,\RR)$ invariance,
we can then write
\eqn\chBaj{
\eqalign{
& \langle A^i(z_i) A^j(z_j) A^k(z_k) A^l(z_l) \rangle =
\cr
& G \left( \matrix{ j & i \cr k & l \cr} \right)(x) \,
{ (z_j-z_l)^{-h_j-h_k-h_l+h_i} (z_i-z_j)^{-h_i-h_j+h_k+h_l}
\over (z_i-z_k)^{2\, h_k} (z_i-z_l)^{-h_i-h_l+h_j+h_k} } \ ,
\cr}
}
where
\eqn\chBak{
G \left( \matrix{ j & i \cr k & l \cr} \right)(x) \,
= \langle i | A^j(1) A^k(x) | l \rangle , \quad
x = {(z_i-z_j)(z_k-z_l) \over (z_i-z_k)(z_j-z_l)} \ ,
}
and we defined (compare with \chBjII)
\eqn\chBaka{
| i \rangle  = \lim_{z_i \rightarrow 0} \, A^i(z_i) | 0 \rangle \ ,
\qquad
\langle i | = \lim_{z_i \rightarrow 0} \, z_i^{-2 h_i}
\langle 0 | A^i(1/z_i)\ .
}
[The expressions \chBak\ and \chBaka\ refer to the operator formalism
and define the 4-point-function for $|x|<1$. The result for $|x|\geq 1$
is obtained by analytic continuation.] The crossing symmetry conditions
that correspond to a change of channels are given by
\eqn\chBal{
\eqalign{
G \left( \matrix{ j & i \cr k & l \cr} \right) (x) & =
(-1)^{h_i+h_j+h_k+h_l} \, x^{-2 h_k}
G \left( \matrix{ j & l \cr k & i \cr} \right) ({1 \over x})
\cr
G \left( \matrix{ j & i \cr k & l \cr} \right) (x) & =
(-1)^{h_i+h_j+h_k+h_l}
G \left( \matrix{ l & i \cr k & j \cr} \right) (1-x) \ .
\cr }}

The connection between the characterization of associativity via
the Jacobi identity for modes and via the condition for crossing
symmetry of four-point functions of currents has been discussed in
\cite{\Bow}. The two are believed to be equivalent, but a direct
proof of this has (to our knowledge) not been given.

There exists a third way to characterize associativity, which was first
discussed in \cite{\Bob}. It uses the following property of the normal
ordered field commutators which we introduced above
\eqn\chBam{
[A,[B,C]](z) + [C,[A,B]](z) + [B,[C,A]](z) = 0 \, .
}
(This property holds despite the fact that the normal ordered
product by itself is not associative!) In many examples \cite{\Bob,\ASS}
it has been found that this condition, which is certainly a
necessary condition for associativity of the OPA, allows one to quickly
derive associativity constraints on the central charge of a
proposed extended algebra. We expect that the condition, when
imposed for all currents in the algebra, is also a sufficient
condition for associativity, but this has not been proven.

\vskip 8mm

\subsec{Auxiliary field theories}

\noindent {\sl 2.3.1. Free fields}

\vskip 3mm

In this section we briefly discuss the conformal field theory of a
single free massless scalar field $\phi (z,\zb)$ and the Feigin-Fuchs
construction. We shall also indicate two alternative free field
constructions for the affine Kac-Moody algebra $A_1^{(1)}$, both of which
can be generalized to other algebras. (See Section 2.3.2 for the
definition of affine Kac-Moody algebras.)

We first consider a single free scalar field $\phi (z,\zb)$ which
is compactified on a circle $\RR /2\pi R$ of radius $R$.
The action of this field theory is given by
\eqn\chBba{
S[\phi] = -\frac{1}{2} \int d^2 z \,
\partial^{\al}\phi\partial_{\al}\phi
}
and the corresponding equation of motion is the 2-dimensional Laplace
equation. The most general solution is given by
\eqn\chBbb{
\phi(z,\bar{z}) = \vph(z) + \bar{\vph}(\bar{z})\ ,
}
where $\vph$ and $\bar{\vph}$ are single-valued functions on the complex
plane. In terms of a mode expansion we have
\eqn\chBbc{
i\partial\vph = \sum_{n\in \ZZ}\ \al_n z^{-n-1}\ ,
}
and an analogous expression for $\overline{\partial\vph}(\bar{z})$.
Canonical quantization gives the following commutation relations
\eqn\chBbd{
\left[ \al_m,\al_n \right]  =  m \, \de_{m+n,0}\ ,
}
which are equivalent to the contraction
\eqn\chBbe{
\partial\vph \underhook{(z)\partial\vph}(w) = {-1 \over (z-w)^2} \ .
}
This contraction, or equivalently the commutation relations
\chBbd, defines the $U(1)$ affine Kac-Moody algebra.

The stress energy tensor is given by
\eqn\chBbf{
T(z) = - {1 \over 2} (\partial\vph\partial\vph)(z)\ ,
}
with the parentheses denoting normal ordering as in Section 2.2.
In terms of modes we have in particular
\eqn\chBbg{
L_0 = {1 \over 2} \al_0^2 + \sum_{n\geq 1}\ \al_{-n}\al_n\ .
}
By invoking the Wick theorem, we can compute the OPA of $T(z)$,
which is precisely given by the result \chBc\ with $c=1$.

The spectrum of the scalar field model is constructed as a Verma module
over a highest weight vector $|\la,\bar{\la}\rangle$. Explicitly
\eqn\chBbh{
\eqalign{
\al_0 |\la,\bar{\la}\rangle = \la |\la,\bar{\la}\rangle\ ,\qquad
\bar{\al}_0 |\la,\bar{\la}\rangle = \bar{\la} |\la,\bar{\la}\rangle
\cr
\al_n |\la,\bar{\la}\rangle = \bar{\al}_n |\la,\bar{\la}\rangle =0\ ,
\quad n>0\ , \cr}
}
and a basis is given by the states
\eqn\chBbi{
\eqalign{
\al_{-k_1}\ldots\al_{-k_m}\bar{\al}_{-l_1}\ldots\bar{\al}_{-l_n}
|\la,\bar{\la}\rangle\ ,\qquad & k_1 \geq k_2 \geq \ldots >0 \ , 
\cr
& l_1 \geq l_2 \geq \ldots >0 \ .  \cr}
}
{}From \chBbg\ we find that the highest weight vector
$|\la,\bar{\la}\rangle$ has conformal dimension $(h,\bar{h})
= ({1 \over 2} \la^2,{1 \over 2} \bar{\la}^2)$.

To determine the set $(\la,\bar{\la})$ that occur in the theory one
observes that, by  using \chBjII, the highest weight
vectors $|\la,\bar{\la}\rangle$ are obtained, from the vertex operators
\eqn\chBbj{
V_{\la,\bar{\la}}(z,\bar{z}) = V_{\la}(z) V_{\bar{\la}}(\bar{z})\ ,
}
where
\eqn\chBbk{
V_{\la}(z) = ({\rm e}^{i\la\vph}) (z)\ ,\qquad
V_{\bar{\la}}(\bar{z}) = ( {\rm e}^{i\bar{\la}\bar{\vph}})
(\bar{z})\ .
}
Locality requires that the spin $h-\bar{h}$ is an integer. Together with
the invariance under $\phi\rightarrow\phi + 2\pi R$ this restricts the
momenta $(\la,\bar{\la})$ to be on the lattice
\eqn\chBbl{
\bGa = \left\{ (\la,\bar{\la}) = (\frac{n}{R} + {1 \over 2}mR,
\frac{n}{R} - {1 \over 2}mR )\ |\ n,m\in\ZZ \right\}\ .
}
The partition function is now easily computed
\eqn\chBblI{
Z = { \sum_{(\la,\bar{\la}) \in \bGa} q^{\frac{1}{2}\la^2}
\bar{q}^{\frac{1}{2}\bar{\la}^2} \over \left| q^{\frac{1}{24}}
\prod_{n\geq 1}(1-q^n)\right| ^2}\ .
}

A special situation occurs for $R=\sqrt{2}$, where there are three
fields of dimension $(1,0)$ namely
\eqn\chBbm{
J^-(z) = \left( {\rm e}^{-i\sqrt{2}\vph} \right)(z) \ , \quad
J^0(z) = i \partial\vph (z)\ , \quad {\rm and} \quad
J^+(z) = \left( {\rm e}^{i\sqrt{2}\vph} \right)(z) \ .
}
The OPA of these three currents closes; this is the so-called vertex
operator construction for the affine Kac-Moody algebra $A_1^{(1)}$
at level 1 \cite{\Halb,\BHN}.

\vskip 6mm

Let us now briefly discuss a small modification of the scalar field
model, known as the Feigin-Fuchs construction \cite{\FFa}.
This construction
is most useful in the study of minimal models, both of the Virasoro
algebra, and, as we shall see later, of minimal models of
$\cW$-algebras. The modification consists of putting a background charge
$Q$ at $z=\infty$. This does not affect the OPE \chBbe, but it does
modify the stress energy tensor to
\eqn\chBbn{
T(z) = -{1 \over 2} (\partial\vph\partial\vph )(z)
+ iQ\partial^2\vph (z)\ .
}
The modified stress energy tensor \chBbn\ again satisfies the
OPE \chBc, but now with a central charge
\eqn\chBbo{
c=1-12 Q^2 \ .
}
The effect of the background charge is thus to screen the central charge
of the original system. Another consequence of the background charge is
that the vertex operators
\eqn\chBbp{
V_{\la}(z) = \left( {\rm e}^{i\la\vph} \right) (z)\ ,
}
now become primary fields of conformal dimension
\eqn\chBbq{
h(\la ) = {1 \over 2} \la (\la - 2Q)\ .
}
Using this modification it is possible to give an explicit construction
of null-states, {\it i.e.}\ the zero's of the Kac determinant, and thus to
prove \chBo. The application of the Feigin-Fuchs construction to minimal
models has been further worked out by Felder in \cite{\Fe}.
In our discussion of $\cW$-algebras in Chapters 6 and 7, we will be
using multi-component extensions of the Feigin-Fuchs construction.

\vskip 6mm

In order to define an alternative free field realization of $A_1^{(1)}$,
we introduce bosonic first order fields $\beta(z)$ and $\gamma(z)$ through
the contraction
\eqn\chBbr{
\ga \underhook{ (z) \be} (w) = {1 \over z-w} \ .
}
One easily checks that the currents
\eqn\chBbs{
\eqalign{
J^-(z) & = -(\ga\ga\be)(z) - \sqrt{2(k+2)}\,
(\ga i\del\vph)(z) - k \del \ga(z) \cr
J^0(z) & = 2(\ga\beta)(z) + \sqrt{2(k+2)}\, i\del\vph(z) \cr
J^+(z) & = \beta(z)
\cr}
}
satisfy the OPA of the level $k$ affine Kac-Moody algebra $A_1^{(1)}$
\cite{\Wak}. The similar construction for more general Lie algebras
\cite{\FFrc,\BMPa,\BMPb} will be used in Chapter 6.

\vskip 8mm

\noindent {\sl 2.3.2. Affine Kac-Moody algebras and WZW models}

\vskip 3mm

The theory of affine Kac-Moody algebras (or AKM algebras in brief)
has played a crucial role in the mathematical analysis of various
CFT's and $\cW$-algebras. It is actually possible to
use representation spaces of AKM algebras to construct a Hilbert
space for a conformal field theory. The symmetry currents in the
CFT (such as the stress energy tensor) are then constructed from
currents taking values in the underlying finite-dimensional Lie
algebra.
Examples for this are the Sugawara and coset constructions and
generalizations that one obtains by solving the Virasoro master
equation \cite{\Hal}. In the unitary case,
these constructions all start from
integrable representations of AKM algebras, which occur at
integer level $k$. In the quantum Drinfeld-Sokolov reduction scheme,
which we discuss in Chapter 6, representations of AKM algebras of
fractional level form the starting point.

Before we say more about the applications, we introduce current
fields $J(z)$ and $\bar{J}(\bar{z})$ (satisfying $\partial_{\bar{z}}
J=\partial_z \bar{J}=0$), that take values in a finite-dimensional
Lie algebra $\bfg$. (We will mostly restrict ourselves to the left
moving components $J(z)$ only.) Choosing a set of anti-hermitean
generators $\{ T_a, a=1,\ldots,{\rm dim}(\bfg)\}$
\foot{In later chapters we will mainly use the Chevalley basis
$\{ e^{\al}, e^{-\al}, h^i\} $.},
\eqn\chBca{
\left[ T_a, T_b \right] = f_{ab}^{\ \ c} T_c   \ ,
}
the components $J^a(z)$ satisfy the OPE
\eqn\chBcb{
J^a(z) J^b(w) = {k\, d^{ab} \over (z-w)^2} + f^{ab}_{\ \ c}
{J^c(w) \over z-w} + \ldots \ .
}
The Cartan-Killing metric
\eqn\chBcc{
d_{ab} = {\rm Tr}(T_aT_b)
}
is used to raise and lower indices.%
\foot{Our normalizations are chosen such that $|\al |^2=2$ for a long
root $\al$ of $\bfg$. The trace is taken over the fundamental
representation of $\bfg$.}
The c-number $k$ is called the central charge or level.
The following commutation relations for the Laurent modes $J_m^a$,
$m\in\ZZ$, of $J^a(z)=\sum_{m\in\ZZ} J_m^a\ z^{-m-1}$ are equivalent
to the OPE \chBcb\
\eqn\chBcd{
\left[ J_m^a, J_n^b \right] = f^{ab}_{\ \ c} J_{m+n}^c +
km d^{ab}\de_{m+n,0}\ .
}
The algebra \chBcd, which is denoted by $\whg$ or $\bfg^{(1)}$,
is called an (untwisted) affine Kac-Moody (AKM) algebra,
or affine Lie algebra in brief. In the
mathematics literature these algebras were first discussed in
\cite{\Kaa,\Moo}, in the physics literature they made their first
appearance in \cite{\BH}; see \cite{\Kad,\KP} for their mathematical
aspects and \cite{\GO} for an introduction from a physicist's point
of view. Notice that the zero modes $J_0^a$ satisfy the commutation
relations of the underlying finite dimensional Lie algebra $\bfg$.

The most interesting representations of an AKM algebra \chBcd\ are the
irreducible highest weight modules (HWM's), which are characterized by
a highest weight $\hat\La$. The projection $\La$ of the weight
$\hat\La$ onto the weight lattice of the underlying Lie algebra $\bfg$
characterizes the $\bfg$-representation of the highest weight state.
A special class of these highest weight representations,
which occur for integer level $k$, are the so-called integrable HWM's
for which $\La$ is an integral dominant weight.
For given integer level $k$ only a finite number of integrable HWM's
$L(\La)$ exist. For fractional levels the more general notion
of `modular invariant representations' was introduced in \cite{\KaWa}.

\vskip 6mm

The most direct application of AKM current algebra to CFT is through
the so-called Sugawara construction, which gives the following
expresion for a CFT stress-energy tensor $T(z)$ in terms of AKM
currents $J^a(z)$
\eqn\chBce{
T(z) = {1 \over 2(k+ \htil)} d_{ab} (J^aJ^b)(z)\ ,
}
The c-number $\htil$ is the dual Coxeter number of $\bfg$ \cite{\Kad}.
Using \chBcb\ one can verify that $T(z)$ satisfies the Virasoro OPE
\chBc, with central charge
\eqn\chBcf{
c=c(\whg,k) = {k({\rm dim}\ \bfg) \over k+\htil}   \ .
}
To each highest weight $\La$ of a HWM $L(\La )$ of the
AKM algebra one can now associate a Virasoro primary field
$\phi_{\La}$, which has conformal dimension
\eqn\chBcg{
h_{\La} = {c_\La \over 2(k+ \htil)}    \ ,
}
where $c_{{\La}}= (\La,\La + 2\rh)$
is the eigenvalue of the second order Casimir
in the $\bfg$-representation characterized by the highest weight
$\La$.

The CFT's that correspond to the form \chBce\ of the stress-energy
tensor are the so-called Wess-Zumino-Witten (WZW) conformal field
theories \cite{\Wia,\WZ}. A WZW model is a nonlinear sigma model on a
compact group manifold $G$, which contains a topological term, the
Wess-Zumino term \cite{\WZ}, in its action.
The presence of this Wess-Zumino
term requires the level $k$ to be an integer. The model is conformally
invariant provided the relative coefficient of the Wess-Zumino
term is chosen in a specific way \cite{\Wia}. The chiral
stress-energy tensor precisely takes the form \chBce, where $\bfg$ is
now the Lie algebra of the group $G$.

The spectrum of the WZW models was determined by Gepner and Witten
\cite{\GW}, who found that only integrable HWM's can occur in
the spectrum. Knizhnik and Zamolodchikov \cite{\KZ}
showed how to apply the techniques of
\cite{\BPZ} to the WZW model, and derived differential equations for the
correlation functions. These so-called
Knizhnik-Zamolodchi\-kov equations in principle solve the WZW model.

WZW models defined on a two-dimensional torus are characterized by
their modular invariant partition function. For the case when the
underlying Lie algebra is $A_1^{(1)}$ all possible modular invariants
have been classified in \cite{\Gea,\CIZa,\CIZb,\Kat}. For higher rank
algebras such complete results have, to our knowledge, not been
obtained.

\vskip 6mm

One can thus construct HWM's of the Virasoro algebra from HWM's of
AKM algebras by means of the Sugawara construction. Clearly, the
scope of this construction is limited since it assumes the presence of
an affine Kac-Moody symmetry in the CFT. There exist, however, more
general Virasoro constructions. One of those is the so-called
Goddard-Kent-Olive coset construction \cite{\GKOa,\GKOb},
which associates a Virasoro algebra to a coset pair $(\whg,\whg')$,
$\whg'\subset\whg$ of AKM algebras%
\foot{The central charge $k'$ of the AKM subalgebra $\whg'$ is
determined by $k'=jk$, where $j$ is the Dynkin
index of the embedding $\bfg'\subset\bfg$.}.
Examples of the coset construction were already given in
\cite{\BH,\Halc}. The coset construction starts from the
Sugawara tensors $T(z)$ and $T'(z)$
associated to $\whg$ and $\whg'$, which generate Virasoro algebras of
central charges $c(\whg,k)$ and $c(\whg',k')$, respectively. One then
constructs the operator
\eqn\chBch{
\widetilde{T}(z) = T(z) - T'(z)\ ,
}
which generates a so-called `coset' Virasoro algebra, of central charge
$c(\whg,\whg',k)$ given by%
\foot{A special situation arises when $c(\whg,k)=c(\whg',k')$,
in which case the embedding $\whg'\subset\whg$ is called a
{\it conformal} embedding. For such embeddings, which were
classified in \cite{\SW,\BBa}, the coset Virasoro algebra is trivial.}
\eqn\chBci{
c(\whg,\whg',k) =  c(\whg,k) - c(\whg',k') \ .
}
An important property of this coset Virasoro algebra is that
it commutes with the AKM subalgebra $\whg'$. This implies that the
$\whg$ HWM's can naturally be interpreted as HWM's of the direct sum
of $\whg'$ and the coset Virasoro algebra. In particular, the coset
vector space obtained from a $\whg$ HWM by identifying states in the
same $\whg'$ HWM, is a (not necessarily irreducible) HWM of the coset
Virasoro algebra.

The coset construction makes it possible to relate modular invariant
partition functions for the AKM algebra $A_1^{(1)}$ and for the
Virasoro algebra \cite{\Boa,\BoGa}.

General Virasoro constructions that are quadratic in the currents
of an AKM $\whg$ have been studied in a systematic fashion (see
\cite{\Hal} for a recent review and references to the original
literature). These constructions correspond to solutions of the
so-called Virasoro master equation. Some of the solutions that have
been found have irrational Virasoro central charge and are thus
non-rational CFT's (compare with Section 4.1). The study of the
Virasoro master equation has revealed an interesting connection with the
theory of graphs and generalized graphs.

\vskip 6mm

In later sections we will argue that the Sugawara and coset constructions
as discussed here are not complete if the Lie algebra $\bfg$ involved
has rank greater than 1, {\it i.e.}\ if it is not $A_1$. The extension of
the Sugawara construction to the so-called Casimir construction, to be
discussed in Section 3.2, will naturally lead to a
class of $\cW$-algebras, the so-called Casimir algebras.
A similar extension of the coset-construction,
to be discussed in Chapter 7, will allow us to construct unitary HWM's
and modular invariant partition functions for these $\cW$-algebras.
More general constructions of $\cW$-algebras from the currents
of AKM algebras have also been considered \cite{\DSc}.

\vfill\eject

%-----------------------------------------------------------------------------

\newsec{$\cW$-algebras and Casimir algebras}

\subsec{$\cW$-algebras: definitions and the example of $\cW_3$}

In the previous chapter we discussed some basics of conformal symmetry
and its elementary realizations in terms of free fields and of the
currents of affine Kac-Moody algebras. We will now focus on the
central topic of this paper: $\cW$-algebras and related structures.

Our first concern is to make more precise what we mean by a
$\cW$-algebra. Since our main interest is in the quantum
$\cW$-algebras that occur in CFT's, we will give a definition,
following \cite{\Go}, that is tailored for this purpose. However,
we would like to stress that the mathematical notions of both
classical and quantum $\cW$-algebras are purely algebraic concepts
which exist independently from the specific context of CFT.

Following \cite{\Go}, we shall first define the notion of a meromorphic
conformal field theory. A quantum $\cW$-algebra can then be defined
to be a meromorphic conformal field theory of a special type.

\vskip 6mm

A meromorphic conformal field theory (mcft) consists of a ``characteristic
Hilbert space'' $\cal H$ (we will sometimes refer to $\cal H$ as the
``vacuum module'') and a map $|\ps\rangle \to V(|\ps\rangle,z)$,
the so-called ``vertex operator map,''
from $\cal H$ into the space of fields.

Furthermore, there should be a distinguished state $|L\rangle$, whose
corresponding vertex operator $T(z) = V( |L\rangle,z)$ is
the stress-energy
tensor of the theory. Its modes $T(z) = \sum_{n\in\ZZ} L_n z^{-n-2}$
satisfy the Virasoro algebra.\smallskip

The vertex operator map has to satisfy the following properties:
\item{(i)} There exists a unique state $|0\rangle \in {\cal H}$
such that $V(|\ps\rangle, z) |0\rangle = e^{zL_{-1}}|\ps\rangle$
\item{(ii)} $ \vev{ \ps_1| V(|\ps\rangle,z) | \ps_2}$ is a meromorphic
function of $z$
\item{(iii)} $\vev{ \ps_1 | V( |\ps\rangle, z) V( |\ch\rangle ,w)|\ps_2}$
is a meromorphic function for $|z|>|w|$
\item{(iv)} $\vev{ \ps_1 | V( |\ps\rangle, z) V( |\ch\rangle ,w)|\ps_2} =
\ep_{\ps\ch} \vev{ \ps_1 | V( |\ch\rangle, w) V( |\ps\rangle ,z)|\ps_2}$
by analytic continuation. Here $\ep_{\ps\ch}=-1$ if both $\ps$ and $\ch$
are fermionic, and $1$ otherwise.\smallskip

\noindent {}From these axioms it follows that the vertex operator map
$|\ps\rangle \to V( |\ps\rangle, z)$ is in fact an isomorphism.

Let ${\cal H} = \oplus_h {\cal H}_h$ denote the decomposition
of $\cal H$ into $L_0$ eigenspaces of eigenvalue $h$ (``conformal
dimensions''). The above axioms imply that only integer or
half-odd-integer conformal dimensions $h$ can occur in a mcft.
The fields of integer dimension are bosonic and those of half-odd-integer
dimension are fermionic. When $|\ps\rangle\in{\cal H}_h$, we will use
the following mode expansion for the corresponding field (compare with
\chBag)
\eqn\chCaa{
V (|\ps\rangle,z) = \sum_n \ps_n z^{-n-h}
}
where $n \in \ZZ$ for $h$ integer and $n \in \ZZ+\half$ for $h$
half-odd integer. It follows that
\eqn\chCab{
\eqalign{
\ps_{-h} |0\rangle & = |\ps\rangle \cr
\ps_n |0\rangle & = 0\, \qquad {\rm for\ } n\geq -h+1 \cr}
}
Moreover, the operator product expansion of two fields $A(z)$
and $B(w)$ of conformal dimensions $h_A$ and $h_B$ may be shown
to equal \chBaa, where
\eqn\chCac{
\{ AB\}_r(w) = V( A_{-r-h_A} B_{-h_B}|0\rangle, w)
}
In particular, the normal ordered product $(AB)(z)$ corresponds to the state
$A_{-h_A}B_{-h_B}|0\rangle$.
\smallskip

A quantum $\cW$-algebra can now be defined to be a meromorphic
conformal field theory of a special type. This will directly imply
that the $\cW$-algebra satisfies a number of CFT consistency
conditions; in particular, it guarantees that the operator product
algebra of the $\cW$-currents will be associative.

\vskip 6mm

\noindent{\bf Definition}\hfill\break\medskip

\noindent{\sl
A quantum $\cW$-algebra is a meromorphic conformal field
theory whose characteristic Hilbert space $\cal H$ contains
a finite number
of distinguished states $|i\rangle$, including the state $
|L\rangle$, whose corresponding vertex operators $W^{(s_i)}(z) = V(
|i\rangle ,z)$ ($T(z) = W^{(2)}(z)$)
are quasiprimary fields of integer conformal dimension $s_i$.
Furthermore, it is required that the entire space of fields is spanned
by normal ordered products of the fields
$W^{(s_i)}(z)$ and their derivatives.}

\vskip 3mm

\noindent It can be shown from the definition that $\cal H$ is spanned
by lexicographically ordered states
\eqn\chCad{
W^{(s_{i_1})}_{-m_1-s_{i_1}}\ldots W^{(s_{i_n})}_{-m_n-s_{i_n}}|0\rangle
}
where $s_{i_j}\geq s_{i_{j+1}}\,, i_j=i_{j+1} \Rightarrow m_j\geq m_{j+1}$
and $m_j\geq0$.
Conversely, if $\cal H$ is spanned by states of this form then all the
fields can be written as normal ordered products of the $W^{(s_i)}(z)$ and
their derivatives.

\vskip 6mm

\noindent {\sl Remarks}

\vskip 3mm

\noindent {\it 1. Generalizations. }\
It is of course possible to relax this definition in various
directions. If we allow an infinite set of extra currents, we
can include important examples such as $\cW_\infty$ (see Section
5.2.1). Also, one can easily define a graded version of
$\cW$-algebras by allowing the spins of the generating currents
to become half-integer (see Section 3.3 for an example).

\vskip 3mm

\noindent {\it 2. Modes. }\
As we discussed in Chapter 2, an OPE algebra of currents is
equivalent to the commutator algebra of their Laurent modes, which
will be $L_m, (TT)_m, \ldots$, $W_m^{(s_i)}, \ldots$\ in the case
of a $\cW$-algebra. The Jacobi identity for this commutator algebra is
equivalent to the condition of crossing symmetry for four-point
functions of currents.

In some cases it is possible to assign half-integer (or, in one case,
$1/3$-integer) modes to some of the currents in a $\cW$-algebra
\cite{\HZa,\HZb,\HZc}. The corresponding twisted sectors of the algebra
can be compared to the Ramond sectors of the superconformal algebras.
The twisting changes the structure of the representation theory and
leads to new modular invariant partition functions. In the Chapters
5 and 6 we will say more about twisted $\cW$-algebras.

It is thus possible to carry out the analysis of $\cW$-algebras entirely
on the level of modes. In this report we will mostly use the formulation
in terms of currents instead.

\vskip 3mm

\noindent {\it 3. Generic vs exotic algebras.}\
The OPE $T$-$T$ (see \chBc)
contains the real parameter $c$, called the central charge.
In certain cases it turns out to be possible
to define a one-parameter family
of consistent $\cW$-algebras, the parameter being $c$.
In such cases various properties of the algebra,
and the representation theory, can be studied as a
function of $c$. We will refer to these algebras by saying that they are
of {\it generic} type. In contrast to this there is the case where a
$\cW$-algebra with a certain set of currents only exists for
some isolated values of the central charge $c$. We will
denote the latter type as {\it exotic}.
These notions will become more clear when we discuss examples
of both possibilities in later sections.%
\foot{In a recent paper \cite{\BW}, the terminology `deformable' and
`non-deformable' is proposed for what we call `generic' and `exotic'
$\cW$-algebras, respectively. As further pieces of terminology, the
notions of `positive-definiteness' and `reductivity' of a $\cW$-algebra
are introduced in \cite{\BW}.}

\vskip 6mm

\noindent
The prototype example of all $\cW$-algebras is the $\cW_3$ algebra%
\foot{For a more systematic nomenclature for $\cW$-algebras, we refer
to later chapters and to Appendix B.}
introduced by A.B.~Zamolodchikov in \cite{\Za}.
It has generators $T(z)$ and $W(z)$, where $W(z)$ is primary of
spin 3 with respect to $T(z)$. The singular OPE of the spin-3
currents reads
\eqn\chCc{
\eqalign{
W\underhook{(z)W}(w) = & \frac{c/3}{\zw^6}
+ \frac{2\,T(w)}{\zw^4}
+ \frac{\partial T(w)}{\zw^3}
\cr
& + \, \frac{1}{\zw^2}\left[ 2\,\beta\La(w)+
\frac{3}{10}\partial^2 T(w) \right]
\cr
& + \, \frac{1}{\zw}\left[ \beta\partial\La(w)+
\frac{1}{15}\partial^3 T(w) \right] \, ,
\cr} }
where
\eqn\chCcI{
\La(w) = (TT)(w)-\frac{3}{10}\partial^2 T(w) \ ,
}
and $\beta$ is given by
\eqn\chCcII{
\beta = {16 \over 22+5c} \,.
}
The central charge parameter $c$ is arbitrary; in the terminology
of `generic' vs.\ `exotic' $\cW$-algebras
the $\cW_3$ algebra is thus an algebra of generic type. Zamolodchikov
showed that these OPE's are the most general compatible with the
requirement of crossing symmetry. The four-point function of the
currents $W(z)$ is given by (compare with \chBaj, \chBak)
\eqn\chCd{
\langle W(z_i) W(z_j) W(z_k) W(z_l) \rangle =
G(x) \, (z_i-z_k)^{-6} (z_j-z_l)^{-6}
}
where
\eqn\chCdI{
\eqalign{
G(x) = & \, {c^2 \over 9} \left[ {1 \over x^6} + {1 \over (1-x)^6} + 1 \right]
\cr
& + 2c \left[ {1 \over x^4} + {1 \over (1-x)^4}
+ {1 \over x^3} + {1 \over (1-x)^3}
- {1 \over x} - {1 \over (1-x)} \right]
\cr
& + c \left( {9 \over 5} + {32 \over 5}{16 \over (22+5c)} \right)
\left[ {1 \over x^2} + {1 \over (1-x)^2}
+ {2 \over x} + {2 \over (1-x)} \right]
\cr}
}
with $x$ as in \chBak. It can easily be checked that the conditions
\chBal\ for crossing symmetry are indeed satisied.

For completeness, we now give the commutator algebra of the Laurent
modes $L_n$ and $W_n$, which are defined as in \chBag.
The commutators $[L_m,L_n]$ and $[L_m,W_n]$ are as in
\chBh\ and \chBagII\ (with $h_W=3$) and we have
\eqn\chBdII{
\eqalign{
[W_m, W_n]
= & \frac{c}{360} \, m (m^2-1)(m^2-4) \delta_{m+n,0}
\cr
& + (m-n) \left\{\frac{1}{15} (m+n+3)(m+n+2) - \frac{1}{6} (m+2)(n+2)
\right\} L_{m+n}
\cr
& + \beta (m-n) \La_{m+n} \,,
\cr}
}
where
\eqn\chBdIII{
\La_m = \sum_n (L_{m-n} L_n) - \frac{3}{10} (m+3) (m+2) L_m \ .
}
Later in this review we will come back to the $\cW_3$ algebra
a number of times.

\vskip 6mm

With the example of $\cW_3$ at hand, we would like to make a few further
remarks on the general structure of $\cW$-algebras. Our definition makes
it clear that the regular terms in any OPE do not contain any independent
information, since they can easily be expressed in terms of the original
currents. However, we would like to stress that the entire set of singular
and regular terms in any OPE forms a representation of conformal symmetry
({\it i.e.}\ the Virasoro algebra), which is in general infinitely
reducible. Concretely, this implies that the set of
all terms in the OPE of two primary fields can
be split as a sum of primary fields and descendant fields that are related
to primary fields that appeared in more singular terms in the expansion
(as in \chBeII). Thus, in general OPE's, an infinite number of primary
fields of the Virasoro algebra are present.

Let us illustrate this with the OPE $W$-$W$ in the $\cW_3$ algebra.
We consider the first regular term in this OPE, which is simply given
by $(WW)(w)$. Clearly, this term is not just a Virasoro descendant of the
identity. Instead, we can `split' it in terms of a descendant field,
given by
\eqn\chCe{
\eqalign{
& \alpha  \left( T \del^2 T - \del ( T \del T) \right)(z)
+ \gamma \, \left( T (TT) \right)(z)
+ \delta \, \del^4 T(z) \cr
& + \epsilon \, \del \left( {1 \over 15}\del^3 T + \beta
\del \left( T^2 - {3 \over 10} \del^2 T \right) \right)(z)
\cr} }
{\it plus} a new Virasoro primary field $\Phi^{(6)}(z)$, which is given by
\eqn\chCf{
\eqalign{
\Phi^{(6)}(z) = & \, (WW)(z)
- \alpha \, \left( T \del^2 T - \del ( T \del T) \right)(z)
- \gamma \left( T (TT) \right)(z) \cr
& - \delta \del^4 T(z)
- \epsilon \del \left( {1 \over 15}\del^3 T + \beta
\del \left( T^2 - {3 \over 10} \del^2 T \right) \right)(z) \, .
\cr} }
In here, $\alpha$, $\gamma$, $\delta$ and $\epsilon$
are functions of $c$ given by
\eqn\chCfII{
\eqalign{
& \alpha = {43c^2-1266c-3784 \over 2(22+5c)(68+7c)(2c-1)} \ , \quad
\gamma = {16(191c+22) \over 3(22+5c)(68+7c)(2c-1)} \ , \cr
& \delta = {-189 c^3 + 812 c^2 - 18900 c - 12320
\over 336 (22+5c)(68+7c)(2c-1)} \ , \quad
\epsilon = {67 c^2 + 178 c - 752 \over 16 (68+7c)(2c-1)}
\cr} }
and $\beta$ is as in \chCcII. We mention here that $\Phi^{(6)}(z)$, which
is primary for all values of $c$, is actually a null field if we choose
$c$ to be $4/5$, $-2$, $-114/7$ or $-23$. For these special values one
expects the existence of $\cW_3$ invariant minimal CFT models. We will
later be discussing the example of $c=4/5$, for which a unitary
$\cW_3$ invariant CFT exists.

The decomposition of all currents in a $\cW$-algebra in terms of
quasi-primary currents and derivatives thereof can also be illustrated
with the case of $\cW_3$. The first few (in terms of conformal dimension)
quasi-primary currents are
\eqn\chCfIII{
T(z)\ , \quad (TT)(z)- {3 \over 10}\del^2 T(z) \ , \quad
W(z)\ , \quad (TW)(z)- {3 \over 14}\del^2 W(z) \ , \quad {\rm etc.}
}

\bigskip

\subsec{Casimir algebras}

In this section we present a generalization of the
traditional Sugawara construction \cite{\Su}, see Section 2.3.2, which
includes higher-spin generators in addition to the stress-energy tensor.
This construction, which was first presented in \cite{\BBSSa}, leads to
so-called Casimir algebras, which are special examples of $\cW$-algebras.
In particular, the Casimir construction provides a realization of
the $\cW$-algebras associated with $X_\ell$, where $X_\ell$ is one of the
simply laced AKM algebras $A_\ell^{(1)}$, $D_\ell^{(1)}$ or
$E_\ell^{(1)}$, with central charge given by $c={\rm rank}(X_\ell)=\ell$.
More details will be provided in Chapter 7.

Our starting point is a conformal field $J(z)$, which takes values
in a Lie algebra $\bfg$ (see Section 2.3.2 for our notations and
conventions). We consider the so-called Casimir operators
\eqn\chCg{
W^{(s_i)}(z) = \frac{1}{s_i!} \eta^{(s_i)}(\bfg,k)\,\sum_{a,b,c,\ldots}
d_{abc\ldots}
(J^a(J^b(J^c(\ldots))))(z) \ ,
}
where $\eta^{(s_i)}(\bfg,k)$ is some normalization constant and
$d_{abc\cdots}$ is a completely symmetric traceless $\bfg$-invariant
tensor of rank $s_i$. The index $i$ labels a basis for
the invariant tensors of
$\bfg$; if $\bfg$ has rank $\ell$ we have $i=1,2,\ldots,\ell$.
The numbers
$s_i$ are equal to the so-called exponents of the Lie algebra $\bfg$
plus one. A list of exponents is provided in Appendix A.
Note that $T^{(s_i)} \equiv d^{abc\cdots}T_aT_bT_c\ldots\,$,
with $T_a$ as in \chBca, is the $s_i$-th Casimir operator of the
Lie algebra $\bfg$.

The first Casimir operator $T(z) \equiv W^{(2)}(z)$ is the Sugawara
stress-energy tensor \chBce, which we discussed in Section 2.3.2.
It satisfies the Virsoro OPE \chBc\ with central charge $c(\whg,k)$
as given in \chBcf. The fact that the $d$-symbols have been chosen
to be traceless guarantees that the remaining Casimir operators
$W^{(s_i)}(z)$, $i=2,3,\ldots,\ell$, are primary fields
of dimension $s_i$ with respect to $T(z)$.

In the remaining part of this section we focus on the relatively
simple example $\whg=A_2^{(1)}$, where two independent Casimir invariants
of orders 2 and 3 exist. The third order Casimir operator takes the form
\eqn\chCj{
W^{(3)}(z) = {1 \over 6} \eta^{(3)}(k)\,d_{abc}(J^a(J^bJ^c))(z)\ ,
}
where
\eqn\chCk{
\eta^{(3)}(k)=\frac{1}{k+3} \sqrt{\frac{6}{5(2k+3)}} \ .
}
In addition to the aforementioned results that $T(z)$ satisfies the
Virasoro operator algebra and that $W^{(3)}(z)$ is a primary field of
dimension 3, we obtain the following contraction of $W^{(3)}$ with itself
\eqn\chCl{
\eqalign{
W^{(3)} \underhook{ (z)W^{(3)} }(w) = & \frac{c/3}{\zw^6}
+ \frac{2\,T(w)}{\zw^4}
+ \frac{\partial T(w)}{\zw^3}
\cr
& + \, \frac{1}{\zw^2}\left[ 2\,\beta\La(w)+
\frac{3}{10}\partial^2 T(w)+
R^{(4)}(w) \right]
\cr
& + \, \frac{1}{\zw}\left[ \beta\partial\La(w)+
\frac{1}{15}\partial^3 T(w)+
\half \partial R^{(4)}(w) \right]
\cr} }
where $\La(z)$ and $\beta$ are given in \chCcI\ and \chCcII\ and the central
charge is equal to $ c = c(A_2^{(1)},k)= 8k/(k+3)$. The field
$R^{(4)}(z)$ is a new primary field which cannot be expressed
in terms of the second and the third order Casimir operators
$T(z)$ and $W^{(3)}(z)$. For general $k$ the field $R^{(4)}(z)$
does not vanish, which means that the operator algebra does not close
on the space spanned by $T(z)$ and $W^{(3)}(z)$ (compare with our
definition in Section 3.1).

This situation improves if the level $k$ is chosen to be 1. It was
shown in \cite{\BBSSa} that in that case the troublesome
field $R^{(4)}(z)$
is a null field (corresponding to a null state in the Hilbert space),
which decouples from the algebra. This then leads to the result that
the $A_2^{(1)}$ Casimir algebra for $k=1$ is actually {\it identical}\
to the $\cW_3$ algebra (with $c=2$) which we introduced in Section 3.1!

In Chapter 7 we shall show that the identification of the $c=2$\ $\cW_3$
algebra with the level-1 $A^{(1)}_2$
Casimir algebra carries over to all of the simply-laced classical
Lie algebras $A_\ell$, $D_\ell$ and $E_\ell$ and the corresponding
$\cW$-algebras,
which we will denote by $\cW_c[X^{(1)}_\ell/X_\ell,1]$ (see Appendix B).
(Clearly, explicit calculations will not be possible in those cases and
we will resort to different techniques.) We will then extend
the Casimir construction to a coset construction, which will allow us to
study the unitary minimal models of various $\cW$-algebras.

\subsec{$\cW$-superalgebras; the example of super-$\cW_3$}

In this section we will take a first look at supersymmetric
extensions of $\cW$-algebras. Rather surprisingly, we will find that
the `minimal' supersymmetric extension of the $\cW_3$ algebra
is only consistent for two values, $c={10 \over 7}$ or
$c=-{5 \over 2}$, of the central charge, so that the minimal
super-$\cW_3$ algebra is what we call an `exotic' algebra.
The original derivation of this result was given in \cite{\IMYa},
where the associativity was checked by considering
the crossing symmetry of 4-point functions. Here we will take the
opportunity to illustrate (as in \cite{\ASS}) a different
technique, which uses the Jacobi identity for graded commutators
of field operators, see Section 2.2.

Before we turn to the `minimal' super-$\cW_3$ algebra,
we make some remarks
about the superconformal algebra. The $N=1$ superconformal algebra is
generated by the super stress-energy tensor $\Th(Z)$ of dimension $3/2$,
\eqn\chCn{
\Th(Z)=\half G(z)+\theta\, T(z) .
}
We follow the conventions of \cite{\FMS,\Fra}:
$Z=(z,\theta)$ is a complex supercoordinate, $T(z)$ is the ordinary
stress energy tensor of dimension $2$
and $G(z)$ is its fermionic superpartner of dimension $3/2$. The
superconformal algebra is represented by the OPE
\eqn\chCo{
\Th(Z_1)\Th(Z_2)=\frac{1}{z_{12}^{3}}\frac{c}{6}+
\frac{\th_{12}}{z_{12}^{2}}
\frac{3}{2}\Th(Z_2)+\frac{1}{z_{12}}
{1 \over 2} D\Th(Z_2)+\frac{\th_{12}}{z_{12}}
\partial\Th(Z_2)+\cdots ,
}
where $z_{12}=z_{1}-z_{2}-\th_{1}\th_{2}$, $\th_{12}=\th_{1}-\th_{2}$,
$D=\partial_{\th}+\th\partial_{z}$, and $\partial=\partial_{z}$.

Let us now consider the superalgebra that is obtained by adding
a primary supercurrent $\Wh(Z)$ of dimension $5/2$ to the superconformal
algebra. The fact that $\Wh(Z)$ is primary of dimension
$5/2$ is expressed by the OPE
\eqn\chCq{
\Th(Z_1)\Wh(Z_2)=\frac{\th_{12}}{z_{12}^{2}}\frac{5}{2}
\Wh(Z_2)+\frac{1}{z_{12}}{1 \over 2} D\Wh(Z_2)+\frac{\th_{12}}{z_{12}}
\partial\Wh(Z_2)+\ldots .
}
Let us now consider the OPE $\Wh(Z_1)\Wh(Z_2)$. We suppose that we
are in the `minimal' case, where the operator product algebra
for $\Wh(Z)$ simply reads \cite{\IMYa} (compare with \chBf\ )
\eqn\chCr{
[\Wh]\cdot [\Wh]=C[I]\ ,
}
with $[I]$ denoting the superconformal family of the identity operator.
In order to fix the coefficients in the OPE we consider
the condition of associativity. We will use the Jacobi identities for
`normal ordered graded commutators' of the currents $\Th(Z)$ and $\Wh(Z)$.
These identities will allow us to fix various
coefficients in a relatively easy way. The general graded Jacobi identity
reads (compare with \chBam\ )
\eqn\chCs{
(-1)^{AC} [[A,B\},C\}(Z)+ {\rm cycl.}=0
}
for general currents $A(Z)$, $B(Z)$, and $C(Z)$. With the
the OPE's \chCo\ and \chCq\ the Jacobi identities for
$\Th\,\Th\,\Th$ and $\Th\,\Th\,\Wh$ are guaranteed.
For $\Th\,\Wh\,\Wh$ we have
\eqn\chCt{
[\Th,\{\Wh,\Wh\}](Z)+2 \, [\Wh,\{\Th,\Wh\}](Z)=0 .
}

The first few singular terms in $\Wh(Z_1)\Wh(Z_2)$ are easily fixed
by writing $\Wh(Z)= {1 \over \sqrt{6}} U(z) + \th W(z)$ and
using the OPE's $W(z_1)W(z_2)$ and $U(z_1)U(z_2)$ as given
in \cite{\Za}. In combination with \chCt\ this leads to
\eqn\chCu{
\eqalign{
\Wh(Z_1)\Wh(Z_2) = & \frac{1}{z_{12}^5}\frac{c}{15}+\frac{\th_{12}}{z_{12}^4}
\Th(Z_2)+\frac{1}{z_{12}^3}\frac{1}{3} D\Th(Z_2)+\frac{
\th_{12}}{z_{12}^3}\frac{2}{3}\partial\Th(Z_2)
+\frac{1}{z_{12}^2}\frac{2}{3}\Th^2(Z_2)
\cr
& +\frac{\th_{12}}{z_{12}^2}\frac{1}{(4c+21)}\left[(c+
\frac{5}{2})\partial^2\Th(Z_2)+22\Th D\Th(Z_2)\right]
\cr
& +\frac{1}{z_{12}}\frac{1}{(4c+21)(10c-7)}\left[(36c+2)
D \Th D \Th(Z_2) \right.
\cr
& \quad + \left.(2c^2-c-37) D\partial^2\Th(Z_2)-(4c-166)
\Th\partial \Th(Z_2) \right]
\cr
& +\frac{\th_{12}}{z_{12}}\frac{1}{(4c+21)(10c-7)}
\left[ (112c-160)\Th D\partial \Th(Z_2) \right.
\cr
& \quad + \frac{4}{3} \left. (2c^2-29c+3)\partial^3\Th(Z_2)
+(144c+8) D\Th \partial\Th(Z_2) \right] + \ldots .
\cr} }
(Note that $\Th^2(Z)$ is actually the same as $\frac{1}{4}\partial D\Th(Z)$.)

The remaining associativity condition is the Jacobi identity for the
triple product $\Wh\,\Wh\,\Wh(Z)$. One finds
\eqn\chCw{
[\Wh,\{\Wh,\Wh\}](Z)= \partial^2 \widehat{\Psi}(Z)\ ,
}
where
\eqn\chCx{
\eqalign{
\widehat{\Psi}(Z) = \frac{1}{9(4c+21)(10c-7)} & \left[
3(2c-83) \Th\partial D\Wh(Z)+
12(18c+1)D\Th\partial\Wh(Z) \right. \cr
&  \; -6(2c-83) \partial\Th D\Wh(Z)-15(18c+1)\partial D\Th\Wh (Z)\cr
&  \; \left. +(2c^2-29c+3)\partial^3 \Wh(Z) \right] .
\cr} }
This result shows that the above operator algebra is {\it not}\
associative for generic values of the central charge $c$. However,
one can check that the field $\widehat{\Psi}(Z)$ is superprimary,
and hence null, for $c=10/7$ or $c=-5/2$. We conclude that for these
values of $c$ all graded Jacobi identities are satisfied.

In \cite{\IMYa}, where the condition of associativity was analysed
by considering crossing symmetry of 4-point correlators, it was shown
that the `minimal' super-$\cW_3$ algebra, as given by the OPE's \chCo,
\chCq\ and \chCu, is indeed associative for $c=10/7$ and $c=-5/2$.

{}From the superspace OPE's listed above the OPE's of the component
fields
$T(z)$, $G(z)$, $W(z)$ and $U(z)$ can easily be obtained. One then finds
that the bosonic $\cW_3$ algebra (see Section 3.1) is a subalgebra of the
`minimal' super-$\cW_3$ algebra for $c=10/7$. A CFT model realizing
super-$\cW_3$ symmetry at $c={10 \over 7}$ will be discussed in
Section 4.2.

We will come back to supersymmetric extensions of $\cW$-algebras in
general and of the $\cW_3$ algebra in particular in later chapters.

\vfill\eject

%------------------------------------------------------------------------

\newsec{$\cW$-algebras and CFT}

\subsec{The chiral algebra in RCFT's.}

We will not attempt to give a comprehensive review of the
structure theory of rational conformal field theories (RCFT's), of which
excellent accounts can be found elsewhere
\cite{\MSc}, but rather recall a few basic facts with special
emphasis on the role played by the chiral algebra.

We first define what we mean by a {\it rational} CFT. (This notion
was introduced by Friedan and Shenker in 1987 (unpublished).) A conformal
field theory is called rational if it has the property that the matrix
${\cal N}_{h\hbar}$ which appears in the torus partition function
\chBw\ has finite rank. RCFT's have the property that their correlation
functions on general (punctured) Riemann surfaces take the form of a
finite sum of holomorphic times antiholomorphic
expressions in the modular parameters of punctured surface. It was shown
in \cite{\AM} that in a RCFT the central charge $c$ and the conformal
dimensions $h$ are all rational numbers.

An immediate consequence of the fact that ${\cal N}_{h\hbar}$ has finite
rank is that the partition function $Z$ (see \chBv) of a RCFT can be
written in the following form
\eqn\chDd{
Z=\sum_{(h,\hbar)} (\chi_h+\ldots)(\bar{\chi}_{\hbar}+\ldots)\ ,
}
where the dots stand for additional Virasoro characters
(with increasing dimension) and the sum is over a {\it finite} set
of labels $(h,\hbar)$. Because the unit operator occurs with
multiplicity one in the theory, the label $(h,\hbar)=(0,0)$ occurs
precisely once in the above summation. The characters that are
collected in the holomorphic factor $(\chi_0+\chi_{s_2}+\ldots)$
are all multiplied with $\bar{\chi}_0$. These terms in $Z$ correspond to
representations associated with chiral primary fields of dimension
$(s_i,0)$, which have the interpretation of currents corresponding
to additional (chiral) symmetries in the CFT.
The conformal families $[\phi_{(s_i,0)}]$, together with the
corresponding operator product expansions, define the so-called chiral
algebra ${\cal A}$ of the RCFT. (In those cases where the above prescription
is ambiguous it will be assumed the characters in $Z$ have been regrouped
such that the chiral algebra is maximally extended.) The algebras
${\cal A}$ and $\overline{\cal A}$, which is composed of the anti-chiral
fields of dimension $(0,s_i)$, are extended conformal algebras.
They contain a Virasoro subalgebra and additional currents
which, due to modular invariance, all have integer conformal dimensions.

Let us now look at the other fields in the theory. It can be shown
that these can be grouped into a finite set of representations of
${\cal A}$ and $\overline{\cal A}$. These representations are HWM's.
The highest weight state is annihilated
by all the positive modes of both the Virasoro generators {\it and}
the additional generators of the chiral algebra. The corresponding
conformal field is called primary with respect to the chiral algebra.
(See Chapters 6 and 7 for a more systematic description of the
representation theory of extended conformal algebras.)

It was established in \cite{\Vea,\MSa} that the fusion rules
of the primary fields in a RCFT can be derived from the transformation
properties under modular transformations of the characters $\hat{\chi}$ of
the chiral algebra ${\cal A}$. This result indicates that the chiral
algebra plays a central role: it dictates both the structure of the
Hilbert space and the form of the interactions. These statements were
made more precise in \cite{\DV,\MSb,\MSc} where the following
results for a RCFT with given chiral algebra ${\cal A}$ were derived
(the assumption is made that both chiral sectors have the same chiral
algebra)
\item{i.} all unitary representations of ${\cal A}$ occur with
multiplicity one,
\item{ii.} the coupling of the left and the right chiral sectors is
either diagonal or off-diagonal by an automorphism of the
fusion algebra.

\noindent
These results show that the classification of all RCFT's can in
principle be done in a two-step process, where in the first step all
chiral algebras that allow a finite set of unitary representations are
determined and in the second step for each algebra the automorphisms of
the fusion algebra are obtained.

It was shown in \cite{\Cab,\Cac}
that the chiral algebra of a RCFT contains
an infinite set of Virasoro-primary currents if the central charge $c$
satisfies $c\geq 1$. However, in many cases it has been found that the
chiral algebra is finitely generated in the following sense. If we have
a set of chiral currents we can
always produce additional currents by taking
derivatives and normal ordered products of currents in the set. We call a
chiral algebra finitely generated if all its currents can be formed from a
finite set of generating currents by repeatedly taking derivatives and
normal ordered products. One easily checks that a chiral algebra which is
finitely generated satisfies the defining properties of what we called a
$\cW$-algebra (Section 3.1). Thus we see the close connection between these
two concepts.%
\foot{Caution: according to our definition, the chiral algebra of
a level-1 WZW model for one of the ADE classical Lie algebras
is (the enveloping algebra of) the full AKM algebra and not
the Casimir algebra discussed in Section 3.2.}

Among the main themes of this review paper are the close connections
between certain $\cW$-algebras on the one hand and classical or
affine Lie algebras on the other. This then might suggest that the
problem of classifying RCFTs can be completely solved by exploiting
these relations. However, a classification of RCFTs along the lines
discussed in this section would at least require a proper understanding
of {\it all} finitely generated $\cW$-algebras, which can be of generic
(deformable) or exotic type. In contrast, the DS reduction scheme and
the coset construction usually lead to $\cW$-algebras that can be defined
for generic central charge $c$. There is thus a missing link, which is
a more systematic understanding of exotic $\cW$-algebras. It is
sometimes possible to construct exotic $\cW$-algebras as extensions
or truncations, which can only be defined for special $c$, of
generic $\cW$-algebras. We refer to Section 5.2.3 for some further
comments about this.

\vskip 6mm

The actual construction of the chiral algebras for RCFT's
has been further analyzed in \cite{\SYa,\SYb,\In}.
These papers introduced
a technique based on the notion of {\it simple currents} and,
related to that, of the {\it center} of a RCFT. Simple currents
are special primary fields whose fusion rules with any other
field contain just one term. This means in particular that they
have well-defined monodromy properties with all fields, and this
property can be used to define an abelian symmetry group which
is called the center of the RCFT.

Let us assume that we are given a RCFT which is diagonal with
respect to a certain chiral algebra. If the set of all primary
fields with respect to this chiral algebra contains simple
currents we can usually construct new, off-diagonal modular
invariant partition functions. These new invariants either
correspond to an automorphism of the fusion rules of the original
chiral algebra or to an extension of the chiral algebra. In most
cases, the actual construction of these new invariants resembles
the orbifold constructions that are well-known in CFT and string
theory.

The simple current modular invariants that correspond to
fusion rules automorphisms have been completely classified in
\cite{\GaSa}. For theories with a center $(\ZZ_p)^k$, with $p$ prime,
a complete classification of {\it all} simple current invariants
was presented in \cite{\GaSb}, see also \cite{\GaSc}.

\subsec{Examples: $\cW_3$ and super-$\cW_3$ minimal models}

In the previous section we discussed the general form of the
torus partition function of RCFT's. We already mentioned in Section
2.2 that modular invariant partition functions for the minimal unitary
series \chBr\ of central charges $c$ have been classified in the
so-called ADE classification \cite{\Gea,\CIZa,\CIZb,\Kat}.
In a similar way, all unitary superconformal models with central charge
$c<3/2$ have been classified in \cite{\Ca,\Ka}. Their central charges are
in the discrete series $c_m=\frac{3}{2}(1-\frac{8}{m(m+2)})$,
$m=3,4,\ldots$.

We will now take a closer look at two specific models which we take from
the classifications just cited. By looking at their chiral algebra
(or chiral superalgebra) we will argue that they actually possess extended
symmetries, which will be $\cW_3$ symmetry for the first model and
super-$\cW_3$ symmetry for the second.
We present these explicit examples to
illustrate the general structure that we discussed in Section 4.1.

\vskip 6mm

For our first example we take a CFT of central charge $c=4/5$, which
corresponds to $m=5$ in the unitary discrete series \chBr. For this
central charge, two modular invariant partition functions exist.
Here we focus on the so-called exceptional modular invariant, which
corresponds to a CFT theory which is related to the 3-states Potts model
at criticality. This partition function contains only a subset of the
primary fields that are allowed by unitarity
\eqn\chDg{
Z=\mid\chi_0+\chi_3\mid^2 + \mid\chi_{2/5}+\chi_{7/5}\mid^2
+ \,2\,\mid\chi_{1/15}\mid^2 + \,2\,\mid\chi_{2/3}\mid^2,
}
where the subscripts denote the conformal dimensions. Of particular
interest in this model is the occurence of
primary fields with dimensions $(h,\hbar)$ given by $(3,0)$ and $(0,3)$,
respectively. The other primary fields in the model are local with
respect to these spin-3 fields. Explicitly we have the following operator
product expansions \cite{\BPZ}
\eqn\chDh{
\eqalign{
\phi_{(3,0)} \cdot \phi_{(2/5,*)}     =[\phi_{(7/5,*)}]\ , & \qquad
\phi_{(3,0)} \cdot \phi_{(7/5,*)}     =[\phi_{(2/5,*)}]\ ,
\cr
\phi_{(3,0)} \cdot \phi_{(2/3,2/3)}   =[\phi_{(2/3,2/3)}]\ , & \qquad
\phi_{(3,0)} \cdot \phi_{(1/15,1/15)} =[\phi_{(1/15,1/15)}]\ ,
\cr} }
where the $*$ stands for $2/5$ or $7/5$ and $[\phi_{(2/3,2/3)}]$
and $[\phi_{(1/15,1/15)}]$ denote either one of the two conformal
families with identical conformal dimensions.

These relations clearly show the existence of extended chiral symmetries
in the model, which are generated by the spin-3 primary fields. The true
symmetry algebra of this particular model is thus an extension of the
conformal algebra, which is generated by the Virasoro generators and the
spin-3 primary fields $\phi_{(3,0)}(z)$ and $\phi_{(0,3)}(\zb)$.
One expects that the partition function \chDg\ can be reexpressed
in terms of characters $\hat{\chi}$ which are defined with respect to
the extended chiral algebra ({\it i.e.}\ the subalgebra generated by $T(z)$
and $\phi_{(3,0)}(z)$). The action \chDh\ of the spin-3 fields on the other
primary fields suggests that the partition function is the `diagonal' sum
of squared extended characters $\hat{\chi}$
\eqn\chDi{
Z=\mid\hat{\chi}_0\mid^2+ \mid\hat{\chi}_{2/5}\mid^2+
\mid\hat{\chi}_{1/15,+}\mid^2
+ \mid\hat{\chi}_{1/15,-}\mid^2 + \mid\hat{\chi}_{2/3,+}\mid^2
+ \mid\hat{\chi}_{2/3,-}\mid^2,
}
where $+$ and $-$ refer to the eigenvalues of the spin-3 operators.

Of course, the chiral algebra of this CFT is precisely the $\cW_3$
algebra which we introduced in Section 3.1, and the structure sketched
above has been confirmed by a detailed analysis of its representation
theory and of the stucture of its modular invariants. The value
$c=4/5$ is actually the lowest central charge that admits unitary
$\cW_3$ invariant CFT's, and the modular invariant \chDi\ is the first
in a long list of $\cW_3$ modular invariants that are known by now.

\vskip 6mm

For our second example we consider a minimal superconformal field
theory of central charge $c=10/7$. We pick this value, which corresponds
to $m=12$ in the superconformal unitary series $c=c_m$ cited above,
since we saw in Section 3.3 that precisely for this value the `minimal'
super-$\cW_3$ algebra can consistently be defined. As we will see later,
this value is also consistent with bosonic $\cW_3$ symmetry,
since it is the $m=6$ position in the unitary
$\cW_3$ series $c_m=2(1-\frac{12}{m(m+1)})$.

It was proposed in \cite{\Bia} that the superconformal field theory with
partition function
\eqn\chDj{
\eqalign{
Z^{(N=1)}_{E_6,D_8} = & \frac{1}{4} \sum^{13}_{s=1,{\rm odd}}
\big( \mid \chi^{NS}_{1s}+\chi^{NS}_{5s}+\chi^{NS}_{7s}
+\chi^{NS}_{11s} \mid^2
+ \, ( \chi^{NS} \rightarrow \widetilde{\chi}^{NS} )
\cr & \qquad \qquad \quad
+ \, \mid \chi^R_{4s} + \chi^R_{8s} \mid^2 \big)
\cr}
}
is a diagonal modular invariant of the minimal super-$\cW_3$
algebra at $c=10/7$. In here we write $\chi^{NS}$ and
$\widetilde{\chi}^{NS}$ for the characters (without and with
the $(-1)^F$ insertion) in the Neveu-Schwarz sector and $\chi^R$
for the characters in the Ramond sector of the $N=1$ superconformal
algebra. The labels $(rs)$ label the various highest weight states in
both sectors.

In later papers \cite{\HR,\SS}, the representation theory of the
super-$\cW_3$ algebra and the branching rules from the super-$\cW_3$
characters to ordinary superconformal characters have been worked out
(see also \cite{\Cas}). In \cite{\SS} some
controversy surrounding this model was resolved and it was established
that the partition function \chDj\ can indeed be written as follows
in terms of characters of the super-$\cW_3$ algebra
\eqn\chDk{
\eqalign{
2 \, Z^{(N=1)}_{E_6,D_8} = &
\left( \mid ch^{NS}_0 \mid^2 + \mid ch^{NS}_{1/14} \mid^2 +
\mid ch^{NS}_{5/14} \mid^2 + \mid ch^{NS}_{1/7,+} \mid^2 +
\mid ch^{NS}_{1/7,-} \mid^2 \right)
\cr
& +  (ch^{NS} \rightarrow \widetilde{ch}{}^{NS})
\cr
& + \frac{1}{4} \left( \mid ch^R_{1/14} \mid^2 + \mid ch^R_{5/14} \mid^2
+ \mid ch^R_{3/2} \mid^2
+ \mid ch^R_{9/14,+} \mid^2 + \mid ch^R_{9/14,-} \mid^2 \right) .
\cr}
}
This combination is precisely a `diagonal' combination of all
characters of the $c=\frac{10}{7}$ super-$\cW_3$ algebra! (We wrote
$ch^{NS}_h$, $\widetilde{ch}_h{}^{NS}$ and $ch^R_h$ for the super-$\cW_3$
characters in the Neveu-Schwarz and Ramond sectors, respectively, and
indicated the sign of the $W_0$ eigenvalue $w$ by $\pm$.) For the actual
derivation of this result some detailed knowledge about characters and
modular invariants for the bosonic $\cW_3$ algebra at $c=10/7$ was used.

Thus we see that this second example, although technically more involved,
is on the same footing as the simple $c=4/5$ example discussed above. In
both cases it turned out to be possible to learn about extensions
of conformal symmetry by careful inspection of known modular invariant
partition functions. These observations are independent from the
detailed study of the operator algebra of the currents involved in the
extended algebra, and they easily go beyond current algebras that can be
constructed by hand. For example, it is straightforward to extend the
examples discussed above to the bosonic $\cW_N$ algebra at
$c=2\frac{N-1}{N+2}$ and to the minimal super-$\cW_N$ algebra at
$c_N=\frac{(3N+1)(N-1)}{2(2N+1)}$ \cite{\SS,\Hoc,\Hod}.
Clearly, the latter algebras cannot easily be obtained in closed form.

\vfill\eject

%------------------------------------------------------------------------

\newsec{Classification through direct construction}

\subsec{The method}

In this chapter we discuss a variety of extended Virasoro
algebras for which the algebra (in the form of (anti-)commutators
or OPE's) is explicitly known. A small number of those have
been found by hand and an additional number have been constructed
with the help of computer power. In the Sections 3.1 and 3.3 we
already showed two explicit examples, which were the $\cW_3$ and
super-$\cW_3$ algebras, respectively.

In later chapters, where we will discuss systematic methods such
as Drinfeld-Sokolov reduction and the coset construction, we will recover
some of the algebras listed below. However, these systematic approaches
at best give existence proofs for some of the algebras, and they certainly
do not lead to explicit results for OPE's. Although these are not always
needed, it is definitely useful to have available explicit and rigorous
constructions of some of the simplest $\cW$-algebras.

We mentioned before that the technical difficulty in constructing an
extended Virasoro algebra with a given set of extra higher-spin fields,
is to make sure that the resulting algebra is associative. In Section 2.2
we discussed three alternative characterizations of associativity, which are
believed to be equivalent. In Section 3.1 we introduced the important
distinction between `generic' algebras (which are
associative for all values of the central charge $c$), and
`exotic' algebras, which are associative for
a finite number of $c$ values only.
Below we will discuss examples of both types.

Before we come to an overview of the results obtained, we would like to
say more about the method of analysis. Let us first consider the analysis
using crossing symmetry of 4-point functions. This method was first
applied in \cite{\Za}, where it was used to fix the coefficients in the
spin-5/2 and spin-3 extended conformal algebras.
In \cite{\Bob}, the behavior
of four-point functions under crossing symmetry was analyzed by using
counting arguments based on the conformal block decomposition of
the 4-point
functions of the currents in extended algebras of the type $\cW(2,s)$.
The analysis of crossing symmetry was further systematized in
\cite{\Bow} (see also \cite{\Hob}).

A second possibility is to do the analysis by using Jacobi identities for
modes of the quasi-primary fields that constitute the $\cW$-algebra.
To streamline these computations some general theory was developed in
\cite{\Bow,\BFKNRV,\KWa}.
The starting point for this is the following form
of the commutation relation of the modes $\phi^i_m$ and $\phi^j_n$
of two quasi-primary fields $\phi^i$ and $\phi^j$, which can be derived
\cite{\Bow} from \chBfII, \chBfIII
\eqn\chEaa{
[\phi^i_m,\phi^j_n] = \sum_k C^{ij}{}_k \, P(m,n;h_i,h_j,h_k)
\, \phi^k_{m+n} + \gamma^{ij} \delta_{m+n}
\left( \matrix{ m+h_i-1 \cr h_i+h_j-1 \cr} \right) \ ,
}
where
\eqn\chEab{
\eqalign{
& P(m,n;h_i,h_j,h_k) = \cr
& \sum_{r=0}^{h_i+h_j-h_k-1}
\left( \matrix{ m+h_i-1 \cr h_i+h_j-h_k-1-r \cr} \right)
{(-1)^r(h_i-h_j+h_k)_r (n+m+h_k)_r \over r! \, (2 \, h_k)_r }\ ,
\cr}
}
with $(x)_r = \Ga(x+r) / \Ga(x)$. One then observes that the Jacobi
identities for those quasi-primaries that can be written as composites
of the generating fields are implied by those of the generating fields.
This reduces the analysis to the generating fields of the $\cW$-algebra,
which have been called `simple fields' in \cite{\BFKNRV}. The extension
of the formalism of \cite{\BFKNRV} to the case of
$N=1$ $\cW$-superalgebras was presented in \cite{\Bl}.

The most extensive and systematic work on the explicit construction
of $\cW$-algebras was presented in the
papers \cite{\BFKNRV} and \cite{\KWa}.
These two papers, which largely overlap, give explicit results
for a large number of $\cW$-algebras with one or two higher-spin
generators in addition to the spin-2 Virasoro generator.

\vskip 6mm

\subsec{Overview of results}

In the list below we restrict ourselves to algebras for which all
operator products (or, equivalently, commutation relations) are
explicitly known and for which associativity has been established.
Algebras whose existence is conjectured on the basis of extrapolation,
general reasoning or wishful thinking are deferred to later chapters.
For completeness, we also mention a few (linear and non-linear)
superconformal extensions of the Virasoro algebra, which are not
$\cW$-algebras in the strict sense, but which fit naturally into
the list.

We recall the notation convention introduced in Appendix B:
by a $\cW$-algebra of type $\cW(2,s_2,s_3,\ldots,s_n)$ we mean an
algebra generated by the Virasoro generator $T(z)$ and additional
primary currents of
spins $s_2,s_3,\ldots,s_n$. For a $\cW$-extension of the $N$-extended
superconformal algebra ($N=$1, 2, 3 or 4) with currents that are
superfields of spins $s_2,s_3,\ldots$, we will write
$\sW^{(N)}(2-\frac{N}{2},s_2,\ldots,s_n)$, where the first entry
denotes the super stress tensor. Notice that these notations only
specify a certain type of algebra; in particular, it is
possible that distinct algebras with the same set of spins of the
generating currents exist.

\vskip 6mm

\noindent {\sl 5.2.1. Generic, linear algebras}

\vskip 3mm

\item{(i)} The Virasoro algebra, given in \chBc\ or \chBh .

\vskip 3mm

\item{(ii)} The $N$-extended superconformal algebras ($N=1,2,3,4$).

\noindent
The classical $N$-extended superconformal algebras, which contain
affine $so(N)$ as a subalgebra, were first given in \cite{\Ad}.
For $N=4$, an additional algebra with only an affine $su(2)$ subalgebra,
exists (the so-called `small' $N=4$ superconformal algebra). The algebras
with $N\leq 3$ and the small $N=4$ algebra possess a unique central
extension. For the $so(4)$ extended $N=4$ algebra two independent central
extensions (corresponding to schwarzian derivatives in $N=4$ superspace)
exist \cite{\Sca}. The $N=4$ quantum algebras are thus parametrized 
by two central charges $c$ and $c^\prime$, or, suppressing one of the
central terms, by $c$ and the value of a deformation parameter $\alpha$
\cite{\Sca,\STPS}. The linear superconformal algebras with $N \geq 5$
do not admit a central extension \cite{\Sca}.

\vskip 3mm

\item{(iii)} $w_\infty$, $\cW_\infty$ and $\cW_{1+\infty}$.

\noindent
We discuss some linear, infinitely generated $\cW$-algebras. Some
early references for these algebras are \cite{\BabII,\Bib,\PRSa,\PRSb,
\PRSc,\Bad,\Mo,\BKa,\BKb}; useful reviews are for example
\cite{\PRSd,\Pop,\Sh}.

The simplest infinitely generated $\cW$-algebra, which
has been named $w_\infty$ \cite{\BabII},
can be viewed as the algebra of area preserving diffeomorphisms of a
two-dimensional cylinder. The algebra
contains generators $w_m^{(s)}$ of spin $s=2,3,4,5,\ldots$. The defining
commutation relations are
\eqn\chEac{
[w_m^{(s)}, w_n^{(t)}] = [(t-1)m-(s-1)n] w_{m+n}^{(s+t-2)}\ .
}
The generators $w_m^{(2)}$ generate a (classical) Virasoro subalgebra.
However, the standard central extension of the Virasoro algebra
can {\it not} be extended to the full algebra $w_\infty$, which should
therefore be viewed as a classical $\cW$-algebra.

The algebra $\cW_\infty$, which was first given in \cite{\PRSa,\PRSb},
is a deformation of $w_\infty$ which is such that the standard central
term in the Virasoro sub-algebra can be extended to the whole algebra.
$\cW_\infty$ can thus be viewed as the quantum version of $w_\infty$.
This has been nicely illustrated
in the context of $\cW$-gravity, where it was shown that the quantization
of the classical theory of $w_\infty$ gravity leads to a quantum theory
based on $\cW_\infty$ \cite{\BHPSSS}.

Following \cite{\PRSa,\PRSb}, we denote the generators of $\cW_\infty$ of
spin $s$ by $V^i_m$, where $s=i+2$, so that the index $i$ ranges from
0 to $\infty$. The defining commutation relations for $\cW_\infty$ can
then be written as
\eqn\chEad{
[V_m^i,V_n^j] = \sum_{l\geq 0} g^{ij}_{2l}(m,n) V_{m+n}^{i+j-2l}
+ c_i(m) \delta^{ij} \delta_{m+n}\ .
}
The structure constants $g^{ij}_{2l}(m,n)$ and the central terms
$c_i(m)$ are completely fixed by the Jacobi identities and take the
following form. For the central terms we have
\eqn\chEae{
c_i(m) = m(m^2-1)(m^2-4) \cdots (m^2-(i+1)^2) c_i\ ,
}
where the central charges $c_i$ are given by
\eqn\chEaf{
c_i ={2^{2i-3} \, i! \, (i+2)! \over (2i+1)!! \, (2i+3)!!} \, c \ .
}
The structure constant $g^{ij}_{2l}(m,n)$ are expressed as
\eqn\chEaf{
g^{ij}_{l}(m,n) = {1 \over 2(l+1)!} \phi^{ij}_l N^{ij}_l(m,n)\ ,
}
where the $N^{ij}_l$ are given by
\eqn\chEag{
N^{ij}_l = \sum_{k=0}^{l+1} (-1)^k \left( \matrix{ l+1 \cr k \cr} \right)
[i+1+m]_{(l+1-k)} [i+1-m]_k [j+1+n]_k [j+1-n]_{(l+1-k)} \ .
}
with $[x]_n = \Ga(x+1) / \Ga(x+1-n)$. Finally, the $\phi^{ij}_l$ are
given by
\eqn\chEah{
\phi^{ij}_l = \sum_{k\geq 0}
{ (-\half)_k ({3 \over 2})_k (-{l \over 2}-\half)_k (-{l \over 2})_k
\over
k! (-i-\half)_k(-j-\half)_k(i+j-l+{5 \over 2})_k } \ ,
}
where $(x)_n = \Ga(x+n) / \Ga(x)$. The $w_\infty$ algebra can be
recovered from $\cW_\infty$ by performing a contraction.

An alternative algebra, which contains a generator of spin 1 in addition
to the spin $2,3,\ldots$ generators of $\cW_\infty$ has been
called $\cW_{1+\infty}$ \cite{\PRSc}. In addition, matrix generalizations
of $\cW_\infty$ and $\cW_{1+\infty}$ have been considered
\cite{\BKb,\OS}.
Supersymmetric extensions of $\cW_\infty$ have been worked out in
\cite{\BPRSS,\BdWVa,\BdWVb}. The papers \cite{\BdWVa,\BdWVb} introduce a
deformation parameter $\lambda$, which is such that the algebra
super-$\cW_\infty(\lambda)$ can be truncated on various subalgebras
for special choices of $\lambda$.

A $c=2$ realization of $\cW_\infty$, which can be viewed as a theory
of $\ZZ_\infty$ parafermions, was discussed in \cite{\BKa}. In
\cite{\BKb} more general unitary representations, of central charge
$c=2p$, $p=1,2,\ldots$, were given. For a systematic discussion of the
representation theory of various infinite $\cW$-algebras,
see for example \cite{\Odb}.

In Section 5.3.4 we will discuss a non-linear extension of $\cW_\infty$,
which can be viewed as a universal $\cW$-algebra.

\vskip 6mm

\noindent {\sl 5.2.2. Generic, non-linear algebras}

\vskip 3mm

\item{(i)} $\cW(2,3)$.

\noindent
This is the $\cW_3$ algebra, which we discussed in
Section 3.1. Its OPE's are given in \chBc, \chBe\ (with $h_W=3$)
and \chCc -\chCcII.

\vskip 3mm

\item{(ii)} $\cW(2,4)$.

\noindent
The explicit OPE's for this algebra,
which was discussed in \cite{\Bob,\HT,\Zh,\BFKNRV,\KWa},
are \chBc, \chBe\ (with $h_W=4$) and
\eqn\chEa{
\eqalign{
 & W(z)W(w) =  \frac{c/4}{\zw^8}
   + \frac{2\,T(w)}{\zw^6} + \frac{\del T(w)}{\zw^5}
\cr
 & \quad + \frac{1}{\zw^4}\left[
   \frac{3}{10} \del^2 T (w) + 2\,\gamma\La(w) \right]
   + \frac{1}{\zw^3}\left[
   \frac{1}{15} \del^3 T (w) + \gamma \del \La (w) \right]
\cr
 & \quad + \frac{1}{\zw^2} \left[
     \frac{1}{84} \del^4 T (w) + \frac{5}{18} \gamma \del^2 \La (w)
     + \frac{24}{\mu} (72c+13) \Omega (w) \right.
\cr
 & \qquad \qquad \qquad
   \left. - \frac{1}{6 \mu} (95 c^2 + 1254 c - 10904) P(w) \right]
\cr
 & \quad + \frac{1}{\zw}\left[
     \frac{1}{560} \del^5 T (w) + \frac{1}{18} \gamma \del^3 \La (w)
     + \frac{12}{\mu} (72c+13) \del \Omega (w) \right.
\cr
 & \qquad \qquad \qquad
   \left. - \frac{1}{12 \mu} (95 c^2 + 1254 c - 10904) \del P(w) \right]
\cr
 & \quad + C_{44}^4 \left\{ \frac{1}{\zw^4} W(w) + \frac{1}{\zw^3}
     \frac{1}{2} \del W (w) \right.
\cr
 & \qquad \qquad + \frac{1}{\zw^2} \left[
     \frac{5}{36} \del^2 W (w) + \frac{28}{3(c+24)} H(w) \right]
\cr
 & \qquad \qquad + \left. \frac{1}{\zw} \left[
      \frac{1}{36} \del^3 W(w) + \frac{14}{3(c+24)} \del H(w)
      \right] \right\} \ ,
\cr}
}
where
\eqn\chEb{
\eqalign{
\Lambda (w) & = (TT)(w) - \frac{3}{10} \del^2 T(w)
\cr
\Omega (w) & = (\Lambda T)(w) - \frac{3}{5} (\del^2 T \, T)(w)
               - \frac{1}{28} \del^4 T(w)
\cr
P(w) & = \frac{1}{2} \del^2 \Lambda(w) - \frac{9}{5} (\del^2 T \, T)(w)
         + \frac{3}{70} \del^4 T(w)
\cr
H(w) & = (TW)(w) - \frac{1}{6} \del^2 W(w)
\cr}
}
and
\eqn\chEc{
 \gamma = \frac{21}{22 + 5c}\ , \qquad
 \mu = (5c+22)(2c-1)(7c+68) \ .
}
The self-coupling constant $C_{44}^4$, which is fixed by the requirement
of associativity, is given by (this value was announced in \cite{\Bob}
and confirmed in \cite{\BFKNRV,\KWa})
\eqn\chEd{
   \left( C_{44}^4 \right)^2 =
    \frac{1}{\mu} 54 (c+24) (c^2 - 172c + 196) \ .
}
(We give this explicit form of the algebra mainly for the purpose of
illustrating how rapidly the complexity of the OPE's increases with
increasing spin of the generating currents. This has been the last algebra
that we display in this explicit form.)
This algebra is related to the Lie algebras $B_2$ and $C_2$ by
Drinfeld-Sokolov reduction
\cite{\Bob,\KWb}. (Warning: the title of the second paper
notwithstanding, the algebra $\cW(2,4)$ is {\it not} directly
relevant for the level-1 $B_2$ WZW models and the
level $(1,k)$ coset models based on this Lie algebra. For
that we need an algebra of type $\cW(2,5/2,4)$ (see below),
which can be viewed as the Casimir algebra (in the sense of Section 3.2)
of the superalgebra $B(0,2)$ \cite{\Wae}.)

\vskip 3mm

\item{(iii)} $\cW(2,6)$.

\noindent The existence of a generic algebra with a spin-6 additional
current was announced in \cite{\Bob}, where it was also shown that
algebras of type $\cW(2,s)$ with $s$ integer or
half-integer and $s>6$ do not exist for generic central charge
$c$. The spin-6 algebra was explicitly constructed in \cite{\FSa};
these results were then confirmed in \cite{\BFKNRV,\KWa}.
It is expected that this algebra is related to the Lie algebra $G_2$
by Drinfeld-Sokolov reduction \cite{\Bob,\BFFOWb}.
\vskip 3mm

\item{(iv)} $\cW(2,3,4)$.

\noindent
This algebra, which is explicitly given in \cite{\BFKNRV,\KWa},
is the third, after Virasoro and $\cW_3$, of the $\cW_N$ algebras,
which are of type $\cW(2,3,\ldots,N)$. Their discovery in
\cite{\FLa,\FLb,\FLc,\BBSSa} was the first systematic extension of
Zamolodchikov's construction of the $\cW_3$ algebra.
(These algebras have not been constructed explicitly for $N\geq 5$).
Of all $\cW$-algebras, these are the ones that are most easily
tractable and that have received the most attention.
We will come back to this series in Chapters 6,7.

\vskip 3mm

\item{(v)} $\cW(2,4,6)$.

\noindent
Solutions to the associativity conditions an algebra
with spins 2,4,6 were found in \cite{\KWa}. One of these has been
identified as the bosonic projection of the $N=1$ superconformal
algebra (see Section 5.3.2.).

\vskip 3mm

\item{(vi)} $\cW(2,5/2,4)$.

\noindent
The existence of this algebra was announced in \cite{\FLc};
the explicit construction was given in \cite{\FST}, see also
\cite{\Aa}. It is the second of in a series of algebras which
are related \cite{\Wab,\Wac} to a coset construction based on
$B^{(1)}_N$ (see Chapter 7) and which have been studied \cite{\Wae}
from the point of view of the quantum Drinfeld-Sokolov reduction of
the superalgebras $B(0,N)$ (see Chapter 6).
The higher algebras in this series are not known explicitly.

\vskip 3mm

\item{(vii)} $\sW^{(1)}(3/2,2)$,\ $\sW^{(1)}(3/2,3/2,2)$.

\noindent
These are examples of generic $\cW$-extensions of the superconformal
algebra \cite{\KMN,\FSc,\Bl}.
The self-coupling of the current $\widehat{W}_2$
in the algebra $\sW^{(1)}(3/2,2)$ vanishes for $c=-6/5$,
in agreement with the finding in \cite{\IMYa} that the
algebra without self-coupling is only associative for $c=-6/5$.
More general $N=1$ supersymmetric $\cW$-algebras
have been proposed \cite{\NM,\EH, \KMN,\II,\FRc, \DRSa,\FGRZ},
but these have not been worked out in closed form.

\vskip 3mm

\item{(viii)} $N=2$ super-$\cW_3$ \cite{\Od,\Roc}.

\noindent
The $N=2$ super-$\cW_n$ algebras were first considered at the classical
level, where they arise from a
Drinfeld-Sokolov reduction of the superalgebras
$A(n-1,n-2)$ see \cite{\NM,\EH, \KMN,\II,\FRb, \Itb,\LPRSW,\NY}.
The quantum algebra $\sW^{(2)}(1,2)$ was first given in explicit
form in \cite{\Od}, see also \cite{\Roc}.

\vskip 3mm

\item{(ix)} Nonlinear extended superconformal algebras.

\noindent
The first examples of these are the $SO(N)$ and $U(N)$
Knizhnik-Bershadsky superconformal algebras \cite{\Kn,\Bea},
which are extended superconformal
algebras with non-linear (quadratic) defining relations. As such they are
very similar to the non-linear $\cW$-algebras. They all admit non-trivial
central extensions. In the papers \cite{\Bowb,\IM} (see also
\cite{\FrLa}) further nonlinear superconformal algebras
were constructed. Among them are two exceptional algebras based on
the superalgebras $G(3)$ and $F(4)$, which have $N=7$ and $N=8$
supersymmetries, respectively.

\vskip 6mm

\noindent {\sl 5.2.3. Exotic algebras}

\vskip 3mm

\noindent
Let us briefly mention some results for these. Algebras of type
$\cW(2,s)$, with $s>2$ integer or halfinteger,
have been studied by a number of
authors \cite{\Za,\Bob,\BFKNRV,\KWa,\Hob}. Among these, only the algebras
with $s=3,4,6$ are generic; exotic algebras with $s=5,7,8$ and
$s=5/2,7/2,\ldots,15/2$ have been constructed. The representation theory
of these algebras was analyzed in \cite{\Varn,\EFHHNV}.
A similar program for $\cW$-superalgebras of type $\sW^{(1)}(3/2,s)$,
with $s$ (half-)integer with $2 \leq s \leq 7/2$, has been carried
out \cite{\FSb} (see also \cite{\Hoa}).
Among these, the only generic algebra
is the one with $s=2$ which we discussed above; of the exotic algebras
the one with $s=5/2$ (which we called the super-$\cW_3$ algebra) was
presented in detail in Section 3.3.

Some further examples: we mention algebras with two higher-spin primaries
(for example, $\cW(2,4,5)$ in \cite{\KWa}) or with a multiplet of
higher spin primaries \cite{\Kau}, for example $\cW(2,4,4)$ at
$c=1$ or $c={656 \over 11}$
(see also \cite{\BFKNRV,\KWa}). $\cW$-superalgebras of type
$\sW^{(2)}(1,(\frac{3}{2})_+,(\frac{3}{2})_-)$ were considered
in \cite{\IMYb}. Further examples of $\cW$-superalgebras can be found in
\cite{\Bera,\Berb,\BEHH}.

\vskip 6mm

Instead of listing still more examples, let us make some general
remarks about the exotic algebras. Their existence can be understood as
follows. If we consider a minimal model, of central charge $c_0$, of some
{\it generic} $\cW$-algebra (which could for example be Virasoro),
it may happen that the model contains chiral primary fields of integer
conformal dimension. Such fields can then be added to the set of currents
in the original $\cW$-algebra, giving rise to an extension of it.
The associativity of the resulting algebra is guaranteed,
but only for central charge $c=c_0$. Thus we should expect that,
in general, this construction gives rise to exotic algebras.

In Section 4.2, we already discussed two examples this enhancement of the
symmetry algebra in specific minimal models, which were a $c=4/5$ minimal
model and a supersymmetric $c=10/7$ minimal model. In the latter example,
the enhanced algebra was the super-$\cW_3$ algebra, which is exotic.
Obviously, the principle can be used to generate many more examples
of exotic algebras.

Taking an opposite point of view, one can try to view exotic $\cW$-algebras
as `truncations' of generic $\cW$-algebras. The idea here is that in certain
minimal models of $\cW$-algebras, one or more of the higher-spin currents
may be `hidden', giving rise to a chiral algebra which is smaller than
the original $\cW$-algebra and which only exists for some specific
values of $c$.

Let us give some examples to explain this idea.
Borrowing some results from Chapter 7, we mention that there exists the
$E_6$ Casimir algebra, which is of type $\cW(2,5,6,8,9,12)$. It is present
in the level-1 $E_6$ WZW model, and in coset models based on
$E_6\oplus E_6/E_6$ at level $(1,k)$ for large enough $k$
\cite{\BBSSb,\Wac}.
However, in the coset model with $k=1$, which has central charge $c=6/7$,
only a truncation of this algebra, of type $\cW(2,5)$ appears. The latter
algebra is exotic, and had been found in the systematic analysis in
\cite{\Bob}. An even more dramatic example is the
$\cW$-algebra for the coset model for $E_8\oplus E_8/E_8$ at level
$(1,1)$, with $c=1/2$, which reduces to the
Virasoro algebra.
(Interestingly, in both these examples the full extended
algebra can be recognized after perturbing the CFT's with a well-chosen
relevant operator \cite{\Zab}.) In \cite{\ASS} a generic extension of the
exotic super-$\cW_3$ algebra of Section 3.3 has been proposed.

Let us conclude this section by remarking that, although we have some
handles for studying the exotic $\cW$-algebras, they are clearly less
tractable than the generic algebras, which can be studied systematically
on the basis of Lie algebra theory.

\vskip 6mm

\subsec{Relating various algebras}

\vskip 3mm

\noindent {\sl 5.3.1. Factoring out spin-${1 \over 2}$ fermions}

\vskip 3mm

In principle, one can consider $\cW$-algebras with
generators of spin $s\leq2$. Adding spin-1 generators to the
Virasoro algebra leads to an algebra which is a semi-direct
product of the Virasoro algebra with an affine Kac-Moody algebra.
Algebras with more than one spin-$3/2$ supercurrent typically include
some spin-1 currents as well (as is the case in the Knizhnik-Bershadsky
algebras cited above).

It is also possible to consider spin-$\half$ fermions as generators
for a (graded) $\cW$-algebra. They occur naturally, for example,
in the $N=3,4$ linear extended superconformal algebras of \cite{\Ad}.
However, it was shown in \cite{\GSb} that the
generators of an extended algebra including spin-$1/2$ fermions can
always be redefined in such a way that the fermions decouple from the
algebra. When applied to the $N=3,4$ extended superconformal algebras,
this leads to the non-linear $SO(3)$ and $SO(4)$ extended superconformal
algebras of Knizhnik-Bershadsky.

\vskip 6mm

\noindent {\sl 5.3.2. Twisted and projected $\cW$-algebras}

\vskip 3mm

In the theory of affine Kac-Moody (AKM) algebras, the idea of twisting
an algebra is well-known \cite{\Kad}. A twisted algebra can be defined if
the underlying finite dimensional
Lie algebra possesses a discrete symmetry
(automorphism), which in the case of the AKM algebras is a $\ZZ_2$ or, in
one case, a $\ZZ_3$ discrete symmetry.
The $\ZZ_2$ twisted AKM algebras are $A_\ell^{(2)}$, $D_\ell^{(2)}$ and
$E_6^{(2)}$, the unique $\ZZ_3$ twisted algebra is $D_4^{(3)}$.

The close relation between AKM algebras on the one hand and
$\cW$-algebras on the other, which will be discussed in the
Chapters 6 and 7, suggests that we can consider twisted versions
of $\cW$-algebras as well. Indeed, we will see that both under
Drinfeld-Sokolov reduction and under the coset
construction the twisting of an AKM algebra can be `pulled back' to a
corresponding $\cW$-algebra. A simple example is a $\ZZ_2$ twisting
of the $\cW_3$ algebra \cite{\HZa}, which is based on the $\ZZ_2$
symmetry
that sends $T(z)$ to $T(z)$ and $W(z)$ to $-W(z)$, and which assigns
half-odd-integer modes to the current $W(z)$. More general twisted
$\cW$-algebras have been discussed in \cite{\HZb,\HZc}.

Once we have a twisted $\cW$-algebra, we can consider the subset of all
currents in the algebra which are integer-moded. In terms of the
automorphism that caused the twisting, this is precisely the invariant
subset. This projected $\cW$-algebra contains the Virasoro algebra,
and forms a new $\cW$-algebra by itself. Let us for example consider
the $\ZZ_2$ projected $\cW_3$ algebra. It will be clear that its
generators include at least $T(z)$ and $\Phi^{(6)}(z)$, where the spin-6
current $\Phi^{(6)}(z)$ is defined in \chCf. By looking at the characters
we can identify additional currents and we find that the $\ZZ_2$
projected $\cW_3$ algebra at least contains independent generators of
spins $2,6,8,10,12$. Similarly, we can identify independent generators
of spins $2,4,6,8,10,12$
for the $\ZZ_2$ projected $\cW_4$ algebra.

The torus partition functions for the twisted $\cW$-algebras considered
in \cite{\HZa,\HZc} are such that the chiral currents that are odd
under the automorphism drop out. As a consequence, the chiral algebras
of these models are, in general,
precisely the projected $\cW$-algebras that we introduced here.

\vskip 6mm

In a very similar way, we can consider the $\ZZ_2$ projection of any
graded $\cW$-algebra. For these there is a natural $\ZZ_2$ automorphism,
which assigns odd $\ZZ_2$ parity to the currents of half-odd-integer
spin and even parity to the integer-spin currents. The projected algebra
contains all currents in the original algebra that have integer spin
(in terms of states this would mean that we select all integer-spin
states in the Neveu-Schwarz vacuum sector). As before we can try to
find a set of generating currents for this set. This gives us a
bosonic $\cW$-algebra, which we call the {\it bosonic projection} of the
original graded algebra. This algebra acts as the chiral algebra of
GSO projected models based on the original graded $\cW$-algebra.

An interesting example is the $N=1$ superconformal algebra, which is
generated by a spin-$3/2$ supercurrent $G(z)$ in addition to the spin-2
stress-energy tensor $T(z)$, with contractions
\eqn\chEf{
\eqalign{
T\underhook{(z) G}(w)
  & = {3/2 G(w) \over (z-w)^2} + {\del G(w) \over z-w} \cr
G\underhook{(z) G}(w) & = {2c/3 \over (z-w)^3} + {2 T(w) \over z-w}\ .
\cr}
}
It was shown in \cite{\Boc} that the bosonic currents
\eqn\chEg{
\eqalign{
W^{(4)}(z) & = (G \del G)(z) + \ldots \cr
W^{(6)}(z) & = (G \del^3 G)(z)
               + {94+5c \over 14+c} (\del G \del^2 G)(z) + \ldots
\cr}
}
(where the dots stand for expressions involving the fields $T(z)$ which
are such that these fields become primary of spin 4 and 6) generate all
integer-spin currents of the $N=1$ superconformal algebra. The bosonic
projection of the $N=1$ superconformal algebra is thus of type
$\cW(2,4,6)$. For completeness we mention that for central charge
$c=7/10$ both currents
$W^{(4)}(z)$ and $W^{(6)}(z)$ are null-fields, so that for that case the
bosonic projection reduces to the Virasoro algebra without any extension.
The $c=1$ case has been discussed in \cite{\DVVV}.

\vskip 6mm

\noindent {\sl 5.3.3. Relations with parafermion algebras}

\vskip 3mm

Certain RCFT's can be conveniently described in terms of chiral
algebras that contain currents of fractional spin.
Such algebras are generally
called parafermion algebras. It is interesting to compare this
description with the purely bosonic formulation, and in particular
to study the (bosonic) chiral algebras of these RCFT's. We will
briefly discuss this for a number of examples.

We first mention the so-called $\ZZ_N$ parafermions \cite{\FZa}, which
are defined in RCFT's of central charge $c_N = 2(N-1)/(N+2)$. It
has been found \cite{\BBSSb},
that the bosonic algebra underlying the model
of $\ZZ_N$ parafermions is the $\cW_N$-algebra. More precisely,
the $\ZZ_N$ parafermion model turns out to be the smallest (in terms of
central charge) unitary CFT with $\cW_N$-symmetry
(this will be made clear
in Chapter 7, where we discuss the discrete series of unitary CFT's
with $\cW_N$-symmetry). The relation between $\ZZ_N$ parafermion algebras
and the $\cW_N$-algebra has been carefully studied in
\cite{\Naa,\Nab,\Db}.

An alternative choice are the $D_N$ parafermions, $N\geq 3$, which
were introduced in appendix A of \cite{\FZa}. They
each allow for a series of unitary CFT's, with central charge given by
\eqn\chEh{
c(k) = (N-1)
 \left\{ 1 - {N(N-2) \over (k+N-2)(k+N) } \right\},
}
with $k$ is positive, half-integer for $N=3$ and positive, integer for
$N>3$. It was observed in \cite{\GSa} that these CFT's
can all be obtained as coset CFT's for the cosets
$\whso(N)_k \oplus \whsu(N)_1 / \whso(N)_{k+2}$, where the subscripts
denote the levels. (For $N=3$ we identify $\whso(3)_k$ with
$\whsu(2)_{2k}$.) Thanks to the formulation of these theories as coset
CFT's, their bosonic chiral algebra can be studied in a straightforward
way. For the case $N=3$ this will be worked out in Chapter 7,
where we will
find that the bosonic $\cW$-algebra underlying the generic $N=3$ model is
of type $\cW(2,3,4,5,6,6)$.

Still other examples of CFT's where parafermionic symmetries can be
identified are the diagonal cosets $\whg \oplus \whg /\whg$ of general
level $(L,l;L+l)$. For the case of $\whg = \whsu(2)$ it has
been found \cite{\BNY,\KMQ,\Raa} that the models
with fixed $L$ (but varying
$l$) can be characterized by a parafermionic current algebra which,
for $L\geq 2$, includes one or more currents of spin $1+2/(L+2)$.
For $L=2$ the extra current is a fermionic supercurrent $G(z)$ of
spin 3/2, with OPE's as in \chEf. For $L=4$ the current algebra contains
two spin-4/3 currents \cite{\FZc}, see also \cite{\Raa,\Rac}.%
\foot{By using the fact that the embedding $\whsu(2)_4 \subset
\whsu(3)_1$ is conformal, it can be seen that the coset CFT's for
$\whso(3)_k \oplus \whsu(3)_1 / \whso(3)_{k+2}$, which were
discussed earlier, arise as special CFT's based on
$\whsu(2)_{2k} \oplus \whsu(2)_4 / \whsu(2)_{2(k+2)}$.}
The bosonic projections of the parafermionic current
algebras for $L\geq 3$ can again be studied by using the
formulation as coset CFT's.

\vskip 3mm

\noindent {\sl 5.3.4. $\whW_\infty(k)$ as a universal structure}

\vskip 3mm

Given the multitude of finitely generated $\cW$-algebras, there
have been attempts to find a universal underlying structure, from
which a number of finitely generated algebras could be obtained
by reduction or truncation. One would expect that a universal
$\cW$-algebra contains all spins $s\geq 2$, and is thus an
infinitely generated $\cW$-algebra (see 5.2.1 (iii)).

In \cite{\LPSW} it was shown that at $c=-2$ the $\cW_N$ algebras can
be constructed by reducing the $\cW_\infty$ algebra. Similar
reductions have been proposed in the context of $d=2$ $\cW$-gravity
and for the $\cW$-constraints in matrix models.

A very interesting proposal was made in the papers \cite{\WYb,\BKe},
which discuss a non-linear deformation of $\cW_\infty$, called
$\whW_\infty(k)$. In \cite{\BKe} the quantum version of this
algebra arises as the chiral algebra of the non-compact coset model
$Sl(2, \RR)_k/U(1)$. By using the connection with (generalized)
parafermions one finds that for $k=-N$ this algebra truncates to the
$\cW_N$ algebra at central charge $c=2(N-1)/(N+2)$. In the limit
$k \rightarrow \infty$, the algebra reduces to $\cW_\infty$.

It has been proposed in \cite{\BKe} that the algebra $\whW_\infty(k)$
is the quantum version of the classical algebra $\cW_{KP}$, which
corresponds to the {\it second} hamiltonian structure in the hamiltonian
formulation of the KP hierarchy \cite{\GDb,\WYb,\FMR}.
In the same spirit, the linear $\cW_{1+\infty}$ algebra has been
identified as the Poisson bracket
algebra for the {\it first} hamiltonian structure of the KP hierarchy
\cite{\WYa,\Yac}.
The KP hierarchy can be reduced to the $N^{th}$ generalized
KdV hierarchy, and the idea is that in this reduction the algebras
$\cW_{KP}$ and $\whW_\infty(k)$ reduce to the classical and
quantum $\cW_N$ algebras, respectively.

It thus appears that the algebra $\whW_\infty(k)$ plays
the role of `universal $\cW$-algebra' for the $\cW_N$-algebras, whose
structure is based on the Lie algebras $A_{N-1}$. It is expected that
similar universal algebras can be obtained for the other series of
Lie algebra based $\cW$-algebras.

As was pointed out in [\EMNa,\EMNb,\BKe],
connections such as the ones described
here may have important applications in string theory. The algebra
$\whW_\infty(k)$ at $k=9/4$ is a symmetry of Witten's black hole
solution to two-dimensional string theory \cite{\Wic} and as such it
provides an infinite number of conserved quantities.
On the other hand, one may expect that the role played by the KdV
hierarchy in the context of matrix models for $d\leq 1$ string
theories, can be extended to a role played by the KP hierarchy,
and the associated $\whW_\infty(k)$ algebra, in a more general context.

\vfil\eject
%\end

%  ch67.tex, corrected 12/22/92

%\input pkjmac
%\input chrefs
%\magnification=1200
%\baselineskip=1.2\baselineskip

\secno=5

%\line{\hfill{\sl ch67.tex, \datum}}

%
%--------------------------------body--------------------------------
%
\newsec{Quantum Drinfeld-Sokolov reduction}

%---------------------------------------------------------------------
\subsec{Introduction}

The most powerful method of constructing $\cW$-algebras is through
the so-called quantum Drinfeld-Sokolov (DS) reduction, in which
one starts with an affine
Lie algebra (or rather a suitable completion of its enveloping
algebra), and reduces this algebra by imposing some constraint
on the generators. At the classical level this procedure, which leads to
the so-called Gelfand-Dikii algebras \cite{\GD}, was pioneered by
Drinfeld and Sokolov \cite{\DS}, and then further explored
by numerous groups \cite{\Ya,\Maa,\Mab,\Baa,%
\Bab,\Bac,\BFFOWa,\BFFOWb,\BFFOWc}.

Quantizing the classical description however poses serious problems.
In a series of papers Fateev and Lukyanov \cite{\FLa,\FLb,\FLc} proposed
to quantize the Gelfand-Dikii algebras by quantizing the free fields
in the classical Miura transformation. However, this approach
works well only for the Lie algebra $A_n$ (and to some extent
also for $D_n$). The observation that for $A_n$
the $\cW$-generators commute with
a set of `screening charges' led to an ad hoc construction of other
$\cW$-algebras as the centralizer of a suitable set of screening charges.

Rather then making a classical detour one
can formulate the reduction as a quantum problem from the outset.
Starting with an affine Lie algebra $\whg$ one imposes a
set of constraints by means of the BRST procedure.
The reduced algebra $\cW[\whg,k]$ is defined as the cohomology of the
BRST operator \cite{\Be,\BOa,\DF,\Fi,\FFra,\Frea,\Freb,\KN}.
This is the approach we will take in the present chapter.
We will argue that from this definition the description of $\cW[\whg,k]$
(for the conventional DS-reduction) as the centralizer of a
set of screening charges follows quite easily, and that, in
preferred circumstances, a generating series for the generators
of this centralizer may be found by proper quantization of the
Miura transformation. Thus making contact with the
work of Fateev and Lukyanov.

The BRST approach not only provides a proper definition of the quantum
$\cW$-algebra, but also a functor that maps modules
of the affine Lie algebra to modules of the corresponding
$\cW$-algebra. Hence, the BRST approach
is a convenient tool for the study of $\cW$-algebra
representations too. The functor was studied in detail in \cite{\FKW}.
\smallskip

In the next section we will
briefly recall some results that were obtained in
the context of Toda field theory, which historically preceded the
systematic study of $\cW$-algebras from the point of view
of DS reduction. The remainder of the chapter is then devoted
to a detailed description of the quantum DS reduction.

\subsec{Lagrange approach: constrained WZW and Toda field theories}

The classical field equations of a Toda field theory are
\eqn\chFaa{
\del \delb \phi = \sum_{i=1}^\ell \beta\alpha_i \exp
  (\beta\alpha_i \cdot \phi)\ .
}
The field $\phi(z,\zb)$ is a vector in the weight space of a (finite
dimensional) semisimple Lie algebra $\bfg$, of rank $\ell$, for which
$\{ \alpha_i \}$ is a set of simple roots. These equations can be
derived from the lagrangian
\eqn\chFab{
{\cal L} = - \half \del \phi \cdot \delb \phi - V_{\beta}\ , \qquad
V_{\beta} = \sum_{i=1}^\ell \exp(\beta \alpha_i \cdot \phi)\ .
}
Toda field theories
are conformally invariant, both at the classical and at the
quantum level. At the quantum level, the conformally improved
energy momentum tensor satisfies the Virasoro current algebra
\chBc, with the central charge given by \cite{\Mana}%
\foot{The minus sign in front of the second term is due to an
unusual choice for the vacuum or, alternatively, to the continuation
$\be\to i\be$ in \chFab. See \cite{\Manb} for a discussion
of this point.}
\eqn\chFac{
c(\beta) = \ell-12 \, | \beta\rho -{1 \over \beta} \rho^{\vee} |^2 \ .
}
For $\bfg=sl(2)$, the Toda field theory reduces to a
Liouville theory for a single scalar field.

In a series of papers
\cite{\BGa,\BGb,\BGc,\BGd} the connection of Toda field
theories to $\cW$-algebras was first observed and worked out.
In \cite{\BGa,\BGb} it was shown that at the classical level a
Toda field theory possesses a set of conserved currents, whose
Poisson bracket algebra forms a classical $\cW$-algebra. In
\cite{\BGc,\BGd} this result was extended to the quantum theory.
Both at the classical and at the quantum level the $\cW$-generators
are multilinear expressions in the fields $\phi$ and their
derivatives. Toda field theories thus lead to free field realizations
of $\cW$-algebras with adjustable central charge $c(\beta)$ as in
\chFac.

For special values of the coupling constant $\beta$, the
central charge can take the value of the one of the minimal RCFT
models of the associated $\cW$-algebra (see Section 6.4 and Chapter
7 for a systematic discussion). It was argued in \cite{\Manb}
that even with this special choice of $\beta$ the Toda field theory
cannot be identified directly with a $\cW$-minimal RCFT,
the problem being that certain projections are needed and that
the Toda spectrum leads only to a subset of the minimal model primary
fields. The papers \cite{\Manb,\MaSp} introduced so-called conformally
extended Toda theories, whose quantum lagrangian
(for $\bfg$ simply-laced) is obtained by replacing the potential
$V_{\beta}$ in \chFab\ by the combination $V_{\beta} + V_{-1/\beta}$.
It is argued in \cite{\Manb,\MaSp} that these conformally extended Toda
theories provide a lagrangian realization of the $\cW$-minimal CFT's.

The papers \cite{\BFFOWa,\BFFOWb} discuss the fact that Liouville
and Toda field theories can be obtained as conformally reduced WZW
theories. This reduction can be viewed as a gauge procedure: the
Toda theory is obtained as the gauge invariant content of a certain
gauged WZW theory. We should stress that the gauging employed here is
different from the gauging of a left-right diagonal subgroup, which leads
to a lagrangian realization of a coset conformal field theory.
Instead, the gauging that leads to a Toda field theory uses
an upper triangular maximal nilpotent subgroup on the left hand side
and a lower triangular one on the right hand side.

It has become clear that under the reduction that takes a WZW
field theory
into a Toda field theory the affine Lie algebra characterizing
the WZW theory reduces to a $\cW$-algebra (see, for example
\cite{\BFFOWc}).
This reduction, which we call a (classical or quantum) Drinfeld-Sokolov
reduction, can be studied in a more algebraic fashion without
making explicit reference to the underlying Lagrange field theories.
This is the approach we follow in Section 6.3.

The Toda field theory approach has been quite instrumental
in the supersymmetric case, where various new $\cW$-superalgebras have
been found as symmetry algebras of supersymmetric Toda field theories
\cite{\NM,\EH,\KMN,\II,\Noh}. In general, a super Toda field theory
based on a basic Lie superalgebra
for which all simple roots can be chosen
to be fermionic is integrable and superconformally invariant.
Its conserved currents generate a $\cW$-superalgebra. Special choices
are $osp(1|2)$, which leads to the (unextended) $N=1$ super-Virasoro
algebra and $sl(2|1)$, which gives the $N=2$ super-Virasoro algebra.
The central charges (as a function of the coupling constant) of a large
class of quantum $\cW$-superalgebras were listed in \cite{\EH}.
The connection between constrained supersymmetric WZW models and
super Toda theories, which is a supersymmetric version of the
DS reduction scheme, was worked out in \cite{\DRSa,\FGRZ}.

Finally we note that Toda field theories
play an important role in the discussion of $\cW$-gravity
(see Section 8.1),
where they arise as effective quantum theories for the
$\cW$-gravity degrees of freedom in the conformal gauge.

%----------------------------------------------------------------

\subsec{Algebraic approach to DS reduction}

\noindent{\sl 6.3.1. $\cW$-algebras from DS reduction}\hfill\medskip

In general the quantum Drinfeld-Sokolov (QDS) reduction consists of
the following steps. One starts with a triple $(\whg,\whg',\ch)$,
consisting of an affine Lie algebra $\whg$, an affine subalgebra $\whg'
\subset \whg$ and a 1-dimensional
representation $\ch$ of $\whg'$.%
\foot{This assumption on $\ch$ may be relaxed.
That involves the use of second class constraints which can, however, be
transformed to first class constraints by introducing a set of auxiliary
fields (see \eg \cite{\BOb}). The first example of this type is the
so-called $\cW_3^{(2)}$-algebra \cite{\Poc,\Beb}
of type $\cW(1,{3\over2},{3\over2},2)$ that can be considered as
the bosonic analogue of the $N=2$ superconformal
algebra. We will not discuss this more general case here, but present
some examples in Section 6.3.3. See also, for example, \cite{\Rob} for
more general results in this direction.}
Next, one imposes the first class constraints $g\sim\ch(g)\,,\forall
g\in\whg'$, by means of the BRST procedure. The cohomology of the BRST
operator $Q$ on the set of normal ordered expressions in currents, ghosts
and their derivatives (or, equivalently, the cohomology of $Q$ on the
Hilbert space associated to this set of fields)
is what is called the Hecke algebra $H_Q(\whg,\whg',\ch)$
of the triple $(\whg,\whg',\ch)$. In particular, the zeroth
cohomology $H^{(0)}_Q(\whg,\whg',\ch)$ is a subalgebra.
This subalgebra is what we would like to call the
$\cW$-algebra $\cW_{DS}[\whg,\whg',\ch]$ associated to the triple
$(\whg,\whg',\ch)$. There is a small subtlety however. For special values
of the level $k$ of $\whg$, the cohomology $H^{(0)}_Q(\whg,\whg',\ch)$
may be slightly bigger than for generic values of $k$ (`generic'
meaning here, and in the sequel, $k\not\in -\dCo + \QQ_+$).
We will therefore define $\cW_{DS}[\whg,\whg',\ch]$ to be
$H^{(0)}_Q(\whg,\whg',\ch)$ for generic values
of $k$. Since the operator product
expansions of the $\cW$-generators are algebraic in $k$ we may
then simply extend the definition of $\cW[\whg,\whg',\ch]$ to
all values of $k$ by means of the acquired OPA
(except, possibly, for a finite set of $c$-values
where the OPE coefficients become singular).
We also would like to remark that, for the conventional QDS reduction to
be discussed below, one can show that the higher cohomologies
$H^{(i)}_Q(\whg,\whg',\ch)$, $i\neq0$ vanish for generic $k$.

In order for $\cW_{DS}[\whg,\whg',\ch]$ to be a $\cW$-algebra in
the sense described in Chapter 3, one has to suitably
choose the triple $(\whg,\whg',\ch)$. We will describe how
a generic class of such triples is
obtained from $sl(2)$ embeddings \cite{\BTV,\FORTW,\ORTW,\Fu}
(for the corresponding hierarchies of integrable differential
equations see \cite{\TV,\Bu,\BDHM,\DM,\DHM,\MO}).%
\foot{It has recently been argued in \cite{\BW} that these
correspond to the
so-called reductive $\cW$-algebras, \ie $\cW$-algebras for which the
classical limit is positive definite. The Lie algebra $\bfg$ corresponds
to a linear truncation, which can be made in the limit $c\to\infty$,
of the vacuum preserving subalgebra (vpa)
generated by the modes $\{ \ph^i_m\,,|m|< h(\ph^i)\}$ for all
quasi-primary fields $\ph^i(z)$. The $sl(2)$ subalgebra corresponds to
the subalgebra of the vpa generated by $\{ L_{-1},L_0,L_1\}$.}

Let us first briefly discuss the classical case. Suppose we have an
$sl(2)$ subalgebra $\{ T^3, T^+, T^-\}$ of $\bfg$. The adjoint
representation of $\bfg$ decomposes into $sl(2)$ representations of
spin $j_k\,,k=1,\ldots,p$, say.\foot{To get first class constraints we
will restrict ourselves here to $sl(2)$ embeddings such that all
$j_k\in \ZZ$.}
Then we may write the current $J(z) =
J_a(z)T^a$ as
\eqn\chFbza{
J(z) = \sum_{k=1}^p \sum_{m=-j_k}^{j_k} U_{k,m}(z) T^{k,m}\,,
}
where $T^{k,m}$ corresponds to the generator of spin $j_k$ and
isospin $m$ under the $sl(2)$ subalgebra. In particular we may put
$T^{1,1}=T^+\,, T^{1,0} = T^3\,, T^{1,-1} = T^-$.
The $sl(2)$ subalgebra $\{T^3, T^+,T^-\}$
can be characterized by a so-called ``defining vector'' $\de$ (which we
can choose to lie in the fundamental Weyl chamber), such that the $sl(2)$
root $\widehat{\al}$ is given by $\widehat{\al} = \de/(\de,\de)$, and
$T^3 = \de\cdot H$.
Take $\whg'$ to be the
affine Lie subalgebra $\widehat{\bf n}_+$ generated by all
$U_{k,m}(z)\,,m>0$.

Now, one may impose the constraint
\eqn\chFbzb{
\ch_{DS}(U_{k,m}(z)) = \cases{  1 & for $(k,m)=(1,1)\,,$ \cr
0 & for all other $(k,m)$ such that $m>0\,.$\cr}
}
Classically, this set of constraints generates enough gauge invariance
to bring the constrained currents in the so-called lowest weight gauge%
\foot{Note that our conventions slightly differ from those of
\cite{\BTV,\TV}.}
\eqn\chFbzc{
J_{fix}(z) = T^+ + \sum_{k=1}^p U_{k,-k}(z) T^{k,-k}\,.
}
The Poisson bracket structure of the current algebra induces a Poisson
bracket structure on the reduced space, and one can show that the algebra
of the $U_{k,-k}(z)\,,k=1,\ldots,p$ closes with respect to this
induced Poisson structure. In particular, the algebra contains a Virasoro
subalgebra generated by $T(z) = \half {\rm Tr\,}(J_{fix}(z)^2)$ with
respect to which the fields $U_{k,-k}(z)$ are primary of conformal
dimension $j_k+1$.

In the quantum set-up very few results have been
obtained so far for $sl(2)$ embeddings other than the principal
embedding.%
\foot{The most extensively studied algebra of this type is the
  Polyakov-Bershadsky algebra $\cW_3^{(2)}$ that arises from the
  so-called $so(3)$ embedding in $sl(3)$ \cite{\Poc,\Beb}.
  For this embedding
  $\underline{8} \to \underline{1} + \underline{2} + \underline{2}
  + \underline{3}$ which indeed leads to a $\cW$-algebra of type
  $\cW(1,{3\over2},{3\over2},2)$ as mentioned before.}
This is clearly an issue that deserves further study.
For this reason, and for the sake of simplicity, from now
on we will be discussing only the `conventional' QDS reduction which
corresponds to the principal $sl(2)$ embedding in $\bfg$.%
\foot{The associated $\cW$-algebra will be denoted by
  $\cW_{DS}[\whg, \widehat{\bf n}_+,\ch_{DS}]$,
  or simply by $\cW[\whg,k]$ if no confusion can arise.}

The defining vector of the principal $sl(2)$ embedding is the
`dual Weyl vector' $\rhv$, \ie for $\al\in \De_+$ we have
$(\rhv,\al)=1$ if and only if $\al$ is a simple root of $\bfg$.
Denoting the currents corresponding to positive roots $\al$ by
$e_\al(z)$, and choosing $T^{1,1} = \sum e^{\alpha_i}$,
the constraint \chFbzb\ is now explicitly given by
\eqn\chFzzz{
\ch_{DS}(e_\al(z)) = \cases{ 1 & for simple roots $\al_i$ \cr
0 & otherwise$\,.$\cr}
}

We introduce pairs of ghost fields
$(b_\al(z),c_\al(z))$, one for every positive root $\al\in\De_+$.
The BRST-operator corresponding to the constraint \chFbzc\ is given by
$Q=\oint j_{BRST}(z) =  Q_0 + Q_1$, where
\eqn\chFbb{
Q_0 = \oint {dz\over 2\pi i} \left(
\sum_{\al\in\De_+} (c_\al e_\al)(z) -\half \sum_{\al\be\ga\in\De_+}
f^{\al\be\ga} (b_\ga c_\al c_\be)(z)\right)
}
is the standard differential associated to $\widehat{\bf n}_+$,
$f^{\al\be\ga}$
are the structure constants of $\bfnp$, and
\eqn\chFbc{
Q_1 = - \oint {dz\over2\pi i} \sum_{\al\in\De_+} \left(
c_\al(z)\ch_{DS}(e_\al(z)) \right) \ .
}
They satisfy
\eqn\chFbd{
Q^2 = Q_0^2 = Q_1^2 = \{ Q_0,Q_1\} = 0 \ .
}
The algebra $\cW[\whg,k]$ will, at least, contain the Virasoro algebra.
An explicit representative is given by
\eqn\chFbe{
T(z) = T^{Sug}(z) + \rh^\vee\cdot\p h(z) + T^{gh}(z) \ ,
}
where $T^{Sug}(z)$ is the Sugawara stress-energy tensor \chBce, and
\eqn\chFbf{
T^{gh}(z) = \sum_{\al\in\De_+} \left(
( (\rh^\vee,\al) -1) (b_\al\p c_\al)(z) + (\rh^\vee,\al) (\p b_\al
c_\al)(z) \right)
}
is the ghost contribution. Note that the choice \chFbe\ assigns conformal
dimensions $(1-(\rhv,\al),(\rhv,\al))$ to the pair $(b_\al(z),c_\al(z))$.
Of course, the `improvement terms' in the
expression \chFbe\ are obtained by the requirement that
the BRST current $j_{BRST}(z) = j_0(z) + j_1(z)$ becomes a primary
field of conformal dimension one, so that $[Q,T(z)] = 0$. (Implying,
at the same time, that the constraints $e_\al(z) \sim
\ch_{DS}(e_\al(z))$ become conformally invariant.)

The central charge of this Virasoro algebra is given by
\eqn\chFbg{
\eqalign{
c & = { k\ {\rm dim\ }\bfg \over k+\dCo} - 12k|\rhv|^2 -
2\sum_{\al\in\De_+} \left( 6(\rhv,\al)^2 - 6(\rhv,\al) +1\right)
\cr
& = \ell - 12 \, |\al_+\rh + \al_-\rhv|^2 \,,\cr}
}
where we have introduced $\al_+\al_-=-1\,, \al_-=-\sqrt{k+\dCo}$.

In the same spirit one would like to construct the other generators
of $\cW[\whg,k]$. This is clearly too complicated in general (see however
\cite{\DF} for $\cW_3$). However,
given the outcome \chFbzc\ of the classical reduction,
we can make the following general observation.

The Virasoro generator \chFbe, which is in the BRST
cohomology, corresponds to the component
$U_{1,-1}(z)$ of the constrained and gauge fixed current $J_{fix}
(z)$ in \chFbzc.
Similarly one expects that the other components $U_{k,-k}(z)$ have
a counterpart in the BRST cohomology, such that the constructed
$\cW$-algebra is generated by a set of currents of conformal dimension
$e_i+1$, where the set $\{e_i, i=1,\ldots,\ell\}$ (the set of `exponents'
of $\bfg$) is the set of $sl(2)$ spins appearing in the decomposition
of the adjoint representation of $\bfg$.
Note that the numbers $e_i+1$ are precisely the orders of the independent
Casimirs of $\bfg$.

We will now argue that the BRST cohomology can be characterized as the
centralizer of a set of screening charges (for more details and precise
statements we refer to \cite{\FFra,\FFrb,\Frea,\Freb}).
This will establish the
connection with the approach of Fateev and Lukyanov
\cite{\FLa,\FLb,\FLc}.
The cohomology of $Q=Q_0+Q_1$ can be calculated by means of the
so-called spectral sequence technique (see \eg \cite{\BT}) applied to the
double complex of $Q_0$ and $Q_1$ (see \chFbd). The spectral sequence is
a systematic approach for calculating the cohomology of $Q$ by starting
from the cohomology of $Q_0$ and making successive higher
order corrections.
In this particular case this procedure stops (\ie the ``spectral sequence
collapses''), for generic level $k$, after the second term, which implies
that $H_Q(*) \cong H_{Q_1} ( H_{Q_0}(*) )$.
It can be proved that the $Q_0$-cohomology is generated
by the fields $\tilde{h}_i(z)$ and $c_{\al_i}(z)$, where
\eqn\chFbi{
\tilde{h}_i(z)  = \al_+ \left( h_i(z) + \sum_{\al\in\De_+}
(\al,\al_i^\vee) (b_\al c_\al)(z) \right) \ .
}
One easily checks, for example, that the OPE of $j_0(z)$ with these
fields is regular. In the next step, to
calculate the cohomology of $Q_1$ on the space of
fields generated by $\tilde{h}_i(z)$ and $c_{\al_i}(z)$, it is convenient
to bosonize $\tilde{h}_i(z)= \al_i^\vee
\cdot i\p\ph(z)$. Then,
in $Q_0$-cohomology, one can identify
\eqn\chFbk{
c_{\al_i}(z) = \exp{( -i\al_+ \al_i\cdot\ph(z))} = \tilde{s}_i^+(z) \ .
}
In particular, since there is no $Q_0$-cohomology at negative ghost
numbers, we have
\eqn\chFbka{
H^{(0)}_{Q_1} = {\rm Ker\ } Q_1 =
\cap_{i=1}^\ell {\rm Ker\ } \oint \tilde{s}_i^+(z) \ ,
}
where the kernels are taken on the fields generated by $\tilde{h}_i(z)$.
Therefore we reach the conclusion that, for generic level $k$,
$\cW[\whg,k]$ can be identified with the centralizer of the set of
`screening charges' $\oint \tilde{s}_i^+(z)\,, (i=1,\ldots,\ell)$
on the space of polynomials (and derivatives thereof) in $i\p\ph^i(z)$.
We would like to stress that, although we set out as defining the
algebra $\cW[\whg,k]$
independent of any particular realization of $\whg$,
that---as follows from the above discussion---the $\cW$-algebra
$\cW[\whg,k]$ comes naturally equipped with a realization
in terms of ${\rm rank\,}\bfg =\ell$ scalar fields $\ph^i(z)$.
Let us, for future use, denote the Fock space of these scalar fields by
$\cF_\La$, where $\La$ labels the eigenvalue of the scalar zero modes
$\al_0^i = p^i$ on the Fock space vacuum $|\La\rangle$.
\bs

It is not hard to construct the Virasoro generator
from this description. This simply amounts to the construction
of a conformal dimension two operator under which all screening
operators \chFbk\ become primary of dimension one. One finds
\eqn\chFbl{
T(z) = -\half (\p\ph\cdot\p\ph)(z) - (\al_+\rh + \al_-\rhv)\cdot i\p^2
\ph(z) \ .
}
It is a well-known fact that for $\bfg =sl(2)$, at generic
level $k$, this is a complete description of the centralizer
(see \eg \cite{\KWc} for a simple proof).

It follows that the eigenvalue $h_\La$ of $L_0$ (conformal dimension)
of the Fock space vacuum $|\La\rangle$ is given by
(compare with \chBbq)
\eqn\chFbla{
h_\La = \half (\La , \La + 2(\al_+\rh + \al_-\rhv) ) \ .
}
At this point we would like to mention one crucial difference between
the simply-laced case and the non simply-laced case. In the simply laced
case, where $\rhv=\rh$, there exists a special point,
namely $\al_\pm= \pm1$,
for which the background charge term in \chFbl\ vanishes. This is exactly
the point where we will make contact with the corresponding
coset construction of the $\cW$-algebra (see Chapter 7). For non
simply-laced Lie algebras there does not exist a point for which the
background charge term vanishes, and there is no known coset construction
that directly corresponds to $\cW[\whg,k]$.
\medskip

The form \chFbl\ of the stress-energy tensor, and the corresponding
value \chFbg\ of the central charge are identical to the result
obtained from a Toda field theory for the scalar fields $\ph^i(z,\zb)$
(see Section 6.2), where we identify $\beta=\alpha_+$. This illustrates
the close connection, which we mentioned earlier,
between the DS reduction scheme and Toda field theories.
\medskip

Not all Lie algebras $\bfg$ give rise to distinct $\cW$-algebras. We
would now like to discuss a remarkable duality which implies, for
instance, that the $\cW$-algebras corresponding to $B_n$ and $C_n$
are essentially the same \cite{\FFrb} (see also \cite{\KWc}).
Given a finite dimensional (semi-simple) Lie algebra $\bfg$, let
$\bfg'$ be its `Langlands dual' (obtained from $\bfg$ by inverting the
arrows in the Dynkin diagram,
\eg $B_n'\cong C_n$). We can identify the Cartan subalgebras of
$\bfg$ and $\bfg'$ by $\al_i' = \al_i^\vee / \sqrt{r^\vee}$, where
$r^\vee$ is the `dual tier number' of $\bfg$ (the maximal number of
edges connecting two vertices of the Dynkin diagram of $\bfg$). This
identification preserves the inner products between the roots.
Now, the centralizer of the screening charge $\oint \exp{(-i\al_+\al_i
\cdot\ph(z))}$ is generated by the Virasoro element
\eqn\chFbn{
T_i(z) = - { 1\over 2(\al_i,\al_i)} (\al_i\cdot\p\ph\, \al_i\cdot
\p\ph)(z) - \half (\al_+\al_i + \al_-\al_i^\vee)\cdot i\p^2\ph(z)
}
and the Cartan subalgebra elements $\be\cdot i\p\ph(z)$ orthogonal to
$\al_i\cdot i\p\ph(z)$. However, one easily checks that this centralizer
coincides with the centralizer of the charge
\eqn\chFbo{
\oint \exp{(-i\al_-\al_i^\vee
\cdot \ph(z))} = \oint \exp{(-i\al'_+\al'_i\cdot\ph(z))} \ ,
}
where $\al'_+ = \al_-\sqrt{r^\vee}$.
This proves that the $\cW$-algebras related
to $\bfg$ and its dual $\bfg'$ are isomorphic, provided one identifies
$r^\vee (k+ \dCo(\bfg)) = (k' + \dCo(\bfg'))^{-1}$.
Moreover, if one applies this theorem to a simply-laced Lie algebra $\bfg$
(for which $\bfg \cong\bfg'\,, r^\vee=1
\,, \al_i = \al_i^\vee$) one discovers that the $\cW$-generators, which
by definition commute with the charges corresponding to the screening
operators $\tilde{s}_i^+(z)$, automatically commute with the charges
corresponding to the `dual screening operators'
$\tilde{s}_i^-(z) = \exp{(-i\al_-\al_i\cdot
\ph(z))}$.\smallskip

Another extremely important property of $\cW[\whg,k]$, which also follows
easily by examining the various $sl(2)$ subalgebras as in \chFbn, is the
Weyl group invariance of the above construction.%
\foot{The invariance under the finite Weyl group $W$ can be
considered a justification for the terminology $\cW$-algebra \cite{\FKW}.}
We claim that the eigenvalues $w_{s_i}(\La)$ of the zero modes of the
generators $W^{(s_i)}(z)$ on the highest weight vector $|\La\rangle$
of $\cF_\La$ are invariant under the (shifted) action of the Weyl group
$W$ of $\bfg$, \ie under (see for example \chFbla)
\eqn\chFboa{
\La \to  w(\La+\al_+\rh + \al_-\rhv) - (\al_+\rh+\al_-\rhv)
\,,\qquad \forall w\in W \ .
}
To prove this, note that for every simple root $\al_i$ of $\bfg$, the
eigenvalue $h_i$ of $T_i(z)$ in \chFbn\
\eqn\chFbob{
{\textstyle{1\over4}}
(\La + \al_+\al_i + \al_-\al_i^\vee, \al_i^\vee) (\La, \al_i)
}
as well as the numbers $(\be,\La)$ for $(\be,\al_i)=0$, are invariant
under the shifted action of $w=r_i$, \ie the reflection in the simple
root $\al_i$. We conclude that the $w_{s_i}(\La)$ are invariant under
the group generated by all these simple reflections, which is precisely
the Weyl group of $\bfg$.\bs

To make the previous discussion somewhat more explicit we will now
illustrate the above in the context of a free field realization of the
affine Kac-Moody algebra $\whg$.
Introduce a set of bosonic first order fields $(\be^\al(z),\ga^\al(z))$
of conformal dimension $(1,0)$ for every positive root $\al\in\De_+$
of $\bfg$, and a set of $\ell={\rm rank}(\bfg)$ scalar fields
$\ph^i(z)$. Introduce the bosonic Fock spaces $\cF^{\ph\be\ga}_\La =
\cF_{\al_+\La}^\ph\otimes \cF^{\be\ga}$, where the vacuum
$|\La\rangle$ is labelled by the scalar zero modes as
$p^i|\La\rangle = \al_+ \La^i |\La\rangle$. Then we have a realization
of the affine Kac-Moody algebra $\whg$ on $\cF^{\ph\be\ga}_\La$ with
highest weight $\La$ and level $k$ \cite{\Wak,\FFrc,\BMPa,\BMPb}
(see Section 2.3.1 for the case of $sl(2)$).
For our purposes it suffices to recall the explicit expression for the
Cartan subalgebra generators in the Chevalley basis
\eqn\chFbaa{
h_i(z) = -\al_- (\al_i^\vee \cdot i\p\ph(z) ) +
\sum_{\al\in\De_+} (\al,\al_i^\vee) (\ga^\al\be^\al)(z) \ .
}
The Virasoro algebra acts on $\cF^{\ph\be\ga}_\La$ by means of the
Sugawara construction, which in this specific realization takes the form
\eqn\chFbab{
T^{Sug}(z) = -\half (\p\ph\cdot\p\ph)(z) - \al_+\rh\cdot i\p^2\ph(z)
- \sum_{\al\in\De_+} (\be^\al\p\ga^\al)(z) \ .
}
This free field realization comes naturally equipped with a
set of screening operators
\eqn\chFbp{
s_i^+(z) = (\be^{\al_i}(z) + \ldots ) \exp{(-i\al_+ \al_i\cdot\ph(z))}\,,
\qquad\qquad i=1,\ldots,\ell
}
where the dots stand for terms of higher order in $\be\ga$-fields.
These screening operators are primary fields of conformal dimension
one (under \chFbab) and satisfy the important property that their
operator product expansion with the affine currents is at most
a total derivative. Consequently, they can be used to construct
intertwining operators between Fock space modules.
Specifically, the operator
\eqn\chFbq{
Q = \int_{{\cal C}} dz_1\ldots dz_n\, s_{i_1}^+(z_1)
\ldots s_{i_n}^+(z_n)
}
will be an intertwining operator, provided the contour is closed in the
homology of the local system determined by the multivalued integrand
of \chFbq. We will come back to this point in Section 6.4.\smallskip

In terms of the above free field realization the modified stress energy
tensor \chFbe\ takes the following form
\eqn\chFbac{
T(z) = T^\ph(z) + T^{\be\ga b c}(z) \ ,
}
where
\eqn\chFbad{ \eqalign{
T^\ph(z) & = -\half (\p\ph\cdot\p\ph)(z) - (\al_+\rh + \al_-\rhv)\cdot
i\p^2\ph(z) \cr
T^{\be\ga bc}(z) & = \sum_{\al\in\De_+} \left( ( (\rhv,\al)-1) (b^\al\p
c^\al + \be^\al\p\ga^\al)(z) + (\rhv,\al)(\p b^\al c^\al +
\p\be^\al\ga^\al)(z) \right)\ .  \cr}
}
Similarly the field $\tilde{h}_i(z)$, in the cohomology of $Q_0$,
takes the form
\eqn\chFbae{
\tilde{h}_i(z) = \al_i^\vee\cdot i\p\ph(z) + \al_+
\sum_{\al\in\De_+} (\al,\al_i^\vee) (b^\al c^\al + \be^\al\ga^\al)(z) \ .
}
One can show, again by using a spectral sequence argument, that
\cite{\BOa,\DF,\Fi,\FFra}
\eqn\chFbaf{
H_Q(\cF^\ph_{\al_+\La} \otimes \cF^{\be\ga} \otimes \cF^{bc} ) \cong
\cF^\ph_{\al_+\La} \otimes H_Q(\cF^{\be\ga}\otimes \cF^{bc}) \cong
\cF^\ph_{\al_+\La}
}
where, in the first equality we have used the fact that $Q$ does not
depend on the scalar fields $\ph^i(z)$ and in the second that
$H_Q(\cF^{\be\ga} \otimes \cF^{bc}) \cong \CC$ which is,
essentially, the familiar quartet decoupling mechanism.
Using this isomorphism one can argue that, in fact, in $Q$-cohomology
we have the following equivalences (see \eg \cite{\BOa})
\eqn\chFbag{
T(z) \sim T^\ph(z)\,,\qquad\qquad \tilde{h}_i(z) \sim \al_i^\vee\cdot
i\p\ph(z) \,,\qquad\qquad s_i^+(z) \sim \tilde{s}_i^+(z) \ .
}
This explains in more detail the origin of the equations \chFbl\ and
\chFbk.

Finally, it follows from \chFbad, or equivalently from \chFbe,
that the conformal dimension of
the Fock space representation $\cF_{\al_+\La}$ obtained by the reduction
\chFbaf\ is given by
\eqn\chFbah{
h_{\al_+\La} = \half \al_+^2 (\La,\La+2\rh) - (\La,\rhv) =
\half (\al_+\La,\al_+\La + 2(\al_+\rh + \al_-\rhv)) \ ,
}
in agreement with \chFbla.

The description of the $\cW$-algebra as the centralizer
of a set of screening charges can now also be proved
without making use of
the spectral sequence corresponding to the decomposition $Q=Q_0+Q_1$
\cite{\Freb}. To this end one uses
the isomorphism of fields and states in the characteristic
Hilbert space (or `vacuum module')
$\cal H$ of the $\cW$-algebra (compare with Section 3.1).
Then one constructs,
for generic level $k$, a resolution $(C^{(i)}{\cal H}^g, d^{(i)})$ of
the characteristic Hilbert
space ${\cal H}^g$ of $\whg$ in terms of free field Fock spaces.
Recall that a resolution $(C^{(i)}L,d^{(i)})$ of a
$\whg$-module $L$ is a complex
$d^{(i)}: C^{(i)}L\to C^{(i+1)}L\,, d^{(i+1)}d^{(i)}=0$
of $\whg$-modules $C^{(i)}L$, where the
differentials $d^{(i)}$ commute with the action of $\whg$, and is such
that the cohomology of this complex is exactly the module $L$,\ ie
\eqn\chFbaj{
H_d^{(i)}( L) \cong \cases{ L & if $\ i=0$ \cr 0 & otherwise . \cr}
}
The terms in the resolution of ${\cal H}^g$
are labelled by elements of the Weyl group $W$ of $\bfg$
\eqn\chFbai{
C^{(i)}{\cal H}^g \cong \bigoplus_{\{ w\in W| \ell(w)=i\} }
\cF_{w\rh-\rh}^{\ph\be\ga}
}
and the differentials $d^{(i)}$ are operators of the type \chFbq.

Applying the functor $H_Q$ to the resolution \chFbai, using \chFbaf,
one obtains a resolution $(C^{(i)}{\cal H},\tilde{d}^{(i)})$
of the characteristic Hilbert space $\cal H$ of $\cW[\whg,k]$
\cite{\Freb} (see also \cite{\Nic}), with terms
\eqn\chFbak{
C^{(i)}{\cal H} \cong \bigoplus_{ \{ w\in W | \ell(w) = i\} }
\cF^\ph_{ \al_+( w\rh-\rh)}
}
and differentials $\tilde{d}^{(i)}$ as in the complex \chFbai,
but with $s_i^+(z)$ replaced by $\tilde{s}_i^+(z)$ (see \chFbag).
In particular the differential $\tilde{d}^{(0)}$ will be the
collection of
screening charges $\oint \tilde{s}_i^+(z)$ acting from $\cF_0^\ph
\to \cF_{-\al_+\al_i}^\ph$, and the intersection of its kernels is
isomorphic to $\cal H$. So, again, we are led to the identification
of $\cW[\whg,k]$ with the centralizer of the set of screening charges
$\oint \tilde{s}_i^+(z)$.\bs

%---------------------------------------------------------------------

\noindent{\sl 6.3.2. Character technique}\hfill\bs

In the above we have seen that, for generic values of the level $k$,
the algebra $\cW[\whg,k]$ can be identified  with the centralizer
of a set of screening charges $Q_i^+ = \oint \exp{(-i\al_+ \al_i\cdot
\ph(z))}$. We have shown that $\cW[\whg,k]$ contains at least the
Virasoro algebra, and have given an explicit expression for the
generating field (see \chFbl). We now want to discuss a useful technique
(which we will also often employ in Chapter 7) to get some insight into
the other generators of $\cW[\whg,k]$. This technique will be referred to
as the ``character technique'' \cite{\Bod,\Boc}.

Suppose we have some $\cW$-algebra of rank $\ell$
and type $\cW(2,s_2,\ldots, s_\ell)$.
The corresponding Verma modules $M(\{h\},c)$ are completely specified by
the eigenvalues $h^{(s_i)}\,,i=1,\ldots,\ell$ of the zero modes
$W^{(s_i)}_0$, and have a character (we will abbreviate $h^{(2)}=h$)
\eqn\chFbba{
{\rm ch\,}_{M}(q) = {\rm Tr\,}_M\ q^{L_0 - {c\over24}} =
{q^{h- {c\over24}} \over \left( \prod_{k\geq1} (1-q^k) \right)^\ell } \ .
}
Conversely, given a character \chFbba, we can anticipate the existence
of a $\cW$-algebra of rank $\ell$. Equation \chFbba, however, does
not give us any information on the conformal dimensions $s_i$ of the
generators. In general, the Verma modules $M(\{h\},c)$ will not be
irreducible,
due to the presence of singular vectors. The singular vector structure
might
be extremely complicated. However, there is one case where one knows
some singular vectors beforehand. That is when we consider the Verma
module built on the vacuum $|0\rangle$ of the conformal field theory
(\ie $h^{(s_i)}=0\,,\forall i$). In that case, all the vectors
\eqn\chFbbb{
W^{(s_i)}_n |0\rangle\,,\qquad\qquad n\geq -s_i+1\,,\ i=1,\ldots,\ell
}
will be singular (see \chCab). These singular vectors generate a
submodule $SM(0,c)$. Clearly, the factor module $\widetilde{M} =
M(0,c)/SM(0,c)$ will have a character
\eqn\chFbbc{
{\rm ch\,}_{\widetilde{M}} (q) = q^{-{c\over24}} \prod_{i=1}^\ell \left(
{(1-q)\ldots (1-q^{s_i-1}) \over \prod_{k\geq1} (1-q^k) }\right) =
{ q^{-{c\over24}} \over \prod_{i=1}^\ell (F_{s_i}(q))}
}
where we have introduced
\eqn\chFbbd{
F_{s}(q) = \prod_{k\geq s} (1-q^k) \ .
}
It may of course happen that the factor module
$\widetilde{M} = M(0,c)/SM(0,c)$ still
contains singular vectors, in which case the `true' (\ie irreducible)
vacuum module $\cal H$ (\ie the characteristic Hilbert space)
is even smaller. If, however, for generic $c$-values the
generators $W^{(s_i)}(z)$ are independent, then \chFbbc,
for generic $c$-values, will be the character of the vacuum module
$\cal H$.\foot{It can happen that, even for generic $c$-values, the
vacuum Verma module contains additional singular vectors beyond those
displayed in \chFbbb. Suppose the lowest one occurs at $L_0$-level $N$.
Then ${\rm ch\,}_{\cal H}$ will have the form \chFbbc\
multiplied by a factor $(1-a_N q^N - a_{N+1} q^{N+1} -\ldots)$ where
$a_k\in\ZZ_+$. This happens for instance in the bosonic projection of
the $N=1$ superconformal algebra \cite{\Boc} (see Section 5.3.2).}
Clearly, given the character \chFbbc, we can anticipate an
underlying $\cW$-algebra of type $\cW(2,s_2,\ldots,s_\ell)$.

Let us now apply these considerations to the QDS reduction. In the
Section 6.3.1 we have mentioned the existence of a resolution of
the characteristic Hilbert space
$\cal H$ in terms of Fock space modules (see \chFbak).
This resolution
allows us to compute the character of $\cal H$ by means of the
Euler-Poincar\'e principle. For $w\in W$ we have (see \chFbah)
\eqn\chFbbe{
h_{\al_+(w\rh-\rh)}  = - (w\rh-\rh,\rhv) \ .
}
Thus, we obtain
\eqn\chFbbf{ \eqalign{
{\rm ch\,}_{\cal H} (q) & = \sum_{i} (-1)^i {\rm ch\,}_{C^{(i)}
{\cal H}} (q) =  q^{-{c\over24}} \sum_{w\in W} \ep(w)
{\rm ch\,}_{\cF_{\al_+(w\rh-\rh)}}(q) \cr
& =
{q^{-{c\over24}}\over \left( \prod_{k\geq1} (1-q^k)\right)^\ell }
\sum_{w\in W} \ep(w) q^{-(w\rh-\rh,\rhv)} \cr}
}
which, by the Weyl-Kac denominator formula (see Appendix A)
\eqn\chFbbg{
\sum_{w\in W} \ep(w) e^{w\rh-\rh} = \prod_{\al\in\De_+} (1-e^{-\al}) \ ,
}
can be rewritten as
\eqn\chFbbh{ \eqalign{
{\rm ch\,}_{\cal H} (q)
& = q^{-{c\over24}} { \prod_{\al\in\De_+} (1- q^{(\rhv,\al)})
\over \left( \prod_{k\geq1} (1-q^k) \right)^\ell } =
q^{-{c\over24}}{ \prod_{i=1}^\ell \left( \prod_{k=1}^{e_i} (1-q^k)
\right)\over \left( \prod_{k\geq1} (1-q^k) \right)^\ell } \cr
& =
{ q^{-{c\over24}} \over F_{e_1+1}(q) \ldots F_{e_\ell+1}(q) } \ .
\cr}
}
Here we have used the fact that $\rhv$ is precisely the defining vector
of the principal $sl(2)$ subalgebra of $\bfg$.
Under this subalgebra the adjoint
representation of $\bfg$ decomposes into $sl(2)$ representations of spin
$e_i\,,i=1,\ldots,\ell$, where $\{ e_i\}$ are the exponents of $\bfg$.
{}From the above discussion, we would thus expect $\cW[\whg,k]$ to be a
$\cW$-algebra of type $\cW(s_1,\ldots,s_\ell)$ where the dimensions
$s_i=e_i+1$ of the generating fields are exactly the orders of the
independent
Casimirs of $\bfg$ (see appendix A for a list), in agreement with the
classical result \cite{\DS,\BFFOWb} and our intuitive reasoning in
Section 6.3.1. Moreover, if we manage to construct a set of fields of
conformal dimensions
$s_i =e_i +1$ in the intersection of the kernels of the screening charges
and succeed in proving that, for generic values of $k$, they are
independent, then the above analysis will prove the completeness of this
set as well as the closure of the algebra that they generate.
This is the strategy we will employ in the examples of
Section 6.3.3.

In the application of this `character technique' to coset models of CFT,
to be discussed in Chapter 7, we do not have a priori knowledge about the
irreducible vacuum modules of the presupposed $\cW$-algebras.
Nevertheless, one can often still get useful information
from the known characters of the
irreducible highest weight modules of the affine Lie algebras $\whg$.
As an illustration, consider again the QDS reduction. For
$\bfg$ simply-laced, and at the special point $\al_\pm=\pm1$, we can
identify the screening operators $s_i^\pm(z)$ with the expressions for
the simple root generators (in the Chevalley basis) of the
vertex operator realization of $\whg$ at level $k=1$.\foot{Of course,
this identification holds up to cocycle factors. These, however, do not
influence the structure of the centralizer.}
Thus, for this particular value of $\al_\pm$, the centralizer of the
screening charges $Q_i^\pm = \oint \tilde{s}_i^\pm(z)$ can be identified
with the subset of normal ordered products of the affine
currents and their derivatives
that are singlets under the horizontal algebra $\bfg$.
This set of `singlets' can be studied by decomposing the irreducible
$\whg$ characters under $\bfg$. This, for simply laced $\bfg$, again
leads to the expression \chFbbh\ \cite{\Bod,\Boc}.
We will come back to this point in Chapter 7.
\bs

%----------------------------------------------------------------------

\noindent{\sl 6.3.3. Examples}\medskip

In this section we will present some general examples of $\cW$-algebras
obtained through the quantum Drinfeld-Sokolov reduction.
Although the previous discussion has been restricted to bosonic
Lie algebras,
most of the results easily carry over to the case of Lie superalgebras.
Instead of developing this here,
we will just present two examples at the end of this section and refer to
the original papers for more details.
\bs

\centerline{--- $\cW[A_n^{(1)},k]$ ---}\medskip

Since the results for the Lie algebra $A_n \simeq sl(n+1)$ are the most
complete we will treat this example in detail and use it as a
reference for the other examples to be discussed later.

Let $\{ \ep_i\,,i=1,\ldots,n+1\}$ be the set of weights of the vector
representation of $A_n$, normalized such that $\ep_i\cdot\ep_j =
\de_{ij} - {1\over n+1}$. They satisfy the constraint $\sum \ep_i =0$.
The simple roots of $A_n$ are given by $\al_i=\ep_i-
\ep_{i+1}$. Consider a set of currents $\{ U_k(z)\}$, of conformal
dimension $k$, defined through the generating expression (``quantum
Miura transformation'').
\eqn\chFca{
\eqalign{
R_{n+1}(z) & = - \sum_{k=0}^{n+1} U_k(z) (\al_0\p)^{n+1-k} \cr
& = ((\al_0\p_z - \ep_1\cdot i\p\ph(z))\ldots (
\al_0\p_z - \ep_{n+1}\cdot i\p\ph(z))) \ .
\cr}
}
We have for example
\eqn\chFcc{
\eqalign{
U_0(z) & = 1 \cr
U_1(z) & = \sum_i \ep_i\cdot i\p\ph(z) = 0 \cr
U_2(z) & = -\sum_{i<j} ((\ep_i\cdot i \p\ph)(\ep_j\cdot i\p\ph))(z)
+ \al_0 \sum_i (i-1) \ep_i\cdot i\p^2\ph(z) \cr
& =- \half( \p\ph\cdot \p\ph)(z) - \al_0 \rh\cdot i\p^2\ph (z) =
T^\ph(z)\cr
U_3(z) & = \sum_{i<j<k} ((\ep_i\cdot i\p\ph)(\ep_j\cdot i\p\ph)
(\ep_k\cdot
i\p\ph))(z) - \al_0 \sum_{i<j} (i-1) \p( (\ep_i\cdot i\p\ph)(\ep_j\cdot
i\p\ph))(z) \cr
& - \al_0 \sum_{i<j} (j-i-1) ( (\ep_i\cdot i\p\ph) (\ep_j\cdot
i\p^2\ph))(z)
+ \half \al_0^2 \sum_i (i-1)(i-2) (\ep_i\cdot i\p^3\ph)(z) \cr}
}
where we have used $\sum_i \ep_i =0$,
$\sum_{i<j} \ep_i\otimes\ep_j = -\half 1\!\!1$, and
$\sum_i i\ep_i = -\rh$.

Now consider the singular part in the OPE $R_{n+1}(z)
\tilde{s}^\pm_i(w)$.
We will show that it is a total derivative for each $i\in\{1,\ldots,n\}$,
which clearly implies that the fields $U_k(z)$ are in the centralizer
of the screening charges $Q^\pm_i$.
Since $\ep_j\cdot\al_i\neq0$ only for $j\in\{ i,i+1\}$, the only terms
contributing to the singular part of the OPE are
\eqn\chFdcf{\eqalign{
(\al_0\p_z & - \ep_i\cdot i\p\ph(z))(\al_0\p_z - \ep_{i+1}\cdot
i\p\ph(z)) e^{-i\al_\pm\al_i\cdot\ph(w)} \cr
& = \left( {-\al_\pm^2 \over (z-w)^2} + (\al_0\p_z  -
\ep_i\cdot i\p\ph(z))
\left( {-\al_\pm\over z-w} \right) \right. \cr
& \ \ \left. + \left( {\al_\pm\over z-w} \right)
(\al_0\p_z  - \ep_{i+1}\cdot i\p\ph(z)) \right)
e^{-i\al_\pm\al_i\cdot\ph(w)} \cr
& = \left( {-\al_\pm^2+\al_0\al_\pm \over (z-w)^2}
+ { i\al_\pm(\ep_i-\ep_{i+1})\cdot\p\ph(w)\over z-w} \right)
e^{-i\al_\pm\al_i\cdot\ph(w)} \cr
& = -\p_w \left( {1\over z-w}  e^{-i\al_\pm\al_i\cdot\ph(w)} \right) \cr}
}
provided $\al_\pm^2 - \al_0\al_\pm =1$, \ie$ \al_0=\al_++\al_-$.
This proves that
\eqn\chFcd{
R_{n+1}(z) \tilde{s}^\pm_i(w) = -\p_w \left( {1\over z-w} R^i_{n+1}(z)
e^{-i\al_\pm\al_i\cdot\ph(w)} \right) \ ,
}
where $R^i_{n+1}(z)$ is defined as in \chFca\ but with the $\ep_i$ and
$\ep_{i+1}$ terms removed.

We would now like to argue that, for generic values of $\al_0$, the
fields $U_k(z)$ are independent. This would, in view of the results
of Section 6.3.2, establish that the $U_k(z)$ form a complete set of
generators for $\cW[A_n^{(1)},k]$ as well
as the closure of the algebra of
$U_k(z)$'s. To this end let us compute the eigenvalues $u_k(\La)$ of the
zero modes $U_{k,0}$ on the highest weight state of the Fock space
$\cF_\La$. If we can show that the resulting polynomials in $\th_i =
(\La,\ep_i)$ are independent then this evidently will imply the
independence of the generators $U_k(z)$. Applying the operator
$R_{n+1}(z)$ (see \chFca) to the highest weight state $|\La\rangle$ we
find
\eqn\chFcda{
-\sum_{k=0}^{n+1} u_k(\La) z^{-k} (\al_0\p)^{n+1-k} =
( \al_0\p - {1\over z}(\La,\ep_1))\ldots ( (\al_0\p - {1\over z}(\La,
\ep_{n+1})) \ .
}
By applying this differential operator to the monomials
$z^j\,,j=0,\ldots n+1$ we find the following recurrence relation for the
eigenvalues $u_k(\La)$
\eqn\chFcdb{
\sum_{k=0}^j {j!\over(j-k)!} \al_0^k u_k(\La) =
(-1)^n \prod_{k=1}^{n+1} ( (\La+\al_0\rh, \ep_k) +
(\half n -j)\al_0 ) \ ,
}
where we have used $(\rh,\ep_k) = \half (n+2-2k)$.
Its solution is (see \eg \cite{\Bic})
\eqn\chFcdc{
u_k(\La) = (-1)^{k-1} \sum_{i_1<\ldots <i_k} \prod_{j=1}^{k}
\left( (\La,\ep_{i_j}) + (k-j)\al_0 \right) \ .
}
Concretely, we have for example
\eqn\chFcde{ \eqalign{
u_2(\La)  & = -\sum_{i_1< i_2} \th_{i_1}\th_{i_2} -
{\textstyle{1\over4}} \left( \matrix{n+2\cr 3\cr}\right) \al_0^2
= \half (\La,\La+2\al_0\rh)\,, \cr
u_3(\La) & = \sum_{i_1<i_2<i_3} \th_{i_1}\th_{i_2}\th_{i_3} +
(n-1) \al_0 \sum_{i_1<i_2} \th_{i_1}\th_{i_2} + \left( \matrix{
n+2\cr 4\cr} \right) \al_0^3\,,\cr}
}
where $\th_i \equiv (\La+\al_0\rh,\ep_i)$.
In general, as one can see from \chFcdb, the eigenvalues $u_k(\La)$
are symmetric polynomials in the variables $\th_i$.
Since the Weyl group of $A_n$ acts as
the permutation group $S_{n+1}$ on the vectors $\ep_i$,
the numbers $u_k(\La)$ are invariant under the (shifted) action of the
Weyl group $W$ of $A_n$, \ie $u_k(\La) = u_k( w(\La+\al_0\rh) -
\al_0\rh)$.

In particular, for $\al_0=0$ we discover that $u_k(\La)$ is simply the
standard (homogeneous)
symmetric polynomial of order $k$ in the variables $\th_i$.
These are well-known to be independent.
This fact can be used to prove
the independence of the generators $U_k(z)$ in an open
neighborhood of $\al_0=0$, and hence the independence for generic
$\al_0$ (\ie for a dense subset of $\CC$)
\cite{\Na,\Wac}. This proves,
as remarked in Section 6.3.2, the closure of the algebra generated
by the fields $U_k(z)\,,k=2,\ldots,n+1$.

By explicit calculation it has been shown that the currents
$U_k(z)$ satisfy an algebra with quadratic defining relations
\cite{\Lu,\FLc}
\eqn\chFcdI{
U_k(z) U_l(w) = \sum_{\ka\geq2} {1\over (z-w)^\ka}
\sum_{p+q=k+l-\ka} C_{kl}^{pq}(\ka)( U_p(z)U_q(w)) \,,
}
where the coefficients
$C_{kl}^{pq}(\ka)$ are algebraic in $\al_0$ and therefore do not have
any singularities.

Note, however, that
the generators of the algebra in the Miura basis are not
primary nor quasiprimary, in general. A basis for the $\cW$-algebra
consisting of (quasi)-primary fields  $\widetilde{U}
_k(z)$ should be obtainable by deformation  of the fields
$U_k(z)$. The projection of $U_k(z)$ onto a quasi-primary field
$\widetilde{U}_k(z)$ is easily accomplished (using \eg
\cite{\BFKNRV}),
and only involves coefficients which are algebraic in $\al_0$.
One has \eg
\eqn\chFce{
\widetilde{U}_3(z) = U_3(z) - \left( {n-1\over2} \right)
   \al_0\p U_2(z) \,,
}
with eigenvalue (use \chFcde)
\eqn\chFcey{
\tilde{u}_3(\La) = \sum_{i_1<i_2<i_3} \th_{i_1}\th_{i_2}\th_{i_3}\,.
}
In this case the resulting field $\widetilde{U}_3(z)$
is, in fact, primary. In general, the
possibility of further projection onto a set of primary field generators
is still an important open problem, and has only been checked in low
spin cases.
The  analogous problem in the classical case has been solved
in \cite{\BFFOWb}. Explicit formulas were obtained in \cite{\DIZ}.
Clearly, in the quantum case, the primary field projection may brake
down for isolated values of $c$, as manifested by the occurrence of
singularities in the operator product expansion coefficients
in Chapter 4 \cite{\Nic,\Nid}.
This does not yet explain the singularity at $c=-22/5$ in the OPE
of two dimension three fields (see \chCc, \chCcII),
since the deformation
\chFce\ is purely algebraic. However, there is another potential source
of singularities, namely, the ones that arise
when the normalization factor required to
bring the fields $\widetilde{U}_k(z)$ to their standard normalization
$\vev{\widetilde{U}_k|
\widetilde{U}_k} = c/k$ becomes singular. Indeed, the normalization
factor required for \chFce\ becomes singular for $c=-22/5$.
As a final remark we note that in terms of the primary fields
$W^{(k)}(z)$ the defining relations of the algebra are no longer
quadratic in general. Explicit operator product expansions in the
$\cW[A_2^{(1)},k]$-case were given in \chCc, and those for
$\cW[A_3^{(1)},k]$ can be found in \cite{\BFKNRV,\KWa}.
\smallskip

Let us, for definiteness, display the generators of the $\cW_3 \cong
\cW[A_2^{(1)},k]$ algebra even more explicitly. Choose thereto an
orthonormal basis with respect to which the simple roots
have the following coordinates: $\al_1 = \sqrt2(1,0)\,,\al_2 =
\sqrt2 (-\half,\half\sqrt3)$. Then
\foot{The expressions are related to those in
\cite{\FZb} by a Weyl reflection.}
\eqn\chFcez{\eqalign{
T(z) & = U_2(z)= \half (H^1H^1 + H^2H^2) - \half \sqrt2 \al_0
\left(  \p H^1 +\sqrt3\p H^2 \right) \cr
\widetilde{U}_3
(z) & = {1\over3\sqrt6} ( 3H^1H^1H^2 - H^2H^2H^2) - \al_0 (
\half H^1\p H^1 + \onethird\sqrt3 H^2\p H^1 - \half H^2\p H^2) \cr
& \ \   + \textstyle{{1\over4}} \sqrt2
\al_0^2 ( \p^2 H^1 - \textstyle{{1\over3}} \sqrt3 \p^2 H^2) \ , \cr}
}
where we have introduced $H^i(z) = i\p\ph^i(z)$. The normalized generator
$W(z) = \sqrt{3\be} \widetilde{U}_3(z)$, where $\be = 16/ (5c+22) =
2/(4-15\al_0^2)$, satisfies the algebra of \chCc.
Note that normalization factor indeed becomes singular for $c = -22/5$.
\bs

\centerline{ --- $\cW[D_n^{(1)},k]$ ---}\bs

For the Lie algebra $D_n\simeq so(2n)$ we proceed similarly.
Let $\{ \pm\ep_i\,,i=1,\ldots,n\}$ denote the weights of the vector
representation of $D_n$, normalized such that $\ep_i\cdot
\ep_j = \de_{ij}$. The simple roots are given by $\al_i=\ep_i-\ep_{i+1}$
for $i=1,\ldots,n-1$ and $\al_n=\ep_{n-1}+\ep_n$.
Define
\eqn\chFcf{
R_n(z) = \ ((\al_0\p_z - \ep_1\cdot i\p\ph(z)) \ldots
(\al_0\p_z - \ep_{n}\cdot i\p\ph(z))) \ .
}
A similar calculation
as for $\cW[A_n^{(1)},k]$ shows that the singular part of the OPE between
$R_n(z)$ and
$\tilde{s}_i^\pm(w)$ is a total derivative for $i\in\{1,\ldots,n-1\}$.
However,
\eqn\chFcfa{
\eqalign{
& (\al_0\p_z - \ep_{n-1}\cdot i\p\ph(z))(\al_0\p_z - \ep_n\cdot
i\p\ph(z)) e^{-i\al_\pm\al_n\cdot\ph(w)}
\cr
& \qquad = \p_w \left( {1\over z-w} e^{-i\al_\pm\al_n\cdot\ph(w)} \right)
+ {2\al_\pm\over z-w} e^{-i\al_\pm\al_n\cdot\ph(w)} (\al_0\p_z) \ .
\cr}
}
So, we conclude that for $i=n$ only the highest component $V_n(z)
=R_n(z)\cdot 1$ is in the centralizer of all the screening charges.
Upon taking the OPE of $V_n(z)$ with itself one clearly generates other
currents in this centralizer. Explicitly \cite{\FLc,\FLd},
\eqn\chFcg{
V_n(z) V_n(w) = {a_n\over (z-w)^{2n}} + \sum_{k=1}^{n-1}
{a_{n-k}\over (z-w)^{2(n-k)} } \left( U_{2k}(z) + U_{2k}(w) \right)
}
for some fields $U_{2k}(z)$. It is convenient to choose $a_k=
\prod_{j=1}^{k-1} (1-2j(j+1)\al_0^2)$. Then, for example,
\eqn\chFch{
U_2(z) = -\half (\p\ph\cdot\p\ph)(z) - \al_0 \rh\cdot i\p^2\ph(z)
= T^\ph(z) \ .
}
The eigenvalue $v_n(\La)$ of the zero mode $V_{n,0}$ on the highest
weight vector $|\La\rangle\in\cF_\La$ follows from \chFcf
\eqn\chFcha{
v_n(\La) = (-1)^n \prod_{k=1}^n (\La+\al_0\rh,\ep_k)
}
and is again invariant under the (shifted) action of the Weyl group
$\La\to w(\La+\al_0\rh)- \al_0\rh$.
By using \chFcg\ one may derive similar formulas for the eigenvalues
$u_{2k}(\La)$, and show explicitly their invariance under the shifted
action of the Weyl group \cite{\FLc,\FLd}. By exploiting these
expressions  one establishes the independence of the
generators $\{ V_n(z)\} \cup \{ U_{2k}(z)\,,k=1,\ldots,n-1\}$,
exactly as in the case of $\cW[A_n^{(1)},k]$.
Noting that the set of dimensions $\{ 2,4,\ldots,2n-2,n\}$ indeed agrees
with the orders of the independent Casimirs of $D_n$, this establishes
the closure of the algebra generated by $V_n(z)$ and $U_{2k}(z)$.
\bs

\centerline{--- $\cW[B_n^{(1)},k]$ ---}\bs

In the case of the Lie algebra $B_n\simeq so(2n+1)$ one may again try to
proceed similarly. Introduce vectors $\ep_i\,,i=1,\ldots,n$ as in the
case of $D_n$, in terms of which the simple roots of $B_n$ are given by
$\al_i=\ep_i-\ep_{i+1}$ for $i=1,\ldots,n-1$, and $\al_n=\ep_n$. Introduce
$R_n(z)$ as in \chFcf. As before, the singular part of the
OPE of $R_n(z)$
with $\tilde{s}_i^\pm(w)$ is a total derivative for
$i=1,\ldots,n-1$, but
\eqn\chFci{
(\al_0\p_z - \ep_n\cdot i\p\ph(z)) e^{-i\al_\pm\al_n\cdot \ph(w)} =
{ \al_\pm\over z-w}  e^{-i\al_\pm\al_n\cdot \ph(w)} \ .
}
Obviously, none of the components in $R_n(z)$ commutes with all
the screening charges and it is not clear how to proceed.
As argued in Section 6.3.2 one expects a $\cW$-algebra of the type
$\cW(2,4,6,\ldots,2n)$, realized in terms of $n$ scalar fields.
In particular, for $n=2$, one would expect to
recover the (unique) algebra of type $\cW(2,4)$ (see Chapter 5). This
has recently been confirmed by explicit construction of the two
generators and verification of their operator product expansion
\cite{\KWb}. For $n=3$ it has been shown by explicitly constructing
all fields in the centralizer
of the screening charges of dimension less or equal to seven
that the algebra is of type $\cW(2,4,6,\ldots)$ \cite{\KWc}.
Their OPEs seem
to lead to the structure constants of the third $\cW(2,4,6)$ algebra
of \cite{\KWa},
but no definite answer is available at the time of writing.%
\foot{This particular $\cW(2,4,6)$ algebra is however
claimed to be inconsistent in \cite{\Flo,\Kli}.}

Inspection of \chFci\ suggests another interesting possibility, to be
discussed below.\bs

\centerline{--- $\cW[B(0,n)^{(1)},k]$ ---}\bs

The free field realization of the
Lie superalgebra $B(0,n)^{(1)}$ leads, similar to the bosonic case of
Section 6.3.1, to a description of the $\cW$-algebra
$\cW[B(0,n)^{(1)},k]$ as the centralizer of a set of screening
charges $s_i(z)$ on the Fock space of $n$ scalar fields $\ph^i(z)$
and one fermionic field $\ps(z)$ \cite{\Ita}.
The screening operators $s_i(z)\,,i=1,\ldots,
n-1$ are as for $B_n^{(1)}$, while $s_n(z) = \ps(z)e^{-i\al_\pm
\ep_n\cdot\ph(z)}$.
The occurrence of the fermion in $s_n(z)$  has an important consequence.
Considering the OPE
\eqn\chFcja{
\eqalign{
& (\al_0\p_z - \ep_n\cdot i\p\ph(z))\ps(z)
  \ps(w)e^{-i\al_\pm \ep_n\cdot\ph(w)} \cr
& \qquad = \p_w \left( {-\al_\mp\over z-w}
 e^{-i\al_\pm \ep_n\cdot\ph(w)}\right)
  + {1\over z-w} e^{-i\al_\pm \ep_n\cdot \ph(w)} (\al_0\p_z)\ ,
\cr}
}
we observe that the operator $U_{n+{1\over 2}}(z) =
R_n(z)\cdot \ps(z)$ ($R_n(z)$ as in \chFcf),
of conformal dimension $n+\half$, is part of the centralizer. By taking
the OPE of $U_{n+{1\over2}}(z)$ with itself one generates a set of fields
$\{ U_{2k}(z)\,,k=1,\ldots,n\}$ analogous to the $\cW[D_n^{(1)},k]$-case.
In particular,
\eqn\chFcj{
U_2(z) = -\half (\p\ph\cdot\p\ph)(z) -\al_0 \rh\cdot i \p^2\ph(z)
- (\ps\p\ps)(z) \ .
}
It has been conjectured that $\{ U_{n+{1\over2}}(z)\} \cup
\{ U_{2k}(z)\,,k=1,\ldots,n\}$ is a complete set of generators of
$\cW[B(0,n)^{(1)},k]$
\cite{\FLc} (in this ref. the algebra was denoted by $\cW B_n$).
For additional details see also \cite{\Ita,\Wab,\Wad,\BOb,\FST,\Aa}.
\bs

\centerline{--- $\cW[A(n,n-1)^{(1)},k]$ ---}\bs

Another interesting (super) $\cW$-algebra is obtained from the
Drinfeld-Sokolov reduction of the Lie superalgebra $A(n,n-1)\simeq
sl(n+1,n)$. At the classical level this reduction was studied in
\cite{\Ina,\FRa,\Frb,\EH,\KMN,\IKa,\IKb} and was shown to give rise to an
$N=2$ $\cW$-algebra. The simplest quantum
case, $n=1$, was analyzed in detail in
\cite{\BOb} and was shown to lead to the $N=2$ superconformal algebra.
The general quantum case was analyzed
in \cite{\Itb,\Itc,\NY}. For $n=2$ this algebra was explicitly
constructed in \cite{\Od,\Roc} by imposing the
associativity constraints. This $\cW$-algebra is supposedly the
chiral algebra of the $\CC P_n\simeq SU(n)/SU(n-1)\times U(1)$
Kazama-Suzuki model.

The affine super algebra $A(n,n-1)^{(1)}$ has a realization in terms
of $2n$ free scalar fields $\ph^i(z)$,
and conjugate pairs of spin $(0,1)$
bosonic (fermionic) first order fields $(\ga^\al(z),\be^\al(z))$
($ (\xi^\al(z),\et^\al(z))$) corresponding to the positive even (odd)
roots $\al\in\De^0_+$ ($\al\in\De^1_+$) of $A(n,n-1)$.
As in the case of $B(0,n)^{(1)}$ the constraint is second class.
It can be treated as a first class constraint upon introducing $2n$
additional free (Majorana) fermionic fields $\ps^i(z)$.
The $2n$ screening operators of $A(n,n-1)^{(1)}$
turn out to be BRST equivalent to the
screening operators $s_i(z) = \al_i\cdot \ps(z)
e^{i\al_+\al_i\cdot\ph(z)}$,
where $\al_1,\ldots,\al_{2n}$ is a set of (fermionic) simple roots for
$A(n,n-1)$. The corresponding $\cW$-algebra can be identified with the
centralizer of the corresponding screening charges on the
$(\ph^i,\ps^i)$-Fock space.

Analogous to the bosonic case $\cW[A_n^{(1)},k]$,
the centralizer can be found
by means of a (super) Miura transformation. The corresponding Lax
operator is most easily described in $N=1$ superfield language.
Introduce a Grassmann variable $\th$, and a set of $2n$ superfields
$\Ph^i(Z) = \ph^i(z) + i\th\ps^i(z)$, $Z=(z,\th)$ as well as a
superderivative $D_Z= {\p_\th} + \th {\p_z}$. Let
\eqn\chFck{
\eqalign{
R_n(Z) & = \sum_{k=0}^{2n+1} U_{{k\over 2}}(Z) (\al_- D)^{2n+1-k} \cr
& = \left( (\al_- D + \ep_1\cdot iD\Ph(Z)) \ldots
(\al_- D + \ep_{2n+1}\cdot iD\Ph(Z) ) \right) \ . \cr}
}
The $\ep_i$ are given by $\ep_i = (-1)^i(\la_i-\la_{i-1})$, where
the $\la_i$, $i=1,2,\ldots, 2n$, are the fundamental weights satisfying
$\la_i \cdot \ep_j = \de_{ij}$ and $\la_0=\la_{2n+1}=0$.
In the superfield language the screening operators take
the form $e^{ i\al_+ \al_j \cdot \Phi}$. The singular terms in the
OPE of the Lax operator $R_n(Z)$ with one of the screening operators
arise from
\eqn\chFckb{
\eqalign{
& (\al_- D_1 + \ep_j\cdot iD\Ph(Z_1))(\al_- D_1 +
\ep_{j+1}\cdot iD\Ph(Z_1))
e^{ i\al_+ \al_j \cdot \Phi(Z_2)} \cr
& \qquad = (-1)^{j+1} D_2 \left(
{\theta_{12} \over Z_{12}} e^{ i\al_+ \al_j \cdot \Phi(Z_2)} \right)\ ,}
}
where we wrote $\th_{12}=\th_1-\th_2$, $Z_{12}=z_1-z_2-\th_1\th_2$
and we used $\al_+ \al_- =-1$. We see that all $U_{{k/2}}(Z)$,
$k=0,\ldots, 2n+1$ are in the centralizer of the screening charges.
The generators $U_0(Z)$ and $U_{1/2}(Z)$ are constant and it can be
verified that $U_1(Z)$ and $U_{3/2}(Z)$ generate an $N=2$ superconformal
algebra of central charge
\eqn\chFcka{
c = 3n \left( 1-(n+1) \al_-^2 \right)\ .
}
We thus obtain a $\cW$-algebra of type
$\sW^{(1)}(1,{3\over2},\ldots,{2n+1\over2})$ or, in terms
of its $N=2$ content, an algebra of type $\sW^{(2)}(1,2,\ldots,n)$.
\bs

\centerline{ --- $\cW[\bfg^{(r)},k]$ ---} \bs

The twisted affine Kac-Moody algebra $\bfg^{(r)}\,,r=2,3$ can be obtained
as the invariant sector under an outer automorphism $\tau$ of $\bfg$
of order $r$ \cite{\Kad}. The corresponding QDS reductions
$\cW[\bfg^{(r)},k]$ can be determined from the commutative diagram
\comdiag{
\eqn\chFcl{ \matrix{
\bfg^{(1)} & \mapright{\tau} & \bfg^{(r)} \cr
\mapdown{d} & & \mapdown{d} \cr
\cW[\bfg^{(1)},k] & \mapright{\tau} & \cW[\bfg^{(r)},k] \cr}
}}
\ie they are obtained by twisting the algebra $\cW[\bfg^{(1)},k]$.

As an example consider $\bfg = A_2$. The outer automorphism $\tau$
of order two acts on the Cartan subalgebra generators in the Cartan-Weyl
basis as
\eqn\chFcm{
\tau(H^1) =   -\half H^1 + \half\sqrt3 H^2\,,\qquad
\tau(H^2) =\half\sqrt3 H^1 + \half H^2 \ .
}
By using \chFcez\ one can explicitly check
\eqn\chFcn{
\tau( U_2(z) ) = U_2(z)\,,\qquad
\tau( \widetilde{U}_3(z) ) = - \widetilde{U}_3(z) \ .
}
The subset of invariant fields under the automorphism $\tau$ is thus
exactly the projected $\cW_3$-algebra mentioned in Section 5.3.2.
This algebra, and its generalizations, were studied in detail in
\cite{\HZa,\HZb,\HZc}.

%---------------------------------------------------------------------

\subsec{Representation theory}

\noindent{\sl 6.4.1. The Kac determinant}\hfill\bs

An important tool in the investigation of the structure of
Verma modules and their irreducible quotients is the so-called
Kac determinant \cite{\Kac}.

The hermiticity requirement $W^{(s)\dagger}_n = W^{(s)}_{-n}$, together
with the normalization $\vev{\La|\La}$, uniquely define a symmetric
sesquilinear form (``Shapovalov form'') on the Verma module $M(\La,c)$.
If $\{ |v_i\rangle\,, i=1,\ldots,p_\ell(N) \}$ is a basis of
$M(\La,c)_{(N)}$ then the determinant of the $p_\ell(N)\times p_\ell(N)$
matrix $\left( {\cal M}_{(N)} \right)_{ij} = \vev{v_i|v_j}$ is called the
Kac determinant (compare with the discussion in Section 2.1). Here,
$p_\ell(N)$ is the number of partitions of $N$ on $\ell$ colors,
\ie $\sum p_\ell(N)t^N = ( \prod(1-t^k))^{-\ell}$.

Since the Kac determinant is clearly a polynomial in $\La$, it is
determined (up to a $\La$ independent factor) by its zero's,
which correspond to singular vectors in $M(\La,c)$.
The strategy now is to
construct explicitly a sufficient number
of singular vectors. Instead of constructing singular vectors
in the Verma
modules directly, it is much easier to construct them in the Fock space
representations. Since to every singular vector in $\cF_\La$ corresponds
a singular vector in $M(\La)$ of the same weight, this also tells us
something about the singular vectors in $M(\La)$.

Singular vectors in $\cF_\La$ are constructed by the following trick
\cite{\Tho}.
Suppose we have an intertwiner $Q: \cF_\La \to \cF_{\La'}$ (\ie
an operator commuting with the action of the $\cW$-generators) then
the image of $|\La\rangle$ under $Q$ is either zero or a singular vector
in $\cF_{\La'}$. Intertwiners $Q$ can be constructed out of properly
integrated products of screening operators $\tilde{s}_i^+(z)$ as
\eqn\chFbdb{
Q = \int_{\cal C} dz_1\ldots dz_n \tilde{s}^+_{i_1}(z_1)\ldots
\tilde{s}^+_{i_n}(z_n) \ .
}
Since the OPE's of the $\cW$-generators with $\tilde{s}^+_i(z)$ are at
most a total derivative (by the definition of the $\cW$-algebra), the
expression \chFbdb\ gives an intertwiner provided the contour $\cal C$
is closed in the homology of the local system determined by the
integrand of \chFbdb\ (see \eg \cite{\BMPd} for more details).
For the purpose of determining the Kac determinant it suffices to
consider intertwiners built out of a single screening operator $\tilde{s}
_i^+(z)$ only.
Doing this, one finds that for weights $\La$ of the form $\La =\al_+\Lap
+\al_-\Lam$, where $\Lap\in P_+\,,\Lam\in P_+^\vee$, the Fock space
$\cF_\La$ has a singular vector at level $N = (\Lap+\rh,\al_i^\vee)
(\Lam+\rhv,\al_i)$ for each $i=1,\ldots,\ell$. This singular vector and
its $\cW$ descendants contribute the following factor to the
Kac determinant ${\rm det\,}{{\cal M}_{(N)}}$
\eqn\chFbdc{
\left( (\La+ \al_+\rh + \al_-\rhv ,\al_i) - ( \half (\al_i,\al_i)m\al_+
+ n\al_-) \right) ^{p_\ell(N-mn)} \ ,
}
where $m=(\Lap+\rh,\al_i^\vee)$ and $n=(\Lam+\rhv,\al_i)$ are both
positive integers. Using Weyl group invariance \chFboa, and
comparison of the order
in $\La$ of the expression thus obtained with the a priori known order
(see \eg \cite{\Boc}), we conclude that the Kac determinant is given,
up to a $\La$ independent factor, by
\eqn\chFbda{
{\rm det\ }{\cal M}_N \sim
\prod_{\al\in\De} \prod_{mn\leq N} \left(
(\La + \al_+\rh + \al_-\rhv, \al) - ( \half(\al,\al)m\al_+ + n\al_-)
\right) ^{p_\ell(N-mn)} \,.
}

For $\bfg = A_1$ this expression reduces to \chBo\
(see also \cite{\Mia,\Boc,\Waa} for various
simply-laced cases, and \cite{\KWb} for $\bfg=B_2$.
The Kac determinant for the supercase $\bfg =B(0,n)$ is discussed
in \cite{\Wab}).

The Kac determinant can be used to determine the structure of the Verma
modules and its quotients. This will be the topic of the next section.
Moreover, from the Kac determinant one can in principle infer what the
unitary representations are. Unfortunately, direct analysis of unitarity
from the Kac determinant is rather cumbersome and has only been completed
in the case of the Virasoro algebra \cite{\FQSb,\FQSc}.
Some results have been obtained however for the $\cW_N$
algebras \cite{\Mia,\Mib,\Mic}. Since these results
can best be understood in the coset approach to $\cW$-algebras
we will postpone this discussion until Chapter 7.
\bs

\noindent{\sl 6.4.2. Completely degenerate representations and
minimal models}\hfill\medskip

The complete structure of Verma modules (or of any of its quotients)
can in principle be determined from the Kac determinant \chFbda.
Instead of giving a complete catalogue of modules of different types,
let us restrict the discussion to the
most important
class as far as physical applications are concerned. This is the class
of irreducible highest weight representations---with highest
weight $\La$,
say---for which the corresponding Verma module $M_\La$ contains ``as many
singular vectors as possible,'' or is ``completely degenerate''
\cite{\FZb,\FLa}.
More precisely, for which the Verma module $M_\La$ has $\ell$
independent singular vectors, one in every simple root direction.
{}From the discussion in Section 6.4.1 it follows that such completely
degenerate weights $\La$ can be parametrized as
\eqn\chFda{
\La = \al_+\Lap + \al_-\Lam
}
with $\Lap\in P_+$ and $\Lam\in P_+^\vee$. The $\ell$ independent
singular vectors occur at levels $(\Lap+\rh,\al_i^\vee)
(\Lam+\rhv,\al_i)\,, i=1,\ldots,\ell$.
The conformal dimension $h_\La$ (see \chFbah) can, in this case,
be written as
\eqn\chFdb{
h_\La - {c\over24} = -{\ell\over24} + \half |\al_+(\Lap+\rh) +
\al_-(\Lam +\rhv) |^2 \ .
}
If, in addition, $\al_+^2$ is a positive rational number, such that
\eqn\chFdc{
\al_+^2 = (k+\dCo)^{-1} =  {p'\over p}\,,\qquad {\rm gcd}(p,p')=1\,,
\qquad {\rm gcd}(p',r^\vee)=1 \ ,
}
it follows from the Kac determinant that the Verma module $M_\La$ in fact
contains an infinite number of singular vectors.

In particular, if we restrict the dominant integral weights $\Lap\,,\Lam$
to the following set
\eqn\chFdd{
\Lap\in P_+^{p-\dCo}\,,\quad \Lam\in P_+^{\vee\,p'-{\rm h}} \ ,
}
where, in addition, $p\geq\dCo\,,p'\geq{\rm h}$, we obtain the class
of so-called ``minimal model representations'' (see \cite{\FZb}
for $A_2$, \cite{\FLa,\FLc} for $A_n$ and $D_n$ and \cite{\BBSSb}
for all $\bfg$ simply-laced).
It can be argued that the (finite set of) primary fields
$\Ph_{\Lap,\Lam}(z)$ corresponding to the
set \chFdd\ form a closed operator product algebra, and hence
constitute a $\cW$-RCFT.

For these minimal models, the labelling of $\La$ by
the pair $(\Lap,\Lam)$ is actually slightly redundant. In general,
we have to make a further field identification. Because of the relation
$p\al_+ + p'\al_-=0$, the symmetry \chFboa\ under the Weyl
group of $\bfg$
is in fact extended to a symmetry under the affine Weyl group, \ie under
\eqn\chFdf{\eqalign{
\Lap & \to w(\Lap + \rh) - \rh + p\be \cr
\Lam & \to w(\Lam + \rhv) - \rhv + p'\be \ , \cr}
}
where $w\in W$ and $\be$ is an arbitrary element of the long root
lattice. The subgroup of transformations \chFdf\
which leaves invariant the fundamental
alcove \chFdd\ is isomorphic to the center of $\bfg$
(\eg $\ZZ_{n+1}$ for $A_n$), hence
we have to identify fields $\Ph_{(\Lap,\Lam)}(z)$ that are related by the
action of the center of $\bfg$. (In the case when odd integer
spin generators
in the $\cW$-algebra are present one sometimes has to make a further
field identification related to the $\ZZ_2$ automorphism
$W^{(s_i)}(z) \to (-1)^{s_i} W^{(s_i)}(z)$.)
\smallskip

For various applications it is useful to have a resolution of an
irreducible module $L_\La$ in terms of free field Fock spaces.
One can try to construct such resolutions directly in the $\cW$ setting
by using the intertwiners \chFbdb. It is however more convenient
to obtain these resolutions from their affine Lie algebra counterparts.
This is achieved by applying the functor $H_Q$ to all the terms in
a resolution of some irreducible highest weight module of $\whg$ and
then making use of \chFbaf. This functor was studied extensively in
\cite{\FKW}. For the case of minimal representations it leads
to a conjecture that there exists a resolution of $L_{\La}$ with terms
\eqn\chFde{
C^{(i)}L_{\La} \cong \bigoplus_{ \{ w\in\widehat{W}\,|\,
\tilde{\ell}(w) =i\} } \cF_{\al_+ w*\Lap + \al_- \Lam} \ ,
}
where $w*\la = w(\la+\rh)-\rh$ \cite{\FKW,\BMPd}.

To conclude this section let us specialize some of the above formulae to
the minimal model case of the $\cW_3$ algebra (the formulae in the case
of $\cW_2$, \ie the Virasoro algebra, can be found in Section 2.1).
The relevant central charge is parametrized by two relatively prime
positive integers
$p$ and $p'$ ($p,p'\geq3$) through (see \chFbg\ and \chFdc)
\eqn\chFdg{
c(p,p') = 2 \left( 1 - {12(p-p')^2\over pp'} \right) \,.
}
Now, parametrizing $\Lap$ and $\Lam$ in \chFdd\ each by two non-negative
integers
\eqn\chFdh{ \eqalign{
\Lap & = r_1\La_1 +r_2\La_2\,,\qquad\qquad 0\leq r_1+ r_2\leq p-3\,,\cr
\Lam & = s_1\La_1+s_2\La_2\,,\qquad\qquad 0\leq s_1+ s_2\leq p'-3\,,\cr}
}
where $\La_i\,,i=1,2$ are the fundamental weights of $sl(3)$, equation
\chFdb\ yields
\eqn\chFdi{
h^{(2)} \left( \matrix{ r_1 & s_1\cr r_2 & s_2\cr} \right) =
{ (p'(r_i+1)- p(s_i+1)) G_{ij} (p'(r_j+1)-p(s_j+1)) - 2(p-p')^2 \over
2pp'}\,,
}
where
$G_{ij} = {\textstyle{1\over3}} \left( \matrix{ 2&1\cr 1&2\cr} \right)$
is the inverse Cartan matrix of $sl(3)$.
The eigenvalue of the $W_3$ generator on the highest weight vector of
$L_{\Lap\Lam}$ follows easily from \chFcey
\eqn\chFdj{ \eqalign{
h^{(3)}  \left( \matrix{ r_1 & s_1\cr r_2 & s_2\cr} \right) = &
{1\over9pp'} \sqrt{ {2\over 3(5p-3p')(5p'-3p)} }
\left( p'(r_1-r_2)-p(s_1-s_2) \right) \cr
& \times \left( p'(2r_1+r_2) - p(2s_1+s_2) \right)
\left( p'(r_1+2r_2) - p(s_1+2s_2) \right)\,. \cr}
}
As argued above, the eigenvalues $h^{(i)}$ in \chFdi\ and \chFdj\
are invariant under a $\ZZ_3$ symmetry. One finds that the
generator $S\,,S^3=1$, acts on the labels by
\eqn\chFdk{
S\ : \ \left( \matrix{ r_1&s_1\cr r_2&s_2\cr} \right) \ \to\
\left( \matrix{ (p-3)-(r_1+r_2) & (p'-3)-(s_1+s_2) \cr
r_1 & s_1 \cr} \right) \,.
}
The $\ZZ_2$ operation $R\,,R^2=1$, that acts by
\eqn\chFdl{
R\ : \ \left( \matrix{ r_1&s_1\cr r_2&s_2\cr} \right) \ \to\
\left( \matrix{ r_2&s_2 \cr r_1 & s_1 \cr} \right) \,,
}
leaves $h^{(2)}$ invariant while it changes the sign of $h^{(3)}$.
In other words, the eigenvalues $h^{(2)}$ inside the fundamental Kac
table \chFdh\ are six-fold degenerate for $h^{(3)}=0$ and three-fold
degenerate otherwise.
\bs

\noindent{\sl 6.4.3. Character formulae}\hfill\bs

Explicit formulae for the characters of $\cW[\whg,k]$ irreducible highest
weight modules can be derived straightforwardly from a free field
resolution of $L_\La$. In this section we want to discuss some explicit
formulae for the case of the minimal
representations discussed in Section 6.4.2. The required free field
resolution, obtained by applying the QDS-functor $H_Q$ to a resolution
of an irreducible $\whg$ highest weight module with properly chosen
admissible weight, was discussed in Section 6.4.2.

Parametrizing $\La = \al_+\Lap + \al_-\Lam$, where $\al_+\,,\Lap$ and
$\Lam$ are as in \chFdc\ and \chFdd, and using equation \chFdb, a simple
application of the Euler-Poincar\'e lemma gives
\eqn\chFbeb{
{\rm ch\,}_{L_\La}(q) =
{1\over \et(\ta)^\ell} \sum_{w\in\widehat{W}} \ep(w)
q^{ {1\over 2pp'} |p'w(\Lap+\rh)-p(\Lam+\rhv)|^2 } \ .
}
In the simply-laced case this character formula was first conjectured
in \cite{\FLc}. The above derivation was given in \cite{\BMPd,\FKW}.
{}From the above
construction it is clear that the characters \chFbeb\ can
be written as certain residues of characters of $\whg$ irreducible
modules with admissible highest weight \cite{\FKW}. This explains an
observation made in \cite{\MP}. This description also,
evidently, relates the modular matrix $S$ and the set of modular
invariants to that of the admissible Kac-Moody representations.

In Chapter 7 we will see that equation \chFbeb\ coincides,
for simply laced Lie algebras $\bfg$, with the branching function
associated with a certain coset. This strongly supports the claim of
equivalence of the two $\cW$ constructions, for which so far (apart
from the $\cW_3$ case) no direct proof is available.
\vfill\eject

%--------------------CHAPTER 7-------------------------------------

\newsec{Coset constructions}

\subsec{Introduction}

The second generic method to obtain examples of $\cW$-algebras from
(a pair of) affine Lie algebras,
which at first sight is completely unrelated to the QDS
reduction scheme of the previous chapter, is through the so-called coset
construction \cite{\GKOa}. We have already seen a glimpse of it
in Section 2.3.2, where we reviewed the affine Sugawara construction
$T^{\whg}(z)$ of
the Virasoro generator and its GKO generalization $T^{\whg/\whg'}(z)$
to coset pairs. We recall that
\eqn\chGaa{
T^{\whg/\whg'}(z) = T^{\whg}(z) - T^{\whg'}(z)\,,
}
with corresponding central charge
\eqn\chGab{
c(\whg,\whg',k) = c(\whg,k) - c(\whg',k')\,.
}
The level $k'$ is determined by $k'=jk$ where $j$ is the Dynkin index of
the embedding $\bfg'\subset\bfg$ \cite{\Dy}.
Moreover, we have the important
property that $T^{\whg/\whg'}(z)$ commutes with the currents of $\whg'$.

The basic idea that generalizes this construction of the
coset Virasoro generator is the following.

Suppose we are given a coset pair $(\bfg,\bfg')\,, \bfg'\subset\bfg$
of finite-dimensional Lie algebras.
Consider their (untwisted) affinizations $\whg'\subset\whg$.
Then define the coset algebra $\cW_c[\whg/\whg',k]$ as
the set of all normal ordered products of $\whg$-currents and their
derivatives that commute (\ie have regular operator product expansions)
with the currents of $\whg'$, modulo the set of null-fields
at this particular value of $k$.%
\foot{Since the level $k'$ of $\whg'$ is
determined through $k'=jk$ we have suppressed $k'$ in the notation
$\cW[\whg/\whg',k]$.}
By definition this algebra closes and satisfies
the other axioms of a $\cW$-algebra except, possibly,
the requirement that it is finitely generated. This is not the
case in general, but
depends on the specific $\whg$-modules one is considering as we will see.
Another issue is that the coset $\cW$-algebra defined in this way depends
crucially on the level $k$ and is in general  ``exotic'' in the
terminology of Section 3.1. Varying $k$ will produce $\cW$-algebras of
different type. For a specific set of cosets the algebra is however
generic. This happens for instance
for the so-called diagonal coset pairs $(\whg\oplus\whg,\whg_{diag})$.

Every coset algebra $\cW_c[\whg/\whg',k]$ comes naturally equipped with a
set of highest weight modules. Let thereto $L_{\La}^{\whg}$ be an
irreducible highest weight module (not necessarily integrable!) of
$\whg$ with highest weight $\La$ and level $k$.
Define $L_{\La,\La'}^{\whg/\whg'}$ as
the space of states in $L_{\La}^{\whg}$ which are highest weight under
$\whg'$. It is clear that $L_{\La,\La'}^{\whg/\whg'}$ is a highest
weight module of the coset algebra $\cW_c[\whg/\whg',k]$.%
\foot{$L_{\La,\La'}^{\whg/\whg'}$ will be infinite-dimensional unless
$\whg'\subset \whg$ is conformal (see Section 2.3.2).}
We have the following decomposition of the $\whg$-module
$L_{\La}^{\whg}$ under the action of $\whg'$
\eqn\chGac{
L_{\La}^{\whg} = \bigoplus_{\La'} \left( L_{\La,\La'}^{\whg/\whg'}
\otimes L_{\La'}^{\whg'} \right) \,,
}
where the sum runs over a set of weights $\La'$ of $\whg'$ at level
$k'=jk$ with corresponding irreducible $\whg'$ highest weight modules
$L^{\whg'}_{\La'}$.
The corresponding equality for the characters reads
\eqn\chGad{
{\rm ch\,}_{L_{\La}}^{\whg}(q,z) =
\sum_{\La'} b_{\La'}^{\La}(q)\ {\rm ch\,}_{L_{\La'}}^{\whg'}(q,z')\,,
}
where $z'$ is determined as a function of $z$ through the embedding
$\bfg'\subset \bfg$.
The branching function $b_{\La'}^{\La}(q)$ is defined through
\eqn\chGae{
b_{\La'}^{\La}(q) = {\rm Tr\,}_{L_{\La,\La'}} \ q^{L_0 - c/24}\,,
}
where $L_0$ and $c$ are the coset expressions given in \chGaa\ and
\chGab.%
\foot{In the sequel we will often suppress the superscripts
$\whg\,,\whg'$ and $\whg/\whg'$ when no confusion can arise.}

In general the modules $L_{\La,\La'}$ will not be irreducible under the
coset algebra $\cW_c[\whg/\whg',k]$. The problem of decomposing
$L_{\La,\La'}$ into irreducible coset representations has not been
completely solved, and will not be discussed in this
report. We will only mention that in the case of
diagonal coset algebras the modules $L_{\La,\La'}$ are believed to be
irreducible. However, one can prove that the modules $L_{0,0}$ {\it are}
always irreducible \cite{\Doa}, implying that $L_{0,0}$ serves as a
characteristic Hilbert space for $\cW_c[\whg/\whg',k]$. In particular we
can study the coset algebras through the character technique
(as in Section 6.3.2).

We would also like to remark that by using the character decomposition
\chGad\ one can derive modular invariant combinations of
$\cW_c[\whg/\whg',k]$-characters from those of the affine Lie
algebras $\whg$ and $\whg'$ [\Boa,\BoGa,\BBSSb,\CRa].\medskip

In this chapter we will mainly restrict our attention to the diagonal
coset algebras $\cW_c[\whg\oplus\whg/\whg,(k_1,k_2)]$ for reasons
alluded to in the above and also simply because not much is known about
generic coset models. Our analysis will be most complete in the case
when one of the levels, say $k_1$, equals one \cite{\BBSSb}, although
most of the analysis goes through
for generic $k_1$. The results for integer $k_1>1$ are
usually presented in the context of parafermionic algebras which fall
outside the scope of this report
(see \eg \cite{\KMQ,\BNY,\Geb,\Raa,\CRa,\SY,\Naa,\Nab}).

In the last section we will briefly touch upon the
subject of more general coset models
(see \eg \cite{\Doa,\Dob,\BY,\GSa,\BoGb,\Db}).

Before we start discussing diagonal coset models it turns out to be
convenient to first discuss a slight generalization of the above
construction. The crucial observation is that the Virasoro algebra
related to $\whg$ as given by the affine Sugawara construction in fact
commutes with the finite-dimensional ``horizontal algebra'' $\bfg$. We
may thus interpret the Sugawara tensor as an element of the coset
algebra $\cW_c[\whg/\bfg,k]$. We will first study this so-called Casimir
algebra \cite{\Fr,\BBSSa,\Thb},
and subsequently argue that the diagonal coset algebras
can be interpreted as its deformation \cite{\BBSSb}.

One of our main themes in this Chapter will be to
establish, for simply-laced $\bfg$, an isomorphism (up to null-fields)
\eqn\chGaf{
\cW_c[\whg\oplus\whg/\whg,(1,k_2)] \cong \cW_{DS}[\whg,k] \,,
}
where $k_2$ and $k$ are algebraically related. This will be done
indirectly by showing that a sufficiently large set of characters (and
thus the Kac determinants) of these algebras are identical.
It is still an important open problem to prove this isomorphism by
direct means and/or to have an a priori understanding why certain coset
$\cW$-algebras are equivalent to certain QDS $\cW$-algebras.
\bs

\subsec{Casimir algebras}

\noindent{\sl 7.2.1. Generalities}\hfill \medskip

Casimir algebras were already introduced in Section 3.2.
We will briefly recall the construction, and give a reinterpretation in
terms of a particular coset. We will indicate that
the Casimir algebra is not only a convenient starting point  for more
general coset constructions, but also provides (at least for a particular
$c$-value) a direct link to the QDS reduction discussed in the last
Chapter.

Our starting point is the so-called affine Sugawara construction (see
Section 2.3.2).
Consider thereto the following composite operator
\eqn\chGbaa{
T(z) = \half\et^{(2)}(\bfg,k) \sum_{a,b} d_{ab}(J^aJ^b)(z)\,.
}
It is well-defined on highest weight modules of $\whg$ and,
for properly chosen normalization constant
$\et^{(2)}(\bfg,k) = 1/(k+\dCo)$
($k\neq-\dCo$), satisfies the Virasoro algebra of central charge
$c = c(\whg,k) = k\ {\rm dim\,}\bfg / (k+\dCo)$. In other words,
every highest weight module of $\whg$ ($k\neq-\dCo$) can be extended to
a highest weight module of the semi-direct sum of a Virasoro algebra and
$\whg$. The currents $J^a(z)$ become primary fields of conformal
dimension one with respect to $T(z)$, \ie
\eqn\chGbab{
T\underhook{(z) J^a}(w) = { J^a(w)\over (z-w)^2} + {\p J^a(w)\over z-w}\,,
}
which, in particular, implies
\eqn\chGbac{
[J_0^a, T(z) ] = 0\,,
}
\ie $T(z)$ is a singlet under the horizontal (finite-dimensional)
subalgebra $\bfg$ of $\whg$.
We conclude that, in the terminology of Section 7.1, the Sugawara tensor
$T(z)$ is an element of the $\cW$-algebra $\cW[\whg/\bfg,k]$ related to
the coset pair $(\whg,\bfg)$. To construct other singlet fields under
$\bfg$ is straightforward. Consider thereto a generic field
\eqn\chGbad{
W^{b_1,\ldots,b_n}_{i_1,\ldots,i_n}(z) = (\p^{i_1}J^{b_1}\ldots
\p^{i_n}J^{b_n})(z)\,.
}
Although the full operator product expansion between the
currents $J^a(z)$ and
$W^{b_1,\ldots,b_n}_{i_1,\ldots,i_n}(w)$ is hard to evaluate, it is
easily seen that the simple pole term only receives contributions
from the simple pole term in (2.69). Therefore
\eqn\chGbae{
[ J^a_0, W^{b_1,\ldots,b_n}_{i_1,\ldots,i_n}(z)] =
\sum_{i=1}^n \sum_c f^{ab_i}{}_c W^{b_1,\ldots,
b_{i-1},c,b_{i+1},\ldots,b_n}_{i_1,\ldots,i_n}(z)\,.
}
This implies that the field $W_{i_1,\ldots,i_n}(z) =
\sum_{b_1,\ldots,b_n} d_{b_1\ldots b_n}
W^{b_1,\ldots,b_n}_{i_1,\ldots,i_n}(z)$ commutes with $\bfg$ if and only
if $d_{b_1\dots b_n}$ is an invariant tensor under $\bfg$, \ie if and
only if $\sum_{b_1,\ldots,b_n} d_{b_1 \ldots b_n} T^{b_1}\ldots T^{b_n}$
is an element in the center of the universal enveloping algebra of
$\bfg$, \ie a Casimir operator. For this reason we will also refer to the
singlet algebra $\cW[\whg/\bfg,k]$ as the Casimir algebra of $\bfg$
at level $k$.

Notice that we may assume that the invariant tensors $d_{a_1\ldots a_n}$
are  completely symmetric, since for any anti-symmetric pair of indices
the field can be reduced by using
\eqn\chGbaf{
(J^aJ^b)(z) - (J^bJ^a)(z) = f^{ab}{}_c (\p J^c)(z)\,.
}
It is clear that the algebra of all singlet fields
$W_{i_1,\ldots,i_n}(z)$ closes. To establish that they give
rise to a $\cW$-algebra as defined in Section 3.1, one has to find a
subset consisting of a finite number of primary fields which---together
with their derivatives and normal ordered products thereof---form a
closed operator product algebra. The other conditions for this to be
a $\cW$-algebra are automatically fulfilled.
The above procedure may (and will) clearly depend on the value of the
level $k$.

The simplest fields%
\foot{In fact, since $d_{a_1\ldots a_M}$ is assumed to be completely
symmetric there is no need to normal order this expression.
The factor $M!$ is inserted for convenience.}
\eqn\chGbag{
T^{(M)}(z) = {1\over M!} \sum_{a_1,\ldots,a_M} d_{a_1\ldots a_M}
(J^{a_1} \ldots J^{a_M})(z)\,,
}
are usually, by abuse of language, referred to as the Casimir operators
of $\whg$. It is easily shown that the operators $T^{(M)}(z)$
are primary with respect to \chGbaa\ of dimension $M$, provided the
$d$-symbols are chosen to be traceless. In general the operators
$T^{(M)}(z)$ will not form a complete set
of generators of $\cW[\whg/\bfg,k]$. It can however be argued
\cite{\Na} that they do generate the complete singlet algebra for 
$\bfg$ simply-laced and $k=1$.
\bs

\noindent{\sl 7.2.2. Character technique} \hfill \medskip

In Section 6.3.2 we have seen that the character technique provides a
very powerful tool for analysing the set of generators of a
$\cW$-algebra. In this section we apply this technique to the Casimir
algebras of the previous section.

Characters of the coset algebra $\cW[\whg/\bfg,k]$ are constructed,
according to the general philosophy of Section 7.1, by decomposing the
characters of $\whg$ with respect to the characters of $\bfg$.
Consider the branching of an integrable highest weight module $L_\La$ of
$\whg$ at level $k$ in terms of irreducible highest weight modules
$L_\la$ of $\bfg$
\eqn\chGbba{
{\rm ch\,}_{L_\La}^{\whg}(q,z) = \sum_{\la\in P_+}
b^\La_\la(q) {\rm ch\,}_{L_\la}^{\bfg}(z)\,.
}
For convenience we will also use $\Ph^\La_\la(q)\equiv
q^{-s_\La}b^\La_\la(q)$
where
\eqn\chGbbb{
s_\La = { |\La+\rh|^2\over2(k+\dCo)} - {|\rh|^2\over2\dCo}
= {c_\La\over2(k+\dCo)} - {1\over24} {k\ {\rm
dim\,}\bfg\over k+\dCo}\,,
}
so that $\Ph^\La_\la(q)$ has an
expansion in terms of integer powers of $q$. Explicitly \cite{\KP},
\eqn\chGbbc{
b_\la^\La(q) \equiv {\rm Tr\,}_{L_{\La,\la}}\ q^{L_0 -c/24}
= \sum_{w\in W} \ep(w) c^\La_{w(\la+\rh)-\rh + k\La_0}(q)
q^{ {1\over2k} |w(\la+\rh)-\rh|^2}\,,
}
where the $c^\La_\la(q)$ are the so-called Kac-Peterson string
functions.

We will discuss several examples
\item{(i)} $\bfg$ simply-laced, $k=1$.\hb
In this case there is a unique string function $c^{\La_0}_{\La_0}(q)=
\left( q^{{1\over24}} \prod (1-q^n) \right)^{-\ell}$. Applying the
denominator formula (9.3) to \chGbbc\ one finds
\eqn\chGbbd{
b_\la^\La(q) = { q^{\half|\la|^2} \prod_{\al\in\De_+}
(1-q^{(\la+\rh,\al)}) \over \left( q^{ {1\over24}}
\prod_{n\geq1} (1-q^n)\right)^{\ell} }\,,
}
for $\la\in P_+\cap (Q+\La)$ and zero otherwise.
In particular, the vacuum character is given by
\eqn\chGbbe{ \eqalign{
\Ph^{\La_0}_0(q) & = { \prod_{\al\in\De_+} (1- q^{(\rh,\al)})
\over \prod_{n\geq1} (1-q^n)^\ell } = { \prod_{i=1}^\ell
\left( \prod_{n=1}^{e_i} (1-q^n)  \right) \over \prod_{n\geq1} (1-q^n)
^\ell } \cr
& = { 1\over \prod_{i=1}^\ell\left( F_{e_i+1}(q)\right) }\,, \cr}
}
where $F_s(q)$ is defined in (6.39).
Since this is precisely the vacuum character of a $\cW$-algebra of type
$\cW(e_1+1,\ldots,e_\ell+1)$ with independent generators, the
above analysis suggests that the singlet algebra
$\cW_c[\whg/\bfg,k=1]$ is of this type.
Exactly the same result \chGbbe\ was obtained in (6.43) for the
vacuum character of the algebra $\cW_{DS}[\whg,k]$ obtained by the
quantum DS-reduction. The main difference, however, is that the analysis
in Section 6.3.2 is valid for generic values of $k$, while the above
analysis is restricted to a particular value for $k$, namely $k=1$.
Another difference is that while the analysis in
Section 6.3.2 also applies to non simply-laced Lie algebras, the
above analysis is restricted to simply-laced ones. For non
simply-laced Lie algebras $\bfg$ the singlet algebra is
essentially different from the QDS reduction based on $\bfg$.
In particular the singlet algebra contains additional fields beyond those
corresponding to the Casimirs of $\bfg$. We discuss
two examples of this type.
\medskip

\item{(ii)} $\whg = B_\ell^{(1)}$ at level $k=1$.\hb
Using the explicit formulae for the string functions in
\cite{\KP} one arrives,
after a straightforward calculation, at the following formula for the
vacuum character of the singlet algebra
\eqn\chGbbp{
\Ph_0^{\La_0}(q) =
{\prod_{i=1}^\ell
\prod_{n=1}^{e_i} (1-q^n) \over \prod_{n\geq1} (1-q^n)^\ell}\times\half
\left( \prod_{n\geq\ell} (1+q^{n+\half}) +
\prod_{n\geq\ell} (1-q^{n+\half}) \right)\,,
}
where $\{ e_i\} = \{ 1,3,5,\ldots,2\ell-1\}$ is the set of exponents of
$B_\ell$. The last term in \chGbbp\ equals the
vacuum character of a fermionic dimension $\ell+\half$ field
projected onto
the $\ZZ_2$ even sector. So, we conclude that the singlet algebra in this
case is the bosonic projection (compare with Section 5.3.2)
of an algebra of type $\cW(2,4,\ldots,2\ell,
\ell+\half)$ (see also \cite{\Wab,\Wae} and \cite{\Boc} for $\ell=1$).
The corresponding QDS reduction is the one based on the affine
Lie superalgebra $B(0,\ell)^{(1)}$ (see Section 6.3.3).
\medskip

\item{(iii)} $\whg = G_2^{(1)} $ at level $k=1$.\hb
A straightforward computation, using the string functions in
\cite{\KP}, gives
\eqn\chGbbw{
\eqalign{
\Ph_0^{\La_0}(q) & =
(1-q)(1-q^2)(1-q^3) \left( (1+q^2+q^3)c_{\La_0}^{\La_0}
- q^{{1\over3}} (1+q+q^3) c_{\La_2}^{\La_0} \right) \cr
% & = {1\over \prod (1-q^n)^3} (1-q)(1-q^2)(1-q^3) \left\{ (1+q^2+q^3)
& = {1\over F_2F_6F_8F_{10}F_{12}^2F_{14}^2F_{15}F_{16}F_{17}F_{18}^2}
( 1-3q^{22} - 7 q^{23} - 14 q^{24} + {\cal O}(q^{25}) )\,, \cr}
}
which suggests that the singlet algebra of $G_2^{(1)}$ at $k=1$ is of
type $\cW(2,6,8,10,12_2, 14_2,$ $15,16,17,18_2)$.
So, apart from the Casimir operators of spin 2 and 6,
we find other integer spin operators.
They can presumably be understood
as composites of (among others) a spin $8/3$ operator, whose existence
can be anticipated in the level-1 non-simply laced vertex operator
realization of $G_2^{(1)}$ \cite{\GNOS,\BTh}. In \cite{\Wad}
it is argued that, in fact, all generators can be found by taking
successive OPE's of the dimension $2$ and
$6$ fields and an additional field of dimension $17/5$.
It is not known whether there exists a QDS reduction leading to the same
algebra.
\bs

\noindent{\sl 7.2.3. 3rd order Casimir example}\hfill\medskip

In this section we present a detailed
example. The easiest nontrivial illustration
of the above considerations is the algebra
of a third order Casimir operator. These exist only for the algebras
$A_{N-1} \cong sl(N)\,, N\geq3$. We have
\eqn\chGbca{ \eqalign{
d_{ab} & = {\rm  Tr\,}(T_aT_b) \,,\cr
f_{abc} & = {\rm Tr\,}( [T_a,T_b] T_c)\,, \cr
d_{abc} & = {\rm Tr\,}(\{ T_a,T_b\} T_c) \,.\cr}
}
Group indices are raised and lowered by means of the Killing metric
$d_{ab}$ (and its inverse $d^{ab}$).
The tensors $d_{ab}\,,d_{abc}$ are completely symmetric while $f_{abc}$
is completely antisymmetric in its indices.
With respect to an antihermitian basis $\{ T_a\}$ the tensors
$d_{ab}\,,f_{abc}$ are real while $d_{abc}$ is pure imaginary.
Let us, for future use, list a set of useful contraction identities
\cite{\MSW,\BBSSa}
\eqn\chGbcaa{
\eqalign{
d^{ab}d_{abc} & = 0 \,,\cr
f_{a}{}^{bc} f_{dbc} & = (-2N) d_{ad} \,,\cr
d_a{}^{bc} d_{dbc} & =2\left( { N^2-4\over N}\right) d_{ad} \,,\cr
f_{ad}{}^b f_{be}{}^c f_{cf}{}^{a} & = N f_{def}\,, \cr
d_{ad}{}^b f_{be}{}^c f_{cf}{}^{a} & = N d_{def} \,,\cr
d_{ad}{}^b d_{be}{}^c f_{cf}{}^{a} & =
\left( {N^2-4\over N}\right) f_{def}\,, \cr
d_{ad}{}^b d_{be}{}^c d_{cf}{}^{a}
& = \left( {N^2 -12\over N}\right) d_{def}\,. \cr}
}
In addition we have the Jacobi identities
\eqn\chGbcab{ \eqalign{
f_{ad}{}^e f_{ebc}&+ f_{bd}{}^e f_{eca} + f_{cd}{}^e f_{eab}  = 0\,,\cr
f_{ad}{}^e d_{ebc}&+ f_{bd}{}^e d_{eca} + f_{cd}{}^e d_{eab}  = 0\,,\cr
f_{ab}{}^c f_{cde}&= {4\over N}(d_{ae}d_{bd}-d_{ad}d_{be})
+(d_{bd}{}^c d_{cae} - d_{ad}{}^c d_{cbe})\,,\cr}
}
and a relation valid for $N=3$ (\ie $sl(3)$) only
\eqn\chGbcac{
d_{ab}{}^c d_{cde} = {\textstyle{2\over3}} (d_{ad}d_{be} + d_{ae}d_{bd}
- d_{ab}d_{de} ) - {\textstyle{1\over3}} (f_{ad}{}^c f_{cbe} + f_{ae}{}^c
f_{cbd})\,.
}
As an intermediate step in the calculation of the OPE $T^{(3)}(z)
T^{(3)}(w)$, it is convenient to define
\eqn\chGbcb{
T^{(2,1)}_a(z) = \half d_{abc}(J^bJ^c)(z)\,.
}
It is straightforward, using the Wick theorem (2.37), to verify that
$T^{(2,1)}_a(z)$ is a  primary field of conformal dimension 2.
The OPE's of $T^{(2,1)}_a(z)$
and $T^{(3)}(z)$ with the elementary fields $J_a(z)$ are given by
\eqn\chGbcc{ \eqalign{
J_a(z) T^{(2,1)}_b(w) & = { k+\half N
\over (z-w)^2} d_{ab}{}^cJ_c(w) + {1\over z-w}
f_{ab}{}^cT^{(2,1)}_c(w)+ \ldots \,,\cr
J_a(z) T^{(3)}(w) & = { k+N\over (z-w)^2} T^{(2,1)}_a(w) + \ldots \,.\cr}
}
Here we have made repeated use of the identities \chGbcaa\ and \chGbcab.
Note that the $(z-w)^{-1}$-terms imply that $T^{(2,1)}_a(z)$
and $T^{(3)}(z)$
transform under the adjoint and singlet representation of the horizontal
algebra $A_{N-1}$, respectively.

After a tedious calculation, using rearrangement lemmas such as (2.38),
one arrives at the following result
\eqn\chGbcd{ \eqalign{
W(z) W(w) = & { c/3 \over (z-w)^6} + { 2T(w) \over (z-w)^4} +
{ \p T(w) \over (z-w)^3}
+ {1\over (z-w)^2} \left( 2\be\La + \textstyle{3\over10} \p^2T
+ R^{(4)}\right)(w) \cr
& + {1\over z-w} \left( \be \p\La + \textstyle{1\over15}\p^3T
+ \half \p R^{(4)} \right) (w)
+ (WW)(w) + \ldots\,. \cr}
}
where
\eqn\chGbce{
c  = c(A_{N-1}^{(1)},k) = { k(N^2-1)\over k+N}\,,
}
$\La(z)$ and $\be$ are defined in equations (3.6), (3.7),
respectively, and
\eqn\chGbcf{
W(z) = \et^{(3)}(A_{N-1},k)\,T^{(3)}(z)\,,\qquad \et^{(3)}(A_{N-1},k) =
{1\over k+N} \sqrt{ {N\over (k+ \half N)(N^2-4)} } \,.
}
Apart from the identity field and its descendants, an additional primary
field $R^{(4)}(z)$ of dimension 4, and its first descendant
$\p R^{(4)}(z)$, enter in \chGbcd. Explicitly,
\eqn\chGbcg{
R^{(4)}(z) =\left(  -2b^2\La -{4\over5}\p^2T + {1\over 2(k+N)} (\p J^a
\p J_a) + (k+N) \left( \et^{(3)} \right)^2 (T^{(2,1)a}T^{(2,1)}_a)
\right)(z) \,.
}
The term $(T^{(2,1)}_aT^{(2,1)a})(z)$
involves the contraction of two 3rd order $d$-symbols.
For $N\geq4$ this will produce an independent 4th order $d$-tensor,
corresponding to the 4th order Casimir of $sl(N)$. For $N=3$, however, no
such independent 4th order Casimir exists. Using the explicit $d$-tensor
contraction \chGbcac\ one can show that for $N=3$
\eqn\chGbch{
R^{(4)}(z) = f(k) \left( (TT) - {8+c\over 12}\p^2T + {22+5c\over 24(k+3)}
(\p J^a \p J_a) \right) (z)\,,
}
where
\eqn\chGbci{
f(k) = - {36 (k+3)^2 \over 5(31k+33) (2k+3) }\,.
}
Clearly, if the field $R^{(4)}(z)$ had not been present
in \chGbcd\
we would have reproduced exactly the $\cW_3$ algebra of (3.5). So, we
are led to the question under which circumstances we can consistently
put $R^{(4)}(z)$ equal to zero or, in other words, if there are cases
for which $R^{(4)}(z)$ is a null-field.
There are various ways of studying this question. One could for instance
attempt to show that the field $R^{(4)}(z)$---together with the fields
generated by taking further OPE's with $R^{(4)}(z)$---form an ideal in
the full operator product algebra. As a first step in this direction we
have (for $N=3$)
\eqn\chGbcj{
T^{(3)}(z)R^{(4)}(w) = f(k) \left\{ {3(1-\half c)\over (z-w)^4} \left(
T^{(3)}(w) + \ldots \right) + {1\over (z-w)^2} \left( R^{(5)}(w) + \ldots
\right) \right\} + \ldots\,,
}
where $R^{(5)}(z)$ is yet another primary field (of dimension 5), and the
dots stand for descendant fields.
{}From this we learn that only for $c=2$ (\ie $k=1$)
it is possible for $R^{(4)}(z)$ to be a null-field.
One would now have to take OPE's with $R^{(5)}(z)$ and continue,
a priori,
ad infinitum. However, knowing that $R^{(4)}(z)$ can only decouple for
$N=3\,,k=1$ we may try to shortcut this
calculation by using a specific realization in which the
null-fields vanish automatically.

For simply laced Lie algebras $\bfg$ at level one, such a realization is
provided by the vertex operator realization
\cite{\Halb,\BHN,\FK,\Seg} in
terms of ${\rm rank\,}\bfg = \ell$ free bosonic scalar fields $\ph^i(z)$.
Specifically, in the Cartan-Weyl basis, the generators are realized by
\eqn\chGbck{ \eqalign{
H^i(z) & = i\p\ph^i(z)\,, \cr
E^\al(z) & = c_{-\al} e^{i\al\cdot\ph(z)} \,,\cr}
}
where the $c_{-\al}$ are certain cocycle factors whose specific form is
irrelevant for the present discussion (see \cite{\GO},
and references therein,
for more details). The Fock space $\cF^\ph_\La$ of the scalar fields
$\ph^i(z)\,,i=1,\ldots,\ell$
is irreducible, hence isomorphic to $L_\La$ (see \eg
\cite{\GO} and references therein).
Inserting the realization \chGbck\ in \chGbag\ one finds
(see \cite{\BBSSa} for more details)
\eqn\chGbcl{ \eqalign{
T(z) & = \half (H^1H^1 + H^2H^2) \,,\cr
W(z) & = -{\textstyle{1\over6}} (3H^1H^1H^2 - H^2H^2H^2) \,,\cr}
}
while $R^{(4)}(z)$ vanishes upon insertion of \chGbck.\foot{In fact one
can write $R^{(4)}(z) \sim (J_aR^{(3)a})(z)$ for some primary
field $R^{(3)a}(z)$ (see Section 7.3.2), and it is even
easier to show the vanishing of $R^{(3)a}(z)$.}
Alternatively, one could compute the OPA of $W(z)$ directly from \chGbcl,
in which case one would obtain \chGbcd\ (for $c=2$) without the field
$R^{(4)}(z)$ being present.

Yet another method for showing that $R^{(4)}(z)$ is a null-field for
$N=3\,,k=1$ is to make use of the character technique of Section 7.2.2.
For $sl(3)\,,k=1$ equation \chGbbe\ gives
\eqn\chGbcm{
\Ph^{\La_0}_0(q) = 1 + q^2 + 2q^3 + 3q^4 +3q^5 + \ldots\,.
}
In particular there should be three independent states at
energy level four.
By calculating the determinant of the inner product matrix of the
three states $\{ L_{-2}L_{-2}|0\rangle\,,L_{-4}|0\rangle\,,W_{-4}
|0\rangle\}$ it is straightforward to show that there are no
nullstates contained in this
set. This also proves that $R^{(4)}(z)$ has to be null.
\smallskip

At this point one can make a remarkable observation that if one
considers the expression \chGbag\ ($M=2,3$) were the sum over the indices
is restricted to the CSA of $sl(3)$, the result is, in fact,
proportional to \chGbcl. This is of course
well-known for the Virasoro generator (``quantum equivalence''
\cite{\GO}), where it is a consequence of the fact that
for simply laced $\bfg$ at level $k=1$ the embedding $u(1)^\ell \subset
\bfg$ is conformal. The expressions \chGbag\ restricted to the
CSA are precisely the Weyl invariant polynomials on $\bfh$.
\smallskip

In Section 6.3.2 we have already argued that the
$\cW$-algebra $\cW_{DS}[\whg,k]$ at the specific point
$\al_\pm=\pm1$ is
in fact isomorphic to $\cW_c[\whg/\bfg,1]$
for $\bfg$ simply-laced. We can
now easily verify that the equations \chGbcl\ and (6.55) are, in fact,
form identical. The above quantum equivalence now also explains,
in concrete
terms, why there exists the Weyl invariance (6.21)
(at the point $\al_\pm= \pm1$).

At level $k=-N$ the fields $T^{(2)}(z)$ and $T^{(3)}(z)$ commute
with the affine currents (see \eg (7.26)) and thus generate
an abelian algebra. More generally, one can prove that
$\cW(\whg/\bfg,-\dCo)$ is isomorphic to the center of the universal
enveloping algebra of $\whg$ at $k=-\dCo$ \cite{\FFrb}.
This can be used---and
in fact was the original motivation in the mathematics literature
for studying $\cW$-algebras---to prove the Kac-Kazhdan
conjecture \cite{\KK,\Hay,\GoW,\Ma}.
\bs

\subsec{$G\times G/G$ coset conformal field theories}

\noindent{\sl 7.3.1. Deforming the singlet algebra}\hfill\medskip

In the previous section we have examined the coset algebra
$\cW_c[\whg/\bfg,k]$ and, for simply-laced $\whg$
at the specific value $k=1$, established
the equivalence with the QDS algebra $\cW_{DS}[\whg,k]$. The latter
algebra was however shown to exist for a continuum of
$c$-values so the question that arises naturally is
whether there exists a coset algebra that corresponds to the generic
case. In this section we will argue
that the relevant coset algebra is $\cW_c[\whg\oplus\whg /
\whg,(1,k_2)]$. In fact, we will establish that the coset algebra
$\cW_c[\whg/\bfg,k_1]$ can be interpreted
as the $k_2\to\infty$ limit of the coset algebra
$\cW_c[\whg\oplus\whg / \whg,(k_1,k_2)]$. In the subsequent section we
will elaborate on this by explicit construction of the spin-3 generator 
of $\cW_c[A_{N-1}^{(1)}\oplus A_{N-1}^{(1)}/ A_{N-1}^{(1)},(1,k_2)]$ 
as a deformation of that of $\cW_c[A_{N-1}^{(1)}/A_{N-1},1]$. 
This will generalize the GKO construction \chGaa\ of the coset Virasoro 
algebra.

Let us distinguish the two algebras in $\whg\oplus\whg$ by
subscripts, \ie $\whg_{(1)}\oplus \whg_{(2)}$, and denote the
respective generators by $J_{(1)}^a(z)$ and $J_{(2)}^a(z)$.
The diagonal subalgebra is generated by $J_{diag}^a(z) = J_{(1)}^a(z)
+ J_{(2)}^a(z)$.

Let $\cH_{(i)}$ be the characteristic Hilbert space of $\whg_{(i)}$
and let $|\ph\rangle \in \cH_{(1)}\otimes\cH_{(2)}$ be a singlet
under $\whg_{diag}$, then
\eqn\chGcaa{
\left( J_{(1)n}^a + J_{(2)n}^a \right) |\ph\rangle = 0\,,\qquad
\forall n\geq0\,.
}
Denote by $P_N$
the orthogonal projection onto the eigenspace $L_0(\whg_{(2)}) = N$.
We have
\eqn\chGcab{
|\ph\rangle = \sum_{0\leq N\leq h_{\ph}} P_N|\ph\rangle\,.
}
By applying $P_N$ to \chGcaa\ one can argue (see
\cite{\BoGb} for details)
that the map $|\ph\rangle \to P_0|\ph\rangle$ is injective, and that
moreover $J_{(1)0}^a P_0|\ph\rangle = 0$. We thus have an injective
map
\eqn\chGcac{
P_0\ :\ \cW[\whg\oplus\whg/\whg,(k_1,k_2)]\ \to\ \cW[\whg/\bfg,k_1]\,.
}
By rescaling $J_{(2)n}^a$ by $\sqrt{k_2}$ one can furthermore argue
that in fact $|\ph\rangle \to P_0|\ph\rangle$ in the limit
$k_2\to\infty$.

To establish that the map $P_0$ in \chGcac\ is actually an isomorphism
in the limit $k_2\to\infty$ we consider the branching function
(see \eg \cite{\KaWa,\KaWb})
\eqn\chGcad{
b^{\La_1,\La_2}_{\La_3} (q) = \sum_{w\in\widehat{W}}
\ep(w) c_{\La_3 - w*\La_2}^{\La_1}(q)\, q^{ {|w(\La_2+\rh)(k_1+k_2+\dCo)
-(\La_3+\rh)(k_2+\dCo)|^2 \over 2k_1(k_2+\dCo)(k_1+k_2+\dCo) } }\,.
}
Here $\La_1\,\La_2$ are integrable weights at levels $k_1$ and $k_2$,
respectively, and $\La_3$ is an integrable weight at
level $k_1+k_2$ such that $\La_1 + \La_2 - \La_3 \in Q$, where
$Q$ is the long root lattice of $\bfg$.

In the limit $k_2 \to \infty$ only the elements $w\in W$ in the sum
\chGcad\ will contribute since the others are of ${\cal O}(q^{k_2})$.
We thus have
\eqn\chGcae{
b^{\La_1,\La_2}_{\La_3}(q) \ \ \mapright{k_2\to\infty}\ \
\sum_{w\in W} \ep(w) c^{\La_1}_{\La_3-w*\La_2}(q)\,
q^{ {|w*\La_2-\La_3|^2
\over 2k_1} }\,.
}
In particular, for the vacuum character $\La_1=k_1\La_0\,,
\La_2=k_2\La_0\,,\La_3=(k_1+k_2)\La_0$, this equation
reduces to the expression \chGbbc\ for $\La=k\La_0\,,\la=0$,
\ie the vacuum character for the
coset pair $(\whg,\bfg,k_1)$. This implies that
the map \chGcae\ becomes surjective in the limit $k_2\to\infty$, and that
for large enough $k_2$
the generators of $\cW[\whg/\bfg,k_1]$ can be deformed to
elements of $\cW[\whg\oplus\whg/\whg,(k_1,k_2)]$ (see also \cite{\Wac}).

The above results imply that the generators of 
$\cW[\whg\oplus\whg/\whg,(1,k_2)]$ are independent
in a neighbourhood of $k_2=\infty$, \ie for $k_2$ large enough,
and thus have to form a closed algebra for a continuum of $k_2$
values and hence, because
the operator expansion coefficients are algebraic in $k_2$, for all $k_2$
values except for those for which the OPE coefficients
develop singularities \cite{\Wac}.
\bs

\noindent{\sl 7.3.2. Concrete example}\hfill\medskip

In this section we show, by explicit construction, that the
Casimir generator $T^{(3)}(z)$ of $\cW[A_{N-1}^{(1)}/A_{N-1},k_1]$
(see \chGbag) can be deformed (for generic values of $(k_1,k_2)$) 
to an element in $\cW[A_{N-1}^{(1)}\oplus A_{N-1}^{(1)}/
A_{N-1}^{(1)},(k_1,k_2)]$ as we have argued in the previous section. 
We will also determine the algebra of this dimension-three
operator. \smallskip

We introduce the following mixed contractions
\eqn\chGcca{
\widetilde{T}^{(i,j)}(z) =
{1\over i!\,j!} \sum_{a_1,\ldots,a_3}
d_{a_1\ldots a_i a_{i+1}\ldots a_3}
\left( J^{a_1}_{(1)} \ldots J_{(1)}^{a_i} J_{(2)}^{a_{i+1}}\ldots
J_{(2)}^{a_3} \right)(z)\,,
}
and take the following ansatz for the coset generator
\eqn\chGccb{
\widetilde{T}^{(3)}(z) = \sum_{i+j=3} a^{(i,j)}(A_{N-1},k_1,k_2)
\widetilde{T}^{(i,j)}(z)\,.
}
It is easy to check that the requirement that \chGccb\
is a singlet under the diagonal $\whg$, \ie that the OPE with the
currents $\widetilde{J}^a = J_{(1)}^a + J_{(2)}^a$ is regular,
uniquely fixes the coefficients $a^{(i,j)}(A_{N-1},k_1,k_2)$
up to an overall multiplicative constant
\eqn\chGccc{
a^{(i,j)}(A_{N-1},k_1,k_2) = (-1)^j \prod_{p=i+1}^3
\left( k_1 + (p-1){N \over 2} \right) \prod_{q=j+1}^3 \left(
k_2 + (q-1) {N \over 2} \right)\,.
}
The coefficient $\tilde{\et}^{(3)}(A_{N-1},k_1,k_2)$ needed to bring the
field $\widetilde{W}^{(3)}(z) = \tilde{\et}^{(3)} \widetilde{T}^{(3)}(z)$
to its standard normalization $\vev{\widetilde{W}^{(3)} | 
\widetilde{W}^{(3)} } = \tilde{c}/3$, where 
\eqn\chGccd{ \eqalign{
\tilde{c} = c(A_{N-1}^{(1)} \oplus A_{N-1}^{(1)},A_{N-1}^{(1)},(k_1,k_2)) &  
= c(A_{N-1}^{(1)},k_1) + c(A_{N-1}^{(1)},k_2) - c(A_{N-1}^{(1)},k_1+k_2) 
\cr & =
{ k_1k_2(k_1+k_2+2N)\, (N^2-1) \over (k_1+N)(k_2+N)
(k_1+k_2+N)} \,,\cr}
}
is given by
\eqn\chGcce{ \eqalign{
\left( \tilde{\et}^{(3)} \right) ^{-2} = &
{N^2-4 \over N} (k_1+N)^2 (k_2+N)^2 (k_1+k_2+N)^2 \cr
& (k_1+{N \over 2}) (k_2+{N \over 2}) (k_1+k_2+{3N \over 2})
\,.\cr}
}
The results \chGccb\ -- \chGcce\ were first given in \cite{\BBSSb}.
The field $\widetilde{W}^{(3)}(z)$ is primary of dimension $3$ 
under the coset Virasoro generator $\widetilde{T}(z)$.

It is now a straightforward excercise to check that by taking the limit
$k_2\to\infty$ in \chGccb\ -- \chGcce\ one recovers the
expression \chGbcf\ for the spin-3 Casimir operator.
This illustrates the claim of Section 7.3.1 that every
generator in $\cW[\whg/\bfg,k_1]$ can be deformed
(with the possible exception of a finite number of $(k_1,k_2)$--values)
to an element of $\cW[\whg\oplus \whg/\whg,(k_1,k_2)]$.

By taking the OPE of the generators $\widetilde{W}^{(3)}$ 
one recovers the result \chGbcd, in terms of the (tilded)
coset fields \chGccb\ and central charge $\tilde{c}$ given by
\chGccd. The field $\widetilde{R}^{(4)}(z)$ is a (coset)
primary of dimension 4. It is given by an involved expression
in terms of the currents $J_{(1)}^a(z)$
and $J_{(2)}^a(z)$ which however drastically simplifies in the case of
$N=3$ and one of the levels (say $k_1$) is put equal to one.
One finds in this case (see \cite{\BBSSb})
\eqn\chGccf{
\widetilde{R}^{(4)}(z) =
c_1(k_2) (R^{(2)}_{(1)ab}R^{(2)ab}_{(2)})(z) + c_2(k_2) (J_{(1)a}R^{(3)a}
_{(1)})(z) + c_3(k_2)(J_{(2)a}R^{(3)a}_{(1)})(z)\,,
}
where
\eqn\chGccg{ \eqalign{
R^{(3)}_a(z) & = 3(J_aT)(z) + f_{abc}(J^b\p J^c)(z) - 2\p^2J_a(z) \,,\cr
R^{(2)}_{ab}(z) & = (3d_{ab}d_{cd} + 24d_{ac}d_{bd} + 8 f_{ace}
f_{bd}{}^e) (J^cJ^d)(z) \,,\cr}
}
are primary fields of dimension $3$ and $2$, transforming in the
$\underline{8}$ and $\underline{27}$ of the horizontal $sl(3)$,
respectively.
The coefficients $c_i(k_2)$ have the property that $c_i(1)=0$.
By the same methods as employed in Section 7.2.3, \ie by using the
explicit vertex operator construction or by studying the
character, one now shows that both $R^{(3)a}(z)$ and $R^{(2)ab}(z)$ are
vanishing on the integrable (irreducible) highest weight modules $L_\La$
at level one. This establishes the existence of a $\cW_3$-algebra
on the coset
modules constructed from the $A_2^{(1)}\oplus A_2^{(1)}$ modules
$L_\La\otimes
M$ at level $(k_1,k_2)=(1,k_2)$ where $M$ is an {\it arbitrary}
$A_2^{(1)}$-module with highest weight.
\medskip

We finish this section by briefly mentioning the possibility
of `supersymmetrizing' specific coset algebras $\cW_c[\whg/\whg',k]$.
By this, we mean adding halfinteger dimension operators, constructed
out of the $\whg$ currents {\it and} $\whg$ primary fields, commuting
with $\whg'$ and such that the combined set again forms a closed
algebra.

The standard example occurs in the context of diagonal cosets
$\cW_c[\whg\oplus\whg/\whg,(k_1,k_2)]$, $\bfg$ simply-laced,
by taking $k_1=\dCo$.
It is easily checked, using (2.74),
that the adjoint representation at level $\dCo$ has conformal dimension
$\half$. The corresponding highest weight fields $\ps^a(z)$ thus
correspond
to a set of dim$(\bfg)$ free fermions. Let us normalize them such that
\eqn\chGdca{
J_{(1)}^a(z) \ps^b(z) = f^{ab}{}_c {\ps^c(w)\over z-w}\,,\qquad\qquad
\ps^a(z)\ps^b(w) = {d^{ab}\over z-w}\,.
}
Then, a dimension $3/2$ operator that commutes with $\whg_{diag}$ is
explicitly given by \cite{\Doa,\BBSSc,\GSa}
\eqn\chGdcc{
G(z) = {\textstyle{1\over3}}(J_{(1)a}\ps^a)(z) - {\dCo\over k_2}
(J_{(2)a} \ps^a)(z)
}
For $\bfg = su(2)$ this leads to a well-known coset construction
of the $N=1$ superconformal algebra \cite{\GKOb}. For higher
rank algebras there will, in general,  be additional generators
of half-integer dimension. The case $\bfg=su(3)$, in particular,
leads to the super-$\cW_3$ algebra discussed in Section 4.2
\cite{\HR,\ASS,\SS}.
More general supersymmetrizable coset algebras have been discussed
in \cite{\BoGb}.\bs

\noindent{\sl 7.3.3. Representation theory}\hfill\medskip

Since the coset algebras $\cW_c[\whg\oplus\whg/\whg,(k_1,k_2)]$ come
naturally equipped with a set of characters (\ie branching functions)
$b^{\La_1,\La_2}_{\La_3}(q)$ (see Section 7.1) we can, in principle,
study their representation theory by examining the set of branching
functions.

Since our main goal will be to argue, for simply-laced $\bfg$, the
equivalence of the coset algebra $\cW_c[\whg\oplus\whg/\whg,(1,k_2)]$ to
the QDS algebra $\cW_{DS}[\whg,k]$ we will restrict the following
discussion to the case $k_1=1$ and $\bfg$ simply-laced.
The general situation can be treated analogously.

The branching functions $b^{\La_1,\La_2}_{\La_3}(q)$ for the case of
integrable weights were already presented in \chGcad. We present here a
slight generalization \cite{\KaWa,\KaWb}.
Let $\wLa_1\in P^1_+$, and suppose
$k_2+\dCo = p/(p'-p)\equiv p/u$. To $\wLa^{(+)}\in P_+^{p-\dCo}$ we
associate the principal admissible weight $\wLa_2 = \wLa^{(+)} -
(u-1)(k_2+\dCo)\La_0$ of level $k_2$. Then the branching function for
the occurrence of $L_{\La_3}$, $\wLa_3 = \wLa^{(-)} -
(u-1)(k_2+\dCo+1)\La_0$, $\wLa^{(-)}\in P_+^{p'-\dCo}$ in the
decomposition of $L_{\La_1}\otimes L_{\La_2}$ under $\whg_{diag}$ is
given by
\eqn\chGcba{
b^{\La_1,\La_2}_{\La_3}(q) = {1\over \et(\ta)^\ell} \sum_{w\in
\widehat{W}} \ep(w) q^{ {1\over2 pp'} |p'w(\Lap + \rh) - p(\Lam+\rh)|^2}
\,.}
provided $\La_1+\La_2-\La_3\in Q$ (and vanishing otherwise).
In particular, for $p'=p+1$ we recover \chGcad. Now observe that, in
fact, \chGcba\  is identical to (6.85), provided we identify
\eqn\chGcbb{
{p'-p\over p} =
{1\over k_2+\dCo} = {1\over k+\dCo} -1\,.
}
This finally proves that $\cW_c[\whg\oplus\whg/\whg, (1,k_2)] \cong
\cW_{DS}[\whg,k]$ (up to null-fields)
provided we identify $k_2$ and $k$ through \chGcbb.
In particular, we read off from \chGcba\ that we have a set of
completely degenerate highest weight modules (\ie nullvectors in all
simple root directions) of $\cW_c[\whg\oplus\whg/\whg,(1,k_2)]$ labelled
by two integrable weights $\Lap\in P_+^{p-\dCo}\,, \Lam\in
P_+^{p'-\dCo}$ of conformal dimension $h_{\Lap\Lam}$ given by (see also
(6.74))%
\foot{We suppress the dependence on $\La_1\in P_+^1$ since $\La_1$
adjusts itself in such a way that $\La_1+\La_2-\La_3\in Q$.}
\eqn\chGcbc{
h_{\Lap\Lam} - {c\over24} = -{\ell\over24} + {1\over 2pp'}
|p'(\Lap+\rh) - p(\Lam+\rh)|^2\,,
}
where, using \chGccd,
\eqn\chGcbd{
c= {k_2(k_2+1+2\dCo)\, {\rm dim\,}\bfg\over
(1+\dCo)(k_2+\dCo)(k_2+1+\dCo)} =
\ell \left( 1 - {\dCo(\dCo+1) (p-p')^2 \over pp'} \right) \,.
}
This is, of course, in agreement with (6.13), for $\al_+^2 = p'/p$.

Having established the equivalence of the coset and QDS $\cW$-algebra we
now want to make some comments on the set of unitary representations for
this algebra as promised in Section 6.4.1. Clearly, the realization of
$\cW_c[\whg\oplus\whg/\whg,(k_1,k_2)]$ on $L_{\La_1\otimes\La_2,\La_3}$
will be unitary when the $\whg$ modules $L_\La$ are unitary
themselves, \ie when they are integrable.
This means $p=k_2+\dCo=p'-1$, $\Lap\in P_+^{p-\dCo}\,,
\Lam\in P_+^{p+1-\dCo}$, $p\geq\dCo$ in \chGcbc, \chGcbd.
The conjecture is that these are {\it all} unitary modules in
the range $c<\ell$, and furthermore that for $c\geq\ell$ there is no
restriction on $(\Lap,\Lam)$ from unitarity.
This conjecture is known to be true for
$\bfg = sl(2)$ \cite{\FQSa,\FQSb,\FQSc,\GKOb},
and also not in contradiction with the analysis in \cite{\Mia,\Mib}.

Modular invariant combinations of the $\cW$-characters $b_{\La_3}^
{\La_1,\La_2}(q)$ \chGcba\ can be constructed out of those for
the $\whg$-characters by using the decomposition \chGad.
The simplest example, for $\cW_3$, is the partition function
of the 3-state Potts model which has been discussed in Section 4.2.
\bs

\noindent{\sl 7.3.4. Kac determinant}\hfill\medskip

In this section we will derive the Kac determinant (see Section 6.3.1)
for the coset
$\cW$-algebra $\cW_c[\whg\oplus\whg/\whg, (1,k)]$, purely from
the coset data, \ie without presupposing a relation to a $\cW$-algebra
arising from the QDS reduction. For simplicity we will restrict
our attention to the case of simply-laced $\bfg$.

We recall the branching function \chGcba
\eqn\chGcda{ {\rm ch\,}_{L_{\Lap\Lam}} (q) =
b^{\La_1,\La_2}_{\La_3}(q) = {1\over \et(\ta)^\ell} \sum_{w\in\widehat{W}}
\ep(w) q^{ {1\over 2pp'} |p' w(\Lap+\rh) - p(\Lam+\rh)|^2 }\,.
}
To derive the Kac determinant we exploit the fact that the branching
function \chGcda\ contains sufficient information on the existence
and whereabouts of singular vectors in the $\cW[\whg\oplus\whg/\whg]$
Verma modules. To this end we expand the expression given in \chGcda\
in powers of $q$ in the limit $k\to \infty$
\eqn\chGcdb{ {\rm ch\,}_{L_{\Lap\Lam}} =
{ q^{h_{\Lap\Lam} - c/24} \over \prod
(1-q^n)^\ell } \left( 1 - \sum_i q^{(\Lap+\rh,\al_i)(\Lam+\rh,\al_i)}
+ \ldots \right)\,,
}
where we have depicted only the leading order correction terms in the
limit $k\to\infty$, which arise from the Weyl group elements $w=r_i\,,
i=1,\ldots,\ell$.
{}From \chGcdb\ we conclude that, for $k$ sufficiently
large, there exists an $i\in\{ 1,\ldots,\ell\}$ such that the
module $L_{\Lap,\Lam}$ has a singular vector at degree
$(\Lap + \rh,\al_i)(\Lam+\rh,\al_i)$. When we parametrize, as usual,
$\La = \al_+\Lap + \al_-\Lam$ the above reasoning shows that
${\rm det\,}{\cal M}_N$ will have a vanishing line
\eqn\chGcdc{
(\La + \al_0\rh,\al_i) - (m\al_+ + n\al_-)\,,
}
for all $m,n\in\ZZ_+$ such that $N=mn$. Including the contribution of
descendant states and using Weyl invariance we conclude, as in Section
6.3.1, that the Kac determinant is given by
\eqn\chGcdd{
{\rm det\,}{\cal M}_N \sim \prod_{\al\in\De} \prod_{mn\leq N}
\left( (\La + \al_0\rh,\al) - (m\al_+ + n\al_-)
\right) ^{p_\ell(N-mn)}\,.
}
This indeed agrees with (6.72) for simply-laced $\whg$.
\bs

\noindent{\sl 7.3.5. The limiting $\cW$-algebra of
diagonal coset models}\hfill\medskip

We have seen that to every diagonal coset pair $(\whg\oplus
\whg, \whg_{diag})$ one can associate a $\cW$-algebra
$\cW[\whg\oplus\whg/\whg,(k_1,k_2)]$. This
$\cW$-algebra ``stabilizes'' for fixed $k_1$ in the limit
$k_2\to \infty$,
where it reduces to the `Casimir algebra'' of $\whg$, \ie the algebra
$\cW[\whg/\bfg, k_1]$. One might pose the question whether this
Casimir algebra in turn ``stabilizes'' in the limit $k_1\to\infty$. This
appears to be the case. To be able to consider the limiting expression
for $b_0^{k\La_0}(q)$ for $k\to\infty$ in \chGbbc, let us recall the
following explicit expression for the Kac-Peterson string functions
(see \eg \cite{\BMPc})
\eqn\chGcga{
c^\La_\la(q) = {q^{-c/24}\over \prod(1-q^n)^{{\rm dim\,}\bfg} }
\sum_{w\in \widehat{W}} \ep(w) q^{h_{w*\La,\la} }
\sum_{ \{ n_\be\in\ZZ | \sum_{\be\in\De_+} n_\be\be = w*\La -\la\} }
\left(\prod_{\be\in\De_+} \ph_{n_\be}(q) \right)\,,
}
where we have introduced
\eqn\chGcgb{
\ph_n = \sum_{m\geq0} (-1)^m q^{\half m(m+1) +nm}\,,\qquad\qquad
\ph_{-n}(q) = q^n \ph_n(q)\,,
}
and
\eqn\chGcgc{
h_{\La,\la} = { (\La,\La+2\rh) \over 2(k+\dCo)} - {(\la,\la)\over 2k}
\,.
}
In the limit $k\to\infty$ only the finite part of the Weyl group
contributes in the sum \chGcga. This inserted in \chGbbc\
produces the required limit. As an example consider $\bfg = A_1$. We have
\eqn\chGcgd{
c_{l\La_1+ (k-l)\La_0}^{k\La_0}(q) \ \ \mapright{k\to\infty}\ \
{ \ph_{-l/2}(q) - \ph_{-1-l/2}(q) \over \prod(1-q^n)^3}\,.
}
Hence
\eqn\chGcge{ \eqalign{
\Ph_0^{k\La_0} \ \ \mapright{k\to\infty}\ \ &
{ \ph_0(q) -2q\ph_1(q) + q^2\ph_2(q) \over \prod(1-q^n)^3 } \cr
& = {1\over F_2F_4F_6^2F_8^2F_9F_{10}^2F_{12} } (1-q^{13}-3q^{14}-7q^{15}
+ {\cal O}(q^{16}) )\,.\cr}
}
This suggests the existence of a ``limiting $\cW$-algebra'' of type
$\cW(2,4,6_2,8_2,9,10_2,12)$ (see also \cite{\Blb}), which
could serve as a ``universal $\cW$-algebra'' in the sense that
it contains all the $A_1$ coset algebras at finite $k$-values.
In fact, the limiting field content is already reached for a low value of
$k$, namely $k=6$.%
\foot{For $k=1,\ldots,5$ the $\cW$-algebras are
of type $\cW(2)$, $\cW(2,4,6)$, \hfill\break
$\cW(2,4,6_2,8,9)$,
$\cW(2,4,6_2,8_2,9,10)$, $\cW(2,4,6_2,8_2,9,10_2)$, respectively.}

For $\bfg =A_2$ one finds similarly a limiting $\cW$-algebra
of type $\cW(2,3,4,5,6_4,7_2,8_7,9_9,$ $10_{12}, 11_{16},12_{26},
13_{26},14_{33},15_{33},16_{12})$. The limiting field content is reached
at $k=8$ \cite{\Blb}.

The relevance of the above considerations remains to be investigated.

\subsec{Other cosets}

\noindent{\sl 7.4.1. $G\times G'/G'$ coset conformal field
theories}\hfill\medskip

The most straightforward generalization of the
$\cW_c[\whg\oplus\whg/\whg]$
construction of Section 7.3 is the construction of
$\cW_c[\whg\oplus\whg'/\whg', (k_1,k_2)]$, where $\bfg'\subset\bfg$ is
an embedding of Dynkin index $j$. The diagonal subalgebra
$\whg'_{diag}\subset \whg\oplus \whg'$ will thus have level $jk_1 +k_2$.
Most of the results of Section 7.3 can easily be carried over
to this case.
In particular, if $|\ph\rangle$ is a $\whg'_{diag}$ singlet in the
characteristic Hilbert space of $\whg\oplus\whg'$ and $P_N|\ph\rangle$
denotes its orthogonal projection onto the eigenspace $L_0(\whg')=N$ then
$P_0|\ph\rangle$ is a $\bfg'$-singlet. The corresponding map
\eqn\chGdac{
P_0\ : \ \cW_c[\whg\oplus\whg'/\whg',(k_1,k_2)] \ \to\
\cW_c[\whg/\bfg',k_1]\,,
}
is injective and turns into an isomorphism in the limit $k_2\to\infty$
\cite{\BoGb}.

The most interesting case occurs when $\whg'\subset\whg$ is conformal
(which requires at least that $k_1=1$) \cite{\BoGb}.
Of course, since a conformal coset
$\whg/\whg'$ is `trivial', \ie has finite branching rules, one can
interpret any $\cW_c[\whg\oplus\whg'/\whg']$
model as a particular case of
a $\cW_c[\whg'\oplus\whg'/\whg']$ model.
Detailed examples of such coset algebras are
discussed in \cite{\GSa,\BoGb}. For instance,
by taking the conformal embedding $so(N)\subset su(N)\,,N\geq3$
of Dynkin index $j=2$,%
\foot{For $N=3$ one rather takes $su(2)\subset
su(3)$ which has Dynkin index $j=4$.}
one obtains a class of $\cW$-algebras
$\cW_c[su(N)\oplus so(N)/so(N), (1,m)]$ that are related to a
class of parafermions with dihedral $D_N$-symmetry \cite{\FZa}
(see also the remarks in Section 5.3.3). In particular,
for $N=3$, where we have $D_3\cong S_3$, we obtain the
Fateev-Zamolodchikov $S_3$ (or spin-$4/3$) parafermionic model
\cite{\FZc,\BNY,\KMQ,\BN,\Rac,\Raa}.
Let us discuss this particular example in somewhat more detail.
That is, we consider the algebra $\cW_c[A_2^{(1)}\oplus A_1^{(1)}/
A_1^{(1)}, (1,m)]$, where $su(2)$ is principally embedded in $su(3)$.
For the coset central
charge we find $c(m) = 2\left( 1- {12\over (m+2)(m+6)}\right)$.
This indeed agrees with the series for the
Fateev-Zamolodchikov spin-$4/3$ parafermionic
algebra which is of type $\cW(2,{4\over3},{4\over3},\ldots)$ \cite{\FZc}.
Let us look at the underlying (bosonic) $\cW$-algebra. First we study the
$m\to\infty$ limit $\cW_c[A_2^{(1)}/A_1,1]$. To count the number
of $su(2)$ singlets $\widetilde{\Ph}^{\La_0}_0(q)$
in the level-1 $\widehat{su}(3)$ vacuum module $\widetilde{L}_{\La_0}$ we
use the fact that under $\widehat{su}(2)$ we have a decomposition
\eqn\chGdaa{
\widetilde{L}_{\La_0} = L_{4\La_0} \oplus L_{4\La_1}\,.
}
Hence
\eqn\chGdab{ \eqalign{
\widetilde{
\Ph}^{\La_0}_0(q) & = \Ph^{4\La_0}_0(q) + \Ph^{4\La_1}_0(q) \cr
& = q^{ {1\over12} }
\left(c^{4\La_0}_{4\La_0}(q) + c^{4\La_1}_{4\La_0}(q) -
2 q^{ {1\over4} }c^{4\La_0}_{2\La_0+2\La_1}(q) \right) \cr
& = {1\over F_2F_3F_4F_5F_6^2} (1 - 2q^9 - 7q^{10} +
{\cal O}(q^{11}) )\,.\cr}
}
which indicates that $\cW_c[A_2^{(1)}/A_1,1]$ and 
thus $\cW_c[A_2^{(1)}\oplus A_1^{(1)}/A_1^{(1)},(1,m)]$, 
for large enough $m$, is of type
$\cW(2,3,4,5,6_2)$. In fact, analysing the branching functions of the
coset pair
$(A_2^{(1)}\oplus A_1^{(1)},A_1^{(1)})$ itself, one finds that the coset
is of type $\cW(2,3,4,5,6_2)$ for $m\geq3$. For $m=1$, where the coset
has central charge $c={6\over7}$,
the same analysis suggests a $\cW$-algebra
of type $\cW(2,5)$. In fact $c={6\over7}$ is precisely one of
the $c$-values for which there exists an exotic $\cW(2,5)$-algebra
\cite{\Bob} (see also Section
5.2.3). For $m=2$ one finds type $\cW(2,3,4,5)$.
\bs

\noindent{\sl 7.4.2. Dual coset pairs}\medskip

Suppose we have two coset pairs $(\whg,\whk)$ and $(\whg',\whk')$.
One can
ask when their corresponding $\cW$-algebras $\cW_c[\whg/\whk,k]$ and
$\cW_c[\whg'/\whk',k']$ are isomorphic. This will evidently be the
case when all the branching functions are in 1--1 correspondence.
Such coset pairs were termed ``dual'' in \cite{\JMb}.

A generic method to obtain dual coset pairs is by
constructing so-called ``$T$-equivalent'' coset pairs \cite{\BoGb,\Al}.%
\foot{This, however, does not produce {\it all} dual coset pairs.}
Two coset pairs $(\whg,\whk)$ and $(\whg',\whk')$ are called
$T$-equivalent when there exists a Lie algebra $\bfg''$ such
that the embeddings $\whg\oplus\whk'\subset
\whg''$ and $\whg'\oplus\whk\subset\whg''$ are conformal. This implies
in particular that\foot{Note that the level of $\whg''$ is necessarily
equal to one. For clarity we have displayed the various levels
by subscripts on $T^\bfg(z)$. The Dynkin index of
$\bfk\subset\bfg$ is denoted by $j(\bfg:\bfk)$.}
\eqn\chGdba{
T^{\bfg''}_1(z) = T^{\bfg}_{j(\bfg'':\bfg)}(z) +
T^{\bfk'}_{j(\bfg'':\bfk')}(z) = T^{\bfg'}_{j(\bfg'':\bfg')}(z) +
T^{\bfk}_{j(\bfg'':\bfk)}(z)\,.
}
Or, by swapping $T$'s around,
\eqn\chGdbb{
T^{\bfg/\bfk}_{j(\bfg'':\bfg)}(z) = T^{\bfg'/\bfk'}_{j(\bfg'':\bfg')}
(z)\,,
}
\ie $T$-equivalent coset pairs have equivalent stress-energy tensors.

Examples of $T$-equivalent coset pairs were given in \cite{\BoGb,\Al}.
Apart from several exceptional cases there are four basic series:
\item{(I)}
$$(su(2n)_m,sp(2n)_m) \ \cong\ (su(2m)_n,u(m)_n)$$
\item{(II)}
$$(su(N)_m\oplus su(N)_n, su(N)_{m+n})\ \cong\
(su(m+n)_N, su(m)_N\oplus u(n)_N)$$
\item{(III)}
$$(so(N)_m\oplus so(N)_n,so(N)_{m+n}) \ \cong\  (so(m+n)_N,
so(m)_N\oplus so(n)_N)$$
\item{(IV)}
$$(sp(2N)_m\oplus sp(2N)_n, sp(2N)_{m+n}) \ \cong\  (sp(2(m+n))_N,
sp(2m)_N\oplus sp(2n)_N)$$

Here we have denoted the respective levels by subscripts.
The ``master Lie algebra'' $\bfg''$ in these four cases is
given by (I) $so(4mn)_1$, (II) $su((m+n)N)_1$, (III)
$so((m+n)N)_1$ and (IV) $so(4(m+n)N)_1$.

The prototype example of a $T$-equivalent coset pair is obtained
by putting $N=m=1$ in the series (IV). This yields two equivalent
ways of producing the Virasoro unitary series (2.26) \cite{\GKOb}.
Another interesting and useful example occurs by putting $m=n=1$ in
the series (II). This example shows that the ``$su(2)$ parafermionic
coset''
$(su(2)_N,u(1))$ (also called $\ZZ_N$ parafermions) \cite{\FZa,\GQ}
is $T$-equivalent to the $\cW_N$ diagonal coset
$(su(N)_1\oplus su(N)_1, su(N)_2)$. This observation, which was first
made in \cite{\BBSSb}, is very useful, for example,
in constructing free field realizations for the $su(2)$ parafermions
\cite{\GrN}.

By iterative use of the four basic series (I)--(IV) one can generate
a multitude of other interesting $T$-equivalent coset pairs.
For instance, summing (II) for $m=1$, we find
\eqn\chGdbc{
\bigoplus_{k=1}^n (su(N)_1\oplus su(N)_k, su(N)_{k+1})
\ \cong\
\bigoplus_{k=1}^n (su(k+1)_N, su(k)_N\oplus u(1))\,.
}
Cancelling various terms on the right hand side of \chGdbc\ yields
a description of the $su(n+1)$ generalized parafermions
\cite{\Geb} in terms of
sums of $\cW_N$ models, generalizing the $su(2)$ case above
$$
(su(n+1)_N, u(1)^n) \ \cong\ \bigoplus_{k=1}^n (su(N)_1\oplus su(N)_k,
su(N)_{k+1})\,.
$$
This explains a curious relation between the central charges of
the generalized parafermion and $\cW_N$ discrete series \cite{\Fa}.

Cancelling also terms on the left hand side of \chGdbc\ gives
$$
(su(n+1)_N,u(1)^n) \ \cong\
( \underbrace{su(N)_1\oplus \ldots\oplus su(N)_1}_{m+1} , su(N)_{m+1})\,,
$$
which is useful for the construction of a generalized parafermion free
scalar field realization \cite{\GrN,\DHS}.
\bs

As a consequence of the conformal embedding in $\whg''$ the
branching functions
of $T$-equivalent coset pairs satisfy a linear relation of the form
\eqn\chGdbg{
\sum_{\La} n_{\La\la'} b^\La_\la ({\bfg/\bfk})(q)
= \sum_{\La'} m_{\La'\la} b^{\La'}_{\la'}({\bfg'/\bfk'})(q)\,,
}
for some integers $n_{\La\la'}$ and $m_{\La'\la}$. To establish
that these
$T$-equivalent coset pairs give rise to a set of dual coset pairs, \ie
coset pairs having the same set of branching functions, requires an
explicit determination of the integers $n_{\La\la'}$ and $m_{\La'\la}$.
Duality has been established in the cases (I) $m=1$
\cite{\JMa,\Al}, (II) $m=n=1$ \cite{\Al}, (IV) $N=1$
\cite{\KaWc,\BoGb,\Al}.

\bs

\vfil\eject

%\end

% ch89.tex,  last updated 12/22/92
% 
%\input pkjmac
%\input chrefs
%\magnification=1200
%\baselineskip=1.2\baselineskip

\secno=7

%\writelabels

%\line{\hfill{\sl ch89.tex, \datum}}

%---------------------------body------------------------------------
\newsec{Further developments}

\subsec{$\cW$-gravity}

In this section we will briefly review some developments in
(classical and quantum) $\cW$-gravity.
$\cW$-gravity is a higher-spin extension of $d=2$ gravity whose
structure is based on an underlying $\cW$-algebra. This $\cW$-algebra
takes over the role played by the Virasoro algebra in pure $d=2$ gravity.
The main applications of $\cW$-gravity, which are in the area of
string theory, will be discussed in the next section. Review papers
on $\cW$-gravity are for example \cite{\BBS,\Huc,\SSvNf}.

\medskip

It is well-known that classical theories of gravity or supergravity
(in general dimensions) can be constructed in a systematic way
by starting from an algebra of space-time (super)symmetries. This
is done by first constructing a gauge theory for the symmetry algebra
and then imposing a number of Yang-Mills curvature constraints.
Because of these constraints general coordinate transformations become
a local symmetry, and in this way a theory of (super)gravity arises
\cite{\KTN}.
For $d=2$ $\cW$-gravity this procedure was successfully applied
in \cite{\SSvNc}, where a covariant Lagrange formulation of classical
$\cW_3$ gravity coupled to scalar fields was presented.

The gauge algebra used in \cite{\SSvNc} for the construction of covariant
$\cW_3$ gravity is a centerless classical limit of the quantum $\cW_3$
algebra. After solving the Yang-Mills curvature constraints in the
corresponding gauge theory, one finds a gauge multiplet which
contains four
zweibein fields $e_\mu{}^+$, $e_\mu{}^-$ and four $\cW_3$ zweibein fields
$b_\mu{}^{++}$, $b_\mu{}^{--}$, where $\mu=\pm$. There are
eight local gauge symmetries: general coordinate, Weyl and local
Lorentz symmetries, together
with their $\cW_3$ analogues. The coupling of these
$\cW_3$ gravity fields
to scalar matter fields was worked out in \cite{\SSvNc}, where a gauge
invariant kinetic action for $N$ scalar fields $\phi^i$, $i=1,2,\ldots,N$
was presented. We shall not display this action here, but instead discuss
the results in the chiral gauge and in the light-cone gauge, which can
both be obtained by fixing some of the local symmetries.

In the chiral gauge
the coupling of scalar matter fields to $\cW_3$ gravity, which was
first obtained in \cite{\Hua}, is easily
explained. We start from the observation \cite{\Hua} that a free
action for $N$ scalar fields admits a rigid $\cW_3$ symmetry.
The transformations of the scalar fields read
\eqn\chHaa{
\delta_\epsilon \phi^i = \epsilon_+ \, \del_- \phi^i \,, \quad
\delta_\lambda \phi^i = \lambda_{++} \, d^{ijk} \,
\del_- \phi^j \del_- \phi^k \,,
}
where $d^{ijk}$ is a symmetric 3-index tensor. [We write $\del_-$ and
$\del_+$ for $\del_z$ and $\del_{\zb}$, respectively.] One can promote
these rigid
symmetries to local gauge invariances by introducing gauge fields
$h_{++}$ and $b_{+++}$ in the standard way. It turns out \cite{\Hua} that
the scalar field action with only the minimal coupling to these gauge
fields,
\eqn\chHab{
S_{\rm ch} = \frac{1}{\pi} \int d^2 z \; \left[
    - \half \del_+ \phi^i \del_- \phi^i + \half h_{++} \del_+ \phi^i
    \del_+ \phi^i + \frac{1}{3} b_{+++} d^{ijk}
    \del_+ \phi^i \del_+ \phi^j \del_+ \phi^k \right] \,,
}
is gauge-invariant, provided we choose the transformation rules of
$h_{++}$ and $b_{+++}$ appropriately and we have the identity
\eqn\chHac{
d^{k(ij} \; d^{l)mk} = \delta^{(ij} \; \delta^{l)m} \,.
}
The two local symmetries, with parameters $\epsilon_+(z,\zb)$ and
$\lambda_{++}(z,\zb)$, can be viewed as particular linear combinations of
all eight local symmetries present in the covariant formulation.

In the light-cone gauge, with both chiralities present, the coupling of
scalar matter fields to $\cW_3$ gravity is more involved.
The $\cW_3$ gauge
fields are $h_{\pm\pm}$ and $b_{\pm\pm\pm}$, corresponding to local
symmetries with parameters
$\ep_\pm(z,\zb)$ and $\la_{\pm\pm}(z,\zb)$. In \cite{\SSvNb} it
was found that the
light-cone gauge action is non-polynomial in the spin-3 gauge fields.
This
difficulty can be circumvented by introducing auxiliary fields $F_+{}^i$
and
$F_-{}^i$, which will play the role of so-called {\it nested covariant
derivatives}. With these variables, the gauge-invariant action takes the
following form \cite{\SSvNb}
\eqn\chHad{
\eqalign{
S_{\rm lc} = & \frac{-1}{\pi} \int d^2 z \, e \left[
  - \half \nabla_+ \phi^i \,\nabla_- \phi^i - F_+{}^i F_-{}^i +
  F_+{}^i \left( \nabla_- \phi^i - \frac{1}{3} b_{---}
  d^{ijk} F_+{}^j F_+{}^k \right) \right.
\cr
&  \quad \quad \quad \left. + F_-{}^i \left( \nabla_+ \phi^i -
   \frac{1}{3} b_{---} d^{ijk} F_-{}^j F_-{}^k \right) \right] ,
\cr}
}
where $e=(1-h_{++}h_{--})^{-1}$ and $\nabla_{\pm} = \del_{\pm} -
h_{\pm\pm}\del_{\mp}$. The field equation of $F_-{}^i$ is algebraic and
leads to
\eqn\chHae{
F_+{}^i = \nabla_+ \phi^i - b_{+++} d^{ijk} F_-{}^j F_-{}^k
}
(with a similar result for $F_-{}^i$).  When solving $F_\pm^i$ by
iteration one obtains a $\cW_3$ generalization of a covariant derivative,
which is infinitely nonlinear and is appropriately called a nested
covariant derivative.

In a similar way, light-cone gauge and covariant actions for $w_\infty$
gravity coupled to scalar fields were constructed in \cite{\BPRSSS}
and in \cite{\SSvNca}, respectively. A geometrical interpretation of
the full non-linear structure of $w_\infty$ gravity was presented
in \cite{\Huba}. We refer to \cite{\BBS,\Hub,\Mik} for further results on
classical $\cW$-gravity.

\medskip

At the classical level, $\cW$-gravity does not have
degrees of freedom, since all the field components are compensated
for by local gauge symmetries (up to possible moduli). However, upon
quantization one or more of these symmetries become anomalous and
therefore quantum $\cW$-gravity does have degrees of freedom.
If the central charge $c$ of the matter system coupled to
$\cW$-gravity is tuned to a specific value, the quantum anomalies
cancel and the $\cW$-gravity degrees of freedom decouple. This is
similar to what happens in $d=2$ induced gauge theories and
$d=2$ ordinary gravity \cite{\Pob,\SSvNf}.

The dynamics of the quantum degrees of freedom of
$\cW$-gravity are governed by an {\it effective} action.
This action can be studied for various gauge
choices, for example in the chiral gauge or in the conformal gauge.
In the chiral gauge the effective action arises in the following way
\cite{\Mat,\SSvNd,\Hubb,\SSvNe,\GvN}.

One first obtains an {\it induced} action by performing a functional
integral over the matter fields, of total central charge $c$, to which
the $\cW$-gravity fields are coupled.
The non-trivial contributions to the induced action arise from
loop-diagrams with $\cW$-gravity fields on external lines and
matter fields
in the loops. The resulting induced action, which can be viewed as an
integrated anomaly, is non-local. The {\it effective} action is then
obtained by renormalizing this induced action, taking into account
loop-diagrams for the anomalously propagating $\cW$-gravity fields.
The entire computation of the effective action can be done
perturbatively, with $1/c$ as the expansion parameter.

In \cite{\dBGa,\dBGb}
the covariant induced action for $\cW_n$ gravity (in the limit
$c\to\pm\infty$) was presented. It is given in terms of
two (left and right) chiral sectors, which contain fields
$b_\pm^{(i)}$, $i=2,\ldots,n$, $(n-1)$ scalar fields $\ph^{i}$,
$i=2,\ldots,n$, and a number of auxiliary fields which play
the role of nested covariant derivatives as in \chHad, \chHae.
When specializing to the conformal gauge, one finds
\cite{\dBGa} that
this induced action takes the form of a Toda action for the scalar fields
$\ph^i$. This implies that in the conformal
gauge the $\cW$-currents of $\cW$-gravity take the familiar
free-field form (compare with \chFcc, \chFce).
Because of this, explicit computations in quantum $\cW$-gravity, such as
for example the analysis of the BRST cohomology of physical states, are
most easily done in the conformal gauge \cite{\DDR}.

\medskip

In the above we mentioned that the effective action for
$\cW$-gravity can be computed perturbatively in quantum field theory,
with $1/c$ as the expansion parameter. However, it is possible to
exploit once more the relation between $\cW$-algebras and affine
Lie algebras to obtain results exact to all orders for some
of the quantities in $\cW$-gravity! The idea here is that
$\cW$-gravity can be obtained by constraining WZW field
theory or, equivalently, by reducing $d=3$ Chern-Simons gauge
theory. We will now briefly explain how this works for
$\cW_3$ gravity.

Consider $\cW_3$ gravity coupled to
a $\cW_3$ CFT of central charge $c$.
The matter sector can be described in the QDS
reduction scheme (see Chapter 6), where the starting point is the affine
Lie algebra $\widehat{sl}(3)$ at level $k$, with (see \chFbg)
\eqn\chHa{
c= c_k = 2- 24 \left( \sqrt{1\over k+3} - \sqrt{k+3} \right) ^2
= 50 - 24 \left( {1 \over k+3} +k+3 \right) \ .
}
To couple this system to $\cW_3$ gravity, we
must make sure that the central charge of the $\cW_3$
gravity sector compensates the matter central charge $c_k$ {\it plus} the
contribution of the $\cW_3$ ghosts, which equals $c_{gh}=-100$
(see Section 8.2). Observe that $c_k = 100 - c_{-(k+6)}$,
which shows that we obtain the correct contribution to the central
charge if we apply the QDS reduction to an $sl(3)$ WZW model of level
$\kappa = -(k+6)$. With \chHa\ this gives
\eqn\chHb{
\kappa = - {1 \over 48} \left( 50-c + \sqrt{(c-98)(c-2)} \right) -3 \ ,
}
where the sign in front of the square root has been chosen in
accordance with the classical limit $c \rightarrow -\infty$, in which
$\kappa \sim c/24$.

The picture that arises at this point is the following: we can represent
the degrees of freedom of the {\it effective} quantum  $\cW_3$
gravity, induced  from a matter system of central charge $c$
and after renormalization,
by a QDS reduced $sl(3)$ WZW model of level $\kappa$
as in \chHb. This observation explains the presence of a `hidden'
affine $sl(3)$
symmetry in $\cW_3$ gravity, which generalizes the affine $sl(2)$
symmetry in ordinary gravity first observed by Polyakov \cite{\Poa}.
For ordinary gravity the connection with a constrained $sl(2)$ WZW model was
first understood in \cite{\KPZ}; the generalization to $sl(n)$ was worked
out in \cite{\BOa}.

{}From the above picture one expects that the effective action for
$\cW_n$-gravity can be obtained in explicit form by reducing
the $sl(n)$ WZW action.
For $\cW_3$ gravity in the chiral gauge this result has been checked
by explicit perturbative computations.
It was found \cite{\Mat,\BFK,\DHR,\OSSvN} that the induced
action for chiral $\cW_3$ gravity, in the limit $c \rightarrow
\pm\infty$, is governed by local Ward identities which can be
obtained by reducing the Ward identities for the $sl(3)$ WZW model.
For finite $c$, the induced action contains additional
terms that correspond to non-local additional terms in the Ward
identities \cite{\SSvNd}. However, the results of \cite{\SSvNe}
suggest that these extra terms get cancelled if one renormalizes the
induced action. In \cite{\SSvNe} the full effective action was
computed through the first non-leading order in the perturbative
$1/c$ expansion,
and it was found that the result precisely takes the form of a
reduced WZW action with renormalized coefficients. The overall
coefficient of the effective action can be identified with the
level $\kappa$ of the affine $sl(3)$ algebra. The perturbative result
for $\kappa$ thus obtained
agrees with the relation \chHb\ through the first
non-leading order in $1/c$.

In \cite{\OSSvN,\SSvNf} the effective action for chiral
$\cW_3$ gravity was given in a closed form.
In terms of Polyakov type variables $f$ and $g$,
this action is {\it local}, and it generalizes the effective
action for ordinary gravity in terms of Polyakov's variable $f$ \cite{\Poa}.

\subsec{$\cW$-symmetry in string theory}

We would now like to come back to some
remarks we made in Chapter 1, namely the applications
of $\cW$-symmetry in string theory. Many of the important
issues in this field
are still unsettled and the discussion below is intended to merely
give a flavor of the developments that are taking place.

In the standard formulation of string theory, the fields representing
the coordinates of a (first-quantized) string in target space-time,
define a `matter' CFT.
The reparametrization invariance on the
string world-sheet implies that this CFT should be coupled
to two-dimensional (world-sheet) gravity. If the total
central charge of the `matter' CFT
is equal to 25,  the $d=2$ gravity sector decouples.
In that case the gravity sector adds one free scalar
field to the space-time coordinates, and it leads to a number of
constraints (the Virasoro constraints) on physical states in the
string theory. If the
matter central charge differs from 25 the extra scalar field becomes
interacting and the situation is more intricate.

In recent years, considerable progress has been made in understanding
the coupling of $c\leq 1$ (minimal) CFT's to $d=2$ gravity.
It has turned out that these theories
can also be studied from the point of view of discretized world-sheets
(matrix models) or from the point of view of topological field theories.
For $c>1$, which is the regime where the bosonic string develops
tachyonic states, the coupling to $d=2$ gravity has run into
`strong-coupling problems' \cite{\KPZ} and has not yet been understood
properly.

\medskip

In the study of $c<1$ CFT's coupled to gravity, interesting
connections with $\cW$-symmetry have come up.
Namely, it was
found, both in the matrix model formulation and in the topological
approach, that the partition function of the theory (as a function of
a set of coupling constants) can be characterized by so-called
$\cW$-constraints \cite{\DVVb,\FKNa,\Goe}. The appearance of these
constraints is closely related to the fact that these models can be
analyzed in terms of generalized KdV hierarchies \cite{\Doc}.
The paper \cite{\FKNb} shows that the $\cW_n$ constraints can
be viewed as reductions of more general $\cW_{1+\infty}$ constraints,
which arise naturally when one views the generalized KdV hierarchies
as reductions of the KP hierarchy.

The study of $c=1$ strings, both in terms of matrix models
and as a continuum theory, has revealed an interesting symmetry
structure. In the case of a flat, uncompactified background the
(chiral) symmetries in the continuum theory have been identified
\cite{\KPa,\Wid} with the area preserving polynomial vector-fields,
generating (the wedge of)
a $w_{\infty}$ algebra as in \chEac.

\medskip

In the cases just mentioned $\cW$-symmetries arise as `bonus'
symmetries in specific string theories that were constructed
without any reference to $\cW$-symmetry. It is certainly
interesting to see if one can construct string theories
with manifest $\cW$-symmetries built in.
A first step in this direction is to study $\cW$-extensions
of $d=2$ gravity and their couplings to CFT. In Section 8.1 we
briefly reviewed some results in this area. We will now
further explore the possibilities to construct $\cW$-extensions
of string theories.

A first remark concerns the $\cW$-constraints in matrix models
and topological field theories mentioned above. It has been proposed
\cite{\AQ,\DDR} that the coupled systems of CFT plus gravity, for which
these constraints occur, are actually closely related to theories
of {\it pure} $\cW$-gravity. Physical states in the spectrum of pure
$\cW$-gravity can sometimes be viewed as CFT matter states `dressed'
by the fields of pure $d=2$ gravity \cite{\DDR}. However, full equivalence
of both theories cannot be claimed, and the connection remains rather
mysterious.

Going one step further, one can consider $\cW$-invariant CFT's coupled
to the corresponding $\cW$-gravity. One interesting observation
is that in this context the critical central charge $c=1$ for
ordinary gravity gets shifted to a higher value. For example, 
for the $\cW$-algebras
related to the $ADE$ simply-laced Lie algebras the threshold value
for the central charge would be the rank $\ell$ of the Lie algebra.
[For $\cW_3$, where $\ell=2$, this can be read off from equation
\chHb; other cases can be treated similarly.]
This makes clear that certain CFT's with $c>1$, whose coupling to
ordinary gravity is problematic, can consistently be coupled to
$\cW$-gravity. We expect that such can be done for {\it any} RCFT,
where the $\cW$-algebra for the $\cW$-gravity theory should be closely
related to the chiral algebra of the RCFT.

Systems of $\cW$-invariant CFT coupled to $\cW$-gravity can tentatively
be interpreted as (critical or non-critical) $\cW$-strings. For the case
of $c<\ell$ minimal models of one of the $ADE$ $\cW$-algebras coupled to
the corresponding theory of $\cW$-gravity, one expects behavior
which is qualitatively similar to that of the standard $c<1$ strings.
In particular, one expects connections with generalized matrix models
and topological field theories, which still remain to be worked out.

The direct generalization of lower critical ($c_1=1$) and upper critical
($c_2=26$) bosonic strings to $\cW$-strings is problematic.
On the level of the algebra the numerology is clear: for example, for the
$ADE$ $\cW$-algebras the lower critical dimension is $c_1=\ell$, and
the upper critical dimension has been found to be
$c_2=2\ell (2{\rm h}^2+2{\rm h}+1)$
\cite{\BBSSa}, where ${\rm h}$ is the Coxeter number
of the Lie algebra. The latter
value is the one for which a nilpotent BRST charge is expected to exist.
For a general (generic) $\cW$-algebra, the critical central charge is
given by
\eqn\chHba{
c = \sum_s \, 2\, (-1)^{2s}\, (6 s^2 - 6 s +1) \ ,
}
where the summation runs over the spins $s$ of the independent generators
of the $\cW$-algebra. However, the existence of a
nilpotent BRST charge is not always guaranteed
(due to complications with non-linearity, see \cite{\SSvNa}) and
should be checked in each individual case.
For the $\cW_3$ algebra, where $c_2=100$, a nilpotent quantum
BRST charge has been constructed in \cite{\Tha,\SSvNa}.
[For infinite $\cW$-algebras the sum \chHba\ needs to be
regularized. It was argued in \cite{\Yab} that for the
$\cW_\infty$ algebra a nilpotent BRST charge exists for $c=-2$.]

The facts
that the notion of critical central charges $c_1=1$ and $c_2=26$ can be
generalized to (at least) the $ADE$ $\cW$-algebras, and that nilpotent
BRST charges can presumably
be constructed, do not by themselves
imply the existence of interesting new string theories. In addition,
a string theory requires that (part of)
the matter conformal field theory
can be viewed as a set of coordinates on a target space-time. In practice
one should therefore consider realizations of (critical or non-critical)
$\cW$-algebras in terms of scalar matter fields. For the $\cW_3$ algebra
realizations in terms of an arbitrary number of scalar fields
(and with adjustable central charge) were discussed in \cite{\Roa}.
The anomaly-free coupling of such matter systems with central
charge $c=100$ to $\cW_3$ gravity was discussed in \cite{\PRSta}.

In general, scalar field realizations of $\cW$-algebras involve
background charges, which lead to mass-shifts in the spectrum of physical
string states. Because of that, the original expectation
that the spectrum of a $\cW$-string might contain states with space-time
spin greater than two \cite{\BGe} is probably not justified.
Also, the fact that the equations determining the BRST cohomology of
physical states involve higher ($\geq 2$) order polynomials in general,
leads to a `branching' of the spectrum of physical states.
As a consequence, in the lower critical dimension $d=\ell$,
$\cW$-strings for the $ADE$ $\cW$-algebras are {\it not}\ free
from tachyons \cite{\DDR}.
In fact, it was argued in \cite{\DDR} that the critical numbers $d_1$ and
$d_2$ of scalar fields in $\cW_n$ gravity (as opposed to the critical
central charges) are $d_1=6/n(n+1)$ and $d_2 = 24 + 6/n(n+1)$.
We would like to stress, however, that the validity of these results
depends crucially on a specific (but
debatable) Ansatz for the matter couplings in the conformal gauge.

The spectrum of critical $\cW_n$ strings has been further analyzed in
\cite{\LPSX}. In \cite{\LPScW} some of these results were generalized
to $\cW$-strings based on more general $\cW$-algebras.

\medskip

String theory is more geometrical than CFT, which seems to be
one of the reasons why the application of finitely generated
$\cW$-algebras to string theory is rather intricate. There have been
a number of proposals for a more geometrical understanding of $\cW$-symmetry,
but their possible applications to string theory have not been clarified.

Some related ideas concerning the geometrical structure of
$\cW$-symmetries have been proposed in the papers \cite{\SSZ,\SoS,\GM},
see also \cite{\Biba,\GLM,\BFK}. The paper \cite{\GM} associates
$\cW$-symmetries
with the extrinsic geometry of the embedding of two-dimensional manifolds
with chiral parametrization into higher dimensional K\"ahler manifolds.
The characteristic equations for such embeddings can be connected to a
Lax pair for certain Toda equations, and these then form the link to
$\cW$-symmetries.

The constructions of $\cW$-gravity based on
Drinfeld-Sokolov reduction and on the connection with $d=3$ Chern-Simons
theory (see Section 8.1) suggest an alternative way to understand the
geometry of $\cW$-symmetries. These ideas have been essential for the
explicit construction of the covariant induced action of
$\cW_n$ gravity \cite{\dBGb}.

\medskip

In conclusion, we would like to stress that the status of $\cW$-symmetry
in the context of CFT is much better understood than the role these
symmetries can play in string theory.
The precise interpretation, in the context of string theory,
of results obtained for $\cW$-gravity and $\cW$-geometry
is not always clear. However, the fact that already `toy models'
of $c<1$ and $c=1$ strings exhibit $\cW$-symmetries certainly suggests
that extended symmetries will eventually also play a role in more
realistic theories of first and second quantized strings.

\vskip 4mm

\noindent {\it Acknowledgements.}\ We would like to thank the editors
of Physics Reports for the invitation to write this review. It is a
pleasure to thank Krzysztof Pilch, Jos\'e Figueroa-O'Farrill,
Alexander Sevrin, Tjark Tjin and Changhyun Ahn for
carefully reading parts of the manuscript and suggesting improvements.
In particular we would like to thank K.~P.
for pointing out our `crimes against the English language', and
J.~F-O'F. for sharing his \TeX nological insights.
P.B. would like to thank the I.T.P. at Stony Brook for their
hospitality and financial support during the course of this work.
Finally,
we would like to express our gratitude to many of our colleagues at
CERN and Stony Brook for their ever continuing encouragement to
complete this paper.
\bs

\noindent {\it Note added.}\ 
An interesting problem that has not been solved is to
generalize the explicit expression for the coset spin-3 
generator, as discussed in section 7.3.2, to the other
coset generators of spin greater than 3. In the preprint
version of this paper we gave some expressions for these
which are however incorrect. The problem can be traced back 
to a higher-spin generalization of (7.26) that was first proposed 
in \cite{\Thb} (equation (2.12)). Inconsistencies in the coset 
expressions derived from this equation show that it cannot be 
correct (see also the erratum to \cite{\Waa}). It is now
understood that the equation fails to hold because it assumes
a tensor identity for higher order $d$-symbols which is simply
not valid. We would like to thank G. Watts for pointing out this 
problem and A. Sudbery for correspondence on the tensor 
identities.

\vfil\eject

%--------------------------------------------------------------------

\newsec{Appendices}

\noindent{\it A. Lie algebra conventions}\hfil\bigskip

\def\Co{{\rm h}}
\def\bfg{{\bf g}}

Throughout the report we use the following conventions (see
\eg [\Kad] for more details):\bs

\settabs 6\columns
\+ $\bfg$ & a finite-dimensional complex semi-simple Lie
(super)algebra\cr
\+ $\bfh$ & the Cartan subalgebra (CSA) of $\bfg$ \cr
\+ $\bfh^*$ & the dual CSA; we will identify $\bfh$ with $\bfh^*$\cr
\+ $\bfg \cong {\bf n}_-\oplus \bfh \oplus {\bf n}_+$ &
\ \ \ \ a Cartan (triangular) decomposition of $\bfg$ \cr
\+ $(\ , \ )$ &  bilinear form on $\bfh^*$ (sometimes also
denoted by a simple dot $\ \cdot$\ ), \cr
\+ & normalized such that $(\al,\al)=2$ for a long root of $\bfg$ \cr
\+ $\ell$ & the rank of $\bfg$ \cr
\+ ${\rm dim\ }\bfg$ & the dimension of $\bfg$ \cr
\+ $\De_+$ & set of positive roots $\al$ of $\bfg$ \cr
\+ $\al_i$ & a simple root of $\bfg$ \cr
\+ $\La_i$ & a fundamental weight of $\bfg$ \cr
\+ $\rh$ & the Weyl vector of $\bfg$ \cr
\+ $\Co$ & the Coxeter number of $\bfg$ \cr
\+ $Q$ & the long root lattice of $\bfg$ \cr
\+ $P_+$ & the set of integral dominant weights $\la\in\bfh^*$\cr
\+ $W$ & the Weyl group of $\bfg$ \cr
\+ $l(w)$ & the length of the Weyl group element $w$\cr
\+ $\ep(w)$ & the determinant (\ie $\pm1$) of the Weyl group element $w$
\cr
\+ $r_\al\ (r_i)$ & reflection in the root $\al\in \De_+$ (/simple root
$\al_i$) \cr
\+ ${\cal U}(\bfg)$ & the universal enveloping algebra of $\bfg$\cr
\+ ${\cal Z}({\cal U}(\bfg))$ & the center of ${\cal U}(\bfg)$\cr
\+ $h^i$ & a basis of the CSA $\bfh$\cr
\+ $e^\al$ & Lie algebra element corresponding to root $\al$\cr
\+ $a_{ij}$ & the Cartan matrix of $\bfg$, \ie $a_{ij} = 2(\al_i,\al_j)
/(\al_i,\al_i)$ \cr
\+ $e_i\,,i=1,\ldots,\ell$ & the exponents of $\bfg$ \cr
\medskip

The dual Lie algebra $\bfg^*$ is the algebra obtained by inverting
the arrows in the Dynkin diagram corresponding to $\bfg$. Its
corresponding characteristics are denoted by a subscript $\vee$, \ie
$\rhv,\,\dCo,\,P_+^\vee$ etc. In particular the roots $\al^\vee$ of
$\bfg^*$ are related to those of $\bfg$ by $\al^\vee = 2\al/(\al,\al)$.
\smallskip

The untwisted affine Lie algebra $\bfg\otimes \CC[t,t^{-1}]
\oplus \CC c$ [\Kad]
corresponding to $\bfg$ will be denoted by either $\whg$ or
$\bfg^{(1)}$. Characteristics of the affine Lie algebra $\whg$ and
the underlying finite-dimensional Lie algebra $\bfg$ will be
distinguished by putting hats on the former. In addition we will
identify affine weights $\widehat{\La}$ with their finite-dimensional
projection $\La$ supplied by the level $k$. We
use the notation $P_+^{(k)}$ for the set of integral dominant weights
of level $k$. The twisted length $\tilde{l}(w)$ of an affine Weyl
group element $w\in \widehat{W}$ is defined in \eg \cite{\BMPb}.
\bs

We have collected some of the characteristics of finite-dimensional
simple Lie algebras $\bfg$ in Table 1.\bs

\vbox{\offinterlineskip
\hrule
\halign{& \vrule# & \strut\quad\hfil#\quad\cr
height2pt&\omit&&\omit&&\omit&&\omit&&\omit&\cr
&$\bfg$\hfil&&dim$\,\bfg$ && $h$ && $\dCo$ && $\{e_i\}$ &\cr
height2pt&\omit&&\omit&&\omit&&\omit&&\omit&\cr
\noalign{\hrule}
height2pt&\omit&&\omit&&\omit&&\omit&&\omit&\cr
& $A_n$ && $n(n+2)$ && $n+1$ && $n+1$ && $1,2,\ldots,n$ &\cr
& $B_n$ && $n(2n+1)$ && $2n$ && $2n-1$&& $1,3,5,\ldots,2n-1$ &\cr
& $C_n$ && $n(2n+1)$ && $2n$ && $n+1$ && $1,3,5,\ldots,2n-1$ &\cr
& $D_n$ && $n(2n-1)$ && $2(n-1)$ && $2(n-1)$ && $1,3,5,\ldots,2n-3,
n-1$ &\cr
& $E_6$ && $78$ && $12$ && $12$ && $1,4,5,7,8,11$  &\cr
& $E_7$ && $133$ && $18$ && $18$ && $1,5,7,9,11,13,17$ &\cr
& $E_8$ && $248$ && $30$ && $30$ && $1,7,11,13,17,19,23,29$ &\cr
& $F_4$ && $52$ && $12$ && $9$ && $1,5,7,11$ &\cr
& $G_2$ && $14$ && $6$ && $4$ && $1,5$ &\cr
height2pt&\omit&&\omit&&\omit&&\omit&&\omit&\cr}
\hrule}
\centerline{\sl Table 1}\bs

\noindent Finally,
we collect here some useful formulae [\Kad]
\eqn\chIaba{ \eqalign{
\rh = \half \sum_{\al\in\De_+} \al\,, & \qquad\qquad
(\rh,\al_i^\vee)=1 \,,\cr
\rh^\vee = \half\sum_{\al\in\De_+} \al^\vee\,,  & \qquad\qquad
(\rh^\vee,\al_i)=1\,,\cr}
}
\eqn\chIac{
{\rm dim\,}\bfg = \ell\,(1+\Co) \,.
}
\noindent The denominator formula:
\eqn\chIaa{
\prod_{\al\in\De_+} (1-e^{-\al}) = \sum_{w\in W} \ep(w) e^{w\rh -\rh} \,.
}
The Freudenthal-de Vries strange formula:
\eqn\chIab{
{ |\rh|^2 \over2\dCo } = { {\rm dim\,}\bfg \over24} \,.
}
And some relations for the exponents:
\eqn\chIad{
\eqalign{
\sum e_i & = \half \ell\,\Co \,,\cr
\sum e_i(e_i+1) & = 4(\rh,\rh^\vee)\,, \cr
& = \third\ell\, \Co(\Co+1) \qquad\qquad {\rm for\ }
\bfg\ {\rm simply\ laced}\,.\cr}
}
\bs\bs

\noindent{\it B. $\cW$-algebra nomenclature}\hfil\bigskip

In this appendix we propose some systematics for naming $\cW$-algebras.
Obviously, this proposal comes somewhat {\it apr\`es la date}, but we
think that it is worthwhile to try and standardize these matters.
We base our notations for $\cW$-algebras on the different approaches to
their construction, which we discuss in the Chapters 5, 6 and 7,
respectively.

\vskip 3mm \noindent
1. We introduce the notion of a $\cW$-algebra of {\it type}
$\cW(2,s_2,s_3, \ldots, s_n)$.
This refers to an algebra that is generated
(in the sense of our definition in Section 3.1) by the
Virasoro generator
$T(z)$ (which is quasi-primary of spin 2) and additional primary
currents of spins $s_2, s_3, \ldots, s_n$. For a $N$-extended
$\cW$-superalgebra
(which contains the $O(N)$-extended superconformal algebra,
$N=1,2,3$ or $4$), we write $\sW^{(N)}(2-{N \over 2},s_2,\ldots,s_n)$.
The first entry refers to the super-Virasoro generator
(which is quasi-primary
of spin $2-{N \over 2}$) and the additional entries $s_2,s_3,\ldots,s_n$
refer to additional generating currents, which are all superfields
in $N$-extended chiral superspace \cite{\Sca}.

Clearly, the type of a $\cW$-algebra does not always completely
fix the algebra. There can be free parameters (the central charge or
still others) and there is a possibility of entirely distinct
algebras with the same set of spins of the generating currents.

\vskip 3mm \noindent
2. In the most general version of the Drinfeld-Sokolov scheme
(see Section 6.3), a $\cW$-algebra is completely determined by a
triple $(\whg,\whg',\ch)$, consisting of an affine Lie algebra
$\whg$, an affine subalgebra $\whg' \subset \whg$ and a 1-dimensional
representation $\ch$ of $\whg'$. We denote the corresponding
$\cW$-algebra by $\cW_{DS}[\whg,\whg',\ch]$. In many cases one
makes a special choice for the defining triple (namely
$\whg'=\widehat{\bf n}_+$, $\ch=\ch_{DS}$) which is completely
determined by the embedding of an $sl(2)$ subalgebra in $\bfg$.
The $\cW$-algebra corresponding to that situation will be called
$\cW_{DS}[\whg,k,\delta]$, where $k$ is the level of $\whg$ and
the vector $\delta$ specifies how the $sl(2)$ subalgebra is embedded
in $\bfg$.

\vskip 3mm \noindent
3. In the coset construction, discussed in Chapter 7, a $\cW$-algebra is
specified by a coset pair $\whg' \subset \whg$, where $\whg$ is an affine
Lie algebra of level $k$ and $\whg'$ is an affine sub-algebra, or by
a pair $\bfg \subset \whg$, where $\bfg$ is the finite dimensional
horizontal subalgebra of $\whg$. (In the latter case the coset
construction
reduces to what we called the Casimir construction or extended Sugawara
construction.) The $\cW$-algebras for such coset pairs will be denoted
by $\cW_c[\whg/\whg',k,*]$, where the $*$ specifies the embedding
$\whg' \subset \whg$, and by $\cW_c[\whg/\bfg,k]$, respectively.

\vskip 6mm

In many cases these notations can be simplified. The subscripts $DS$
and $c$ can be dropped if the context is clear. The embedding data
$\delta$ and $*$ can be defaulted for `obvious' choices, such as the
principal $sl(2)$ embedding for $\delta$, or, for $*$, the diagonal
embedding $\whg \subset \whg \oplus \whg$.

Furthermore, there are `nicknames' for the most familiar algebras.
In particular, there are the $\cW_n$ algebras, of type
$\cW(2,3,\ldots,n)$, which can be realized as
$\cW_{DS}[A_{n-1}^{(1)},k]$,
as $\cW_c[A_{n-1}^{(1)}/A_{n-1},1]$ for $c=n$, or as
$\cW_c[A_{n-1}^{(1)}\oplus A_{n-1}^{(1)}/A_{n-1}^{(1)},(1,k)]$ for
the central charges in the minimal series.
We used the name super-$\cW_3$ algebra for the algebra of type
$\sW^{(1)}(3/2,5/2)$ discussed in Section 3.3.
Similarly, there are the $N=2$ super-$\cW_n$ algebras,
of type $\sW^{(2)}(1,2, \ldots, n)$, which can be realized
as $\cW_{DS}[A(n,n-1)^{(1)},k]$ (see Section 6.3.3).

Algebras that are obtained by applying the DS scheme to various
embeddings of $sl(2)$ in $A_{n-1}$ are sometimes called
$\cW_n^{(l)}$ algebras, where the superscript indicates the embedding
that is used (with $l=1$ denoting the principal embedding).

In the literature the notation $\cW X_\ell$ algebra is often used for
a $\cW$-algebra that one can associate with the Lie algebra $X_\ell$.
This notation works fine for the simply-laced Lie algebras
$X_\ell=A_\ell,D_\ell$
or $E_\ell$ but it is confusing in other cases and we tried as much as
possible to avoid it. We can explain our concern with the example
$X_\ell=B_\ell$.
DS reduction of the affine algebra $B_\ell^{(1)}$ leads to the
$\cW$-algebra $\cW_{DS}[B_\ell^{(1)},k]$, which is of type
$\cW(2,4,6,\ldots,2\ell)$. On the other hand, the coset $\cW$-algebras
$\cW_c[B_\ell^{(1)}/B_\ell,1]$ and
$\cW_c[B_\ell^{(1)} \oplus B_\ell^{(1)} / B_\ell^{(1)}, (1,k)]$ are all
of type $\cW(2,4,6,\ldots,2\ell,\ell+1/2)$, which is the spin content
that one would get by applying DS reduction to the superalgebra
$B(0,\ell)^{(1)}$ rather than to $B_\ell^{(1)}$ itself.

\vfil\eject
%\end

\listrefs

\vfill\eject\end
\bye